\documentclass[11pt]{article}

\usepackage{geometry}
\usepackage{amsmath,amsthm,amsfonts,amssymb,epsfig,color}
\newcommand{\mysymbol}{{\sf invsgn}}
\newcommand{\varpsi}{{\psi}}
\newcommand{\remove}[1]{}

\newcommand{\dist}{{\mathbf{dist}}}

\newcommand{\tw}{{\mathbf{tw}}}

\renewcommand{\int}{{\mathbf{int}}}

\newcommand{\eg}{{\mathbf{eg}}}

\newcommand{\wpmem}{{\sc $p$-min/eq/max-CMSO}}
\newcommand{\myemptyset}{\epsilon}
\newcommand{\pmin}{{\sc $p$-min-CMSO[$\psi$]}}
\newcommand{\peq}{{\sc $p$-eq-CMSO[$\psi$]}}
\newcommand{\pmax}{{\sc $p$-max-CMSO[$\psi$]}}
\newcommand{\pmem}{{\sc $p$-min/eq/max-CMSO[$\psi$]}}
\newcommand{\pmm}{{\sc $p$-min/max-CMSO[$\psi$]}}

\newcommand{\pmeq}{{\sc $p$-min/eq-CMSO[$\psi$]}}

\newcommand{\cov}{\mbox{\bf cov}}
\newcommand{\pack}{\mbox{\bf pack}}
\newcommand{\h}[1]{\end{document}}

\newcommand{\mycirc}{\diamond}

\newcommand{\mar}[1]{#1}
\newcommand{\term}[1]{#1}

\newcommand{\myparagraph}[1]{\vspace{3mm}\smallskip\noindent{\textbf{\sffamily #1} \ }}

\usepackage[russian,spanish,greek,english]{babel}
\usepackage[utf8x]{inputenc}

\usepackage{titling}

\usepackage{graphicx}  
\usepackage{enumerate}

\usepackage[colorlinks,citecolor=green,linkcolor=red,urlcolor=blue,linktoc=magenda,pdftitle="(Meta) Kernelization",pdfauthor="Hans L. Bodlaender Fedor V. Fomin Daniel Lokshtanov Eelko Penninkx Saket Saurabh Dimitrios M. Thilikos"]{hyperref}

\newtheorem{theorem}{Theorem}[section] 
 
\newtheorem{corollary}[theorem]{Corollary} 
\newtheorem{proposition}[theorem]{Proposition} 
\newtheorem{lemma}[theorem]{Lemma} 
\newtheorem{definition}[theorem]{Definition} 
\newtheorem{remark}{Remark}
\newtheorem{observation}{Observation}

\newtheorem{claim}{Claim}

\begin{document}
\thanksmarkseries{arabic}
\title{{\bf (Meta) Kernelization}\thanks{A preliminary  version of this article appeared in  \href{http://ieeexplore.ieee.org/xpl/login.jsp?tp=&arnumber=5438590&url=http\%3A\%2F\%2Fieeexplore.ieee.org\%2Fstamp\%2Fstamp.jsp\%3Ftp\%3D\%26arnumber\%3D5438590}{Proceedings of the 50th Annual IEEE Symposium on Foundations of Computer Science  (FOCS 2009), IEEE, 2009, pp. 629--638.}}
%~$^,$\thanks{%
%{Emails:  Hans L. Bodlaender: {\sf H.L.Bodlaender@uu.nl},
%Fedor V. Fomin: {\sf fomin@ii.uib.no},
%Eelko Penninkx: {\sf penninkx@cs.uu.nl},
%Daniel Lokshtanov: {\sf daniello@ii.uib.no}, 
%Saket Saurabh: {\sf saket@imsc.res.in}, 
%Dimitrios  M. Thilikos: {\sf sedthilk@thilikos.info}.}%
%}
}
\author{\href{http://www.staff.science.uu.nl/~bodla101/}{Hans L. Bodlaender}\thanks{Utrecht University,  Utrecht, the Netherlands. Email: 
{\sf \{H.L.Bodlaender$\mid$penninkx\}@uu.nl}}
\and \href{http://www.ii.uib.no/~fomin/}{Fedor V. Fomin}\thanks{University of Bergen,   Bergen,  Norway. Email: {\sf \{fomin$\mid$daniello\}@ii.uib.no}}~$^,$\thanks{Supported by ``Rigorous Theory of Preprocessing", ERC Advanced Investigator Grant 267959.  }
\and \href{http://www.ii.uib.no/~daniello/}{Daniel Lokshtanov}$^{3}$\medskip
\and \href{http://www.cs.uu.nl/staff/penninkx.html}{Eelko Penninkx}$^{2}$
\and \href{http://www.imsc.res.in/~saket/}{Saket Saurabh}$^{3,}$\thanks{The Institute of Mathematical Sciences, CIT Campus, Chennai, India. Email: {\sf saket@imsc.res.in}. Supported by ``Parameterized Approximation", ERC Starting Grant 306992. } 
\and \href{http://www.thilikos.info}{Dimitrios  M. Thilikos}\thanks{Department of Mathematics, National and Kapodistrian University of Athens, Athens, Greece and AlGCo project-team, CNRS, LIRMM, France. Email:  {\sf sedthilk@thilikos.info}. 
Co-financed by the E.U. (European Social Fund - ESF) and Greek national funds through the Operational Program ``Education and Lifelong Learning'' of the National Strategic Reference Framework (NSRF) - Research Funding Program: ``Thales. Investing in knowledge society through the European Social Fund''.}}

\date{\vspace{3mm}September 2013\vspace{-.6cm}}

\maketitle
\begin{abstract}
\noindent 
In a parameterized problem, every instance $I$ comes with a positive integer $k.$ The 
problem is said to admit a {\em polynomial kernel} if, in polynomial time, one can reduce the size 
of the instance $I$ to a polynomial in $k,$ while preserving the answer. In this work we give two 
meta-theorems on kernelzation. The first theorem says  that all problems expressible in Counting Monadic 
Second Order Logic and satisfying a coverability property admit a polynomial 
kernel on graphs of bounded genus. Our second result is that all problems that have finite 
integer index and satisfy a weaker coverability property admit a linear kernel 
on graphs of bounded genus.  These theorems unify and extend { all} previously 
known kernelization results for planar graph problems. 
\end{abstract}
%
%
%\noindent \textbf{Categories:} {F.2.2}, {Analysis of Algorithms and Problem Complexity}{Nonnumerical Algorithms and Problems}; 
%{G.2.2}, {Graph Theory}{Graph Algorithms};
%{G.2.2}, {Graph Theory}{Graph Algorithms};
%{G.2.2}{Discrete Mathematics}{Nonnumerical Algorithms and Problems}
%\noindent \textbf{Terms:}{Algorithms, Design, Theory}


\newpage

\tableofcontents

%\newpage

%%%%%%%%%%%%%%%%%%%%%%%%%%%%%%input{meta_intro}
\section{Introduction}
\label{sec:intro}
%\listoftodos
%\todo{ADD BACK THE BOTTOMSTUFF}

Preprocessing (data reduction or kernelization)
%\sef{Please un-Unicode everything!}
as a strategy of coping with hard problems is
universally used in almost every implementation.
The history of preprocessing, like applying reduction rules to 
simplify truth functions, can be traced back to the 1950's
\cite{Quine52}.   A natural question in this regard is how to measure the quality of the 
preprocessing rules proposed for a specific problem. For a long time 
the mathematical analysis of polynomial time preprocessing algorithms 
was neglected.  The basic reason for this anomaly was that if we start with an instance $I$ of an 
{\sf NP}-hard problem and can show that, in polynomial time, we can replace this with an 
equivalent instance $I'$ with $|I'|<|I|$ then that would imply  {\sf P}={\sf NP} in classical 
complexity. 
%lack of tools in classical complexity
%in t traditional complexity to be ab
%However, for a long time the mathematical analysis
%of preprocessing algorithms was almost neglected.
The situation changed drastically with advent of parameterized complexity. Combining tools from 
parameterized  and classical complexities 
%by the beginning  of this century.  With the new tools appearing
%{From} the parameterized complexity of Downey and Fellows
%\cite{DowneyF98} 
it has become possible  to derive
upper and lower bounds on the sizes of reduced instances, or so called
{\em kernels}. 
% Importance of preprocessing and the mathematical challenges it poses is beautifully 
% expressed in the following quote by Fellows from \cite{Fellows06}:  
% ``It has become clear, however, that far from being trivial and
% uninteresting, that pre-processing has unexpected practical power
% for real world input distributions, and is mathematically a much
% deeper subject than has generally been understood."  

%\paragraph*{Kernelization}.

\myparagraph{Kernelization.} In parameterized complexity each problem instance comes with a parameter $k$ and the parameterized problem is said to admit a {\it polynomial kernel} if there is a polynomial time algorithm (the degree of polynomial is independent of $k$), called a {\em kernelization} algorithm, that reduces the input instance down to an instance with size bounded by a polynomial $p(k)$ in $k,$ while preserving the answer. This reduced instance is called a {\em $p(k)$ kernel} for the problem. If $p(k) =O(k),$ then we call it a {\em linear kernel} (for a more formal definition, see Subsection~\ref{subsec:paraalgoandkern}). Kernelization has been extensively studied in the realm of parameterized complexity,  resulting in polynomial kernels 
for a variety of problems. Notable examples of kernelization include a $2k$-sized vertex kernel for {\sc Vertex Cover}~\cite{CKJ01}, a $355k$ vertex kernel for {\sc Dominating Set} on planar graphs~\cite{AlberFN04}, which later was improved to a $67k$ vertex kernel~\cite{ChenFKX07}, and an $O(k^2)$ kernel for {\sc Feedback Vertex Set}~\cite{Thomasse09} parameterized by the solution size. 

One of the most important results in the area of kernelization was given by \cite{AlberFN04}. 
They gave the first linear sized kernel  for the {\sc Dominating Set} problem
on planar graphs. The work of \cite{AlberFN04} triggered an explosion 
%\sed{Shall we mention here that no kernels exist in general because the problem is W[2]-hard?}
of papers on kernelization, and in particular on kernelization of problems 
on planar graphs. Combining the ideas of  \cite{AlberFN04} with problem specific data reduction rules, kernels of linear sizes were obtained for a variety of parameterized problems  on planar graphs including
\textsc{Connected
Vertex Cover, Minimum Edge Dominating Set, Maximum Triangle
Packing, Efficient Edge Dominating Set, Induced Matching, 
Full-Degree Spanning Tree, Feedback Vertex Set, Cycle Packing}, 
and \textsc{Connected Dominating Set} ~\cite{Alber:2006yg,AlberFN04,BodlaenderP08,BodlaenderPT08,ChenFKX07,GN07ICALP,GuoNW06,KanjPXS08,LokshtanovMS09,MoserS07}. {\sc Dominating Set}  has received special attention from kernelization view point, leading to a linear kernel on 
 graphs of bounded genus~\cite{FT04ICALP} and polynomial kernel on graphs excluding a fixed graph $H$ as a minor and on  
 $d$-degenerated graphs~\cite{AG08TechReport,PhilipRS09}.  We refer to the survey of 
 \cite{GN07SIGACT} for a detailed treatment of the area of kernelization. 
% While positive kernelization results have appeared regularly over the last two decades, the first results establishing infeasibility of polynomial kernels for specific problems have appeared only recently. In particular, Bodlaender et al. \cite{BDFH08} and Fortnow and Santhanam \cite{FS08} have developed a framework for showing that a problem does not admit a polynomial kernel unless the polynomial hierarchy collapses to the third level,  
%  a collapse which is deemed unlikely. We refer to~\cite{BDFH08,FS08} for further details on kernelization lower bounds.
 %\footnote{Saket: This paper is all about 
 %positive result so I do not think we need to mention negative result in introduction, conclusion may be the right %place for this.}

Most of the papers on linear kernels on planar graphs
have the following  idea in common: find an appropriate region decomposition (essentially a partitioning of  the vertex set into graphs of small diameter) of the input planar graph based on the problem in question,  
%into parts corresponding to some regions of the plane, 
and then perform
\emph{problem specific} rules to reduce the part of the graph
inside each region. The first step towards the general abstraction
of all these algorithms was initiated by  \cite{GN07ICALP},
who proved a general decomposition theorem for all problems with 
a specific distance property. Combining this decomposition theorem 
with problem specific reduction rules yields linear kernels for 
various problems on planar graphs.  Thus all previous work on
 kernelization was strongly based on the design of reduction rules particular to the problem in question.  
 In this paper we step aside and  find
properties of problems,  such as {\em expressibility in Counting Monadic Second Order Logic (CMSO)}, which allows these reduction
rules to be automated.
%
%\paragraph{Relevant Work.}
%It is worth to mention here some previous relevant works on 
%  solving  computationally hard problems on some special  
%inputs by just using data reduction.  Arnborg et al.~\cite{ArnborgCPS93}  have shown that each problem in Monadic
%Second Order Logic on graphs of bounded treewidth can be solved in linear time, by
%using only reductions, that is, no tree decomposition is built.  There are similar ideas even in the earlier papers 
%of Fellows and Langston, see  for an example~\cite{FellowsL89,FellowsL92,FellowsL94}. 

%\paragraph*{Algebraic reduction techniques}  
\myparagraph{Algebraic reduction techniques.}   
%One of the main techniques used in this work is the extension of the
%protrusion theory employed in \cite{H.Bodlaender:2009ng,F.V.Fomin:2010oq} for
%obtaining meta-kernelization theorems for problems on sparse graphs like
%planar and $H$-minor-free graphs, to general graphs. Bodlaender et
%al.~\cite{H.Bodlaender:2009ng} were first to use protrusion techniques (or
%rather graph reduction techniques) to obtain kernels, but the idea of using
%graph replacement for algorithms has been there for long time. 
The idea of
graph replacement for algorithms dates back to Fellows and
Langston~\cite{FellowsL89}. 
Arnborg et al.~\cite{ArnborgCPS93} proved that every set of graphs of bounded treewidth that is definable by a Monadic Second Order Logic (MSO) formula is also definable by reduction.
By making use of algebraic reductions, Arnborg et al.~\cite{ArnborgCPS93}  obtained a linear time algorithm for MSO expressible
problems on graphs of bounded treewidth. Bodlaender and
de Fluiter~\cite{BodlaenderF96a,BodlaendervA01a,Fluiter97} generalized these
ideas in several ways---in particular, they applied it to a number of  optimization
problems. It is also important to mention the work of Bodlaender and
Hagerup~\cite{BodlaenderH98}, who used the concept of graph reduction to
obtain parallel algorithms for MSO expressible problems on graphs of bounded treewidth.

%\paragraph*{Algorithmic meta-theorems} 
\myparagraph{Algorithmic meta-theorems.} 
Our results can be seen as what Grohe and Kreutzer
 call 
\emph{algorithmic meta-theorems} \cite{Grohe07logic,Kreutzer2011}. 
 Meta-theorems bring out the deep relations
between logic and combinatorial structures, which is a fundamental issue
of computational complexity. Such theorems also 
yield a better understanding of the scope of general algorithmic
techniques and the limits of tractability.
A typical example of meta-thoerem is the  celebrated Courcelle's theorem
\cite{Courcelle92a} which states that all graph
properties definable in MSO  can be decided
in linear time on graphs of bounded treewidth. More recent
examples of such meta-theorems state that all first-order 
definable properties on planar graphs can be decided in linear
time~\cite{FrickG01-dec} and that all first-order definable optimization problems
on classes of graphs with excluded minors can be approximated in
polynomial time to any given approximation ratio \cite{DawarGK07}.
Our meta-theorems not only give a uniform and natural
explanation for a large family of known kernelization results but also
provide a variety of new results. In what follows we build up towards 
our theorems. We first give necessary definitions %and introduce various structures
 needed  to formulate our results. 
% 
% \medskip
% \noindent{\sf Started here 29/01/2013 (SED)}
%\hrule

%\paragraph*{Parameterized graph problems}
\myparagraph{Parameterized graph problems.}
A parameterized graph problem  $\Pi$ in general can be seen as a subset of $\Sigma^{*}\times \Bbb{Z}^{+}$
where, in each instance $(x,k)$ of $\Pi,$ $x$ encodes a graph and $k$ is the parameter (we denote by $\Bbb{Z}^{+}$ the set of all non-negative integers). In this paper we extend this definition by permitting the parameter $k$ to be negative 
with the additional constraint that either all pairs with non-positive value of the parameter 
are in $\Pi$ or that no such pair is in $\Pi$. Formally, a parametrized problem $\Pi$
is a subset of $\Sigma^{*}\times \Bbb{Z}$ where for all $(x_{1},k_{1}),(x_{2},k_{2})\in\Sigma^{*}\times \Bbb{Z}$
with $k_{1},k_{2}<0$ it holds that $(x_{1},k_{1})\in\Pi$ if and only if  $(x_{2},k_{2})\in\Pi$.
This extended definition encompasses the traditional one and is being adopted for technical reasons
(see Subsection~\ref{subsec:finiinteginde}).
In many cases, in the pair $(x,k)$, $x$ will encode an {\em annotated graph}, that is a pair $(G,S)$ where $S$ is 
a subset of the vertices of $G,$ i.e., $S$ contains the {\em annotated} vertices of $G.$ 
In this paper, we mostly work on problems restricted to certain graph classes.
For this reason, given a   graph class ${\cal G},$ 
we use notation
 $\Pi\mar{\doublecap}{\cal G}$ for the set of instances of $\Pi$ 
 minus the instances $(x,k)$ where $x$ does not encode a graph in ${\cal G}.$ That way, the new problem 
$\Pi'=\Pi\doublecap {\cal G}$ is a subset of $\Sigma^{*}\times \Bbb{Z}$ that
corresponds to the restriction of $\Pi$ to graphs in ${\cal G}.$  
In this paper we mostly apply such restrictions to bounded genus graphs. We 
denote by $\mar{{\cal G}_{r}}$ the class of  graphs that are $2$-cell embeddable in 
some surface of Euler genus at most $r.$

%\paragraph*{$r$-coverable problems}
\myparagraph{$r$-coverable problems.} 
Let $G=(V,E)$ be  a graph embedded without crossings in a surface.   (For more details on graph embeddings, see Subsection~\ref{sec:comresultsallofthem}.)
The \term{{\em radial distance}}
%\ff{My suggestion is drop the word radial even from the definition} 
between two vertices $x,y$ of $G$ in this embedding is one less than the minimum length of an alternating sequence of vertices and faces starting from $x$ and ending in $y,$ such that every two consecutive elements of this sequence are incident  %\ff{correct is incident with (but adjacent to). I did changes everywhere} 
with each other. 
Given a set $S\subseteq V,$ 
we define $\mar{{\bf R}_{G}^{r}(S)}$ to be the set of all vertices of $G$ whose radial distance from some vertex of $S$ is at most $r$

Let $r$ be a non-negative integer. 
We say that a parameterized graph problem $\Pi$
has the \term{\emph{radial $r$-coverability property} }  if all  
YES-instances of $\Pi$ encode graphs embeddable 
in some surface of Euler genus at most $r$ and there exist such an embedding 
and  a
set $S \subseteq V$ such that $|S|\leq r \cdot k$ and  ${\bf R}_{G}^{r}(S)=V.$
We say that a problem $\Pi$ is {\em  radially {$r$}-coverable} if either $\Pi$ or its 
``complement in ${\cal G}_{r}$'', namely 
$\overline{\Pi}\cap {\cal G}_{r}$
 %$(\Sigma^{*}\times \Bbb{Z}^{+}\setminus \Pi)\doublecap {\cal G}_{r}$ 
 has the radial $r$-coverability property, (here, $\overline{\Pi}=\Sigma^{*}\setminus \Pi$). 
Every problem $\Pi$ that has the radial $r$-coverability property is radially $r$-covervable. However, the converse is not necessarily true. In particular, the {\sc $p$-Independent Set} problem can easily be seen to be radially $r$-coverable 
 but it does not have the radial $r$-coverability property.

%\todo[inline]{Definitions of coverable and quasi coverable are very strangely phrased.}
%
% 
%We say that a parameterized problem on graphs $\Pi$
% is \emph{compact} if there exist a constant $r$ such that one of the following holds:
% \begin{enumerate}
% \item
%For every graph of a YES-instance of $\Pi,$ there is an embedding of $G$ into a surface of Euler-genus  at most $r$ 
%and a set $S \subseteq V$ such that $|S|\leq r \cdot k$ and  ${\bf R}_{G}^{r}(S)=V.$    
%\item For every graph of a NO-instance of $\Pi,$ either there is no embedding of $G$ into a surface  of Euler-genus  at most $r$ or there exist such an embedding and a set $S \subseteq V$ such that $|S|\leq r \cdot k$ and  ${\bf R}_{G}^{r}(S)=V.$   
%\end{enumerate}

%and $k\leq |V|^r.$ 

%\paragraph*{$r$-quasi-coverable problems}
\myparagraph{$r$-quasi-coverable problems.}
A parameterized graph problem $\Pi$ has the  \term{\emph{radial $r$-quasi-coverability property}}  
if  all  
YES-instances of $\Pi$ encode graphs embeddable 
in some surface of Euler genus at most $r$ and there exist such an embedding 
and  a
set $S \subseteq V$ such that $|S|\leq r \cdot k$ and $\tw(G\setminus {\bf R}_{G}^{r}(S))\leq r$ 
 %and $k\leq |V|^r$ 
(by $\tw(G)$ we denote the treewidth of $G,$ for the formal definition, see~Subsection~\ref{subsec:treewidth}). 
We say that a problem $\Pi$ is  {\em radially  $r$-quasi-coverable}, 
if   either $\Pi$  or $\overline{\Pi}\cap {\cal G}_{r}$ has the radial $r$-quasi-coverability property. Every problem $\Pi$ that has the radial $r$-quasi-coverability property is radially $r$-quasi-covervable. Again, the converse is not necessarily true. For an example, the {\sc $p$-Cycle Packing} problem is  radially $r$-quasi-coverable 
 but it does not have the radial $r$-quasi-coverability property. 

%\medskip

Notice that if a problem is $r$-coverable then it is also $r$-quasi-coverable.
 From now on, for simplicity, we drop the terms ``radial'' and ``radially'' and we
 simply use the terms  {\em ``$r$-quasi-coverability property''}
or {\em ``$r$-quasi-coverable''}.

%
%\begin{remark}
%In the definitions of coverability and quasi-coverability we have restricted the upper bound on $|S|$ to a  
%linear function of $k.$ However, if  one replaces this condition with a polynomial function of $k$ 
%our results still are able to obtain polynomial kernels.  
%\end{remark}

%\paragraph*{Counting Monadic Second Order Logic}
\myparagraph{Counting Monadic Second Order Logic.} 
We use CMSO~\cite{ArnborgLS91,Courcelle90,Courcelle97}, an extension of MSO, as a basic tool to express properties of vertex/edge sets in graphs. 
%In fact, it is known that every set $\cal F$ of graphs of bounded treewidth is CMSO-definable if and only if $\cal F$ has finite state~\cite{Lapoire98}. 
As in this section our aim is 
to define a series of CMSO-based problem properties,  
we avoid the formal definitions of CMSO and we postpone them for 
Subsection~\ref{subsec:counmonasecoordelogi}. 
\smallskip

% 
% A CMSO-minimization problem on graphs is a problem where we are given a graph $G=(V,E)$ as input. The objective is to find a set $S$ of vertices or edges of minimum size such that a problem-specific predicate $P(G,S)$ is satisfied, where $P$ is a property of vertex/edge sets in graphs expressible in CMSO. In a CMSO-maximization problem on graphs the objective is to find a set $S$ of maximum size satisfying $P(G,S).$ Our first result concerns a parameterized analogue of CMSO-minimization/maximization problems, and parameterized exact-size CMSO problems on graphs. 
% 

Our first result concerns a parameterized analogue of graph optimization problems where the objective is to find a maximum or minimum sized vertex or edge set satisfying a CMSO-expressible property.
%
%A  \term{{\em {\sc $p$-min-CMSO} problem}} is a parametrized graph problem
%that can be defined as follows given some 
% $\psi$ %whose free variables are a graph and a subset of its vertices
%such that the following holds: $G=(V,E)$ and $k$ are  the graph and the parameter \sed{$k$ here should %be non-negative}
%of a YES-instance of $\Pi$ if and only if  there exist a set $S\subseteq V$ 
%where $|S|\leq k$ such that $G$ and $S$ satisfy $\psi,$ i.e., $(G,S)\models \psi.$
%In this case, we say that {\em $\Pi$ is definable by the sentence $\psi$}.
%
%
We now define a class of parameterized problems, called  \term{{\em {\sc $p$-min-CMSO} problems}}\footnote{We follow the notation given in the book by Flum and Grohe~\cite{FG06} and add ``$p$"  in front of names of problems to emphasize that these are parameterized problems.}, 
with one problem for each CMSO sentence $\psi$ 
on graphs, where $\psi$ has a free vertex set variable $S$. The {\sc $p$-min-CMSO} problem defined by $\psi$ is denoted by \pmin{} and defined 
as follows. 
%Let $\psi$ be a CMSO sentence on graphs, with a free vertex set variable $S$.
%,   free variables of $\psi$  are a graph $G$ and a subset $S$ of the vertices of $G$. \todo{what does this line means?}
%Consider the following parameterized problem.
\begin{center}
\fbox{\begin{minipage}{11cm}
\noindent{\pmin{}}\\
\noindent {\em Input}: A graph $G=(V,E)$ and a non-negative integer $k$ \\
\noindent{\em Parameter:} $k$\\
\noindent{\em Question}: Is there a subset $S\subseteq V$ such that $|S|\leq k$ and $(G,S)\models \psi$?
\end{minipage}}
\end{center}
\medskip
In other words, \pmin{} is  a subset $\Pi$ of $\Sigma^{*}\times \Bbb{Z}$
where for every $(x,k)\in \Sigma^{*}\times \Bbb{Z}^{+}$, 
$(x,k)\in \Pi$ if and only
if  there exists a set $S\subseteq V$ 
where $|S|\leq k$ such that the graph $G$ encoded by  $x$ together with  $S$ satisfy $\psi,$ i.e., $(G,S)\models \psi.$ For 
$(x,k)\in \Sigma^{*}\times \Bbb{Z}^{-}$ we know that $(x,k)\notin \Pi$. 
%already know whether it is in $\Pi$ or not. 
In this case, we say that {\em $\Pi$ is definable by the sentence $\psi$} and that  $\Pi$ is a \pmin{}.

The definition of  \peq{} (resp. \pmax{})  problem is the same as the one for  \pmin{} problem with the difference that now we ask that $|S|=k$ (resp. $|S|\geq k$) and that for any $(x,k)\in \Sigma^{*}\times \Bbb{Z}^{-}$ we have that $(x,k) \in \Pi$.  
We can also extend the notion of  a \pmem{} problems to edge versions.  In these problems  
 $S$ is a subset of edges instead of a subset of vertices. All of our results  can be straightforwardly extended to this alternate setting.  {  In particular, an edge set problem on graph $G=(V,E)$ can be transformed to a vertex subset problem on the edge-vertex incidence graph $I(G)$ of $G$, which is  is a bipartite graph with 
vertex bipartition's  $V$ and $E$ with edges between vertices $v\in V$ and $e\in E$ if and only if $v$ is incident  with $e$ in $G$. Observe that if $G$ can be embedded in surface $\Sigma$ then so does $I(G)$ and even the treewidth of these graphs only differ by a factor of $2$.  To make the translation work throughout the paper, it is sufficient to use the fact that the property of being an incidence graph of a graph $G$ is expressible in MSO. To avoid complications in our proof we omit the details for this. }

%\sef{Think whether we move it in the end!}

%
% In a {\sc $p$-min-CMSO} graph problem $\Pi \subseteq {\cal G}_{g} \times \mathbb{N},$ we are given a graph $G=(V,E)$ and an integer $k$ as input. The objective is to decide whether there is a vertex/edge set $S$ of size at most $k$ such that the CMSO-expressible predicate $P_\Pi(G,S)$ is satisfied. In a {\sc $p$-eq-CMSO} problem the size of $S$ is required to be exactly $k$ and in a {\sc $p$-max-CMSO} problem the size of $S$ is required to be at least $k.$ 

The \term{\emph{annotated version} $\Pi^{\alpha}$} of a \pmem{} problem $\Pi$ is
the parameterized graph problem whose  instances are pairs of the form $((G,Y),k)$ where $(G,Y)$ is an annotated  graph and $k$ is a non-negative integer.  In the {\em annotated version} of a \pmeq{} problem, $S$ is additionally required to be a subset of $Y.$ For the annotated version of a  \pmax{} problem $S$ is not required to be a subset of $Y,$ but instead of $|S| \geq k$ we demand that $|S \cap Y| \geq k.$ A problem is an {\em annotated \pmem{} problem} if it is the annotated version of some \pmem{} problem.
%
%\sfdd{For all problems we just need the certificate to be a subset of the annotated vertices!}
%

%We consider the {\em standard parameterization} of all our problems, that is, $k$ is the parameter.\footnote{Fedor:  Why do we need the term standard parameterization here (and everywhere).   Our problem $\Pi$ is already a pair $(G,k),$ so the parameter is already there. Can we avoid it? Do we need this?}

%\paragraph*{Our results}
\myparagraph{Our results.}
Our first result is the following theorem (the proofs of Theorems~\ref{thm:cmsol}, \ref{thm:cmsolnotanotated}, and~\ref{thm:automata} are  given in Section~\ref{sec:derivourresults}).

\begin{theorem}
\label{thm:cmsol}
If $\Pi$ is an $r$-coverable  \pmm{}  (respectively \peq{}) problem,
then 
the annotated version $\Pi^{\alpha}$  admits 
a quadratic (respectively cubic) kernel.
\end{theorem}

Let us remark that, while  a parameterized graph problem is a special case of its annotated version where all vertices are annotated, the existence of a polynomial kernel for the annotated version does not imply directly that the corresponding (non-annotated) parameterized graph problem admits a polynomial kernel. Indeed, 
a kernelization algorithm for an annotated parameterized graph problem $\Pi^{\alpha}$ is a polynomial time algorithm that, 
given an input $(G=(V,E),Y,k)$ of  $\Pi^{\alpha},$ computes an equivalent instance $(G'=(V',E'),Y',k')$ of
$\Pi^{\alpha}$ such that  $\max\{|V'|,k'\}=k^{O(1)}.$  The point here is that even when $Y=V,$ we cannot guarantee that $Y'=V'.$
However, there is a simple trick resolving this issue, given some additional complexity
conditions.  In particular, Theorem~\ref{thm:cmsol} can be used to prove the following. 

\begin{theorem}
\label{thm:cmsolnotanotated}
If $\Pi$ is an {\sf NP}-hard  $r$-coverable
%\sed{Why $\Pi$ cannot be  just {\sf NP}-hard?}
\pmem{} problem and  $\Pi^{\alpha}$ is in ${\sf NP},$ then $\Pi$ admits a polynomial kernel. 
\end{theorem}

Theorems \ref{thm:cmsol} and~\ref{thm:cmsolnotanotated}  provide polynomial kernels for a 
variety of  parameterized  graph problems. However, many parameterized graph problems in
the literature are known to admit linear kernels on planar graphs. Our next theorem unifies and 
generalizes {\em all} known linear kernels for parametrized  graph problems on surfaces. To 
this end we make use of the notion of having {\em Finite Integer Index} or, in short, {FII}. This 
term first appeared in the works of ~\cite{BodlaendervA01a,Fluiter97} and is similar to the
notion of {\em finite state} \cite{AbrahamsonF93,BoriePT92,Courcelle90}. As the definition of 
the property of having FII is long,  we defer it to Subsection~\ref{subsec:finiinteginde}. Out 
next result is the following.

\begin{theorem}
\label{thm:automata}
If $\Pi$ is an  $r$-quasi-coverable  parameterized graph problem that has FII, then $\Pi$ admits a linear kernel.
%Let $\Pi\subseteq {\cal G}_g\times \mathbb{N}$ be a coverable graph problem with FII. Then $\Pi$ admits a linear kernel.
\end{theorem}

Our theorems are similar in spirit, yet they have a few differences. In particular, not every \pmem{} problem has FII. For  example, the \textsc{Independent Dominating Set} problem is a \pmin{} problem, but it does not have FII. Also the class of parameterized graph problems that have FII does not have a syntactic characterization and hence it may take some more work to apply Theorem~\ref{thm:automata}  than Theorem~\ref{thm:cmsol}. On the other hand, Theorem \ref{thm:automata}   applies to $r$-quasi-coverable problems and 
yields linear kernels. That way, it  unifies and implies results presented  in~\cite{AlberDN06,AlberFN04,BodlaenderP08,BodlaenderPT08,ChenFKX07,FT04ICALP,GN07ICALP,GuoNW06,KanjPXS08,LokshtanovMS09,MoserS07} as a corollary.%\sef{Check if this is a complete list.}

At high level, the proofs of  our theorems consist of    combinatorial decomposition and algebraic reductions. The combinatorial part shows how a graph can be decomposed into pieces with specific properties, and the algebraic reductions part explains how these pieces can be reduced. The important tool in both parts is the notion of \emph{protrusion}, i.e.  a subset of vertices of a graph, inducing a graph of constant treewidth and separated from the remaining part of the graph by a constant number of vertices.
In the algebraic reductions part of the proof, we show that sufficiently large protrusions can be  replaced by  equivalent  protrusions of smaller size.  %While for FII protrusion replacements are not very difficult,
 For CMSO problems algebraic reduction step is much more technical and involved  than for FII. Here we  work with annotated  problems and  perform replacements in several  stages. 
 %In addition, the proof requires  a variant of Courcelle's  Theorem   for    structures. 
 
 In the combinatorial part, the  result concerning quasi-coverable problems  is roughly as follows. Suppose that after deleting  $k$ constant radius  balls 
 from  a bounded-genus graph $G$ the remaining part of $G$ has constant treewidth. 
 Then either $G$ has a  protrusion of sufficiently large size (and in this case we can apply protrusion reduction to reduce the instance), or $G$ has $O(k)$ vertices. The proof of this result is based on a new treewidth-obstruction lemma for graphs embedded on a surface of bounded genus, which is interesting in its own right.  More precisely the lemma states that if 
 a graph of bounded genus has two vertices which are far apart (in the radial distance) and cannot be separated by a small separator, then the treewidth of the graph is large. Concerning coverable problems, we show that every bounded genus graph $G$  whose vertices can be covered by  $k$ balls of constant radius admits a \emph{protrusion decomposition}.  A protrusion decomposition is  a partition of the vertex set into $O(k)$ sets, one of these sets is a set $S$ of size $O(k)$ and the other sets are protrusions separated from each other by $S$. Combined with protrusion replacement rules for CMSO problems, such a decomposition implies the existence of a  polynomial kernel for every coverable CMSO problem. 
 
  \medskip The remaining part of this paper is organized as follows. In the next section (Section~\ref{sec:preliminaries}) we give a series of definitions on basic notions that are necessary to describe our results. In Section~\ref{sec:avariodcourctheo} we give a proof of a variant of the classical Courcelle's Theorem which we use in the proofs of our results.  In 
  Section~\ref{sec:metaalgoframforkern} we present our meta-algorithmic framework for kernelization and 
explain how our main results are derived from a series of algorithmic and combinatorial properties. 
The algorithmic properties  are proved in Section~\ref{sec:redrules} while our combinatorial 
results are proven in Section~\ref{sec:comresultsallofthem}.
Some criterion for proving that a problem in graphs 
has FII are given in Section~\ref{sec:criterialforprovingfii} and in Section~\ref{sec:implication}
we give an extended exposition of how our results 
can be applied to concrete problems. In Section~\ref{sec:conclusion}, we conclude   
with some open problems and further research directions.
At  the end of the paper, we append a short compendium 
of problems for which linear or polynomial kernels are consequences of our results.

\section{Definitions and  Notations}
\label{sec:preliminaries}
In this section we give necessary definitions, set up notations and derive some preliminary results that we make use of 
in proving the main results of the paper. 

\subsection{Preliminaries}
In this section we define some concepts that we use in the rest of this paper. Given a graph $G=(V,E)$ we use the notation $V(G)$ and $E(G)$ for $V$ and $E$ respectively. Given a set $S\subseteq V(G),$ we define $\mar{\partial_G(S)}$ as the set of vertices in $S$ that have a neighbor in $V\setminus S.$ For a set $S \subseteq V(G),$ the  {\em neighbourhood}  of $S$ in $G$ is $\mar{N_G(S)} = \partial_G(V(G) \setminus S).$ We also define 
the {\em closed neighborhood of $S$ in $G$} as $\mar{N_{G}[S]}=S\cup \partial_G(V(G) \setminus S).$ When it is clear from the context, we omit the subscripts. 

 Let~$G=(V,E)$ be a graph.  A graph~$G'=(V',E')$ is a
 \term{\emph{subgraph}} of~$G$ if~$V' \subseteq V$ and~$E' \subseteq E.$
 The subgraph~$G'$ is called an  \term{\emph{induced subgraph} } of~$G$ if~$E'
 = \{\{u,v\} \in E \mid u,v \in V'\}$.  In this case, $G'$~is also called the
 subgraph \emph{induced by~$V'$} and is denoted by~$\mar{G[V']}.$ 
Given a graph $G$ and a set $S\subseteq V,$ we denote by $G\setminus S$
the graph $G[V\setminus S].$ If $S\subseteq E,$ we denote $\mar{G\setminus S}=(V,E\setminus S).$ We also use the term $(x,y)$-path for a path in $G$  that has $x$ and $y$ as endpoints.

Throughout this paper we use $\mar{\Bbb{Z}}$,  $\mar{\Bbb{Z}^{+}}$  and  $\mar{\Bbb{Z}^{-}}$  for the sets of  integers,  non-negative and 
non-positive integers respectively.   Finally,  we use $\mar{\Bbb{N}}$ for the set of positive integers.

\subsubsection{Parameterized algorithms and kernels}
\label{subsec:paraalgoandkern}
An instance of a parameterized problem consists of $(x,k),$ where $k$ is called the parameter. 
Thus 
a parameterized problem $\Pi$ is a subset of $\Sigma^{*}\times \mathbb{Z}$ for some finite alphabet $\Sigma$ such that for all $(x_{1},k_{1}),(x_{2},k_{2})\in \Sigma^{*}\times \mathbb{Z}$
with $k_{1},k_{2}<0$ it holds that $(x_{1},k_{1})\in \Pi\iff (x_{2},k_{2})\in \Pi$.
%We assume that $k$ is {\em  given in unary} (with a sign bit) and hence $|k|\leq |x|.$ 
%\todo[inline]{Why $x$ being give in unary implies that $k\leq |x|$.} 
A central notion in 
parameterized complexity is {\em fixed parameter tractability}, which means, 
for a given instance $(x,k),$ 
solvability in time $f(k)\cdot p(|x|),$ where $f$ is an arbitrary function of $k$ and $p$ is a polynomial in the input size. 
The notion of {\em kernelization} is formally defined as follows. 

\begin{definition}{\rm [\bf Kernelization]} 
Let $\Pi\subseteq \Sigma^{*}\times \mathbb{Z}$ be a parameterized problem
and $g$ be a computable function.  
We say that $\Pi$ {\em admits a kernel of size $g$} if there exists an algorithm
$\mathcal{K},$ called  {\em kernelization algorithm}, or, in short, a \term{{\em kerneliza\-tion}}, 
that given $(x,k)\in \Sigma^{*}\times \mathbb{Z}^+,$ 
outputs, in time polynomial in $|x|+k,$ a pair $(x',k')\in \Sigma^{*}\times \mathbb{Z}^+$ such that
\begin{itemize}
\item[(a)] $(x,k)\in \Pi$ 
if and only if $(x',k')\in \Pi,$ and 
\item[(b)] $\max\{|x'|, k' \}\leq g(k).$
 \end{itemize}
 For every $(x,k)\in \Sigma^{*}\times \mathbb{Z}^-,$  the algorithm outputs a trivial equivalent instance. 
When $g(k)=k^{O(1)}$ or $g(k)=O(k)$ then we say that 
$\Pi$ {\em admits a polynomial} or \term{{\em linear kernel}} respectively.
\end{definition}

%
%A {\em structure} of type $({\sf graph},{\sf vertex\ set})$
%is a pair $(G,S)$ where $G$ is a graph  and $S\subseteq V(G).$

In this paper, we study parameterized problems
on graphs. However, in many cases we have to deal with 
annotated graph problems whose input is a pair 
$(G,S)$, where $S$ is a set of annotated vertices of $G.$
For such problems the task is to find a solution that is contained in $S.$ For this reason, 
we use the term {\em parameterized graph problem} for every subset $\Pi$ of $\Sigma^{*}\times \mathbb{Z},$  
where in each instance $I=(x,k)\in\Sigma^{*}\times \mathbb{Z}$ the string
$x$ is encoding either a graph $G=(V,E)$ or a pair $(G,S)$ with  $S\subseteq V$ and the integer $k$ encodes the parameter.  

%
%Given an instance $I=(x,k)\in\Sigma^{*}\times \mathbb{Z}$ of a parameterized graph problem, the {\em  graph} and the {\em parameter} of $I$  are the graph encoded by $x$ and the integer $k$ respectively.

%The combinatorial tools that we use in this section are based on the notion of treewidth. 
%We denote  the treewidth of a graph $G$ by ${\bf tw}(G)$; its definition can be found 
%at Section~\ref{sec:preliminaries}.\medskip

%

\subsubsection{Tree-width}
\label{subsec:treewidth}

Let $G=(V,E)$ be a graph. A \term{{\em tree decomposition}} of $G$ is a pair $(T, 
\mathcal{ X}=\{X_{t}\}_{t\in V(T)})$ where $T$ is a tree and ${\cal X}$ is a collection of subsets of $V$
such that: 
\begin{itemize}
%%\setlength{\itemsep}{-0.7mm}
%\item   $\bigcup_{u\in V_T}=V,$  
\item $\forall {e=\{u,v\}\in E},~\exists {t\in V(T)} : \{u,v\}\subseteq X_{t}$   and 
\item $\forall {v\in V}, \ T[\{t\mid v\in X_{t}\}]$ is non-empty and connected. 
\end{itemize}
%\begin{itemize}
%\item $\cup_{u\in V_T}=V$
%\item $\forall_{e\in E}\ \exists_{t\in V_T} : e\subseteq X_{t}$ 
%\item $\forall_{v\in V}\ T[\{t\mid v\in X_{t}\}]$ is connected.
%\end{itemize}
We call the vertices of $T$ \term{{\em nodes}} and
the sets in $\mathcal{ X}$ \term{{\em bags}} of the tree decomposition $(T,{\cal X}).$
 The {\em width} of $(T,{\cal X})$ is equal to $\max_{}\{|X_t|-1\mid {t\in V(T)}\}$ and the
 \term{{\em treewidth}} of $G=(V,E)$ is the 
minimum width over all tree decompositions of $G.$  We denote  the treewidth of a graph $G$ by ${\bf tw}(G)$.

A {\em nice tree decomposition} is a triple $(T,{\cal X},r)$ where $(T,{\cal X})$ is a tree decomposition 
where the tree $T$ is rooted on some vertex $r\in V(T)$ and  the following conditions are satisfied: 
\begin{itemize}
\item Every node of the tree $T$ has at most two children; 
\item if a node $t$ has two children $t_1$ and $t_2,$ then $X_t = X_{t_1} = X_{t_2}$ (we call $t$ a {\em join  node}); and
\item  if a node $t$ has one child $t_1,$ then either $|X_t | = |X_{t_1} | + 1$  and $X_{t_1} \subset X_{t}$ (in this case we call $t_1$ {\em introduce node}) or $|X_t| = |X_{t_1} | -1$ and $X_t \subset X_{t_1}$ (in this case we call $t_1$ {\em forget node}). 
\end{itemize}
It is possible to transform a given tree decomposition $(T,{\cal X})$ into a nice tree decomposition $(T',{\cal X}',r)$
where the root $r$ is any vertex of $T$ in time $O(|V|+|E|)$~\cite{Bodlaender96ali}.

%
%\medskip
%\hrule
%\smallskip

\subsection{Boundaried Graphs} 
\label{subsec:boungrap}
Here we define the notion of {\em boundaried graphs} and various operations on them.
\begin{definition}{\rm [\bf Boundaried Graphs]}\label{def:boungraph}
A \term{boundaried graph} is a graph $G$ with a set $B\subseteq V(G)$ 
of  distinguished vertices and an injective labelling $\mar{\lambda}$ 
from $B$  to the set $\Bbb{Z}^{+}$. The set $B$ is called the \term{{\em boundary}} of $G$ and  the vertices in $B$  are called  {\em boundary vertices} or \term{{\em terminals}}. 
Given a boundaried graph $G,$ we denote its boundary by $\mar{\delta(G)},$
we denote its labelling by $\lambda_G$, 
and we define its {\em label set} by $\mar{\Lambda(G)}=\{\lambda_{G}(v)\mid v\in \delta(G)\}$.
Given a finite set $I\subseteq \Bbb{Z}^{+}$, we define 
$\mar{{\cal F}_{I}}$  to denote the class of all boundaried graphs whose label set is $I$.
Similarly, we define ${\cal F}_{\subseteq I}=\bigcup_{I'\subseteq I}{\cal F}_{I'}$.
We also denote by $\mar{{\cal F}}$ the class of all boundaried graphs.
Finally we say that a boundaried graph is a {\em $t$-boundaried} graph if $\Lambda(G)\subseteq \{1,\ldots,t\}$.
\end{definition}

%
%We remark that  in the labelling of the boundary of a $t$-boundaried graph, not all $t$ available labels are necessary used.

%
%For a graph $G=(V,E)$ and a vertex set $S \subseteq V,$ we sometime consider the graph $G[S]$ as the 
%$|\partial(S)|$-boundaried graph with $\partial(S)$ being the boundary.

\begin{definition}{\rm [\bf Gluing by $\oplus$]} Let $G_1$ and $G_2$ be two  boundaried graphs. We denote by $G_1 \mar{\oplus} G_2$ the  graph 
(not boundaried) obtained by taking the disjoint union of $G_1$ and $G_2$ and identifying equally-labeled vertices of the boundaries of $G_{1}$ and $G_{2}.$ In $G_1 \oplus G_2$ there is an edge between two labeled vertices if there is either an edge between them in $G_1$ or in $G_2.$ 
\end{definition}

\begin{definition}
Let $G=G_{1}\oplus G_{2}$ where $G_{1}$ and $G_{2}$ are boundaried graphs.
We define the \term{{\em glued}} set of $G_{i}$ as the set $B_{i}=\lambda_{G_{i}}^{-1}(\Lambda(G_{1})\cap \Lambda(G_{2})), i=1,2$. For a vertex $v\in V(G_{1})$ we define its \term{{\em heir}} $\mar{h(v)}$ in 
$G$ as follows: if $v\not\in B_{1}$ then $h(v)=v$, otherwise $h(v)$ is the result of the identification 
of $v$ with an equally labeled vertex in $G_{2}$. The {\em heir} of a vertex in $G_{2}$ is defined symmetrically. The \term{{\em common boundary}} of $G_{1}$ and $G_{2}$ in $G$ is equal 
to $h(B_{1})=h(B_{2})$ where the evaluation of $h$ on vertex sets is defined in the obvious way.
The {\em heir} of an edge $\{u,v\}\in E(G_{i})$ is the edge $\{h(u),h(v)\}$ in $G$.
\end{definition}

Let ${\cal G}$ be a class of (not boundaried)  graphs.
By slightly abusing notation we say that a boundaried graph {\em belongs to a graph class ${\cal G}$} if the underlying graph belongs to ${\cal G}.$

%\begin{definition}{\rm [\bf Replacement]}
%Let $G$ be a  graph,  $X$ be a subset of $V(G)$, and $\lambda: \partial_{G}(X)\rightarrow \Bbb{Z}^+$.
%Let $G_1$ also  be a boundaried graph.
%The result of \term{{\em $\lambda$-repla\-cing}} {\em  $X$ with $G_1$} is the graph $G^{\star}\oplus G_{1},$
%where $G^{\star}=G\setminus (X\setminus \partial (X))$ and is treated as a boundaried graph, where 
%$\delta(G^{\star})=\partial_{G}(X)$ and  $\lambda_{G^{\star}}=\lambda.$
%\end{definition}
%\todo[inline]{Do we ever use  Replacement in the paper now.}

\subsection{Finite Integer Index}

\label{subsec:finiinteginde}
\begin{definition}{\rm [\bf Canonical equivalence on boundaried graphs.]}
Let $\Pi$ be a parameterized graph problem whose instances are pairs of the form $(G,k).$
 Given two boundaried graphs $G_1,G_2~\in {\cal F},$ we say that \term{$G_1\!\equiv _{\Pi}\! G_2$} if 
$\Lambda(G_{1})=\Lambda(G_{2})$
 and there exist a \term{{\em transposition constant}}
$c\in\Bbb{Z}$ such that 
\begin{eqnarray*}
\forall(F,k)\in {\cal F}\times \Bbb{Z} &&  (G_1 \oplus F, k) \in \Pi \Leftrightarrow (G_2 \oplus F, k+c) \in \Pi.\label{eq:fiidef}
\end{eqnarray*}
 %\sed{$c$ can be a positive or negative integer?}
%\end{itemize}
\end{definition}
Note that  the relation $\equiv_{\Pi}$  is
an equivalence relation. Observe that $c$ could be negative in the above definition. This is the reason we extended the definition of parameterized problems to include negative parameters also.

Next  we define a notion of ``transposition-minimality'' for the members 
of  each equivalence class of $\equiv_{\Pi}.$

\begin{definition}{\rm [\bf Progressive representatives]}
\label{def:progrepr}
Let $\Pi$ be a parameterized graph problem whose instances are pairs of the form $(G,k)$
and let ${\cal C}$ be some equivalence class of $\equiv_{\Pi}$. We say that $J\in{\cal C}$ is a \term{{\em progressive 
representative}}
of ${\cal C}$ if for every $H\in{\cal C}$
there exists $c\in\Bbb{Z}^{-},$ such that 
\begin{eqnarray}
\forall(F,k)\in {\cal F}\times \Bbb{Z} \ \ \  (H \oplus F, k) \in \Pi \Leftrightarrow (J\oplus F, k+c) \in \Pi. \label{eq:progfii}
\end{eqnarray}
\end{definition}

The following lemma guaranties the existence of a progressive representative for each equivalence class of $\equiv_{\Pi}$.

\begin{lemma}
\label{lem:existprog}
Let $\Pi$ be a parameterized graph problem whose instances are pairs of the form $(G,k)$.
%and let $t\in\Bbb{Z}^{+}.$  
 Then each  equivalence class of $\equiv_{\Pi}$ has a progressive representative.
\end{lemma}

\begin{proof}
We first examine the case where every  instance of $\Pi$ with a negative valued parameter is a NO-instance.\smallskip

Let ${\cal C}$ be an equivalence class of $\equiv_{\Pi}$.
We distinguish two cases:\smallskip
%We examine two cases.

\noindent{\em Case 1.}
Suppose first that for every $H\in{\cal C},$  every $F\in{\cal F},$
 and every integer  $k\in\Bbb{Z}$ it holds that $(H \oplus F, k) \not\in \Pi.$
Then we set  $J$ to be an arbitrary chosen graph in ${\cal C}$ and $c=0.$ 
In this case, it is obvious that~\eqref{eq:progfii} holds for every  $(F,k)\in {\cal F}
\times \Bbb{Z}.$\medskip

\noindent{\em Case 2.}
Suppose now that for some $H_{0}\in{\cal C},$ $F_{0}\in {\cal F}$, and $k_{0}\in \Bbb{Z}$ 
it holds that  that  $(H_0 \oplus F_{0}, k_{0}) \in \Pi$. Among all such triples, choose 
the one where the value of $k_{0}$ is minimized. Since every instance of $\Pi$ with a negative valued 
parameter is a NO-instance, it follows that $k_0$ is well defined and  is non-negative.
We claim that $H_0$ is a progressive representative.

Let $H\in{\cal C}.$ As $H_0\equiv_{\Pi} H,$ there is a constant $c$
such that 
\begin{eqnarray*}\forall (F,k)\in {\cal F}
\times \Bbb{Z}\ \ \ 
 (H\oplus F, k) \in \Pi \Leftrightarrow (H_0 \oplus F, k+c) \in \Pi . %\label{eq:equivprogproof}
\end{eqnarray*}
It suffices to prove that $c\leq 0$. Assume for a contradiction that 
$c>0$. Then, by taking $k=k_0-c$ and $F=F_0,$ we have that
\begin{eqnarray*}
 (H\oplus F_0, k_0-c) \in \Pi \Leftrightarrow (H_0 \oplus F_0, k_0-c+c) \in \Pi. 
 \end{eqnarray*}
Since $(H_0 \oplus F_0, k_0) \in \Pi$ it follows that $(H\oplus F_0, k_0-c)\in \Pi$ contradicting the choice of $H_0,F_0,k_0$.\medskip

Suppose now that  every  instance of $\Pi$ with a negative valued parameter is a YES-instance.
The proof of this case is symmetric to the previous one: just replace every occurrence of ``$\in\Pi$''
with a ``$\not\in\Pi$'' and every occurrence of ``$\not\in\Pi$'' 
with ``$\in\Pi$'' and the ``NO-instance'' with ``YES-instance''.
\end{proof}

Notice that two  boundaried graphs with different label sets belong to 
different equivalence classes of $\equiv_{\Pi}.$ Hence for every equivalence 
class ${\cal C}$ of $\equiv_{\Pi}$ there exists some finite set $I\subseteq\Bbb{Z}^{+}$ such that 
${\cal C}\subseteq  {\cal F}_{I}$. We are now in position  to give the following definition.

\begin{definition}{\rm [\bf Finite Integer Index]}
\label{def:deffii}
A parameterized graph problem $\Pi$ whose instances are pairs of the form $(G,k)$
has {\em Finite Integer Index} (or simply has \term{{\em FII}}), if and only if for every finite $I\subseteq \Bbb{Z}^+,$
the number of equivalence classes of  $\equiv_{\Pi}$ that are subsets of ${\cal F}_{I}$
is finite. For each $I\subseteq \Bbb{Z}^{+},$ we define $\mar{{\cal S}_I}$ to be
a set containing exactly one progressive representative of each equivalence class of $\equiv_{\Pi}$
that is a subset of ${\cal F}_{ I}$. We also define $\mar{{\cal S}_{\subseteq I}}=\bigcup_{I'\subseteq I} \mar{{\cal S}_{I'}} $. 
\end{definition} 
%
%The next lemma follows directly by the fact that $\equiv_{\Pi,t}$ is an equivalence relation.
%
%\begin{lemma}
%\label{lem:intersectionfii}
%Let $\Pi_{1}$ and $\Pi_{2}$ be problems whose instances are pairs of the form $(G,k).$ If both $\Pi_{1}$ and $\Pi_{2}$ have FII, then also $\Pi_{1}\cap \Pi_{2}$ has FII.
%\end{lemma}

% in the next subsection.

\subsection{Structures and its properties}
\label{lem:stru}
%Towards this,  
We first define the notions of {\em structure} and {\em arity of a structure}.

\begin{definition}{\rm [{\bf Structure and arity]}}
A  \term{{\em structure}} is a tuple where the first element of the tuple is a graph $G$ and the remaining elements of the 
tuple are either subsets of $V,$ subsets of $E,$ vertices in $G$ or edges in $G.$ The arity of the structure 
is the number of elements in the tuple.
 \end{definition}

Given a structure $\alpha$ of arity $p$ and an integer $i\in\{1,\ldots,p\}$ we let $\alpha[i]$ denote the $i$'th element of $\alpha.$ The graph of a structure $\alpha$ is denoted by $G_\alpha $ and it appears as the first element of the structure, that is, it is $\alpha[1].$ {\em Appending} a subset $S$ of 
$V(G_{\alpha})$ to a structure $\alpha$ of arity $p$ produces a new structure, denoted by $\alpha' = \alpha\, \mar{\mycirc}\, S$, of arity $p+1$ with the first $p$ elements of $\alpha'$ being the elements of $\alpha$ and $\alpha'[p+1] = S.$ Appending an edge set, a vertex, or an edge to a structure is defined similarly. 
For example, consider the structure $\alpha=(G_{\alpha},S,e),$
of arity $3$ where $S\subseteq V(G_{\alpha})$ and $e\in E(G_{\alpha}).$
Let also $S'$ be some subset of $V(G_{\alpha})$ and let $u\in V(G_{\alpha}).$ Appending $S'$ to $\alpha$ results to the structure $\alpha'=\alpha\mycirc S'=(G_{\alpha},S,e,S'),$ while 
appending $u$ to $\alpha'$ results to the structure 
$\alpha''=\alpha'\mycirc u=(G_{\alpha},S,e,S',u).$
 
Next we define the 
notions of {\em type} of a structure and {\em property} of structures. 

\begin{definition}{\rm [{\bf Type  of structure]}}
The type of a structure of arity $p$ is another tuple of arity $p,$ denoted by ${\bf type}(\alpha),$ where  the first element  ${\bf type}(\alpha)[1]$ is {\sf graph}, 
while for every $i\in\{2,\ldots,p\},$  ${\bf type}(\alpha)[i]$ is {\sf vertex}, {\sf edge}, {\sf vertex set} or {\sf  edge set} according to what the $i$'th element of $\alpha$ is. Note that we distinguish between a set containing a single vertex or  edge from just a single vertex or edge. 
\end{definition}

\begin{definition}{\rm [{\bf Properties of structures]}}
%Let $p$ be a positive integer.
A {\em property} of structures is a
function   $\sigma$ that assigns to each structure 
 a  value in $\{\mbox{\sf true, false}\}.$ 
\end{definition}

\subsection{Counting Monadic Second Order Logic and its properties}
\label{subsec:counmonasecoordelogi}
The syntax of Monadic Second Order Logic (MSO) of graphs includes the logical connectives $\vee,$ $\land,$ $\neg,$ 
$\Leftrightarrow ,$  $\Rightarrow,$ variables for 
vertices, edges, sets of vertices, and sets of edges, the quantifiers $\forall,$ $\exists$ that can be applied 
to these variables, and the following five binary relations: 
\begin{enumerate}

\item 
$u\in U$ where $u$ is a vertex variable 
and $U$ is a vertex set variable; 
\item 
 $d \in D$ where $d$ is an edge variable and $D$ is an edge 
set variable;
\item 
 $\mathbf{inc}(d,u),$ where $d$ is an edge variable,  $u$ is a vertex variable, and the interpretation 
is that the edge $d$ is incident with the vertex $u$; 
\item 
 $\mathbf{adj}(u,v),$ where  $u$ and $v$ are 
vertex variables  and the interpretation is that $u$ and $v$ are adjacent; \item 
 equality of variables representing vertices, edges, sets of vertices, and sets of edges.
\end{enumerate}

 In addition to the usual features of monadic second-order logic, if we have atomic sentences testing whether the cardinality of a set is equal 
to $q$ modulo $r,$ where $q$ and $r$ are integers such that $ 0\leq q<r $ and $r\geq 2,$ then 
this extension of the MSO is called the {\em counting monadic second-order logic}. Thus CMSO 
is MSO with the following atomic sentence for a set $S$: 
\begin{quote}
$\mathbf{card}_{q,r}(S) = \mathbf{true}$ if and only if $|S| \equiv q \pmod r.$ 
\end{quote}
We refer to~\cite{ArnborgLS91,Courcelle90,Courcelle97} for a detailed introduction on CMSO. 

\medskip

A CMSO sentence $\varpsi$ where some of the variables are free
%\sed{To say ``bound by quantifiers'' is surrey wrong} 
can be evaluated on a structure $\alpha$ by instantiating the free variables of $\varpsi$ by the 
elements of $\alpha.$ In order to determine which variables of $\psi$ are instantiated 
by which elements of $\alpha$ we need to introduce some conventions.

In a CMSO-sentence $\psi,$  each free variable $x$
has a \term{{\em rank}} $r_x\in\Bbb{N}\setminus\{1\}$ associated to it. Thus a CMSO-sentence $\psi$ can be seen as a string accompanied by a tuple  of integers
containing one integer $r_x$ for each free variable $x$ of $\psi$.

We say that {\bf type}$(\alpha)$  \term{{\em matches}} $\psi$ 
if the arity of $\alpha$ is  at least 
$\max r_x,$ where the maximum is taken  over  each  free variable $x$ of $\psi$ and for each free variable $x$ of $\psi$,  ${\bf type}(\alpha)[r_x]$ corresponds to the kind of the variable $x.$
For an example,
if $x$ is a vertex set variable, then ${\bf type}(\alpha)[r_x]={\sf vertex\ set}.$ 
Finally we say that $\alpha$ {\em matches} $\psi$ if ${\bf type}(\alpha)$ matches $\psi$.
For each free variable $x$ of $\psi$ and a structure $\alpha$ that matches $\psi$
the \term{{\em corre\-spon\-ding element}} of $x$ in $\alpha$ is $\alpha[r_x]$.
 
\begin{definition} {{\rm [{\bf Property $\sigma_{\psi}$}]}}
Each CMSO-sentence $\psi$ defines a property $\mar{\sigma_{\psi}}$ on structures as follows:
For every structure $\alpha$ that does not match $\psi$ the value of $\sigma_{\psi}(\alpha)$ is equal to {\sf false}, otherwise the value of $\sigma_{\psi}(\alpha)$ is the result of  the evaluation of $\psi$ with each free 
variable ${x}$ of $\psi$ instantiated by $\alpha[r_{x}].$
\end{definition}
 
%
%
%A given CMSO sentence  $\varpsi$ can only be evaluated on structures of 
%specific types, since all free variables need to be instantiated and we cannot instantiate an edge set 
%variable by a vertex set, vertex or edge.
%
% Thus, since the arity of the structure also counts the 
%graph, to be able to evaluate $\varpsi$ on $\alpha,$ the arity of $\varpsi$ needs to be at least one 
%more than the number of free variables in $\alpha.$ 

Note that it is not necessary that every element 
of $\alpha$ corresponds to some variable of $\psi$. However, 
 it is 
still possible that the sentence  $\psi$ can be 
evaluated on the structure $\alpha$
and, in this case, the evaluation of the sentence  
does not depend on all the elements of the structure.

A property $\sigma$ is {\em CMSO-definable} if there 
exists a sentence  $\varpsi$ such that $\sigma=\sigma_{\psi}$.
%
%
%, for every structure $\alpha,$ it holds that
%\begin{eqnarray}
% \alpha\models \psi & \Leftrightarrow & \sigma(\alpha)={\sf true} \label{eq:theone_eval}
%\end{eqnarray}
%Where $ \alpha\models \psi$ reads as ``$\alpha$ satisfies  $\psi$'' or  ``$\alpha$ models  $\psi$''.
In this case we say that the CMSO-sentence $\psi$ {\em defines} $\sigma$.
\smallskip

\begin{observation}\label{morgan}
For every CMSO-definable property $\sigma$ there exists a 
CMSO-sentence $\psi$ that defines $\sigma$ and has the following additional
features.
\begin{enumerate}
\item Each  variable of $\psi$ has a unique name.
\item $\psi$ does not use the ${\bf adj}$ operator,
\item  $\psi$ does not have conjunctions,
\item  $\psi$ does not have universal quantifiers.
\end{enumerate}
\end{observation}

\begin{proof}
Let $\psi'$ be a CMSO-sentence defining $\sigma$. We construct another CMSO-sentence
$\psi$ defining $\sigma$ so that  $\psi$ satisfies  
Properies~(1)--(4). For Property (1), we rename each variable so that it has a unique name.  
When we rename a free variable $x$ of $\psi$  of rank $r_x$ to $x'$
we let $x'$ have rank $r_{x'}=r_x$ in $\psi'$.

 For Property (2), we replace each occurrence of ${\bf adj}(x,x')$ by 
$\exists x''\in E: {\bf inc}(x'',x)\wedge {\bf inc}(x'',x').$ For Properties (3) and (4), just 
use the fact that $\wedge$ and $\forall$ can be expressed using $\vee,$ $\exists,$ and $\neg$
by De Morgan's laws.  
\end{proof}
We call CMSO-sentences satisfying Properties (1)--(4) of Observation~\ref{morgan} \term{{\em normalized}} CMSO-sentences.\medskip

\subsection{Boundaried structures}\label{lem:comstru}
In this subsection we extend the notion of boundaried graphs to boundaried structures.

\begin{definition}{\rm [{\bf Boundaried structure]}}
A \term{{\em boundaried structure}} is a tuple where the first element is a boundaried graph $G$ and the remaining 
elements are either subsets of $V(G),$ subsets of $E(G),$ vertices in $V(G),$ edges in $E(G),$ or the symbol $\star.$ 
For a boundaried structure $\alpha,$ $\alpha[i]$ is the $i$'th element of $\alpha$ and $G_\alpha = \alpha[1]$ is always a boundaried graph. 
\end{definition}

\begin{definition}{\rm [{\bf Type of a  boundaried structure]}}
The \term{{\em type}} of the boundaried structure is defined similarly to the 
type of a structure; for a boundaried structure $\alpha$ of arity $p,$ ${\bf type}(\alpha)$  is a tuple of arity $p,$ 
where the first element of ${\bf type}(\alpha)$ is {\sf boundaried graph}, while for every $i\in\{2,\ldots,p\},$ ${\bf 
type}(\alpha)[i]$ is {\sf vertex}, {\sf edge}, {\sf $\star$}, {\sf vertex set}, or {\sf edge set} according to what  $\alpha[i]$ is. 
\end{definition}

\begin{definition}{\rm [{\bf Type matching}] }
Given a CMSO-formula $\psi,$ we say that ${\bf type}(\alpha)$ {\em matches} $\psi$
if the arity of $\alpha$ is  at least 
$\max r_x,$ where the maximum is taken  over  each  free variable $x$ of $\psi$
and for every free variable $x$ of $\psi$
\begin{itemize}
\item  if $x$ is a vertex variable then ${\bf type}(\alpha)[r_x]\in\{\star,{\sf vertex}\}$
\item if $x$ is a edge variable then ${\bf type}(\alpha)[r_x]\in\{\star,{\sf edge}\}$
\item if $x$ is a vertex set  variable then ${\bf type}(\alpha)[r_x]={\sf vertex\ set}$
\item if $x$ is a edge set  variable then ${\bf type}(\alpha)[r_x]={\sf edge\ set}$
\end{itemize}
We say that $\alpha$ matches $\psi$ if 
${\bf type}(\alpha)$ \term{{\em matches}} $\psi$.
\end{definition}

We denote by ${\cal A}$ the set of all boundaried structures.
Given some $p\in\Bbb{N}$, 
we denote by \term{${\cal A}^{p}$} the set of all boundaried structures of arity $p$ and given 
a finite set $I\subseteq \Bbb{Z}^+$ we denote  by \term{${\cal A}^{p}_{I}$}  the set of all boundaried 
structures of arity $p$  whose boundaried graph has  label set $I$. Notice that 
according to this definition, ${\cal A}^{1}_{I}$ is essentially the same as ${\cal F}_{I}$.  
Finally, we say that a boundaried structure $\alpha$ is a \term{$t$-{\em boun\-daried structure}}
if $\Lambda(G_{\alpha})\subseteq\{1,\ldots,t\}$.

\begin{definition}{\rm [{\bf Compatiblity}] }
For two boundaried structures $\alpha$ and $\beta$ we say that $\alpha$ and $\beta$ are \term{{\em compatible}},
we denote this by $\alpha\mar{\sim_c}\beta$, if the following conditions are satisfied.
\begin{itemize}%\setlength\itemsep{-.3mm}
 \item $\alpha$ and $\beta$ have the same arity $p.$
 \item For every $i \leq p,$  ${\bf type}(\alpha)[i]={\bf type}(\beta)[i]\neq \star$ {\em or} exactly one out of ${\bf type}(\alpha)[i],$ ${\bf type}(\beta)[i]$ is a vertex or edge and exactly one of them is a $\star.$

 \item For every $i\in\{2,\dots,p\}$ such that both $\alpha[i]$ and $\beta[i]$ are vertices, $\alpha[i] \in \delta(G_\alpha),$ $\beta[i] \in \delta(G_\beta)$ and $\lambda_{G_\alpha}(\alpha[i]) = \lambda_{G_\beta}(\beta[i]).$
 
 \item For every $i$ such that both $\alpha[i]$ and $\beta[i]$ are edges, $\alpha[i] \in E(G_\alpha[\delta(G_\alpha)]),$ $\beta[i] \in E(G_\beta[\delta(G_\beta)])$ and  $\lambda_{G_\alpha}(\alpha[i]) = \lambda_{G_\beta}(\beta[i])$ (here we extend the function $\lambda$ 
 to sets in the obvious way).
 \end{itemize}
 \end{definition}
 
 \begin{definition}{\rm [{\bf Gluing of boundaried compatible structures}] }
When two boundaried structures $\alpha$ and $\beta$ are {\em compatible}, the operation of {\em gluing} $\alpha$ and $\beta$ is defined as follows. 
\begin{itemize}
\item $\alpha \oplus \beta$ is a structure $\gamma$ with the same arity, say $p,$ as $\alpha$ and $\beta.$ 
\item $G_\gamma = G_\alpha \oplus G_\beta.$ 
\item For every $i\in\{2,\ldots,p\}$ such that both 
$\alpha[i]$ and $\beta[i]$ are both vertex sets or both edge sets,
we define $\gamma[i]=h(\alpha[i])\cup h(\beta[i]).$
%in the following way: 

%
%notice first that each vertex $v$ of $G_{\gamma}$
%either  originates from $G_{\alpha}$ or from $G_{\beta}$ or is an identified vertex.
%In case $v$ originates from $G_{\alpha}$ (resp. $G_{\beta}$) 
%we put $v$ into $\gamma[i]$ if $v$ originates from a vertex of $\alpha[i]$ (resp. $\beta[i]$).
%If $v$ is an identified vertex that results from identifying $x\in V(G_{\alpha})$ and $y\in V(G_{\beta})$ we put $v$ into $\gamma[i]$ if $x\in \alpha[i]$ or $y\in\beta[i]$.
%
%\item  For every $i\in\{2,\ldots,p\}$ such that both 
%$\alpha[i]$ and $\beta[i]$ are edge sets we define $\gamma[i]$
%in the following way: notice first that each endpoint of an edge $\{v_1,v_2\}$ of $G_{\gamma}$
%either  originates from $G_{\alpha}$ or from $G_{\beta}$ or is an identified vertex.
%
%In case $v_1$ originates from $G_{\alpha}$ (resp. $G_{\beta}$) 
%we put $v$ into $\gamma[i]$ if $v$ originates from a vertex of $\alpha[i]$ (resp. $\beta[i]$).
%If $v$ is an identified vertex that results from identifying $x\in V(G_{\alpha})$ and $y\in V(G_{\beta})$ we put $v$ into $\gamma[i]$ if $x\in \alpha[i]$ or $y\in\beta[i]$.

\item For every $i\in\{2,\ldots,p\}$ such that both 
 $\alpha[i]$ and $\beta[i]$ are vertices  or both are edges we 
 have $h(\alpha[i])=h(\beta[i])$ (by compatibility) and we set $\gamma[i] = h(\alpha[i]) = h(\beta[i]).$ 
 If $\alpha[i]=\star$ we set  $\gamma[i] = h(\beta[i])$ whereas if  $\beta[i]=\star$ we set  $\gamma[i] = h(\alpha[i]).$ By compatibility, exactly one of these cases apply for every $i.$
 \end{itemize}
\end{definition}

\section{A variant of Courcelle's Theorem}
\label{sec:avariodcourctheo}
In this subsection we give a proof of a variant of the classical Courcelle's Theorem~\cite{Courcelle90,Courcelle92a,Courcelle97}, 
which we use in the proofs of our results.

 %The statement of the theorem and its proof is given in Subsection~\ref{lem:canequ}.  Before that, we introduce some basic notions.
We define the \term{{\em compatibility equivalence}} relation $\mar{\equiv_c}$ on boundaried structures as follows. We say that $\alpha \equiv_c \beta$ if for every boundaried structure $\gamma,$ $$\alpha\sim_c \gamma\iff \beta\sim_c\gamma.$$ Clearly $\equiv_c$ is an equivalence relation. We now make the following observation.
%\sed{I find this over-complicated and somehow redundant }
\begin{observation}\label{obs:compequiv} For every arity $p$ and finite set $I\subseteq \Bbb{Z}^{+},$ the relation $\equiv_c$ has a finite number of equivalence classes when restricted to  ${\cal A}_{I}^{p}.$
\end{observation}
\begin{proof}
Define the \term{{\em compatibility signature}} of a boundaried structure $\alpha$ to be a string ${\bf s}(\alpha)$ that encodes the following information about $\alpha$:
\begin{itemize}
\item $\Lambda(G_{\alpha})$
 \item ${\bf type}(\alpha).$
 %\item $G_\alpha[\delta(G_\alpha)].$
 %\item For every $i$ such that $\alpha[i]$ is a vertex set $s$ encodes $\alpha[i] \cap \delta(G_\alpha).$
 %\item For every $i$ such that $\alpha[i]$ is an edge set $s$ encodes $\alpha[i] \cap E(G_\alpha[\delta(G_\alpha)]).$
 \item For every $i$ such that $\alpha[i]$ is a vertex, ${\bf s}(\alpha)$ encodes whether $\alpha[i] \in \delta(G_\alpha),$ and if so, it  encodes $\lambda_{G_{\alpha}}(\alpha[i]).$% (Here $\omega_{J}$ is the order function defined in the preliminaries.)
 \item For every $i$ such that $\alpha[i]$ is an edge, ${\bf s}(\alpha)$ encodes whether $\alpha[i] \in E(G_\alpha[\delta(G_\alpha)]),$ and if so, it also encodes $\lambda_{G_{\alpha}}(\alpha[i]).$
\end{itemize}
Clearly, for every fixed $I$ and $p,$ the compatibility signature ${\bf s}(\alpha)$ can be encoded by a  
number of bits that depends only on $I$ and $p$ and hence there are only finitely many different 
compatibility signatures for boundaried structures in ${\cal A}^{p}_{I}.$ It is easy to verify that
whether a boundaried structure $\alpha\in{\cal A}_{I}^p$ is compatible with 
a  boundaried structure $\gamma\in{\cal A}^p$ can be deduced solely from $\gamma$ and the 
compatibility signature of $\alpha$. Thus, if two boundaried structures $\alpha$ and $\beta$ have the same compatibility signatures then $\alpha \equiv_c \beta.$ This completes the proof.
\end{proof}

%\subsection{The canonical equivalence}
%\label{lem:canequ}

\begin{definition}{\rm [\bf Canonical equivalence on structures.]}
For a property $\sigma$ of structures, we define the corresponding \term{{\em canonical equivalence relation}} $\mar{\equiv_{\sigma}}$ on boundaried structures. For two boundaried structures $\alpha$ and $\beta$ we say $\alpha \equiv_{\sigma} \beta$ if $\alpha \equiv_{c} \beta$ and for all boundaried structures $\gamma$ compatible to $\alpha$ (and thus also to $\beta$), we have 
\begin{eqnarray*}
\sigma(\alpha \oplus \gamma)={\sf true} \Leftrightarrow \sigma(\beta \oplus \gamma)={\sf true}. %\label{eq:theother_eval}
\end{eqnarray*}
\end{definition}
It is easy to verify that $\equiv_{\sigma}$ is an equivalence relation. We say that a 
property $\sigma$ of structures is \term{{\em finite state}} if, for 
every $p\in \Bbb{N}$ and $I\subseteq \Bbb{Z}^{+},$ the 
equivalence relation $\equiv_{\sigma}$ has a finite number of equivalence classes when 
restricted to ${\cal A}^{p}_{I}$.  
Given a CMSO-sentence $\psi,$ we say that $\equiv_{\sigma_{\psi}}$ is the \term{{\em canonical equivalence relation}} corresponding to $\psi$ and we simply denote this relation by  $\mar{\equiv_{\psi}}$.
% is the  canonical equivalence relatio.

In our arguments, the following lemma will be crucial. While it is an implicit consequence
of the results~\cite{ArnborgLS91,Courcelle90,Courcelle97,Courcelle92a,AbrahamsonF93,BoriePT92,DowneyF98}, in the rest of this section, 
we give a complete and self-contained proof.

\begin{lemma}
\label{log_lem}
Every CMSO-definable property on structures has finite state.\end{lemma}
\begin{proof}
Our aim is to prove that for every $p\in \Bbb{N}$ and finite $I\subseteq \Bbb{Z}^{+},$ and 
CMSO-definable property $\sigma$, the equivalence relation $\equiv_{\sigma}$ has a finite number of equivalence classes when restricted to ${\cal A}^{p}_{I}$.  
For this we will define, for every normalized CMSO-sentence $\psi,$ a function ${\sf sgn}_{\psi}$ that takes as input  a boundaried structure and 
outputs a string in  $\{0,1\}^{*}$. To prove the result it suffices to show the following two properties of the function ${\sf sgn}_{\psi}$:

\begin{itemize}
\item[({\bf i})]\label{cond2}  for all $p\in\Bbb{N}$, $J\subseteq \Bbb{Z}^+$, the set ${\sf sgn}_{\psi}({\cal A}_{I}^{p})$ is finite. 
\item[({\bf ii})]\label{cond1} for every two boundaried structures $\alpha$ and $\beta$, if ${\sf sgn}_{\psi}(\alpha)={\sf sgn}_{\psi}(\beta)$ then $\alpha\equiv_{\sigma}\beta.$
\end{itemize}

\noindent We need  the following claim:\\

\noindent{\em Decoder Claim:}  In order to prove Property ({\bf ii}), it is enough to prove that for every  CMSO-sentence 
$\psi$ defining a property $\sigma$, there exist two functions
\begin{eqnarray*}
{\sf dec}_{c}:\{0,1\}^*\times {\cal A}^p & \rightarrow & \{{\sf true},{\sf false}\}\\
{\sf dec}_{\psi}:\{0,1\}^*\times {\cal A}^p & \rightarrow & \{{\sf true},{\sf false}\}
\end{eqnarray*}
 such that 
for every pair $\alpha\in {\cal A}_{I}^p$ and $\gamma\in{\cal A}^p$
we have that  
\begin{eqnarray}
{\sf dec}_{c}({\sf sgn}_{\psi}(\alpha),\gamma)={\sf true}& \iff &  \alpha\sim_c\gamma.\label{lab:comptc}
\end{eqnarray}
and 
for every pair $\alpha\in {\cal A}_{I}^p$ and $\gamma\in{\cal A}^p$ with $\alpha\sim_c\gamma$
it holds that 
\begin{eqnarray}
{\sf dec}_{\psi}({\sf sgn}_{\psi}(\alpha),\gamma)={\sf true} & \iff & \sigma(\alpha\oplus\gamma)={\sf true}.\label{lab:comptc2}
\end{eqnarray}

\noindent{\em Proof of Decoder Claim:} For the proof of the above claim, assume that for some $\alpha,\beta\in{\cal A}_{I}^p$, 
it holds that 
\begin{eqnarray}
{\sf sgn}_{\psi}(\alpha) & = & {\sf sgn}_{\psi}(\beta).\label{leb:thisalso}
\end{eqnarray}
Then for all $\gamma\in{\cal A}^p$, it holds that 
$$\alpha\sim_c \gamma\Leftrightarrow^{\eqref{lab:comptc}} {\sf dec}_{c}({\sf sgn}_{\psi}(\alpha),\gamma)= {\sf true} \Leftrightarrow^{\eqref{leb:thisalso}}
{\sf dec}_{c}({\sf sgn}_{\psi}(\beta),\gamma)={\sf true}\Leftrightarrow^{\eqref{lab:comptc}}  \beta\sim_c \gamma,$$ hence 
$\alpha\equiv_{c}\beta$. Further, for all $\gamma\in{\cal A}^p$
such that $\alpha\sim_c\gamma$ it holds that
\begin{eqnarray*}
\sigma(\alpha\oplus \gamma)={\sf true} & \Leftrightarrow^{\eqref{lab:comptc2}} &  {\sf dec}_{\psi}({\sf sgn}_{\psi}(\alpha),\gamma)= {\sf true} \\ 
& \Leftrightarrow^{\eqref{leb:thisalso}}& 
{\sf dec}_{\psi}({\sf sgn}_{\psi}(\beta),\gamma)={\sf true} \\
& \Leftrightarrow^{\eqref{lab:comptc2}} & \sigma(\beta\oplus \gamma)={\sf true},
\end{eqnarray*}
and thus $\alpha\equiv_{\sigma}\beta,$ as required. This completes the proof of the decoder  claim.\medskip

We start by partially defining the outputs of ${\sf sgn}_{\psi}$ as follows. If 
 $\alpha$ does not match $\psi$ then ${\sf sgn}_{\psi}(\alpha)$
is the null string, denoted by $\myemptyset,$ otherwise, ${\sf sgn}_{\psi}$ 
encodes the compatibility signature of $\alpha$ (as defined in the proof of Observation~\ref{obs:compequiv})
and additional information about $\alpha$ that will be specified later in the proof.

The existence of  a function  ${\sf dec}_{c}$ satisfying~\eqref{lab:comptc} follows directly 
from the proof of  Observation~\ref{obs:compequiv}.

We define the function ${\sf dec}_{\psi}$ such that ${\sf dec}_{\psi}(\myemptyset,\gamma)={\sf false}$
for every boundaried structure $\gamma$. Also ${\sf dec}_{\psi}({\sf sgn}_{\psi}(\alpha),\gamma)={\sf false}$
whenever ${\bf type}(\alpha\oplus\gamma)$ does not match $\psi$. Observe that this can be checked 
using the compatibility signature of $\alpha$ (that is already encoded in ${\sf sgn}_{\psi}(\alpha)$) 
and $\gamma$.
Thus  ${\sf dec}_{\psi}$ satisfies~\eqref{lab:comptc2}
for all pairs $\alpha$, $\gamma$ such that 
 $\alpha\oplus\gamma$ does not 
match $\psi.$  

In the remainder of the proof, we will complete the definition of ${\sf sgn}_{\psi}$
and we will define ${\sf dec}_{\psi}$ for all pairs ${\sf sgn}_{\psi}(\alpha)$, $\gamma$ such that
 $\alpha\oplus\gamma$ match $\psi.$ This should be done in a way such that 
 ({\bf i}) holds for ${\sf sgn}_{\psi}$ and \eqref{lab:comptc2} holds for ${\sf dec}_{\psi}$.\medskip

 We now define  ${\sf sgn}_{\psi}$  and ${\sf dec}_{\psi}$ and prove that they have the claimed properties 
 for the case   where $\alpha$ matches $\psi$ and  
$\psi$ is an atomic  CMSO-sentence. An atomic  CMSO-sentence
is a sentence of the form ``$u \in S$'', ``$e \in S$'', 
``$u = v$'', ``$e = d$'', ``$\mathbf{inc}(d,u)$'', %``$\mathbf{adj}(u,v)$'', 
or ``$\mathbf{card}_{q,r}(S)$'' where $S$ is a set variable, $u$ and $v$ are vertex variables,  $e$ and $d$ are edge variables and $r\in\Bbb{N}\setminus\{1\}$ and $q\in\{0,\ldots,r-1\}$.
 In this case, we append to
${\sf sgn}_{\psi}(\alpha)$   certain information about $\alpha$ that
% In particular,  if 
% $\alpha$ does not match $\psi$ then ${\sf sgn}_{\psi}(\alpha)$
%is the null string, denoted by $\myemptyset,$ otherwise ${\sf sgn}_{\psi}(\alpha)$

\begin{enumerate}[(i)]
%\item\label{sgn1} encodes the compatibility signature of $\alpha$ (as defined in the proof of Observation~\ref{obs:compequiv}).

\item\label{sgn2} encodes $G[\delta(G_{\alpha})],$ 

\item\label{sgn3} encodes $\lambda_{G_{\alpha}},$

\item\label{sgn4} for every vertex variable $x$, encodes whether $\alpha[r_x]=\star$ or not (recall that $r_x$ is the rank of $x$). If $\alpha[r_x]\ne\star$, then ${\sf sgn}_{\psi}(\alpha)$ encodes whether 
$\alpha[r_x]\in \delta(G_{\alpha})$ and, if this is the case, also encodes $\lambda_{G_{\alpha}}(\alpha[r_x]),$

\item\label{sgn5} for every edge variable $x$, encodes whether $\alpha[r_x]=\star$ or not. If $\alpha[r_x]\ne\star$, ${\sf sgn}_{\psi}(\alpha)$ also encodes whether $\alpha[r_x]\subseteq \delta(G_{\alpha})$ and  if this is the case, also encodes $\lambda_{G_{\alpha}}(\alpha[r_x]),$

\item\label{sgn6}  for every vertex set variable $x$, encodes $\lambda_{G_{\alpha}}(\alpha[r_x]\cap \delta(G_{\alpha})),$

\item\label{sgn7}  for every edge set variable $x$, encodes $\lambda_{G_{\alpha}}(\alpha[r_x]\cap E(\delta(G_{\alpha})))$ (here $\lambda_{G_{\alpha}}$ is extended to sets of unordered pairs in the natural way),

\item\label{sgn8} for every vertex variable $x$ such that $\alpha[r_x]\ne \star$  and every vertex set variable $x',$ encodes whether $\alpha[r_x]\in\alpha[r_{x'}]$.

\item\label{sgn9} for every edge variable $x$ such that $\alpha[r_x]\ne \star$  and every edge set variable $x',$ encodes whether $\alpha[r_x]\in\alpha[r_{x'}]$.

\item\label{sgn10} for every pair of vertex variables $x$, $x'$ where $\alpha[r_x]\neq\star\neq\alpha[r_{x'}]$,
encodes whether $\{\alpha[r_x],\alpha[r_{x'}]\}\in E(G_{\alpha}),$ 

\item\label{sgn11} for every vertex variable $x$ and every edge variable $x'$, where $\alpha[r_x]\neq\star\neq\alpha[r_{x'}],$ encodes whether $\alpha[r_x]\in\alpha[r_{x'}]$ (i.e, whether $\alpha[r_{x'}]$ is incident to $\alpha[r_x]$),

\item\label{sgn12} if $\psi$ is ``${\bf card}_{q,r}(x)$'' where $x$  is either a vertex set or an edge set variable,
encodes $|\alpha[r_x]| \pmod r$,

\item\label{sgn13}  for every pair of vertex variables $x$, $x'$ where $\alpha[r_x]\neq\star\neq\alpha[r_{x'}]$,
encodes whether $\alpha[r_{x}]=\alpha[r_{x'}]$,

\item\label{sgn14} for every pair of edge variables  $x$, $x'$ where $\alpha[r_x]\neq\star\neq\alpha[r_{x'}]$,
encodes whether $\alpha[r_{x}]=\alpha[r_{x'}]$,\end{enumerate}

To see that  ${\sf sgn}_{\psi}(\alpha)$ satisfies Property~({\bf i}), it is enough to verify that, for every 
$\alpha\in{\cal A}_{I}^p,$ the length of ${\sf sgn}_{\psi}(\alpha)$ is upper bounded 
by a function depending only the atomic formula $\psi$, the integer $p$, and the set $I$.

We now define ${\sf dec}_{\psi}({\sf sgn}_{\psi}(\alpha),\gamma)$ for the case where $\psi$ is an atomic CMSO-formula
and $\alpha\oplus\gamma$ matches $\psi$ and prove that ${\sf dec}_{\psi}$ satisfies \eqref{lab:comptc2}
for this case.
For this,
%
%
%We now show that ${\sf dec}_{\psi}$  exists for every atomic formula $\psi$. 
%First of all, we define ${\sf dec}_{\psi}(\myemptyset,\gamma)={\sf false}$
%for all boundaried structures $\gamma$. Also ${\sf dec}_{\psi}({\sf sgn}_{\psi}(\alpha),\gamma)={\sf false}$
%whenever ${\bf type}(\alpha\oplus\gamma)$ does not match $\psi$. Observe that this can be checked 
%using ${\sf sgn}_{\psi}(\alpha)$ and $\gamma$ (without any access to $\alpha$).
%Thus  ${\sf dec}_{\psi}$ satisfies~\eqref{lab:comptc2}
%for all pairs $\alpha$, $\gamma$ such that 
% $\alpha\oplus\gamma$ does not 
%match $\psi.$ We next examine the case where  $\alpha\oplus\gamma$  
%matches $\psi$.
%
we distinguish cases depending on the kind of $\psi$. During our case analysis,
 we use quotes $\mbox{``}$  $\mbox{''}$ in order to delimit the string that corresponds 
to a formula and we use the symbol $\mar{\circ}$ to denote the concatenation operation between  strings. For example,
if $\psi=\mbox{``}\exists x \forall y\ \neg \phi(x,y)\mbox{''}$, then $\psi=\mbox{``}\exists x \forall y\mbox{''}\circ\mbox{``}\neg \phi(x,y)\mbox{''}$.

We give a detailed proof in the case where $\psi=\mbox{``}x\in x'\mbox{''}$. We also provide 
a brief description of the proofs for the remaining cases that can all be formalized in a similar fashion.
\medskip

\noindent{\em Case 1:} $\psi=\mbox{``}x\in x'\mbox{''}$ where $x$ is a vertex variable and 
$x'$ is a vertex set variable. Then ${\sf dec}_{\psi}({\sf sgn}_{\psi}(\alpha),\gamma)$ is computed by the  procedure in Table~\ref{table:procedure}:

\begin{table}[ht]
\label{table:procedure}
\begin{center}
\fbox{\begin{minipage}{15cm}
\begin{tabbing}
{\bf if} \= $\alpha[r_x]\ne\star$\ \ \ \ \ \ ({\em using the compatibility signature of $\alpha$})\\
   \> {\bf then} {\bf if} \= $\alpha[r_x]\in \alpha[r_{x'}]$ \ \ \ \ \ \ ({\em using \mbox{\rm (vii)}})\\
   \>            \> {\bf then} {\bf return} {\sf true}			\\
   \>            \> {\bf else}  {\bf if} \= $\alpha[r_x]\in\delta(G_{\alpha})$  \ \ \ \ \ \ ({\em using \mbox{\rm (iii)}}) \\
   \>	          \>            \> {\bf then  if} \= $\lambda^{-1}_{G_{\gamma}}(\lambda_{G_{\alpha}}(\alpha[r_x]))\in\gamma[r_{x'}]$ \ \ \ \ \ \ ({\em using \mbox{\rm (iii)}})\\
   \> 		\> 		\> 		\> {\bf then return} {\sf true} \\
   \> 		\> 		\> 		\> {\bf else return} {\sf false} \\
   \>	          \>            \> {\bf else return} {\sf false}\\
   \> {\bf else\ \  if} \= $\gamma[r_x]\in \gamma[r_{x'}]$ \ \ \ \ \ \ \ \  \ \ \  ({\em notice that $\gamma[r_x]\ne\star$, since $\alpha\sim_{c}\gamma$})\\
     \>	          \>  {\bf then return} {\sf true} \\
     \> 	\> {\bf else if} \= $\gamma[r_x]\in\delta(G_{\gamma})$\\
     \> 	\> \> {\bf then if} \= $\lambda^{-1}_{G_{\alpha}}(\lambda_{G_{\gamma}}(\gamma[r_x]))\in\alpha[r_{x'}]$ \ \ \ \ \ \ ({\em using \mbox{\rm (iii)} and \mbox{\rm (v)}})~~~  \\
   \> 		\> 		\> 		\> {\bf then return} {\sf true} \\
   \> 		\> 		\> 		\> {\bf else return} {\sf false} \\   
   \>	          \>            \> {\bf else return} {\sf false}
\end{tabbing}
\end{minipage}}
\end{center}
\caption{The procedure of the Case 1 in the proof of Lemma~\ref{log_lem}.}
\end{table}
It can be easily verified that the above procedure outputs {\sf true} if and only if 
$(\alpha\oplus\gamma)[r_x]\in (\alpha\oplus\gamma)[r_{x'}]$ that is, if and only if
$\sigma(\alpha\oplus\gamma)={\sf true}.$ Furthermore, every query of the above procedure 
can be answered by inspecting ${\sf sgn}_{\psi}(\alpha)$ and $\gamma$.
The numbers in the parentheses in the above procedure correspond to the items of the encoding 
of ${\sf sgn}_{\psi}(\alpha)$ that are used to answer each query about $\alpha$. This completes the proof of Case 1.\medskip

\noindent{\em Case 2:} $\psi=\mbox{``}x\in x'\mbox{''}$ where $x$ is an edge variable and 
$x'$ is a edge set variable. Here the function ${\sf dec}_{\psi}$ should decide 
whether $\sigma(\alpha\oplus \gamma)$ is true which, in this case, is the same as asking 
whether $(\alpha\oplus \gamma)[r_x]\in(\alpha\oplus \gamma)[r_{x'}]$ is true.
This last question is equivalent to asking whether one of the following holds
\begin{eqnarray}
\alpha[r_x]\in \alpha[r_{x'}]  \label{lab:int1sq}\\
\gamma[r_x]\in \gamma[r_{x'}] \label{lab:int2sq}\\
 \alpha[r_x]\in E(G_{\alpha}[\delta(G_{\alpha})])
 & \mbox{~and~} & \lambda_{G_{\alpha}}(\alpha[r_x])\in  \lambda_{G_{\gamma}}(\gamma[r_{x'}]\cap E(G_{\gamma}[\delta(G_{\gamma})])) \label{lab:int3sq}\\
 \gamma[r_x]\in E(G_{\gamma}[\delta(G_{\gamma})])
 & \mbox{~and~} & \lambda_{G_{\gamma}}(\gamma[r_x])\in  \lambda_{G_{\alpha}}(\gamma[r_{x'}]\cap E(G_{\alpha}[\delta(G_{\alpha})])) \label{lab:int4sq}
\end{eqnarray}
%Where we interpret  $\alpha[r_x]\myin \gamma[r_{x'}]$
%as asking whether $\alpha[r_x]$ is an edge with both endpoints in $\delta(G_{\alpha})$
%and $\lambda_{G_{\alpha}}(\alpha[r_x])\subseteq \lambda_{G_{\gamma}}(\gamma[r_{x'}]\cap \delta(G_{\gamma}))$
%and  $\gamma[r_x]\myin \alpha[r_{x'}]$ is interpreted symmetrically.
Each  query in~\eqref{lab:int1sq}--\eqref{lab:int4sq} can be answered  given $\gamma$ and ${\sf sgn}_{\psi}(\alpha)$
(but no access to $\alpha$ itself).\medskip

\noindent{\em Case 3:} $\psi=\mbox{``}x = x'\mbox{''}$ where both $x$ and $x'$ 
are vertex variables. Here the function ${\sf dec}_{\psi}$ should decide 
whether $\sigma(\alpha\oplus \gamma)$ is true which, in this case, is the same as asking 
whether $(\alpha\oplus \gamma)[r_x]=(\alpha\oplus \gamma)[r_{x'}]$ is true.
This last question is equivalent to asking whether one of the following holds
\begin{eqnarray}
\alpha[r_x]=\alpha[r_{x'}]\ne\star && \label{lab:intr1}\\
 \gamma[r_x]=\gamma[r_{x'}]\ne\star &&\label{lab:intr2}\\
 \alpha[r_x]\in\delta_{G_{\alpha}} \mbox{~and~}   \gamma[r_{x'}]\in\delta_{G_{\gamma}} & \mbox{~and~}  &\lambda_{G_{\alpha}}(\alpha[r_x])=\lambda_{G_{\gamma}}(\gamma[r_{x'}]) \label{lab:intr3}\\
 \alpha[r_{x'}]\in\delta_{G_{\alpha}} \mbox{~and~}   \gamma[r_x]\in\delta_{G_{\gamma}} & \mbox{~and~} & \lambda_{G_{\alpha}}(\alpha[r_{x'}])=\lambda_{G_{\gamma}}(\gamma[r_x]).\label{lab:intr4} \label{lab:int2s4}
\end{eqnarray}
The above is correct because $\alpha\sim_c\gamma$ implies that at most one of $\alpha[r_x]$ and $\gamma[r_x]$ 
is a $\star$ and, whenever neither of them are $\star$'s, it holds that $\alpha[r_{x}]\in \delta_{G_{\alpha}},$ $\gamma[r_{x}]\in \delta_{G_{\gamma}},$ and $\lambda_{G_{\alpha}}(\alpha[r_x])=\lambda_{G_{\gamma}}(\gamma[r_x])$ and  the same holds for  $\alpha[r_{x'}]$ and $\gamma[r_{x'}]$. Again, each  query in~\eqref{lab:intr1}--\eqref{lab:intr4} can be answered  given $\gamma$ and ${\sf sgn}_{\psi}(\alpha).$\medskip

\noindent{\em Case 4:} $\psi=\mbox{``}x = x'\mbox{''}$ where both $x$ and $x'$ 
are edge variables. This case is very similar to the Case 3 and is omitted.\medskip

\noindent{\em Case 5:} $\psi=\mbox{``}\mathbf{inc}(x,x')\mbox{''}$ where $x$ is  an edge variable and
$x'$ is a vertex variable.  Again, here the function ${\sf dec}_{\psi}$ should decide 
whether $\sigma(\alpha\oplus \gamma)$ is true and this   is equivalent to $(\alpha\oplus \gamma)[r_{x'}]\subseteq (\alpha\oplus \gamma)[r_x]$.
This last question is equivalent to asking whether one of the following holds
\begin{eqnarray}
\star\ne\alpha[r_{x'}]\subseteq\alpha[r_x]  & \label{lab:intr1s}\\
 \star\ne\gamma[r_{x'}]\subseteq\gamma[r_x] &\label{lab:intr2s}\\
 \alpha[r_{x'}]\in\delta(G_{\alpha}) & \mbox{~and~} & \lambda_{G_{\alpha}}(\alpha[r_{x'}])\in\lambda_{G_{\gamma}}(\gamma[r_x])\label{lab:intr3s}\\
 \gamma[r_{x'}]\in\delta(G_{\gamma}) & \mbox{~and~} & \lambda_{G_{\gamma}}(\gamma[r_{x'}])\in\lambda_{G_{\alpha}}(\alpha[r_x])\label{lab:intr4s}
\end{eqnarray}
As in Case 3, the above is correct because of the fact that $\alpha\sim_c\gamma$
and it is enough to verify that  each  query in~\eqref{lab:intr1s}--\eqref{lab:intr4s} can be answered  given $\gamma$ and ${\sf sgn}_{\psi}(\alpha).$\medskip

\noindent{\em Case 6:} $\psi=\mbox{``}\mathbf{card}_{q,r}(x)$'' where $x$ is  a vertex set variable.  
The function ${\sf dec}_{\psi}$ should decide 
whether $\sigma(\alpha\oplus \gamma)$ is true which in this case means that $$|(\alpha\oplus \gamma)[r_x]| \equiv q \pmod r.$$
This, in turn, is equivalent to
\begin{eqnarray}
|\alpha[r_x]|+|\gamma[r_x]|-|\lambda_{G_{\alpha}}(\alpha[r_x]\cap \delta(G_{\alpha}))\cap \lambda_{G_{\gamma}}(\gamma[r_x]\cap \delta(G_{\gamma}))| \equiv q \pmod r \label{lab:modulo}
\end{eqnarray}
It is easy to see that~\eqref{lab:modulo} can be evaluated given $\gamma$ and ${\sf sgn}_{\psi}(\alpha).$
This proves Property ({\bf ii}), therefore the statement of the lemma holds when $\psi$ 
is an atomic sentence.\medskip

To complete the proof we now complete the definition of  ${\sf sgn}_{\psi}$ for every non-atomic normalized CMSO-sentence $\psi$ and we will define ${\sf dec}_{\psi}$ for all pairs ${\sf sgn}_{\psi}(\alpha)$, $\gamma$ such that $\alpha\oplus\gamma$ match $\psi.$ As in the case of atomic formulas, this should be done in a way such that 
 ({\bf i}) holds for ${\sf sgn}_{\psi}$ and \eqref{lab:comptc2} holds for ${\sf dec}_{\psi}$.\medskip

By using induction,  we assume that ${\sf sgn}_{\psi'}$ and ${\sf dec}_{\psi'}$ have  
been defined such that ${\sf sgn}_{\psi'}$ satisfies  Property~({\bf i}) and ${\sf dec}_{\psi'}$
satisfies~\eqref{lab:comptc2}  for every  normalized CMSO-sentence  $\psi'$ 
and has  length smaller than $\psi$. This, together with the decoder claim 
implies Property~({\bf ii}) for $\psi'$, namely that  
\begin{eqnarray}
\forall\alpha',\beta'\in{\cal A}\ \  {\sf sgn}_{\psi'}(\alpha')={\sf sgn}_{\psi'}(\beta')\Rightarrow \alpha'\equiv_{{\psi'}}\beta'. \label{lab:resind}
\end{eqnarray}
One of the following cases applies:\medskip

\noindent{\em Case 1.}  $\psi=\mbox{``}\neg\mbox{''}\circ\psi'$, where both $\psi$ and $\psi'$ 
have the same free variables whose rank is the same in $\psi$ and $\psi'$.
From the induction hypothesis, we know that there exist  ${\sf sgn}_{\psi'}$ and ${\sf dec}_{\psi'}$
 such that ${\sf sgn}_{\psi'}$ satisfies  Property~({\bf i}) and ${\sf dec}_{\psi'}$
satisfies~\eqref{lab:comptc2}. 
We define
\begin{eqnarray}
{\sf sgn}_{\psi}(\alpha) & = & {\sf sgn}_{\psi'}(\alpha)\label{lab:decodert34}
\end{eqnarray}
We also define
\begin{eqnarray}
{\sf dec}_{\psi}({\sf sgn}_{\psi}(\alpha),\gamma) &  = &  \neg {\sf dec}_{\psi'}({\sf sgn}_{\psi'}(\alpha),\gamma) \label{lab:decmysymb34}
\end{eqnarray}
Notice that, in~\eqref{lab:decmysymb34}, ${\sf dec}_{\psi}$ is indeed a function of 
${\sf sgn}_{\psi}(\alpha)$ and $\gamma$ because of the definition of ${\sf sgn}_{\psi}(\alpha)$ in~\eqref{lab:decodert34}. By induction hypothesis,  for every $p\in\Bbb{N}$ and $I\subseteq \Bbb{Z}^+$,  ${\sf sgn}_{\psi}({\cal A}_{I}^{p})={\sf sgn}_{\psi'}({\cal A}_{I}^{p})$  is finite, yielding that ${\sf sgn}_{\psi}$ satisfies Property~({\bf i}).

To prove that ${\sf dec}_{\psi}$ satisfies~\eqref{lab:comptc2}, 
let $\alpha\in {\cal A}_{I}^p$ and $\gamma\in{\cal A}^p$ with $\alpha\sim_c\gamma$.
Then  
\begin{eqnarray*}
\sigma_{\psi}(\alpha\oplus\gamma) = \neg\sigma_{\psi'}(\alpha\oplus\gamma)= \neg {\sf dec}_{\psi'}({\sf sgn}_{\psi'}(\alpha)) =^{\eqref{lab:decmysymb34}} {\sf dec}_{\psi}({\sf sgn}_{\psi}(\alpha),\gamma)
\end{eqnarray*}
where the second equation holds because of the induction hypothesis.\medskip

\noindent{\em Case 2.} $\psi=\psi_{1}\circ\mbox{``} \vee \mbox{''}\circ\psi_{2}$\ where  $\psi_1$ and $\psi_{2}$ have the same free variables and the free variables have the same rank in $\psi,$
$\psi_{1},$ 
and $\psi_{2}$.
From the induction hypothesis, we know that there exist  ${\sf sgn}_{\psi_1}$, ${\sf sgn}_{\psi_2},$ ${\sf dec}_{\psi_1},$ and ${\sf dec}_{\psi_2}$ such that ${\sf sgn}_{\psi_1}$ and ${\sf sgn}_{\psi_2}$  both satisfy  Property~({\bf i}) while ${\sf dec}_{\psi_1}$ and ${\sf dec}_{\psi_2}$ both 
satisfy~\eqref{lab:comptc2}. 

We define
\begin{eqnarray}
{\sf sgn}_{\psi}(\alpha) & = & {\sf encode}({\sf sgn}_{\psi_1}(\alpha),{\sf sgn}_{\psi_2}(\alpha))\label{lab:decodert56}
\end{eqnarray}
where ${\sf encode}$ is a function that receives two strings and encodes them as a single string.
We also define two functions ${\sf decode}_{1}$ and ${\sf decode}_{2}$ such that 
$${\sf decode}_i({\sf encode}({\bf s}_{1},{\bf s}_{2}))={\bf s}_{i}, \mbox{~for~} i\in\{1,2\}.$$
We now define
\begin{eqnarray}
{\sf dec}_{\psi}({\sf sgn}_{\psi}(\alpha),\gamma) \phantom{\vee} =   \phantom{\vee}{\sf dec}_{\psi_1}({\sf decode}_{1}({\sf sgn}_{\psi}(\alpha)),\gamma) & \nonumber \\
 \phantom{}\vee{\sf dec}_{\psi_2}({\sf decode}_{2}({\sf sgn}_{\psi}(\alpha)),\gamma) & \nonumber
\end{eqnarray}

From~\eqref{lab:decodert56}, we have that 
  for every $p\in\Bbb{N}$ and $I\subseteq \Bbb{Z}^+$,
\begin{eqnarray}
{\sf sgn}_{\psi}({\cal A}_{I}^{p})\subseteq {\sf encode}({\sf sgn}_{\psi_1}({\cal A}_{I}^p),{\sf sgn}_{\psi_2}({\cal A}_{I}^p))\cup\{\epsilon\}
\label{lab:uinnres}
\end{eqnarray}
By the induction hypothesis, ${\sf sgn}_{\psi_i}({\cal A}_{I}^p)$ is finite, for $i\in\{1,2\}$. This,
together with~\eqref{lab:uinnres}, implies that  ${\sf sgn}_{\psi}$ satisfies Property~({\bf i}).
  
To prove  that ${\sf dec}_{\psi}$ satisfies~\eqref{lab:comptc2}, observe that for all $\alpha\in{\cal A}_{I}^{p},\gamma\in {\cal A}^{p}$ such that $\alpha\sim_c\gamma,$ 
\begin{eqnarray*}
\sigma_{\psi}(\alpha\oplus\gamma) = {\sf true} & \iff & (\sigma_{\psi_1}(\alpha\oplus\gamma) = {\sf true})\bigvee (\sigma_{\psi_2}(\alpha\oplus\gamma) = {\sf true})\\
& \iff & ({\sf dec}_{\psi_1}({\sf sgn}_{\psi_1}(\alpha),\gamma)= {\sf true})\bigvee ( {\sf dec}_{\psi_2}({\sf sgn}_{\psi_2}(\alpha),\gamma)= {\sf true}) \\
& \iff &({\sf dec}_{\psi_1}({\sf decode}_{1}({\sf sgn}_{\psi}(\alpha)),\gamma)= {\sf true})  \\
& & \bigvee ( {\sf dec}_{\psi_2}({\sf decode}_{2}({\sf sgn}_{\psi}(\alpha)),\gamma)= {\sf true})\phantom{....}\\
 & \iff & {\sf dec}_{\psi}({\sf sgn}_{\psi}(\alpha),\gamma)= {\sf true}.
\end{eqnarray*}
The first equivalence holds because of the definition of $\psi$, the second 
by the induction hypothesis, the third by the definition of ${\sf decode}_{i}$, and the last one by the definition of ${\sf dec}_{\psi}$.\medskip

\noindent{\em Case 3.} $\psi=\mbox{``}\exists x\subseteq V(G)\mbox{''} \circ \psi'$,    
where $\psi$ has $p$ free variables and  $\psi'$ has $p+1$ free variables, the ranks 
of the free variables of $\psi$ and $\psi'$ are the same, except for the variable $x$ 
which is a free variable in $\psi'$ but is not free in $\psi$ and the rank of $x$ in $\psi'$ 
is $p+1$.
%
%
%where $\psi$ defines 
%a property $\sigma$ of structures of arity $p$,  $\psi'$ defines a property $\sigma'$ of structures of arity 
%$p+1$, and $x$ is a free variable of $\psi'$ with rank $p+1$, i.e., $r_{x}=p+1$.
From the induction hypothesis, we know that there exist  ${\sf sgn}_{\psi'}$ and ${\sf dec}_{\psi'}$
 such that ${\sf sgn}_{\psi'}$ satisfies  Property~({\bf i}) and ${\sf dec}_{\psi'}$
satisfies~\eqref{lab:comptc2}. 
We define
\begin{eqnarray}
{\sf sgn}_{\psi}(\alpha) & = & {\sf encode}(\{{\sf sgn}_{\psi'}(\alpha\mycirc x)\mid x\subseteq V(G_{\alpha})\})\label{lab:decodert}
\end{eqnarray}
where, given a set ${\cal W}$ of signatures the string ${\sf encode}({\cal W})$ encodes all members of 
${\cal W}.$ We also define the function ${\sf decode}$ that receives as an entry 
a string ${\bf s}$ and outputs the set of strings  that are encoded to it, in particular ${\sf decode}({\sf encode}({\cal W}))={\cal W}.$ We now define
\begin{eqnarray}
{\sf dec}_{\psi}({\sf sgn}_{\psi}(\alpha),\gamma) &  = & \!\!\!\!\!\!\!\!\!\!\!\bigvee_{\mbox{\small ${\bf s}\in\mbox{\sf\small decode $({\sf sgn}_{\psi}(\alpha))$}\atop \mbox{\small $\mbox{$y\subseteq V(G_{\gamma})$}\atop \mbox{\small such that}\ \mbox{$\mysymbol_{\psi'}({\bf s})\sim_{c} (\gamma\mycirc y)$}$}$}}\!\!\!\!\!\!\!\!\!\!\!\sigma_{\psi'}(\mysymbol_{\psi'}({\bf s})\oplus (\gamma\mycirc y)) \label{lab:decmysymb}
\end{eqnarray}
where, given a string ${\bf s}$ encoding a signature,  $\mysymbol_{\psi'}({\bf s})$
returns the lexicographically smallest boundaried structure $\alpha^{\star}$ such that ${\sf sgn}_{\psi'}(\alpha^{\star})={\bf s}$. First observe that the function ${\sf dec}_{\psi}$ is indeed a function 
of ${\sf sgn}_{\psi}(\alpha)$ and $\gamma$.
By the construction of ${\sf sgn}_{\psi},$ for all $p\in\Bbb{N}$ and every finite  $I\subseteq \Bbb{N}$, it holds that 
$${\sf sgn}_{\psi}({\cal A}_{I}^{p})\in {\sf encode}(2^{{\sf sgn}_{\psi'}({\cal A}_{I}^{p+1})})\cup \{\myemptyset\}$$
which proves that ${\sf sgn}_{\psi}$ satisfies Property~({\bf i}) (given a set $X$ we denote by $2^X$
the set of all its subsets). It remains to prove that 
${\sf dec}_{\psi}$ satisfies~\eqref{lab:comptc2}, namely 
that for all $\alpha\in{\cal A}_{I}^p$ and $\gamma\in{\cal A}_{I}$ such that $\alpha\sim_{c}\gamma$, the following hold
\begin{eqnarray}
{\sf dec}_{\psi}({\sf sgn}_{\psi}(\alpha),\gamma)= {\sf true} \Rightarrow  \sigma_{\psi}(\alpha\oplus\gamma) = {\sf true}\label{lab:elast1}\\
{\sf dec}_{\psi}({\sf sgn}_{\psi}(\alpha),\gamma)= {\sf true} \Leftarrow  \sigma_{\psi}(\alpha\oplus\gamma)= {\sf true} \label{lab:elast2}
\end{eqnarray}
To prove~\eqref{lab:elast1}, assume that  ${\sf dec}_{\psi}({\sf sgn}_{\psi}(\alpha),\gamma)= \mbox{\sf true}$.
Thus there exist some $y\subseteq V(G_{\gamma})$ and 
${\bf s}\in {\sf decode}({\sf sgn}_{\psi}(\alpha))$
such that   $\mysymbol_{\psi'}({\bf s})\sim_{c} (\gamma\mycirc y)$ and 
\begin{eqnarray}
\sigma_{\psi'}(\mysymbol_{\psi'}({\bf s})\oplus (\gamma\mycirc y))={\sf true}.\label{lab:thistrue}
\end{eqnarray}
As ${\sf decode}({\sf sgn}_{\psi}(\alpha))=\{{\sf sgn}_{\psi'}(\alpha\mycirc x)\mid x\subseteq V(G_{\alpha})\},$
we may select an $x\subseteq V(G_{\alpha})$  such that 
${\bf s}={\sf sgn}_{\psi'}(\alpha \mycirc x)$. Therefore, the construction 
of $\mysymbol_{\psi'}$ ensures that 
${\sf sgn}_{\psi'}(\mysymbol_{\psi'}({\bf s}))={\bf s}={\sf sgn}_{\psi'}(\alpha\mycirc x)$.
From~\eqref{lab:resind}, $\mysymbol_{\psi'}({\bf s})\equiv_{{\psi'}} \alpha\mycirc x$. This means that
$(\alpha\mycirc x)\sim_c (\gamma\mycirc y)$, 
$\sigma_{\psi'}(\mysymbol_{\psi'}({\bf s})\oplus (\gamma\mycirc y))=\sigma_{\psi'}((\alpha\mycirc x)\oplus (\gamma\mycirc y))$, and, from~\eqref{lab:thistrue}, it follows that 
\begin{eqnarray*}
\sigma_{\psi'}((\alpha\mycirc x)\oplus (\gamma\mycirc y))={\sf true}.%\label{lab:thistrue2}
\end{eqnarray*}
Recall that $(\alpha\mycirc x)\oplus (\gamma\mycirc y)=(\alpha\oplus\gamma)\mycirc(x\cup y)$, therefore
\begin{eqnarray*}
\sigma_{\psi'}((\alpha\oplus\gamma)\mycirc(x\cup y))={\sf true}.%\label{lab:thistrue3}
\end{eqnarray*}
which, by the definition of $\psi,$ implies that  $\sigma_{\psi}(\alpha\oplus\gamma)={\sf true}$ and~\eqref{lab:elast1} follows.\medskip

It now remains to prove~\eqref{lab:elast2}.
Assume  that the value of $\sigma_{\psi}(\alpha\oplus\gamma)=$= {\sf true} . 
Thus, by the definition of $\psi$,
 there exist some $x\subseteq V(G_{\alpha})$ and some $y\subseteq V(G_{\gamma})$ such that 
$ (\alpha\mycirc x)\sim_{c} (\gamma\mycirc y)$ and 
\begin{eqnarray}
\sigma_{\psi'}((\alpha\mycirc x)\oplus (\gamma\mycirc y)) & = & {\sf true} \label{lab:eppggd}
\end{eqnarray}
Let ${\bf s}={\sf sgn}_{\psi'}(\alpha\mycirc x)$ and observe, by~\eqref{lab:decodert}, 
that ${\bf s}\in{\sf decode}({\sf sgn}_{\psi}(\alpha)).$ 
By the definition of $\mysymbol_{\psi'}$ we have that 
${\sf sgn}_{\psi'}(\mysymbol_{\psi'}({\bf s}))={\sf sgn}_{\psi'}(\alpha\mycirc x)={\bf s}.$
By~\eqref{lab:resind}, $\mysymbol_{\psi'}({\bf s})\equiv_{{\psi'}} \alpha\mycirc x.$
Hence, from~\eqref{lab:eppggd}, we obtain that
$\mysymbol_{\psi'}({\bf s})\sim_c (\gamma\mycirc y)$ and 
\begin{eqnarray*}
\sigma_{\psi'}(\mysymbol_{\psi'}({\bf s})\oplus (\gamma\mycirc y)) & = & {\sf true}.
\end{eqnarray*}
Notice that ${\bf s}$ and $y$  certify, in~\eqref{lab:decmysymb},
that ${\sf dec}_{\psi}({\sf sgn}_{\psi}(\alpha),\gamma)={\sf true}$, yielding~\eqref{lab:elast2}.\medskip

\noindent{\em (Multi) case 4.}  $\psi=\mbox{``}\exists x\subseteq E(G)\mbox{''}\circ\psi'$ or
$\psi=\mbox{``}\exists x\in V(G)\mbox{''}\circ\psi'$ or 
$\psi=\mbox{``}\exists x\in E(G)\mbox{''}\circ\psi'$. The proof of the 
first case is the same as the proof of Case 3.
The proof for the remaining two cases differs from the proof of Case 3
only in that when the variables of $x$ an $y$ in the proof are quantified 
as vertices or edges of the vertex or edge set respectively 
of a boundaried structure, they may also 
take the value $\star$.\medskip

As the above case analysis is complete, the proof follows.
\end{proof}

%\section{Derivation of our results}
%\label{sec:derivourresults}

\section{Derivation of our results}
\label{sec:metaalgoframforkern}
\label{sec:derivourresults}

In this section we give 
%a series  of meta-algorthmic concepts  and 
two master theorems 
from which all our results will be derived. We start with fundamental notions of our paper. These are the  notions  of 
 \emph{protrusion}, \emph{protrusion replacement}, and  \emph{protrusion decomposition}.

\begin{definition}{\rm [\bf $t$-protrusion]}
Given a graph $G,$ we say that a set $X \subseteq V$ is an \term{{\em $t$-protru\-sion}} of $G$ if $|\partial(X)|\leq t$ and $\tw(G[X]) \leq t.$
\end{definition}
%
%For an $r$-protrusion $X,$ the vertex set $X' = X \setminus \partial(X)$ is a \emph{restricted $r$-protrusion}. The set $X'$ is the restricted protrusion of $X$ and $X$ is the protrusion of $X'.$ 
\begin{definition}{\rm [\bf $(f,a)$-protrusion replacement family]}
Let $\Pi$ be a parameterized graph problem, let $f:\Bbb{Z}^{+}\rightarrow \Bbb{Z}^{+}$ be a non-decreasing function and let $a\in\Bbb{Z}^{+}.$
An \term{{\em $(f,a)$-protrusion replacement family} } for $\Pi$ is a collection ${\cal A}=\{{\sf A}_{i}\mid i\geq 0\}$ of algorithms, 
such that  algorithm ${\sf A}_{i}$ receives as input a pair $(I,X),$ where 
\begin{itemize}
\item  $I$ is an instance  of $\Pi$ whose graph and parameter are $G$ and $k\in\Bbb{Z},$  
\item $X$ is an $i$-protrusion  of $G$ with at least $f(i)\cdot k^{a}$
vertices, 
\end{itemize}
and outputs
%, in $O(|I|)$ steps, 
an equivalent instance $I^{*}$
such that, if $G^*$ and $k^*$  are the graph and the parameter 
of $I^*,$  then  $|V(G^{*})|<|V(G)|$ and $k^{*}\leq k.$The running time of a $(f,a)$-protrusion replacement family is the running time of ${\sf A}_{i}$.
\end{definition}
%\todo{Why $k$ has to be positive in the defn $3$}

\begin{definition}{\rm [{\bf $(\alpha,\beta)$-Protrusion decomposition]}}
An \term{$(\alpha,\beta)${\em -protrusion decomposition} } of a graph $G$
is  a partition ${\cal P}=\{R_{0},R_{1},\ldots,R_{\rho}\}$ of $V(G)$ 
such that \begin{itemize}
\item $\max\{\rho, |R_{0}|\}\leq \alpha,$  
\item each $R^{+}_{i}=N_{G}[R_i],$ $i\in\{1,\ldots,\rho\},$ is a $\beta$-protrusion of $G,$ and 
\item for every ${i\in\{1,\ldots,\rho\}},\ N_{G}(R_{i})\subseteq R_0.$ 
%SED: actually, this last condition is not necessary! But makes things  more visualizable!
\end{itemize}
We call the sets $R^{+}_{i},$ $i\in\{1,\ldots,\rho\},$ the  {\em protrusions} of ${\cal P}.$
\end{definition}

\subsection{Meta-algorithmic properties}

We define the following two properties for a  
parameterized graph problem $\Pi.$ 
\begin{itemize}
\item[{\bf A}] {\rm [}{\bf Protrusion replacement}:{\rm ]} There exists an  $(f,a)$-protrusion replacement family ${\cal A}$ for $\Pi,$ for some function $f: \Bbb{Z}^{+}\rightarrow \Bbb{Z}^{+}$
and some $a\in\Bbb{Z}^{+}.$
\item[{\bf B}] {\rm [}{\bf Protrusion decomposition:}{\rm ]} There exists a constant $c$ such that, 
if  $G$ and $k\in\Bbb{Z}^+$ are the graph and the parameter of a YES-instance of $\Pi$
then $G$ admits a $(c\cdot k,c)$-protrusion decomposition.
\end{itemize}

We also consider the following weaker version of the  combinatorial property:

\begin{itemize}
\item[{\bf B}$^{*}$] {\rm [}{\bf Weak protrusion decomposition:}{\rm ]}  There exist a constant $c'$ and a non-de\-creasing  
function $g:\Bbb{Z}^{+}\rightarrow \Bbb{Z}^{+}$ such that,  
for every $x\in\Bbb{Z}^{+},$ if  $G$ and $k\in\Bbb{Z}^+$ are the graph and the parameter of a YES-instance of $\Pi$ such that 
all $c'$-protrusions of $G$ are of size at most $x,$ then 
$G$ has a $(g(x)\cdot k,g(x))$-protrusion decomposition.
\end{itemize}

To see that ${\bf B}$  implies {\bf B}$^{*},$ set $c'=1$ and consider the function $g,$ 
with  $g(x)=c,$ where $c$ is the constant in the definition of ${\bf B}.$

\subsection{The meta-algorithm}

All our kernelization algorithms are based on the following procedure that
makes use of  some {$(f,a)$-protrusion replacement family}  ${\cal A}=\{{\sf A}_{i}\mid i\geq 0\}.$ In the following procedure, given a set $R\subseteq V(G)$, we define
${\cal C}_{R}$ as the set of connected components of $G\setminus R$ that have 
treewidth at most $|R|$. Let $X_{R}$ be the set of  vertices that are either in $R$ or 
in some of the connected components of ${\cal C}_{R}$. 

\smallskip
\begin{center}
{\begin{minipage}{11.5cm}
\noindent{\sf Meta-kernelization$(t)$}\\
\noindent{\sl Input}: An instance $I$ of a parameterized graph problem.\\
\noindent{\sl Output}: An equivalent instance $I'.$
\smallskip
\smallskip 

~~~\begin{minipage}{11cm}

\noindent  If $k\geq 0$ and $|I|\leq k$, we return $I$. 
While there exists some $R\subseteq V(G)$ of size at most $2t$ such that 
 $|X_{R}|\geq  f(2\cdot |R|)\cdot k^{a}$,  apply 
algorithm ${\sf A}_{2\cdot|R|}$ with the  pair $(I,X_{R})$ as  input  and replace $I$ by the output $I'$ of this algorithm. In case the parameter $k'$ of $I'$ is negative, then output a trivial YES or NO instance of $\Pi$ depending on whether $(I' ,-1)\in\Pi$ or not.

\end{minipage}
\end{minipage}}
\end{center}
\smallskip

\begin{lemma}
\label{lem:procmetarunsinsteps}
Procedure {\sf Meta-kernelization$(t)$}  runs  in $|I|^{O(t)}$ steps.
Moreover, it outputs an instance with a graph $G$ such that 
for all $i\in\{0,\ldots,t\},$ all $i$-protrusions of $G$ have size at most $f(2i)\cdot k^{a}$.
\end{lemma}

\begin{proof}
Notice that the  while-loop of the procedure will be applied less than  $n=|I|$ times, since 
each iteration decreases the size of the graph by at least one.
In each iteration of the outer loop we have to consider $O(|I|^{2t})$
different choices for $R$. For each choice of $R$ the set $X_{R}$ can
be computed in linear time using the algorithm   of~\cite{Bodlaender96ali}.
That way, the procedure requires $O(|I|^{2t+2})$ steps in total.
To show that the input specifications of the algorithm ${\sf A}_{2\cdot |R|}$ are 
 satisfied when it is called, we argue that every time 
the algorithm $A_{2\cdot |R|}$ is applied to $(I,X_{R})$,
$X_{R}$ is  a $2\cdot |R|$-protrusion of the graph $G$ in the instance of $I$. For this,
notice that $\partial_{G} (X_{R})\subseteq R$ and $\tw(G[X_{R}])\leq \tw(G[X_{R}\setminus R])+|R|\leq 2|R|$.

Let $I'$ be the output of {\sf Meta-kernelization$(t)$} and $G$ be the graph of $I'$.
Assume towards a contradiction that for some $j\in\{0,\ldots,t\},$ 
$G$ contains a $j$-protrusion $X$ of size $>f(2j)\cdot k^{a}.$
Let $R=\partial_{G}(X)$. Observe that $|R|\leq j$ 
and that every connected component $C$ of $G\setminus R$
that contains at least one vertex of $X$ is contained in $X$.
Thus $\tw(C)\leq j$, therefore  $X\subseteq X_{R}$. 
But then, $X_{R}$ is a $2j$-protrusion of $G$ of size $\geq f(2j)\cdot k^{a}$, 
contradicting the fact that $I'$ is the output of  {\sf Meta-kernelization$(t)$}.
\end{proof}

\subsection{Two master theorems}
Our results can be deduced from the following two master theorems. While their proofs are similar in spirit, we present them 
separately in order to illustrate  the way properties ${\bf A},$ ${\bf B},$ and ${\bf B}^{*}$ are combined.

\begin{theorem}
\label{master1}
If a parameterized graph problem $\Pi$ has property {\bf A} for some nonnegative constant $a$ 
and property {\bf B} for some constant $c,$ then $\Pi$ admits a kernel of size $O(k^{a+1}).$ %The running time of the kernelization algorithm on some instance $(I,k)$ is  $O(|I|^{c+2}).$
\end{theorem}
\begin{proof}
Let ${\cal A}=\{{\sf A}_{i}\mid i\geq 0\}$ be an $(f,a)$-protrusion replacement family for $\Pi.$
We claim that the required kernelization algorithm is {\sf Meta-kernelization$(c)$}. 

Suppose that $I$ is a YES-instance  of $\Pi.$   {\sf Meta-kernelization$(c)$}
procedure transforms $I$ to a YES-instance $I^{*}$ of $\Pi.$ Assume that 
$G^{*}$ and $k^{*}$ are the graph and the parameter of $I^{*}$ respectively. First of all 
we assume that $k^{*} \geq 0$ else  {\sf Meta-kernelization$(c)$}  returns a trivial YES or NO instance.  
Let ${\cal P}=\{R_{0},R_{1},\ldots,R_{\rho}\}$ be a $(c\cdot k^{*},c)$-protrusion decomposition of $G^{*}$
for some $\rho \leq c\cdot k^{*},$ whose existence follows from property {\bf B}. Notice that $k^{*}\leq k.$ 
Therefore, from Lemma~\ref{lem:procmetarunsinsteps}, we have  that 
$$|V(G^*)|\leq |R_{0}|+\sum_{i=1}^{\rho} |R_i| \leq c\cdot k + c\cdot k\cdot  f(2c)\cdot k^{a}  = c\cdot k\cdot (f(2c)\cdot k^{a}+1).$$
Hence, if the above procedure outputs an instance whose graph has more than $c\cdot k\cdot  (f(2c)\cdot k^{a}+1)$ vertices,
then the $(I,k)$ is a NO-instance and in this case the algorithm outputs a trivial  NO-instance of $\Pi.$ Otherwise,  by Lemma~\ref{lem:procmetarunsinsteps}, the algorithm outputs, in $O(|I|^{2c+2})$  steps, an equivalent instance
with a graph on $O(k^{a+1})$ vertices, as required. \end{proof}

When $a=0$, we can use the   weaker condition   {\bf B}$^{*}$ and have a linear kernel.

\begin{theorem}
\label{master2}
If a parameterized graph problem $\Pi$ has property {\bf A} for $a=0,$
and property {\bf B}$^{*}$ for some constant $c,$ 
then $\Pi$ admits a linear kernel.
 %Moreover, the running time of the kernelization algorithm is  $O(k^{q}),$ for some constant $q$ that depends on $f,$ $g,$ and $c.$ 
\end{theorem}
\begin{proof}
Let ${\cal A}=\{{\sf A}_{i}\mid i\geq 0\}$ be an $(f,0)$-protrusion replacement family for $\Pi.$ (Notice that in this proof it is important that $a=0.$)

Let also $g:\Bbb{Z}^{+}\rightarrow \Bbb{Z}^{+}$ be a function  such that,  
for every $x\in\Bbb{Z}^{+},$ if  $G$ and $k$ are the graph and the parameter of a YES-instance of $\Pi$ such that 
all $c$-protrusions of $G$ have size at most $x,$ then 
$G$ has a $(g(x)\cdot k,g(x))$-protrusion decomposition.
We claim that the required kernelization algorithm is {\sf Meta-kernelization$(c)$}.  Let $t=g(f(2c)).$ 

Suppose now that $I$ is a YES-instance of $\Pi.$  {\sf Meta-kernelization$(c)$}
procedure transforms $I$ to a YES-instance $I^{*}$ of $\Pi.$ Assume that 
$G^{*}$ and $k^{*}$ are the graph and the parameter of $I^{*}$ respectively.  First of all 
we assume that $k^{*} \geq 0$ else  {\sf Meta-kernelization$(c)$}  returns a trivial YES or NO instance.
By Lemma~\ref{lem:procmetarunsinsteps},  $I^{*}$ has no  $c$-protrusion of size at least $f(2c)$. By applying Condition ${\bf B}^{*}$ for  $x=f(2c),$ we have that 
$G^{*}$ has a $(t \cdot k^*,t)$-protrusion decomposition 
${\cal P}=\{R_{0},R_{1},\ldots,R_{\rho}\}$  for some $\rho \leq t\cdot k^{*}.$ Notice that $k^{*}\leq k.$ 
By Lemma~\ref{lem:procmetarunsinsteps}, we have  that 
$$|V(G^*)|\leq |R_{0}|+\sum_{i=1}^{\rho} |R_i| \leq t\cdot k + t\cdot k\cdot  f(2c)  = t\cdot k\cdot (f(2c)+1).$$
Hence, if the above procedure outputs an instance whose graph has more than $t\cdot k\cdot (f(2c)+1)$ vertices,
then the algorithm outputs a trivial  NO-instance of $\Pi.$ Otherwise, by Lemma~\ref{lem:procmetarunsinsteps}, 
the algorithm outputs, in $O(|I|^{2t+2})$ steps,  an equivalent instance 
on $O(k)$ vertices, as required. 
\end{proof}

We now have all necessary notions to present how the meta-algorithmic theorems mentioned in the introduction
are derived from  Master Theorems~\ref{master1} and~\ref{master2}.

\subsection{Problems  having the algorithmic and combinatorial properties}

Our meta-algorthmic results follow by combining  the following six results.
The  first four imply the  protrusion replacement  property {\bf A}.

 \begin{itemize}
 \item Every annotated  \pmin{}  problem has the  protrusion  replacement  property  {\bf A} for $a=1.$ 
 (Lemma~\ref{lem:annotatedforaequal1}, Subsection~\ref{subsec:protredufamiforanot})
 \item Every annotated  \peq{}  problem  has the  protrusion  replacement  property  {\bf A} for  $a=2.$ 
 (Lemma~\ref{lem:annotatedforaequal2}, Subsection~\ref{subsec:aprotreplfamiforannopro}) 
 \item Every annotated  \pmax{}  has the  protrusion  replacement  property  {\bf A} for $a=1.$ 
 (Lemma~\ref{lem:annotatedforaequal1butmax}, Subsection~\ref{subsec:aprotreplfamiforannopromax}) 
 \item Every parameterized graph problem $\Pi$ that  has {FII} has the  protrusion  replacement  property  {\bf A}  for $a=0.$ 
 (Lemma~\ref{lem:fiiwithaequal0}, Subsection~\ref{subsec:aprotreplfamifii})
 \end{itemize}

%\begin{lemma}
%\label{lem:annotatedforaequal1}
%Every annotated  \pmin{}  problem has the  protrusion  replacement  property  {\bf A} for $a=1.$
%\end{lemma}
%
%\begin{lemma}
%\label{lem:annotatedforaequal2}
%Every annotated  \peq{}  problem  has the  protrusion  replacement  property  {\bf A} for  $a=2.$
%\end{lemma}
%
%\begin{lemma}
%\label{lem:annotatedforaequal1butmax}
%Every annotated  \pmax{}  has the  protrusion  replacement  property  {\bf A} for $a=1.$
%\end{lemma}
%
%\begin{lemma}
%\label{lem:fiiwithaequal0}
%Every parameterized graph problem $\Pi$ that  has {FII} has the  protrusion  replacement  property  {\bf A}  for $a=0.$
%\end{lemma}

\noindent The two last results imply the  protrusion  decomposition  properties
{\bf B} and {\bf B}$^{*}.$

\begin{itemize}
 \item Every $r$-coverable problem has the  protrusion  decomposition  property {\bf B}. 
 (Lemma~\ref{lem:compactcombcondition}, Subsection~\ref{subsec:fortheconditionwithb})  
 \item Every $r$-quasi-coverable problem  has the weak protrusion  decomposition  property {\bf B}$^*\!.$  
(Lemma~\ref{lem:quasicompactcombcondition}, Subsection~\ref{subsec:proofoftheoremquasi}). 

 \end{itemize}

%\begin{lemma}
%\label{lem:compactcombcondition}
%Every $r$-coverable problem has the  protrusion  decomposition  property {\bf B}.
%\end{lemma}
%
%
%\begin{lemma}
%\label{lem:quasicompactcombcondition}
%Every $r$-quasi-coverable problem  has the weak protrusion  decomposition  property {\bf B}$^*\!.$
%\end{lemma}
%
%
%
%
%
%
%The proofs of Lemmata~\ref{lem:annotatedforaequal1}--\ref{lem:quasicompactcombcondition} are postponed 
%until Sections~\ref{sec:redrules} and~\ref{sec:comresultsallofthem}.
%In particular, Section~\ref{sec:redrules} is devoted to the proof of 
%Lemmata~\ref{lem:annotatedforaequal1} (Subsection~\ref{subsec:protredufamiforanot}),~\ref{lem:annotatedforaequal2} (Subsection~\ref{subsec:aprotreplfamiforannopro}),~\ref{lem:annotatedforaequal1butmax} (Subsection~\ref{subsec:aprotreplfamiforannopromax}), and~\ref{lem:fiiwithaequal0} (Subsection~\ref{subsec:aprotreplfamifii}).
%In Section~\ref{sec:comresultsallofthem}
%we prove  Lemmata~\ref{lem:compactcombcondition} (Subsection~\ref{subsec:fortheconditionwithb}) and~\ref{lem:quasicompactcombcondition} (Subsection~\ref{subsec:proofoftheoremquasi}).

\subsection{Derivation of Theorems~\ref{thm:cmsol},~\ref{thm:cmsolnotanotated},~and~\ref{thm:automata}}

All our main results are consequences of   Master Theorems~\ref{master1} and~\ref{master2}. 
Theorem~\ref{thm:cmsol}  follows from  Master Theorem~\ref{master1} and Lemmata~\ref{lem:annotatedforaequal1},~\ref{lem:annotatedforaequal2},~\ref{lem:annotatedforaequal1butmax}, and~\ref{lem:compactcombcondition}.
Moreover, Theorem~\ref{thm:automata} follows from Master Theorem~\ref{master2} and Lemmata~\ref{lem:fiiwithaequal0} and~\ref{lem:quasicompactcombcondition}.
We conclude this section with the proof of Theorem~\ref{thm:cmsolnotanotated}%\sed{I changed this to theorem...} 

\begin{proof}[of Theorem~\ref{thm:cmsolnotanotated}]
Suppose that  $\Pi$ is {\sf NP}-hard and its annotated version $\Pi^{\alpha}$ is in {\sf NP}. Consider an algorithm that, given an instance $I=(G,k)$ of $\Pi,$ applies first  
the kernelization algorithm of Theorem~\ref{thm:cmsol} as a subroutine on the 
annotated instance $((G,V(G)),k),$ that is, all the vertices of $G$ are set to be annotated. 
This subroutine outputs an equivalent {\sl annotated} instance $I'=((G',Y'),k)$ of $\Pi^{\alpha}$ 
where the number of vertices in $G'$ is 
a polynomial function of $k.$ 
The next step of the algorithm is to  
apply a polynomial time many-to-one reduction 
from $\Pi^{\alpha}$ to $\Pi$ on $I'$ and obtain an equivalent instance $I''=(G'',k'')$
where $|I''|$ is a polynomial function of $|I'|.$
This reduction exists from the Cook--Levin theorem, as
$\Pi^{\alpha}\in{\sf NP}$ and $\Pi$ is {\sf NP}-hard.
Then  $|I''|$ is a polynomial function of $k$ and this two-step polynomial-time algorithm 
is the desired kernelization algorithm  for $\Pi.$ The reduction from $\Pi^\alpha$ to $\Pi$ might output an instance 
$I''$ with parameter $k''$ where $k''$ is exponential in $|I''|$ because $k''$ could be encoded in binary. However, since 
$\Pi$ is a \pmem{} problem, $(I'',k'')\in \Pi$ if and only if  $(I'',k''')\in \Pi$, where $k'''=\min\{k'',|I''|+1\}$.  The kenrelization algorithm outputs $(I'',k''')$. 
\end{proof}

\section{Reduction Rules}
\label{sec:redrules}
In this section we prove the existence of protrusion replacement families for \pmem{} graph problems and for parameterized problems that have FII.

\subsection{Model checking on structures}
In order to prove our reduction rules we consider an extension of \wpmem{} problems to a setting where the input is a structure rather than a graph. Specifically we consider the following problems.

\medskip
\begin{center} 
\fbox{\begin{minipage}{11cm}
\noindent{\sc Min/Max-CMSO on Structures}\\
\noindent {\em Input}: A structure $\alpha$ and a CMSO sentence $\psi$.\\
\noindent{\em Output}: A minimum/maximum size subset $S$ of $V(G)$ (or $E(G)$) such\\
\noindent{\phantom{{\em Output}:}} that $(\alpha \diamond S) \models \psi.$
\end{minipage}}
\end{center}
\medskip

\medskip
\begin{center} 
\fbox{\begin{minipage}{11cm}
\noindent{\sc Eq-CMSO on Structures}\\
\noindent {\em Input}: A structure $\alpha$, a CMSO sentence $\psi,$ and an integer $k$.\\
\noindent{\em Output}: A subset $S$ of $V(G)$, (or $E(G)$) $|S|=k$ such that $(\alpha \diamond S) \models \psi.$
\end{minipage}}
\end{center}
\medskip

Observe that in the above problems the CMSO sentence is part of the input and not fixed as in the case of \pmem{} problems. 
We will repeatedly apply the following result from \cite[Theorem~5]{BoriePT92}, see also~\cite{ArnborgLS91}.
\begin{proposition}
\label{lema:borie}
There exists a computable function $f:\Bbb{Z}^{+}\times\Bbb{Z}^{+}\rightarrow\Bbb{Z}^{+}$ 
and an algorithm that solves 
{\sc Min/Max/Eq-CMSO on Structures} in 
$f(\tw(G_\alpha),|\psi|)\cdot |V(G_\alpha)|$ steps.
\end{proposition}

Proposition~\ref{lema:borie} is a slight strengthening of Theorem~$5$ of~\cite{BoriePT92}; what is shown there explicitly is the corresponding version where the input is a graph rather than a structure. Arnborg et al.~\cite{ArnborgLS91} show the variant of Proposition~\ref{lema:borie} for MSO logic rather than CMSO logic. Either of these proofs can be made to work {\em both} on structures and with CMSO logic.

The construction of each  protrusion replacement family depends on whether we are dealing with an annotated \pmin{}, \peq{}, or \pmax{} problem, or whether the problem in question has FII. For the case of annotated %\pmem{} 
problems, the constructions consist  of  three parts. 
In the first two parts, we focus on reducing the set of annotated vertices, and in the last part we  %\sed{I do not like this text!}
reduce the set of vertices. In all cases, we assume that we are given a 
sufficiently large $t$-protrusion.
%\sfdd{In this section, some $r$'s that should be $t$'s may still remain}
In the following discussion we deal with 
annotated \pmem{} problems where the set $S$ in question  
is a set of vertices. The case where $S$ is a set of edges can be dealt with in an identical manner.

%\sef{Check the edge issue in defs...}

\subsection{Protrusion replacement  families for annotated \pmin{} Problems}
\label{subsec:protredufamiforanot}

We start from the existence of a protrusion replacement family for annotated \pmin{} problems. The 
technique employed in this section will act as a template for other types of annotated problems.
%how we 
%handle the annotated \pem{} problems. 
 Recall that in an
 annotated \pmin{} problem $\Pi^{\alpha}$ we are given a structure  $(G,Y)$ and an integer $k.$ The objective is to find a set $S\subseteq Y$ of size 
 at most $k$ such that $(G,S)$ models some CMSO sentence $\psi.$
 For our reduction rule, we are also given a sufficiently large $t$-protrusion $X.$ In the first step of the reduction, we show 
 that the set $Y \cap X$ can be substituted in $O(|X|)$ steps by a new 
 set $Z$ of $O(k)$ vertices such that $((G,Y),k)$ is a YES-instance if and only if $((G,Z\cup (Y\setminus X)),k)$ is a YES-instance.
 In the second step  we show that the $t$-protrusion 
 $X$ can be partitioned into $O(k)$ $t'$-protrusions, where $t'=O(t),$ such that each $t'$-protrusion 
 contains vertices from $Z$ only in its (bounded size) boundary. In the third and 
 final step of the reduction rule, we replace the largest $t'$-protrusion with an equivalent, but smaller, $t'$-boundaried graph. For the case of  \pmin{} 
 problems, these three reduction steps correspond to Lemmata~\ref{lem:annotated},~\ref{lem:split},~and~\ref{lem:red1} respectively.\medskip

%\begin{definition}{\rm [{\bf 
%Canonical equivalence relation on structures of type $({\sf graph},{\sf vertex\ set}).$}]}
%Given a $t\in \Bbb{Z}^{+},$ we use the notation $\mar{{\cal U}_{t}}$ for  the set of all $t$-boundaried structures with type $({\sf graph},{\sf vertex\ set}).$ Let $\psi$ be a CMSO formula whose free variables are a graph and a vertex set.
%Given $\alpha,\beta\in {\cal U}_{t},$ we 
%say that $\alpha\mar{\equiv_{\psi,t}} \beta$ if 
%\begin{eqnarray}
%\forall{\gamma\in{\cal U}_{t}} & & \alpha\oplus \gamma\models \psi\Leftrightarrow \beta\oplus \gamma\models \psi\label{lab:rkkdf}
%\end{eqnarray}
%and call the equivalence relation $\equiv_{\psi,t}$ \term{{\em canonical equivalence relation
%for $\psi$ on ${\cal U}_{t}.$}} 
%\end{definition}

%A direct corollary of Lemma~\ref{log_lem} is the following.
%\begin{lemma}
%\label{log_lem_bis}
%For every $t\in\Bbb{Z}^{+}$ and every CMSO sentence  $\psi$  whose free 
%variables are a graph and a vertex set, the  number of equivalence classes of the canonical equivalence relation
%for $\sigma_{\psi}$ on ${\cal U}_{t}$ is bounded by a function of $t$ and $|\psi|.$
%\end{lemma}

We start by proving a lemma that lets us analyze the interior of a protrusion without bothering about the rest of the graph.
\begin{lemma}
\label{lem:findMinAnnotated}
There is an algorithm that given two boundaried structures $(G_X, Y_X)$ and $(G_R, S_R)$ of type $({\sf graph},{\sf vertex\ set})$ and a CMSO-sentence $\psi$ finds a minimum size set $S_X \subseteq Y_X$ such that $(G_X, S_X) \oplus (G_R, S_R) \models \psi$ in time $|V(G_X \oplus G_R)|\cdot f(|\psi|, \tw(G_X \oplus G_R))$.
\end{lemma}

\begin{proof}
Let $(G', Y', S_R') = (G_X, Y_X, \emptyset) \oplus (G_R, \emptyset, S_R)$. Finding the desired set $S_X \subseteq Y$ now amounts to finding a minimum size set $S_X' \subseteq Y'$ such that $(G', S_X' \cup S_R') \models \psi$. This is easily formulated as {\sc Min-CMSO on Structures} and hence may be solved in the desired running time by Proposition~\ref{lema:borie}.
\end{proof}

%\begin{lemma}
%\label{lem:translate}
%Let $(F,W)$  be a $t$-boundaried structure, $\tw(F)\leq t$, and let $\beta_{1},\ldots,\beta_{\rho}$ be $t$-boundaried structures such that the graph of each $\beta_i$ has at most $\nu$ vertices. Then, given a CMSO sentence $\psi,$ there is a graph $G^{+}$ and a series of CMSO sentences  $\psi_{1}^+,\ldots,\psi_{\rho}^+$ such that 
%\begin{itemize}
%\item For all $j\in\{1,\ldots,\rho\},$
%$(F,W)\oplus \beta_{j}\models \psi\iff (G^{+},W)\models \psi_{j}^+.$
%\item  the sizes of $\psi_{j}^+, i\in\{1,\ldots,\rho\}$ 
%are bounded by some function of $|\psi|,\nu,$ and $t.$
%\item $\tw(G^+)\leq t+\nu.$

%\item $|V(G^{+})|\leq |V(F)|+\nu-t.$
%\sed{fix the subgraph issue!}

%\end{itemize}
%This is an old NULL lemma.
%\end{lemma}

%: STOP HERE 
%We now provide the reduction rule for annotated \pmin{} problems.

\medskip\noindent\textbf{Reducing the set of annotated vertices}. The first step of our reduction rule is based on the following lemma.

\begin{lemma}
\label{lem:annotated}
Let $\Pi^{\alpha}$ be an annotated \pmin{} problem 
%defined by some CMSO-sentence $\psi$
 and let $t$ be an integer.
Then there exists an algorithm that given an instance $((G,Y),k)$ of $\Pi^{\alpha}$ and a $t$-protrusion $X$
of $G,$ outputs  in   $O(|X|)$ steps  an equivalent instance  $((G,Y'),k)$ of $\Pi^{\alpha},$ where
$|Y'\cap X|=O(k)$ and $Y' \subseteq Y.$
%
%Let $((G,Y),k)$ be an
%instance of $\Pi^{\alpha}$ and let  $X$ be an $r$-protrusion 
%of $G.$ Then there is an  $O(|X|)$ time algorithm, that 
%
%
%
%computes a set of vertices 
%$Z \subseteq X \cap Y$ with $|Z|= O(k),$ such that if there exist $S \subseteq Y$ with $(G,S)\models \psi$ and 
%$|S|\leq k,$ then there exist   $S'\subseteq Y$ where $(G,S')\models \psi,$ $|S'|\leq k,$ and $S' \cap X \subseteq Z.$
\end{lemma}

%\begin{remark}
 %Let us remark that 
 We remark that the constants hidden in the ``$O$''-notation of the complexity of the algorithm and the size of its output depend only on the length of the CMSO-sentence $\psi$ defining  $\Pi^{\alpha}$ and the constant $t.$ From now onwards, we will not explicitly mention this. 
%\end{remark}

\begin{proof}
Let $\psi$ be the CMSO-sentence mentioned in the definition of $\Pi^{\alpha}.$  Lemma~\ref{log_lem} implies 
that the canonical equivalence relation $\equiv_{\sigma_\psi}$ has finitely many equivalence classes
on the 
set of boundaried structures of arity two with label set $\{1,\ldots,t\}$. Let  ${\bf MinRep}(\psi,t)$  be  a set containing a representative (a boundaried structure of arity two)  for each equivalence class of $\equiv_{\sigma_\psi}$ with  the minimum number of  vertices in the graph of a structure.
%Let ${\bf MinRep}(\psi,t)$ 
%be a set containing a representative with the minimum number of  vertices for each 
%equivalence class of $\equiv_{\sigma_\psi}$. 
Given $G$, $Y$ and $X$ we define the 
sets $B = \partial_G(X)$, $R = (V(G) \setminus X) \cup B$ and the boundaried 
structures $(G_X, Y_X)$ and $(G_R, Y_R)$ as follows. The boundaried 
graphs $G_X$ and $G_R$ are just $G[X]$ and $G[R]$ respectively. Both 
have boundary $B$, with labels from $\{1,\ldots,t\}$ such that $G_X \oplus G_R = G$. 
Similarly $Y_X = Y \cap X$ while $Y_R = Y \setminus X$, such that $(G,Y) = (G_X,Y_X) \oplus (G_R,Y_R)$.

For every structure $\alpha = (G_R^\alpha, S_R^\alpha) \in {\bf MinRep}(\psi,t)$ we find using Lemma~\ref{lem:findMinAnnotated} a minimum size set $S_X^\alpha \subseteq Y_X$ such that $(G_X, S_X^\alpha) \oplus \alpha \models \psi$. Since $|{\bf MinRep}(\psi,t)|$ and the size of each structure in ${\bf MinRep}(\psi,t)$ depends only on $|\psi|$ and $t$, and the treewidth of $G[X]$ is at most $t$, this takes time $O(|X|)$. Now, define 
\begin{equation*}
Y_X' = \bigcup_{\alpha \in {\bf MinRep}(\psi,t)}\left\{
\begin{array}{rl}
S_X^\alpha & \text{if } |S_X^\alpha| \leq k,\\
\emptyset & \text{otherwise}.
\end{array} \right.
\end{equation*}
We set $Y' = Y_X' \cup Y_R$ (formally $Y_X'$ and $Y_R$ are vertex sets of different graphs, so actually 
$Y'$ is the second element of the 2-tuple of  $(G_X,Y_X') \oplus (G_R,Y_R)$,
i.e., $Y'=((G_X,Y_X') \oplus (G_R,Y_R))[2]$, but this is just semantics). Since  $|{\bf MinRep}(\psi,t)|$ depends only on $|\psi|$ and $t$ the construction of $Y'$ implies $|Y' \cap X| = O(k)$.

To complete the proof, it remains to show that $((G,Y'),k) \in \Pi^{\alpha}$ if and only if $((G,Y),k) \in \Pi^{\alpha}$. For the forward direction we have that $Y' \subseteq Y$ and hence feasible solutions to $((G,Y'),k)$ are also feasible for $((G,Y),k)$. We now turn to proving the reverse direction. Let $S \subseteq Y$, $|S|\leq k$ be such that $(G,S) \models \psi$. Let $S_X = X \cap S$ and $S_R = S \setminus X$. Observe that $(G_X,S_X) \oplus (G_R, S_R) = (G, S)$ and that $|S_X| + |S_R| = |S| \leq k$. Choose $\alpha = (G_R^\alpha, S_R^\alpha) \in {\bf MinRep}(\psi,t)$ such that $\alpha \equiv_{\sigma_\psi} (G_R, S_R)$. Let $S_X^\alpha \subseteq Y_X$ be the set computed for $\alpha$ in the previous paragraph. Since 
$$(G_X, S_X) \oplus \alpha \models \psi \iff (G_X, S_X) \oplus (G_R, S_R) \models \psi \iff \mbox{\sf true}$$
it follows that  $|S_X^\alpha| \leq |S_X| \leq k$. Thus $S_X^\alpha \subseteq Y_X'$. Let $S' = S_X^\alpha \cup S_R$ (again, formally $S_X^\alpha$ and $S_R$ are vertex sets of different graphs, so actually $S' = ((G_X,S_X^\alpha) \oplus (G_R,S_R))[2]$). We have that $S' \subseteq Y'$, $|S'| \leq |S_X^\alpha| + |S_R| \leq |S_X| + |S_R| = |S| \leq k$. Finally we observe that
\begin{align*}
(G,S') \models \psi \\
\iff (G_X,S_X^\alpha) \oplus (G_R, S_R) \models \psi \\
\iff (G_X,S_X^\alpha) \oplus \alpha \models \psi\\
\iff \mbox{\sf true.}
\end{align*}
This concludes the proof.
\end{proof}

%In kernelization algorithm we use Lemma~\ref{lem:annotated}  to change the set $Y$ to $(Y \setminus X) \cup Z.$ 
%The next lemma exploits the fact that $Z$ contains at most $O(k)$ 
%\sed{fix!}
%vertices.

\medskip\noindent\textbf{Partitioning Protrusions}.
In the second step of the reduction rule, the $t$-protrusion $X$ is partitioned into $O(k)$ smaller $t'$-protrusions for some $t'=O(t).$

\begin{lemma}
\label{lem:split}
Let $G$ be a graph, $Y$ be a subset of its vertices, and $k$ be an integer.  Let 
also $X$ be a $t$-protrusion and $Z = X \cap Y$ such that $|Z|\leq k.$  There 
is an $O(|X|)$ step algorithm that outputs a  collection ${\cal Q}$ of $(4t+2)$-protrusions
 such that $X = \bigcup_{Q\in{\cal Q}}Q,$ $|{\cal Q}|=O(k),$  and for every 
$Q\in{\cal Q},$ $Z \cap Q \subseteq \partial_{G}(Q).$ 
\end{lemma}

\begin{proof}
We assume that $G[X]$ is connected, otherwise we work independently on its connected components.
We find a nice tree decomposition of $G[X]$ and then we add $\partial_{G}(X)$ to all its  bags.
We denote  the resulting tree decomposition by $(T,{\cal X})$ and, clearly,
it has width most $2t.$
 
The decomposition  $(T,{\cal X})$  can be constructed in  $O(|X|)$ steps, see e.g. \cite{Bodlaender96ali}.
Now we mark a subset of the nodes of $T.$ 
For each vertex $z\in Z$ we mark, if exists, the forget node $t_{z}$
with the property that $\{z\}=X_{t_{z}}\setminus X_{t_{z}'},$  where $t_{z}$ is the child of  $t_{z}'$ in $T.$
%
%
%where a 
%\sed{The proof (as well as its spirit) is not hard to understand for somebody that is familiar with nice tree-%decompositions. We may have problems with a reviewer that is not so acquainted.}
%vertex in $Z$ is forgotten. 
As each vertex is forgotten at most once in a nice tree 
decomposition, so far we have marked at most  $|Z|+1$  nodes of $T.$ Now, as long as this is possible, we  keep 
marking each bag that is the lowest common ancestor of two already marked nodes. 
Using a standard counting argument for trees, it follows that, in the worst case, 
this operation   doubles the number of marked nodes. Hence, 
there are at most $O(|Z|)$ marked nodes;  we denote this set by $M.$
We say that two nodes $t_{1},t_{2}\in M$ are {\em linked} if these nodes are the only marked nodes of the  
$(t_1,t_2)$-path in $T.$  
%exists in the path connecting $t_{1}$ to $t_{2}$ in $T.$ 
We  define the set
\begin{eqnarray*}
P  & = & \{(t_1,t_2)\mid \mbox{$t_1$ and $t_2$ are linked nodes of $M$ and $t_1$ is a predecessor of $t_2$}\}. 
\end{eqnarray*} 
We observe that $|P|=O(|Z|)$ and  each marked node belongs to some pair in $P.$
Let ${\cal C}$ be the set of the connected components of 
$G[X]\setminus \bigcup_{t\in M}X_{t}.$
By the construction of $M,$ the neighborhood of 
a connected component $C$ in ${\cal C}$ may intersect
either a single bag $X_{t}$ of $T,$ or two bags $X_{t_{1}},X_{t_{2}}$ of $T$
such that $(t_{1},t_{2})\in P.$
In the first case, we define $R(C)$ to be some pair in $P$ that contains $t$ as an endpoint (if there are 
many such pairs, we make an arbitrary choice). 
In the second case, we define $R(C)=\{t_{1},t_{2}\}.$
Given a pair $p$ of $P,$ we use the 
notation $L^{-1}$ to denote the union of the vertex sets of all the connected 
components of ${\cal C}$ that map to $p.$
It is now easy to see that that ${\cal R}=\{L^{-1}(p)\mid p\in P\}$
is a partition of $G[X]\setminus \bigcup_{t\in M}X_{t}.$
As each vertex from  $Z$ is in some  bag corresponding to a  marked node,
none of the sets in ${\cal R}$ intersects $Z.$ 
Moreover the neighborhood in $G$ of each set in ${\cal R}$ is
a subset of at most two bags of $(T,{\cal X})$ and 
%as these bags correspond to forget nodes 
thus its neighborhood has at most $2(2t+1)$ vertices.
We now define the set ${\cal Q}=\{V(R)\cup \partial_{G}(V(R))\mid R\in {\cal R}\}.$
Then each member $Q$ of ${\cal Q}$ is 
an $(4t+2)$-protrusion of $G$ where $Z \cap Q \subseteq \partial_{G}(Q).$
Moreover, $\bigcup_{Q\in {\cal Q}}=X$ and the lemma follows as $|{\cal Q}|=|P|=O(k).$
\end{proof}
%\begin{remark}
%The assumption that $Y$ is the set of ``black vertices'' is irrelevant here. One could prove 
%Lemma~\ref{lem:split} for any set $Y'$ resulting in $O(|Z'=X\cap Y'|)$ $r'$-protrusions 
%$X_1, X_2, \ldots ,X_{\ell'}$ such that $X = X_1 \cup X_2 \cup \ldots \cup X_{\ell'}$ and for every 
%$i \leq \ell',$ $Z' \cap X_i \subseteq \partial(X_i).$ 
%\end{remark}

We will also need the following simple decomposition lemma for $t$-protrusions.
\begin{lemma}
\label{lem:stplittreedec}
 If a graph $G$ contains a $t$-protrusion $X$ where $|X|>c>0,$ then it also contains a $(2t+1)$-protrusion $Y$ where $c<|Y|\leq 2c.$ Moreover, given a tree-decomposition of $X$ of width at most $r,$ a tree decomposition of $Y$ of width at most $2t$ can be found in $O(|X|)$ steps.
\end{lemma}
\begin{proof} If $|X|\leq 2c,$ we are done.
Assume that $|X|>2c$ and let  
%$$({\cal X}=\{X_{t}\}_{t\in V(T)}),T,t)$$  
\[(T, \mathcal{ X}=\{X_{t}\}_{t\in V(T)},t)\] 
be a nice tree-decomposition of $G[X],$ rooted at  some, arbitrary chosen, node $t$ of $T.$ Given a vertex $x$ of the rooted tree $T,$  we denote by ${D}(x)$ the 
subset of $V(T)$ containing $x$ and all its descendants in $T$ and by $T_x$ the subtree of $T$ rooted at $x$.   
Let $B\subseteq V(T)$ be the set containing  each vertex $x$ of $T$ with 
the property that the vertices appearing in $\bigcup_{y\in {D}(x)}X_y$ (i.e. the vertices of the nodes 
corresponding to $x$ and its descendants) are more than $c.$ As $|X|\geq 2c,$ $B$ is a non-empty set. We choose $b$ to be  a member of $B$ 
whose descendants in $T$ do not belong in $B.$ The choice of $b$ and the fact that $T$ is a binary tree ensure that $c<|\bigcup_{y\in {D}(b)}X_y|\leq 2c.$
We define $Y=\partial_{G}(X)\cup \bigcup_{y\in {D}(b)}X_y$ and observe that 
\begin{eqnarray}
(T_b,  {\cal X}'=\{\partial_{G}(X)\cup X_{t}\}_{t\in D(b)} , b) 
%({\cal X}'=\{\partial_{G}(X)\cup X_{t}\}_{t\in D(b)}),{D}(b))
\label{eq:newdecnonis}
\end{eqnarray}
is a tree decomposition of $G[Y].$ As $|\partial_{G}(X)|\leq t,$   the width of the tree decomposition in~\eqref{eq:newdecnonis}  is at most $2t.$ Moreover, 
it holds that $\partial_{G}(Y)\subseteq  \partial_{G}(X) \cup X_{b},$ therefore $Y$ is a $(2t+1)$-protrusion of $G.$\end{proof}

\medskip\noindent\textbf{Reducing Protrusions}.
In the third phase of our reduction rule, we find a protrusion to replace, and perform the replacement.

\begin{lemma}
\label{lem:red1}
Let $\Pi^{\alpha}$ be an annotated \pmeq{} problem. 
% definable by some CMSO-sentence $\psi.$
%and let $\psi'$ be a CMSO-sentence on graphs. 
Then for every integer $t$ there is a $c_{1}\in\Bbb{Z}^{+}$ (depending only on $|\psi|$ and $t$) and  an algorithm that given 
an instance $((G,Y),k)$ of $\Pi^{\alpha}$ and a $t$-protrusion $X$ of $G,$ where $c_1 <|X| \leq 2c_{1}$ and $X\cap Y\subseteq \partial_{G}(X),$ outputs, 
in  $O(|X|)$  steps, an equivalent instance  $((G^*,Y^*),k)$ of $\Pi^{\alpha}$ such that $|V(G')| < |V(G)|.$
\end{lemma}

\begin{proof}  
We define an equivalence relation between boundaried structures of type $({\sf graph},{\sf vertex\ set})$ as follows:
Let $\alpha_{1}=(G_{1},Y_{1})$ and $\alpha_{2}=(G_{2},Y_{2})$ be two boundaried structures with labelling functions $\lambda_{1}: \delta(G_1) \rightarrow \{1,\ldots,t\}$ and $\lambda_{2}: \delta(G_2) \rightarrow \{1,\ldots,t\}$ respectively, such that $Y_{1}\subseteq \delta(G_1)$ and $Y_{2}\subseteq \delta(G_2)$.

We say that $\alpha_{1}\approx\alpha_{2}$ if the following conditions are satisfied:
\begin{enumerate}\setlength\itemsep{-.7mm}
\item $\Lambda(G_{1})=\Lambda(G_{2})$
\item $\lambda_{1}(Y_{1})=\lambda_{2}(Y_{2})$
\item for every $S_{1}\subseteq Y_{1}$ and $S_{2}\subseteq Y_{2}$ such that $\lambda_{1}(S_{1})=\lambda_{2}(S_{2}),$ it follows that 
%$G_{1}\mycirc S_{1}\equiv_{\psi,t}G_{2}\mycirc S_{2}$ (or, in other words,
 $(G_{1}, S_{1})\equiv_{\sigma_{\psi}}(G_{2}, S_{2})$.
\end{enumerate}

%
%
% Given some $I\subseteq \{1,\ldots,t\}$, we 
%say that $G_{1}\approx_{I} G_{2}$ if 
%%\sfdd{Unst this!}
%\begin{eqnarray}
%\forall_{I'\subseteq I\cap (\lambda_{1}(B_{1})\cap \lambda_{2}(B_2))} \ (G_{1},\lambda^{-1}_{1}(I'))\equiv_{\psi,t} (G_{2},\lambda_{2}^{-1}(I')).\label{eq:lambdaissue}
%\end{eqnarray}
%%\begin{eqnarray}
%%\forall_{S_{1}\subseteq B_{1}} \ \forall_{S_{2}\subseteq B_{2}} \ (G_{1},S_{1})\equiv_{\psi,t} (G_{2},S_{2}).\label{eq:lambdaissue}
%\end{eqnarray}
Notice that $\approx$ is an equivalence relation. Because, in the above definition, 
the sets $S_{1}$ and $S_{2}$  cannot have more than $t$ vertices,
the number of equivalence classes of $\approx$
depends only on $t$ and the number of equivalence classes of $\equiv_{\sigma_{\psi}}$ on boundaried structures of arity two whose label set is a subset of $\{1, \ldots, t\}$. By Lemma~\ref{log_lem} the number of such equivalence classes is finite and upper bounded by a function of $|\psi|$ and $t$. Thus the number of equivalence classes of $\approx$ is also upper bounded by a function of $|\psi|$ and $t$. Let ${\cal S}$ be a set of minimum size representatives of the equivalence classes of $\approx$ 
and let $c_{1} = \max_{\alpha \in {\cal S}} |V(G_\alpha)|$. 
\medskip

Let $G$, $Y$ and $X$ be a graph and vertex sets as in the statement of the Lemma. We now define the sets $B = \partial_G(X)$, $R = (V(G) \setminus X) \cup B$ and the boundaried structures $(G_X, Y_X)$ and $(G_R, Y_R)$ as follows. The boundaried graphs $G_X$ and $G_R$ are just $G[X]$ and $G[R]$ respectively. Both have boundary $B$, with labels from $\{1,\ldots,t\}$ such that $G_X \oplus G_R = G$. Similarly $Y_X = Y \cap X$ while $Y_R = Y \setminus X$, such that $(G,Y) = (G_X,Y_X) \oplus (G_R,Y_R)$. Observe that $|V(G_X)| = |X| > c_1$.

%Our algorithm has in its source code hard-wired a table that for every boundaried structure $(G_X, Y_X)$ with label set from $\{1,\ldots,t\}$ and $|V(G_X)| \leq 2c_1$ contains the $(G_X',Y_X') \in {\cal S}$ such that $(G_X',Y_X') \approx (G_X, Y_X)$. The size of this table is a constant that depends only on $\psi$ and $t$. The algorithm looks up in the table and finds the representative $(G_X',Y_X') \in {\cal S}$ such that $(G_X',Y_X') \approx (G_X, Y_X)$.

Our algorithm has in its source code hard-wired a table that for every boundaried structure $\alpha$ of type $({\sf graph},{\sf vertex\ set})$ with label set from $\{1,\ldots,t\}$ and $|V(G_\alpha)| \leq 2c_1$ contains the $\beta \in {\cal S}$ such that $\beta \approx \alpha$. The size of this table is a constant that depends only on $|\psi|$ and $t$. The algorithm looks up in the table and finds the representative $(G_X',Y_X') \in {\cal S}$ such that $(G_X',Y_X') \approx (G_X, Y_X)$. By construction we have  $|V(G_X')| \leq c_1 < |V(G_X)|$. The algorithm outputs the instance $((G',Y'),k)$ where $(G',Y') = (G_X',Y_X') \oplus (G_R,Y_R)$. Since $|V(G_X')| < |V(G_X)|$ it follows that $|V(G')| < |V(G')|$ and it remains to argue that the instances $((G,Y),k)$ and $((G',Y'),k)$ are equivalent.

Suppose that $((G,Y),k)$ is a YES-instance and let $S \subseteq Y$, $|S|\leq k$ ($|S|=k$ for \peq{}) be such that $(G,S) \models \psi$. Let $S_X = X \cap S$ and $S_R = S \setminus X$. Observe that $(G_X,S_X) \oplus (G_R, S_R) = (G, S)$, $S_X = S_X \cap X \subseteq Y \cap X \subseteq \partial(X)$, and that $|S_X| + |S_R| = |S|$. Let $S_X'$ be the subset of $\delta(G_X')$ such that $\lambda_{G_{X}'}(S_X') = \lambda_{G_{X}}(S_X)$. Since $S_X \subseteq Y_X \subseteq \delta(G_X)$ it follows that $|S_X| = |S_X'|$. Furthermore, property $3$ of $\approx$ yields that $(G_X,S_X) \equiv_{\sigma_\psi} (G_X',S_X')$. Let $S' = S_X' \cup S_R$ (formally $S_X'$ and $S_R$ are vertex sets of different graphs, so we set $S' = ((G_X',S_X') \oplus (G_R,S_R))[2]$). Since $S_R \cap \delta(G_R) = \emptyset$ we have that $|S'| = |S_X'| + |S_R| =  |S_X| + |S_R| = |S|$. Thus, if $|S| \leq k$ then $|S'| \leq k$, while if $|S| = k$ then $|S'| = k$. Finally we observe that
\begin{align*}
(G',S') \models \psi \\
\iff (G_X',S_X') \oplus (G_R, S_R) \models \psi \\
\iff (G_X,S_X) \oplus (G_R, S_R) \models \psi\\
\iff (G,S) \models \psi \iff \mbox{\sf true.}
\end{align*}
This concludes the forward direction of the proof. The reverse direction is symmetric.
\end{proof}

Lemmata \ref{lem:annotated}, \ref{lem:split}, and \ref{lem:red1} together yield a reduction rule for all 
annotated \pmin{} problems.

\begin{lemma} 
\label{lem:CMSOMinReduce}
Let $\Pi^{\alpha}$ be an annotated  \pmin{} problem. 
% definable by some CMSO-sentence $\psi.$
Then for every $t,$ there is a constant $c_{2}>0$ (depending only on  $|\psi|$ and  $t$)
and an algorithm that, given an instance $((G,Y),k)$ of $\Pi^{\alpha}$ and a $t$-protrusion $X$ with $|X|>c_{2}k,$ outputs, 
in $O(|X|)$ steps,  an equivalent instance  $((G^*,Y^*),k)$ of $\Pi^{\alpha}$ such that 
   $|V^{*}|<|V|.$
\end{lemma}

\begin{proof} Let $|\partial_{G}(X)|=t.$
The algorithm starts by applying Lemma \ref{lem:annotated} to $X,$ and producing an equivalent instance $((G,Y'),k)$ where $|Y'\cap X|\leq ak,$ for some constant $a$ depending only on $|\psi|$ and $t.$ Let $Z=Y'\cap X.$ The next step is to apply Lemma~\ref{lem:split} and construct  a  collection ${\cal Q}$ of $(4t+2)$-protrusions
 such that $X = \bigcup_{Q\in{\cal Q}}Q,$ $Z \cap Q \subseteq \partial_{G}(Q)$ for each ${Q\in{\cal Q}}$, and  $|{\cal Q}|\leq bk$ for some constant $b$ depending only on $|\psi|$ and $t.$ Let $c_{1}$ be the constant as guaranteed by Lemma~\ref{lem:red1} when applied on $(8t+4)$-protrusions, and set $c_{2}=c_{1}\cdot b$.
 By the pigeon-hole principle, some $(4t+2)$-protrusion $Q$ in ${\cal Q}$  has size at least $|X|/bk > c_{1}.$ We apply Lemma~\ref{lem:stplittreedec} and obtain a  $(8t+4)$-protrusion $Q' \subseteq Q$ such that $Z \cap Q' \subseteq \partial(Q')$ and $c_1 < |Q'| \leq 2c_1$. Finally we apply the algorithm of Lemma \ref{lem:red1} on $Q'$ and construct an equivalent instance of $\Pi^{\alpha}$ as required.
\end{proof}
%
%We can now restate Lemma~\ref{lem:CMSOMinReduce} as follows:
%
%\begin{corollary}
%Every annotated  \pmin{} problem definable by some CMSO-sentence $\psi,$ satisfies condition {\bf A} for $a=1.$
%\end{corollary}
%

We are now ready to prove the following result. 
% first of the results in Section~\ref{sec:derivourresults}.
\begin{lemma}
\label{lem:annotatedforaequal1}
Every annotated  \pmin{}  problem has the  protrusion  replacement  property  {\bf A} for $a=1.$
\end{lemma}
\begin{proof}
%[of Lemma~\ref{lem:annotatedforaequal1}]
According to the terminology that we introduced in Section~\ref{sec:metaalgoframforkern}, we have to prove that 
there exist an $(f,1)$-protrusion replacement family ${\cal A}$ for $\Pi^\alpha.$ Indeed, this directly follows from Lemma~\ref{lem:CMSOMinReduce} if  we define  $f: \Bbb{Z}^{+}\rightarrow \Bbb{Z}^{+}$ such that for every $r,$ $f(r)$ is the constant $c_{2}$ of Lemma~\ref{lem:CMSOMinReduce}.
\end{proof}

\subsection{Protrusion replacement  for annotated \peq{} Problems}
\label{subsec:aprotreplfamiforannopro}

In this section we give a reduction rule for annotated \peq{} problems. The rule is very similar to the one for the  \pmin{} problems described in the previous section. The main difference between the two problem variants is that we now need to keep track of solutions of {\em every} possible size between $0$ and $k,$ instead of just the smallest one. Because of this, we require the protrusion to contain at least $ck^2$ vertices instead of $ck$ vertices, in order to be able to reduce it. We start by proving adaptations of Lemmata~\ref{lem:findMinAnnotated} and~\ref{lem:annotated}   to \peq{} problems.

\begin{lemma}
\label{lem:findEqAnnotated}
There is an algorithm that given two boundaried structures $(G_X, Y_X)$ and $(G_R, S_R)$ of type $({\sf graph},{\sf vertex\ set})$,a CMSO-sentence $\psi$ and non-negative integer $k$, finds a $S_X \subseteq Y_X$ of size $k$ such that $(G_X, S_X) \oplus (G_R, S_R) \models \psi$ or concludes that no such set exists in time $|V(G_X \oplus G_R)|\cdot f(|\psi|, \tw(G_X \oplus G_R))$.
\end{lemma}

\begin{proof}
Let $(G', Y', S_R') = (G_X, Y_X, \emptyset) \oplus (G_R, \emptyset, S_R)$. Finding the desired set $S_X \subseteq Y$ now amounts to finding a set $S_X' \subseteq Y'$ of size $k$ such that $(G', S_X' \cup S_R') \models \psi$. This is easily formulated as {\sc Eq-CMSO on Structures} and hence may be solved in the desired running time by Proposition~\ref{lema:borie}.
\end{proof}

\begin{lemma}
\label{lem:annotatedEQRed}
Let $\Pi^{\alpha}$ be an annotated \peq{} problem and let $t$ be an integer. Then there exist an algorithm that given an instance $((G,Y),k)$ of $\Pi^{\alpha}$ and a $t$-protrusion $X$ of $G,$ outputs  in $O(k|X|)$ steps  an equivalent instance  $((G,Y'),k)$ of $\Pi^{\alpha},$ where $|Y'\cap X|=O(k^2)$ and $Y' \subseteq Y.$
\end{lemma}

\begin{proof} The proof of the lemma starts exactly as in the proof of Lemma~\ref{lem:annotated}. 
For a CMSO-sentence  $\psi$ defining $\Pi^{\alpha}$, Lemma~\ref{log_lem} implies that the canonical equivalence relation $\equiv_{\sigma_\psi}$ has finitely many equivalence classes on the set of boundaried structures of arity two with label set $\{1,\ldots,t\}$. We denote by  ${\bf MinRep}(\psi,t)$   a set containing a representative (a boundaried structure of arity two)  for each equivalence class of $\equiv_{\sigma_\psi}$ with  the minimum number of  vertices in the graph of a structure. For given $G$, $Y$ and $X$, we define the sets $B = \partial_G(X)$, $R = (V(G) \setminus X) \cup B$ and the boundaried structures $(G_X, Y_X)$ and $(G_R, Y_R)$ as follows. The boundaried graphs $G_X$ and $G_R$ are just $G[X]$ and $G[R]$ respectively. Both have boundary $B$, with labels from $\{1,\ldots,t\}$ such that $G_X \oplus G_R = G$. Similarly $Y_X = Y \cap X$ while $Y_R = Y \setminus X$, such that $(G,Y) = (G_X,Y_X) \oplus (G_R,Y_R)$.

For every structure $\alpha = (G_R^\alpha, S_R^\alpha) \in {\bf MinRep}(\psi,t)$ and every integer $i \leq k$ we use Lemma~\ref{lem:findEqAnnotated} to find a set $S_X^{\alpha,i} \subseteq Y_X$ such that $|S_X^{\alpha,i}|=i$ and $(G_X, S_X^\alpha) \oplus \alpha \models \psi$. If no such set exists we set $S_X^{\alpha,i} = \emptyset$.  Since $|{\bf MinRep}(\psi,t)|$ and the size of each structure in ${\bf MinRep}(\psi,t)$ depends only on $\psi$ and $t$, and the treewidth of $G[X]$ is at most $t$, this takes time $O(k|X|)$. Now, define 
\begin{equation*}
Y_X' = \bigcup_{\substack{\alpha \in {\bf MinRep}(\psi,t) \\ i \leq k}} S_X^{\alpha,i}
\end{equation*}
We set $Y' = Y_X' \cup Y_R$ (formally $Y_X'$ and $Y_R$ are vertex sets of different graphs, so actually $Y' = ((G_X,Y_X') \oplus (G_R,Y_R))[2]$). Since  $|{\bf MinRep}(\psi,t)|$ depends only on $|\psi|$ and $t$ the construction of $Y'$ implies $|Y' \cap X| = O(k^2)$.

To complete the proof, it remains to show that $((G,Y'),k) \in \Pi^{\alpha}$ if and only if $((G,Y),k) \in \Pi^{\alpha}$. For the forward direction we have that $Y' \subseteq Y$ and hence feasible solutions to $((G,Y'),k)$ are also feasible for $((G,Y),k)$. We now turn to proving the reverse direction. Let $S \subseteq Y$, $|S| = k$ be such that $(G,S) \models \psi$. Let $S_X = X \cap S$ and $S_R = S \setminus X$. Observe that $(G_X,S_X) \oplus (G_R, S_R) = (G, S)$ and that $|S_X| + |S_R| = |S| = k$. Choose $\alpha = (G_R^\alpha, S_R^\alpha) \in {\bf MinRep}(\psi,t)$ such that $\alpha \equiv_{\sigma_\psi} (G_R, S_R)$. Set $i = |S_X|$, and let $S_X^{\alpha,i} \subseteq Y_X$ be the set computed for $\alpha$ and $i$ in the previous paragraph. The existence of $S_X^{\alpha,i}$ of size $i$ is guaranteed by the fact that  
$$(G_X, S_X) \oplus \alpha \models \psi \iff (G_X, S_X) \oplus (G_R, S_R) \models \psi \iff \mbox{\sf true}.$$
By construction $S_X^{\alpha,i} \subseteq Y_X'$. Let $S' = S_X^{\alpha,i} \cup S_R$ (again, formally $S_X^{\alpha,i}$ and $S_R$ are vertex sets of different graphs, so actually $S' = ((G_X,S_X^{\alpha,i}) \oplus (G_R,S_R))[2]$). We have that $S' \subseteq Y'$. Further, since $S_R \cap \delta(G_R) = \emptyset$ we have that $|S'| = |S_X^{\alpha,i}| + |S_R| = |S_X| + |S_R| = |S| = k$. Finally we observe that
\begin{align*}
(G,S') \models \psi \\
\iff (G_X,S_X^{\alpha,i}) \oplus (G_R, S_R) \models \psi \\
\iff (G_X,S_X^{\alpha,i}) \oplus \alpha \models \psi\\
\iff \mbox{\sf true.}
\end{align*}
This concludes the proof.
\end{proof}

\begin{lemma}
\label{lem:annotatedeq}
Let $\Pi^{\alpha}$ be an annotated  \peq{} problem.  
%definable by some CMSO-sentence $\psi.$ 
Then for every $t,$ there is a constant $c_{2}\in\Bbb{Z}^{+}$ (depending only on  $|\psi|,$ and  $t$) and an algorithm that, given an instance $((G,Y),k)$ of $\Pi^{\alpha}$ and a $t$-protrusion $X$ with $|X|>c_{2}k^2,$ outputs  in $O(k\cdot |X|)$ steps an equivalent instance  $((G^*,Y^*),k)$ of $\Pi^{\alpha}$ such that $|V^{*}|<|V|.$
\end{lemma}

\begin{proof}
The algorithm starts by applying Lemma \ref{lem:annotatedEQRed} to $X,$ and producing an equivalent instance $((G,Y'),k)$ where $|Y'\cap X|\leq ak^2,$ for some constant $a$ depending only on $|\psi|$ and $t.$ Let $Z=Y'\cap X.$ The next step is to apply Lemma~\ref{lem:split} and construct  a  collection ${\cal Q}$ of $(4t+2)$-protrusions such that $X = \bigcup_{Q\in{\cal Q}}Q,$ $Z \cap Q \subseteq \partial_{G}(Q)$ for each ${Q\in{\cal Q}}$, and  $|{\cal Q}|\leq bk^2$ for some constant $b$ depending only on $|\psi|$ and $t.$ Let $c_{1}$ be the constant as guaranteed by Lemma~\ref{lem:red1} when applied on $(8t+4)$-protrusions, and set $c_{2}=c_{1}\cdot b$.
 By the pigeon-hole principle, some $(4t+2)$-protrusion $Q$ in ${\cal Q}$  has size at least $|X|/bk^2 > c_{1}.$ We apply Lemma~\ref{lem:stplittreedec} and obtain a  $(8t+4)$-protrusion $Q' \subseteq Q$ such that $Z \cap Q' \subseteq \partial(Q')$ and $c_1 < |Q'| \leq 2c_1$. Finally we apply the algorithm of Lemma \ref{lem:red1} on $Q'$ and construct an equivalent instance of $\Pi^{\alpha}$ as required.
\end{proof}

We are now ready to prove the following result.
% in Section~\ref{sec:derivourresults}.
\begin{lemma}
\label{lem:annotatedforaequal2}
Every annotated  \peq{}  problem  has the  protrusion  replacement  property  {\bf A} for  $a=2.$
\end{lemma}

\begin{proof}
%[of Lemma~\ref{lem:annotatedforaequal2}]
According to the terminology that we introduced in Section~\ref{sec:metaalgoframforkern}, we have to prove that there exists an $(f,2)$-protrusion replacement family ${\cal A}$ for $\Pi^\alpha.$ Indeed, this directly follows from Lemma~\ref{lem:annotatedeq} if  we define $f: \Bbb{Z}^{+}\rightarrow \Bbb{Z}^{+}$ such that for every $r,$ $f(r)$ is the constant $c_{2}$ in the proof of the same lemma.
\end{proof}

\subsection{Protrusion replacement  for annotated \pmax{} Problems}
\label{subsec:aprotreplfamiforannopromax}

We now give a reduction rule for annotated \pmax{} problems. The rule is still similar to the ones described in the two previous sections, but differs more from the \pmin{} problems than \peq{} did. We start by proving a variant of lemma~\ref{lem:findMinAnnotated} for \pmax{} problems.

\begin{lemma}
\label{lem:findMaxAnnotated}
There is an algorithm that given two boundaried structures $(G_X, Y_X)$ and $(G_R, S_R)$ of type $({\sf graph},{\sf vertex\ set})$ and a CMSO-sentence $\psi$ finds a set $S_X \subseteq V(G_X)$ such that $(G_X, S_X) \oplus (G_R, S_R) \models \psi$ and $|S_X \cap Y_X|$ is maximized. The running time of the algorithm is $|V(G_X \oplus G_R)|\cdot f(|\psi|, \tw(G_X \oplus G_R))$.
\end{lemma}

\begin{proof}
Let $(G', Y', S_R',V') = (G_X, Y_X, \emptyset,V(G_X)) \oplus (G_R, \emptyset, S_R, \emptyset)$. Finding the desired set $S_X$ now amounts to finding a set $S_X' \subseteq V'$ such that $(G', S_X' \cup S_R') \models \psi$ and $|S_X' \cap Y'|$ is maximized. This is easily formulated as {\sc Max-CMSO on Structures} and hence may be solved in the desired running time by Proposition~\ref{lema:borie}.
\end{proof}

\begin{lemma}
\label{lem:annotatedMax}
Let $\Pi^{\alpha}$ be an annotated \pmax{} problem and let $t$ be an integer. There exists an algorithm that given an instance $((G,Y),k)$ of $\Pi^{\alpha}$ and a $t$-protrusion $X$ of $G,$ outputs  in $O(|X|)$ steps an equivalent instance $((G,Y'),k)$ of $\Pi^{\alpha},$ where $|Y'\cap X|=O(k)$ and $Y' \subseteq Y.$
\end{lemma}

\begin{proof}
By Lemma~\ref{log_lem},  for  a CMSO-sentence $\psi$ defining $\Pi^{\alpha}$,
 the canonical equivalence relation $\equiv_{\sigma_\psi}$ has finitely many equivalence classes on the set of boundaried structures of arity two with label set $\{1,\ldots,t\}$.  As in proofs of 
  Lemmata~\ref{lem:annotated} and \ref{lem:annotatedEQRed}, we define the following objects. 
  We set  ${\bf MinRep}(\psi,t)$  to be  a set containing a representative (a boundaried structure of arity two)  for each equivalence class of $\equiv_{\sigma_\psi}$ with  the minimum number of  vertices in the graph of a structure.
  %We set  ${\bf MinRep}(\psi,t)$ to be a set containing a representative with the minimum number of  vertices for each equivalence class of $\equiv_{\sigma_\psi}$.  
Also for  $G$, $Y$ and $X$, we define  sets $B = \partial_G(X)$, $R = (V(G) \setminus X) \cup B$, and the boundaried structures $(G_X, Y_X)$ and $(G_R, Y_R)$ as follows. 
Again, the boundaried graphs $G_X=G[X]$ and $G_R=G[R]$   have boundary $B$ with labels from $\{1,\ldots,t\}$ such that $G_X \oplus G_R = G$. Similarly $Y_X = Y \cap X$ while $Y_R = Y \setminus X$, such that $(G,Y) = (G_X,Y_X) \oplus (G_R,Y_R)$.

By making use of  Lemma~\ref{lem:findMaxAnnotated}, for every structure $\alpha = (G_R^\alpha, S_R^\alpha) \in {\bf MinRep}(\psi,t)$,  we find  a set $S_X^\alpha \subseteq V(G_X)$ such that $(G_X, S_X^\alpha) \oplus \alpha \models \psi$ and $|S_X \cap Y_X|$ is maximized. Since $|{\bf MinRep}(\psi,t)|$ and the size of each structure in ${\bf MinRep}(\psi,t)$ depends only on $|\psi|$ and $t$, and the treewidth of $G[X]$ is at most $t$, this takes time $O(|X|)$. If $|S_X^\alpha \cap Y_X| \leq k$, let $\hat{S}_X^\alpha = S_X^\alpha \cap Y_X$. On the other hand, if $|S_X^\alpha \cap Y_X| > k$, set $\hat{S}_X^\alpha$ to be a set of arbitrarily chosen $k$ vertices from $S_X^\alpha \cap Y_X$. Now, define 
\begin{equation*}
Y_X' = \bigcup_{\alpha \in {\bf MinRep}(\psi,t)} \hat{S}_X^\alpha.
\end{equation*}
We set $Y' = Y_X' \cup Y_R$ (formally $Y_X'$ and $Y_R$ are vertex sets of different graphs, so actually $Y' = ((G_X,Y_X') \oplus (G_R,Y_R))[2]$). Since  $|{\bf MinRep}(\psi,t)|$ depends only on $|\psi|$ and $t$ the construction of $Y'$ implies $|Y' \cap X| = O(k)$.

To complete the proof, it remains to show that $((G,Y'),k) \in \Pi^{\alpha}$ if and only if $((G,Y),k) \in \Pi^{\alpha}$. For the forward direction we have that $Y' \subseteq Y$, and hence for any set $S \subseteq V(G)$ such that $(G,S) \models \psi$ and $|S \cap Y'| \geq k$ we also have that $|S \cap Y| \geq k$. We now turn to proving the reverse direction. Let $S \subseteq V(G)$, $|S \cap Y|\geq k$ be such that $(G,S) \models \psi$. Let $S_X = X \cap S$ and $S_R = S \setminus X$. Observe that $(G_X,S_X) \oplus (G_R, S_R) = (G, S)$ and that $|S_X \cap Y_X| + |S_R \cap Y_R| = |S \cap Y| \geq k$. Choose $\alpha = (G_R^\alpha, S_R^\alpha) \in {\bf MinRep}(\psi,t)$ such that $\alpha \equiv_{\sigma_\psi} (G_R, S_R)$. Let $S_X^\alpha \subseteq V(G_X)$ be the set computed for $\alpha$ in the previous paragraph. Since 
$$(G_X, S_X) \oplus \alpha \models \psi \iff (G_X, S_X) \oplus (G_R, S_R) \models \psi \iff \mbox{\sf true}$$
it follows that $|S_X^\alpha \cap Y_X| \geq |S_X \cap Y_X|$. Furthermore we have that $|S_X^\alpha \cap Y_X'| \geq |\hat{S}_X^\alpha| \geq \min(|S_X \cap Y_X|,k)$.

Let $S' = S_X^\alpha \cup S_R$ (again, formally $S_X^\alpha$ and $S_R$ are vertex sets of different graphs, so actually $S' = ((G_X,S_X^\alpha) \oplus (G_R,S_R))[2]$). We have that
$$|S' \cap Y'| \geq |S_X^\alpha \cap Y_X'| + |S_R \cap Y_R| \geq \min(|S_X \cap Y_X|,k) + |S_R \cap Y_R| \geq \min(|S \cap Y|,k) \geq k.$$ 
Finally we observe that
\begin{align*}
(G,S') \models \psi \\
\iff (G_X,S_X^\alpha) \oplus (G_R, S_R) \models \psi \\
\iff (G_X,S_X^\alpha) \oplus \alpha \models \psi\\
\iff \mbox{\sf true.}
\end{align*}
This concludes the proof.
\end{proof}

\begin{lemma}
\label{lem:redMax}
Let $\Pi^{\alpha}$ be an annotated \pmax{} problem. 
%definable by some CMSO-sentence $\psi.$ 
Then for every integer $t$ there is a $c_{1}\in\Bbb{Z}^{+}$ (depending only on $|\psi|$ and $t$) and  an algorithm that given an instance $((G,Y),k)$ of $\Pi^{\alpha}$ and a $t$-protrusion $X$ of $G,$ where $c_1 <|X| \leq 2c_{1}$ and $X\cap Y\subseteq \partial_{G}(X),$ outputs, in $O(|X|)$ steps, an equivalent instance $((G^*,Y^*),k)$ of $\Pi^{\alpha}$ such that $|V(G')| < |V(G)|.$
\end{lemma}

\begin{proof}  
Let $\psi$ be the CMSO-sentence mentioned in the definition of $\Pi^{\alpha}.$ 
%Let $\psi$ be a CMSO-sentence defining $\Pi^{\alpha}.$  
By Lemma~\ref{log_lem},   the canonical equivalence relation $\equiv_{\sigma_\psi}$  has finitely many equivalence classes on the set of boundaried structures of arity two with label set $\{1,\ldots,t\}$. 
Let  ${\bf MinRep}(\psi,t)$  be a set containing a representative (a boundaried structure of arity two)  for each equivalence class of $\equiv_{\sigma_\psi}$ with  the minimum number of  vertices in the graph of a structure. 
%Let ${\bf MinRep}(\psi,t)$ be a set containing a representative with the minimum number of  vertices for each equivalence class of 
%$\equiv_{\sigma_\psi}$. 
We now define an equivalence relation $\approx$ between boundaried structures $\alpha = (G_\alpha, Y_\alpha)$ of type $({\sf graph},{\sf vertex\ set})$ that satisfy $Y_\alpha \subseteq \delta(G_\alpha)$. Let $\alpha_{1}=(G_{1},Y_{1})$ and $\alpha_{2}=(G_{2},Y_{2})$ be two boundaried structures with labelling functions $\lambda_{1}: \delta(G_1) \rightarrow \{1,\ldots,t\}$ and $\lambda_{2}: \delta(G_2) \rightarrow \{1,\ldots,t\}$ respectively, such that $Y_{1}\subseteq \delta(G_1)$ and $Y_{2}\subseteq \delta(G_2)$. We say that $\alpha_{1}\approx\alpha_{2}$ if the following conditions are satisfied:
\begin{enumerate}\setlength\itemsep{-.7mm}
\item $\Lambda(G_{1})=\Lambda(G_{2})$
\item $\lambda_{1}(Y_{1})=\lambda_{2}(Y_{2})$
\item for every $S_{1}\subseteq V(G_{1})$ there is a $S_{2}\subseteq V(G_{2})$ such that $\lambda_{1}(S_{1} \cap \delta(G_1))=\lambda_{2}(S_{2} \cap \delta(G_2)),$ and $(G_{1}, S_{1})\equiv_{\sigma_{\psi}}(G_{2}, S_{2})$.
\item for every $S_{2}\subseteq V(G_{2})$ there is a $S_{1}\subseteq V(G_{1})$ such that $\lambda_{1}(S_{1} \cap \delta(G_1))=\lambda_{2}(S_{2} \cap \delta(G_2)),$ and $(G_{1}, S_{1})\equiv_{\sigma_{\psi}}(G_{2}, S_{2})$.
\end{enumerate}

Notice that $\approx$ is an equivalence relation. Further, consider two boundaried structures $\alpha_{1}=(G_{1},Y_{1})$ and $\alpha_{2}=(G_{2},Y_{2})$ such that  $\Lambda(G_{1})=\Lambda(G_{2})$, $\lambda_{1}(Y_{1})=\lambda_{2}(Y_{2})$, and for each subset $L \subseteq \{1,\ldots,t\}$ the sets 
$$\{\beta \in {\bf MinRep}(\psi,t)~:~\exists S_1 \subseteq V(G_1), \lambda_1(S_1 \cap \delta(G_1)) = L \wedge (G_1,S_1) \equiv_{\sigma_\psi} \beta\}$$
and 
$$\{\beta \in {\bf MinRep}(\psi,t)~:~\exists S_2 \subseteq V(G_2), \lambda_2(S_2 \cap \delta(G_2)) = L \wedge (G_2,S_2) \equiv_{\sigma_\psi} \beta\}$$
are the same. It is easy to verify that in this case $(G_{1},Y_{1}) \approx (G_{2},Y_{2})$. Thus the number of equivalence classes of $\approx$ is upper bounded by a function of $|\psi|$ and $t$. Let ${\cal S}$ be a set of minimum size representatives of the equivalence classes of $\approx$ and let $c_{1} = \max_{\alpha \in {\cal S}} |V(G_\alpha)|$. 
\medskip

Let $G$, $Y$ and $X$ be a graph and vertex sets as in the statement of the Lemma. We now define the sets $B = \partial_G(X)$, $R = (V(G) \setminus X) \cup B$ and the boundaried structures $(G_X, Y_X)$ and $(G_R, Y_R)$ as follows. 
The boundaried graphs $G_X=G[X]$ and $G_R=G[R]$   have boundary $B$  with labels from $\{1,\ldots,t\}$ such that $G_X \oplus G_R = G$. 
We define  $Y_X = Y \cap X$ and $Y_R = Y \setminus X$, such that $(G,Y) = (G_X,Y_X) \oplus (G_R,Y_R)$. Observe that $|V(G_X)| = |X| > c_1$.

%The boundaried graphs $G_X$ and $G_R$ are just $G[X]$ and $G[R]$ respectively. Both have boundary $B$, with labels from $\{1,\ldots,t\}$ such that $G_X \oplus G_R = G$. 
%Similarly $Y_X = Y \cap X$ while $Y_R = Y \setminus X$, such that $(G,Y) = (G_X,Y_X) \oplus (G_R,Y_R)$. Observe that $|V(G_X)| = |X| > c_1$.

%Our algorithm has in its source code hard-wired a table that for every boundaried structure $(G_X, Y_X)$ with label set from $\{1,\ldots,t\}$ and $|V(G_X)| \leq 2c_1$ contains the $(G_X',Y_X') \in {\cal S}$ such that $(G_X',Y_X') \approx (G_X, Y_X)$. The size of this table is a constant that depends only on $\psi$ and $t$. The algorithm looks up in the table and finds the representative $(G_X',Y_X') \in {\cal S}$ such that $(G_X',Y_X') \approx (G_X, Y_X)$.

Our algorithm has in its source code hard-wired a table that for every boundaried structure $\alpha$ of type $({\sf graph},{\sf vertex\ set})$ with label set from $\{1,\ldots,t\}$ and $|V(G_\alpha)| \leq 2c_1$ contains the $\beta \in {\cal S}$ such that $\beta \approx \alpha$. The size of this table is a constant that depends only on $|\psi|$ and $t$. The algorithm looks up in the table and finds the representative $(G_X',Y_X') \in {\cal S}$ such that $(G_X',Y_X') \approx (G_X, Y_X)$. By construction we have  $|V(G_X')| \leq c_1 < |V(G_X)|$. The algorithm outputs the instance $((G',Y'),k)$ where $(G',Y') = (G_X',Y_X') \oplus (G_R,Y_R)$. Since $|V(G_X')| < |V(G_X)|$ it follows that $|V(G')| < |V(G')|$ and it remains to argue that the instances $((G,Y),k)$ and $((G',Y'),k)$ are equivalent.

Suppose $((G,Y),k)$ is a YES-instance and let $S \subseteq V(G)$, $|S \cap Y|\geq k$ be such that $(G,S) \models \psi$. Let $S_X = X \cap S$ and $S_R = S \setminus X$. Observe that $(G_X,S_X) \oplus (G_R, S_R) = (G, S)$, $S_X \cap Y_X \subseteq \delta(G_X)$, and that $|S_X \cap Y_X| + |S_R \cap Y_R| = |S \cap Y|$. Let $S_X'$ be a subset of $V(G_X')$ such that $\lambda_{G_{X}'}(S_X' \cap \delta(G_X')) = \lambda_{G_{X}}(S_X \cap \delta(G_X))$ and $(G_{X}', S_{X}')\equiv_{\sigma_{\psi}}(G_{X}, S_{X})$. The existence of such a set $S_X'$ is implied by property $(3)$ of $\approx$. Since $Y_X \subseteq \delta(G_X)$, $Y_X' \subseteq \delta(G_X')$, $\Lambda_{G_{X}}(Y_{X})=\Lambda_{G_{X}'}(Y_{X}')$ and $\Lambda_{G_{X}}(S_{X} \cap \delta(G_X))=\Lambda_{G_{X}'}(S_{X}' \cap \delta(G_X'))$ we have that $|S_X \cap Y_X| = |S_X' \cap Y_X'|$. 
 
Let $S' = S_X' \cup S_R$ (formally $S_X'$ and $S_R$ are vertex sets of different graphs, so we set $S' = ((G_X',S_X') \oplus (G_R,S_R))[2]$). Since $S_R \cap \delta(G_R) = \emptyset$ we have that $|S' \cap Y'| = |S_X' \cap Y_X'| + |S_R \cap Y_R| = |S_X \cap Y_X| + |S_R \cap Y_R| = |S \cap Y|$. Thus, if $|S \cap Y| \geq k$ then $|S' \cap Y'| \geq k$. Finally we observe that
\begin{align*}
(G',S') \models \psi \\
\iff (G_X',S_X') \oplus (G_R, S_R) \models \psi \\
\iff (G_X,S_X) \oplus (G_R, S_R) \models \psi\\
\iff (G,S) \models \psi \iff \mbox{\sf true.}
\end{align*}
This concludes the forward direction of the proof. The reverse direction is symmetric, but using property $4$ of $\approx$ rather than property $3$.
\end{proof}

\begin{lemma}
\label{lem:annotatedmax}
Let $\Pi^{\alpha}$ be an annotated \pmax{} problem.  
%definable by some CMSO-sentence $\psi.$ 
Then for every $t,$ there is a constant $c_{2}>0$ (depending only on  $\psi,$ and  $t$) and an algorithm that, given an instance $((G,Y),k)$ of $\Pi^{\alpha}$ and a $t$-protrusion $X$ with $|X|>c_{2}k,$ outputs, in $O(|X|)$ steps,  an equivalent instance  $((G,Y^*),k)$ of $\Pi^{\alpha}$ such that $|V^{*}|<|V|.$
\end{lemma}

\begin{proof} 
Let $|\partial_{G}(X)|=t.$
The algorithm starts by applying Lemma \ref{lem:annotatedMax} to $X,$ and producing an equivalent instance $((G,Y'),k)$ where $|Y'\cap X|\leq ak,$ for some constant $a$ depending only on $|\psi|$ and $t.$ Let $Z=Y'\cap X.$ The next step is to apply Lemma~\ref{lem:split} and construct  a  collection ${\cal Q}$ of $(4t+2)$-protrusions
 such that $X = \bigcup_{Q\in{\cal Q}}Q,$ $Z \cap Q \subseteq \partial_{G}(Q)$ for each ${Q\in{\cal Q}}$, and  $|{\cal Q}|\leq bk$ for some constant $b$ depending only on $|\psi|$ and $t.$ Let $c_{1}$ be the constant as guaranteed by Lemma~\ref{lem:redMax} when applied on $(8t+4)$-protrusions, and set $c_{2}=c_{1}\cdot b$.
 By the pigeon-hole principle, some $(4t+2)$-protrusion $Q$ in ${\cal Q}$  has size at least $|X|/bk > c_{1}.$ We apply Lemma~\ref{lem:stplittreedec} and obtain a  $(8t+4)$-protrusion $Q' \subseteq Q$ such that $Z \cap Q' \subseteq \partial(Q')$ and $c_1 < |Q'| \leq 2c_1$. Finally we apply the algorithm of Lemma \ref{lem:redMax} on $Q'$ and construct an equivalent instance of $\Pi^{\alpha}$ as required.
\end{proof}

%The third result of Section~\ref{sec:derivourresults} can be proved as follows.

Now we show the following result. 
\begin{lemma}
\label{lem:annotatedforaequal1butmax}
Every annotated  \pmax{}  has the  protrusion  replacement  property  {\bf A} for $a=1.$
\end{lemma}
\begin{proof}[of Lemma~\ref{lem:annotatedforaequal1butmax}]
According to the terminology that we introduced in Section~\ref{sec:metaalgoframforkern}, we have to prove that 
there exists an $(f,1)$-protrusion replacement family ${\cal A}$ for $\Pi.$
Indeed, this directly follows from Lemma~\ref{lem:annotatedmax} if  we define  $f: \Bbb{Z}^{+}\rightarrow \Bbb{Z}^{+}$ such that for every $r,$ $f(r)$ is the constant $c_{2}$ in the statement of the same lemma.
\end{proof}

\subsection{A protrusion replacement family based for problems that have FII}
\label{subsec:aprotreplfamifii}

In the previous sections we gave reduction rules for annotated \pmem{} problems. These reduction rules, together with the results proved later in this article will give quadratic or cubic kernels for the problems in question. However, for many problem a linear kernel is possible. In this section we provide reduction rules for graph problems that have FII. These reduction rules will yield linear kernels. The main reduction lemma is the following.

\begin{lemma}
\label{lem:red2finiteindex}
Let $\Pi$ be a problem that has  FII. 
Then for every $t\in\Bbb{Z}^{+},$ there exists a $c\in\Bbb{Z}^{+}$ (depending on $\Pi$ and $t$), 
and an algorithm that, given an instance $(G,k)$ of $\Pi$
and a $t$-protrusion $X$ in $G$ with $|X|>c,$ outputs, in   $O(|X|)$ steps,
an equivalent instance   $(G^*,k^*)$ of $\Pi$ where $|V(G^*)| < |V(G)|$ and $k^* \leq k.$
\end{lemma}

\begin{proof}
Recall that we denote by  ${\cal S}_{\subseteq [2t+1]}$ a set of (progressive) representatives for $\equiv_\Pi$ restricted to boundaried graphs with label sets from $\{1,\ldots, 2t+1\}$. Let 
$$c=\max_{}\Big{\{}|V(Y)|~\big{|}~ Y\in {\cal S}_{\subseteq [2t+1]}\Big{\}}.$$ 
%We also denote by $\hat{\cal G}^{\leq 2c}$ the set of all $(2t+1)$-boundaried graphs with at most $2c$ vertices. Let $\zeta: \hat{\cal G}^{\leq 2c}\rightarrow \hat{\cal S}$ be the function such that if $R$ is an $i$-boundaried graph in $\hat{\cal G}^{\leq 2c},$ then $\zeta(R)$ is the unique $i$-boundaried  graph in $\hat{\cal S}$ such that  $H \equiv_{\Pi,i} \zeta(R),$ i.e., $\zeta$ maps each such $R$ to a progressive  representative of it.  We also set up the function $\eta:\hat{\cal G}^{\leq 2c}\rightarrow \Bbb{Z}^{-}$ such that if $R\in\hat{\cal G}^{\leq 2c}$ then 
%\begin{eqnarray}
%\forall (F,k)\in {\cal F}_{t}
%\times \Bbb{Z} & &   (R \oplus F, k) \in \Pi  \Leftrightarrow    (\zeta(R) \oplus F, k+\eta(R)) \in \Pi. \label{eq:progresivereplacement}
%\end{eqnarray}%\sed{Shall we say $\Bbb{Z}^{+}$ instead?}
%Such a function $\eta$ can be defined because the values of $\zeta$ are progressive representatives
%(see Lemma~\ref{def:progrepr}).

Our algorithm has in its source code hard-wired a table that stores for each boundaried graph $G_Y$ in ${\cal F}_{\subseteq [2t+1]}$ on at most $2c$ vertices a boundaried graph $G_Y' \in {\cal S}_{\subseteq [2t+1]}$ and a constant $\mu \leq 0$ such that $G_Y \equiv_\Pi G_Y'$, and specifically
\begin{align}\label{eq:progresivereplacement}
\forall (F,k) \in {\cal F} \times \mathbb{Z} ~:~ (G_Y \oplus F, k) \in \Pi \iff (G_Y' \oplus F, k + \mu) \in \Pi.
\end{align}
The existence of such a constant $\mu \leq 0$ is guaranteed by the fact that ${\cal S}_{\subseteq [2t+1]}$ is a set of progressive representatives.

We now apply Lemma~\ref{lem:stplittreedec} and find a $(2t+1)$-protrusion $Y$ of $G$ where $c<|Y|\leq 2c.$ Split $G$ into two boundaried graphs $G_Y = G[Y]$ and $G_R = G[(V(G) \setminus Y) \cup \partial(Y)]$ as follows. Both $G_R$ and $G_Y$ have boundary $\partial(Y)$, and since $|\partial(Y)| \leq 2t+1$ we may label the boundaries of $G_Y$ and $G_R$ with labels from $[2t+1]$ such that $G = G_Y \oplus G_R$. As $c < |V(G_Y)| \leq 2c$ the algorithm can look up in its table and find a $G_Y' \in {\cal S}_{\subseteq [2t+1]}$ and a constant $\mu$ such that $G_Y \equiv G_Y'$ and $G_Y$, $G_Y'$ and $\mu$ satisfy Equation~\ref{eq:progresivereplacement}. The algorithm outputs 
$$(G^*, k^*) = (G_Y' \oplus G_R,k+\mu).$$ 
Since $|V(G_Y')| \leq  c < |V(G_Y)|$ and $k^* \leq k + \mu \leq k$ it remains to argue that the instances $(G, k)$ and $(G^*, k^*) $ are equivalent. However, this is directly implied by Equation~\ref{eq:progresivereplacement}.
%
%
%As $G[Y]\in \hat{\cal G}^{\leq 2c},$
%as in Lemma~\ref{lem:stplittreedec}.
%we can treat $R=G[Y]$ as a $(2t+1)$-boundaried graph with $B=\partial_{G}(Y)$ as boundary and by considering some labelling $\lambda$ of the
%vertices of $B.$
%
%We set  $H = \zeta(R)$ and $k^* = k+\eta(R).$ 
%We construct $G^*$ by replacing   $R$ by $H$ in $G.$ 
%Since $|Y| > c$ and $H$ has at most $c$ vertices, we obtain that $|V(G^*)| < |V(G)|.$  
%As all values of $\eta$ are non-positive, we have that $k^{*}\leq k.$
%
%We now treat $F=G[V(G)\setminus(Y\setminus B)]$
%as  a $(2t+1)$-boundaried graph with $B$ as boundary, labeled by $\lambda.$ Observe that $G=R\oplus F$
%and $G^{*}=H\oplus F.$ As $k^{*}=k+\eta(H),$~\eqref{eq:progresivereplacement} implies  that 
%$(G,k) \in \Pi$ if and only if $(G^*,k^*) \in \Pi.$ 
%
\end{proof}

We are now in position to prove Lemma.
%~\ref{lem:fiiwithaequal0} of Section~\ref{sec:derivourresults}.

\begin{lemma}
\label{lem:fiiwithaequal0}
Every parameterized graph problem $\Pi$ that  has {FII} has the  protrusion  replacement  property  {\bf A}  for $a=0.$
\end{lemma}
\begin{proof}
%[of Lemma~\ref{lem:fiiwithaequal0}]
According to the terminology that we introduced in Section~\ref{sec:metaalgoframforkern}, we have to prove that 
there exists an $(f,0)$-protrusion replacement family ${\cal A}$ for $\Pi.$
Indeed, this directly follows from Lemma~\ref{lem:red2finiteindex} if  we define  $f: \Bbb{Z}^{+}\rightarrow \Bbb{Z}^{+}$ such that for each $r,$ $f(r)$ is the constant $c$ in the statement of the same lemma.
\end{proof}

\section{Combinatorial results}
\label{sec:comresultsallofthem}

We start this section with some necessary definitions from   graph theory. 

\subsection{{Definitions from  graph theory}}\label{subsec:defsGT}
Let  $e=\{u,v\}$  be an edge of a graph $G=(V,E)$. We obtain the  graph  $G/e$ by \emph{contracting}  $e$. This means that
the edge $e$ is removed and its endpoints $u$, $v$,  are merged into a new vertex $v_e$, such that each edge  incident to either $u$ or $v$ is incident to $v_e$. Note that loops and multiple edges can appear as a result of edge contractions.  More formally, 
let $f$ be a function mapping  $u,v$ to $v_e$ and all remaining vertices  in $V\setminus\{u,v\}$ to itself.  The contraction of $e$ results in a new graph $G/e=(V',E')$, where 
$V'=(V\setminus \{u,v\})\cup \{v_e\}$, $E'=E\setminus \{e\}$, and for every $w\in V$, $w'=f(w)\in V'$ is incident with  an edge $e'\in E'$ if and only if, the corresponding edge, $e\in E$ is incident with  $w$ in $G$.
  When we have to remain in the class of simple graphs, loops and multiple edges resulting by contractions are deleted.

{{A graph $H$ is a {\em minor} of a graph $G$, we write $H\leq_{\rm mn} G$, if $H$   can be obtained by contracting some edges of a  subgraph of $G$.
A graph class $\mathcal C$ is {\em  minor-closed} if every minor of every graph
in $\mathcal C$ also belongs to  $\mathcal C$. A minor-closed
graph class $\mathcal C$ is $H${\em -minor-free}    
  if $H \notin \mathcal C.$}}

Given a graph $G=(V,E),$ we define the (normal) {\em distance} 
between two of its vertex sets $X$ and $Y$ as the shortest path distance between them, i.e. the minimum length of a path with endpoints in $X$ and $Y,$  and denote it by  $\dist_{G}(X,Y).$  
Given a set $S\subseteq V$ of vertices, we denote by ${\bf B}_{G}^{r}(S)$ the set of all vertices that are within distance at most $r$ from some vertex of $S$
in $G$. 

We also need some  notions  from topological 
graph theory. All concepts that we do not define here can be found in the book 
\cite{MoharT2001}.
The \emph{Euler genus} $\eg(\Sigma)$ of a nonorientable surface
$\Sigma$ is equal to  the nonorientable genus
$\tilde{g}(\Sigma)$ (or the crosscap number).
The Euler genus $\eg(\Sigma)$ of an orientable   surface
$\Sigma$ is $2{g}(\Sigma),$ where ${g}(\Sigma)$ is  the orientable genus
of $\Sigma.$  
We say that a graph $G$ is \emph{$\Sigma$-embedded} if it is accompanied with an embedding of the graph into $\Sigma.$ We also sometimes refer  to  an embedding as to a drawing of $G$ in $\Sigma$. 
We treat edges and loops (in some proofs we will also allow loops and multiple edges) 
  as subsets of the surface  $\Sigma$  that are homeomorphic to the open interval $(0,1)$.
We define the endpoints  of an edge $e$ as the set of points of $\Sigma$ that are in the closure 
of $e$ but not in $e.$ 
We call by {\em face} of a $\Sigma$-embedded graph $G$ any connected component of
$\Sigma\setminus (E(G) \cup V(G))$. 
All embeddings we consider are {\em $2$-cell embeddings}, which are embeddings 
with each face being homeomorphic to a disk.

%  
% \textbf{TO DO:  check if we use noose at all. } 
%Let $G$ be a $\Sigma$-embedded graph. 
%%Let $G=(V,E)$ be a graph embedded in some surface $\Sigma.$ 
%A {\em noose} in $G$ is a
%closed curve $N$ of $\Sigma$ meeting only the vertices of $G$; we denote 
%these vertices by $V_N=V\cap N.$ 
% In other words, a noose is a subset of $\Sigma$ homeomorphic to a cycle which intersects no edges of the drawing of $G$ in $\Sigma$. We call a noose in $\Sigma$ \emph{contractible} if it is null-homotopic in~$\Sigma$. Thus a contractible noose forms a boundary of some disc in $\Sigma$. We call a noose $N$ in $\Sigma$ \emph{surface separating} (or just separating) if $\Sigma\setminus N$ is disconnected.  See Figures~\ref{fig:mrefl} and \ref{fig:secsp} for examples of separating and non-separating nooses. 
%We also define the {\em
%length}  of $N$ as the number of vertices it meets and we denote it by $|N|,$ that is, 
%$|N|=|V_N|.$ The {\em face-width} of $G$ is the minimum length of a non-contractible {(non null-homotopic in $\Sigma$)} noose of $G$.
%%\ff{Shall we define contractibility?}
%%noose in $G$ (a noose is {\em contractible} if its contraction to a point does not alter the surface). 

Given a $\Sigma$-embedded graph $G$, we define its {\em radial graph} $R_{G}$ as an embedded graph whose vertices are the vertices and the faces of $G$ (each face $f$ of $G$ is represented by a point $v_{f}$ in it). 
Roughly, each point $v_f$ is adjacent to all vertices $v$ incident to $f$. 
However, a face can be incident ``several times" with  the same vertex, and $R_G$ can have multiple edges.  For a point $v_f$ in the face $f$ and vertex $v$ incident with $f$, we draw a maximum number of multiple  edges   in $f$  such that  for every pair of multiple edges $e$ and $e'$ the open disc bounded by these edges   intersects $G$.  
%An edge between a vertex $v$ and a vertex $v_{f}$ is drawn if and only if $v$ is incident with $f.$ 
 Thus $R_G$ is a bipartite multigraph, embedded in the same surface as $G.$  
 Radial graphs provide an alternative way of viewing radial distance defined in Section~\ref{sec:intro}: the radial 
distance of a pair of vertices in $G$ corresponds to their normal distance in $R_G.$ The relation between radial and normal metrics is captured by the following observation. 
\begin{observation}
\label{obs:radialmetric}
If $G$ is a  $\Sigma$-embedded graph, then for every set $S\subseteq V$ and every $r\in\Bbb{Z}^{+}$, it holds that ${\bf B}_{G}^{r}(S)\subseteq {\bf R}_{G}^{2r}(S).$
\end{observation}

\subsection{Decomposition lemma for coverable problems}
%Proof of Lemma~\ref{lem:compactcombcondition}}
\label{subsec:fortheconditionwithb}

In this section we show the following decomposition result. 
\begin{lemma}
\label{lem:compactcombcondition}
Every $r$-coverable problem has the  protrusion  decomposition  property {\bf B}.
\end{lemma}

In order to prove Lemma~\ref{lem:compactcombcondition}, we have to show that every $r$-coverable problem satisfies combinatorial property {\bf B}, i.e. admits a protrusion decomposition. 
  Lemma~\ref{lem:compactcombcondition} follows directly 
from  the following lemma. 
% is a direct consequence of the following graph-theoretic result.

\begin{lemma}  
\label{thm:decomgenus2}
Let $r$ be a positive integer and let $G=(V,E)$ be a graph embedded in a surface ${\Sigma}$  of Euler genus $g$ 
that contains a  set $S$ of vertices, $|S|\leq k,$  
such that ${\bf R}_{G}^{r}(S)=V.$ Then  $G$ has an 
%$(\alpha,\beta)$-protrusion decomposition, where $\alpha=O(r^{2}g+rg^{3/2}+kg)$ and $\beta=O(r).$
$(\alpha k,\beta)$-protrusion decomposition
for some  constants $\alpha$ and $\beta$ than depend only on $r$ and $g$.
%, where $\alpha=O(r^{2}g+rg^{3/2})$ and $\beta=O(r).$
\end{lemma}

{Indeed, since a problem is $r$-coverable, there is a set $S$, $|S|\leq r\cdot k,$ 
such that ${\bf R}_{G}^{r}(S)=V.$ Then    combinatorial 
property {\bf B} holds  for $c=r\cdot \max\{\alpha,\beta\}.$
}

%Indeed, if we set $g=r$ in Lemma~\ref{thm:decomgenus2},  then   combinatorial 
%property {\bf B} holds  for $c=\max\{\alpha,\beta\}.$
%
\medskip

The rest of this subsection is devoted to the proof of Lemma~\ref{thm:decomgenus2}.
We start from  a series of definitions and preliminary results. The first observation follows directly from the definition of protrusion decomposition. 

\begin{observation}
\label{obs:minprot}
If $G$ has  an $(\alpha k,\beta)${-protrusion decomposition},
then the same holds for every subgraph of $G.$
\end{observation}

% 
% \textbf{TO DO !!! Check if need it anymore}
% 

%The next proposition follows from~\cite[(2.7)]{RobertsonS84} that the treewidth of a planar graph of radius $d$ does not exceed $3d+1$.
%%bidimentionality of $r$-domination.%\sed{Is bidimensionality an OK word for this paper?}

%
%\begin{proposition}
%\label{boundtwsurf}
%There exists a %linear
%  function $f_{1}: \mathbb{Z}^+\rightarrow \mathbb{Z}^+$ such that for every  planar graph $G=(V,E)$ and every  set $S\subseteq V$ of vertices 
%such that $V={\bf B}_{G}^{r}(S)$ and $|S|\leq k,$ it holds that $\tw(G)\leq f_{1}(r) {k}.$
%\end{proposition}
% 
%%The best known  estimation  $f_{1}(r)=(2r+1)\sqrt{4.5}$ is  given in~ \cite{Thilikos11}.
% \medskip

%The following proposition is a consequence of the result from \cite{Eppstein00} on the treewidth of graphs with bounded genus and diameter.
%\begin{proposition}
%\label{boundtwsurf2}
%There exists s  function $f_{2}:\Bbb{Z}^{+}\times \Bbb{Z}^{+}\rightarrow \Bbb{Z}^{+}$ such that if 
%$G=(V,E)$ is a graph of Euler genus at most $g$ 
%containing set $S\subseteq V,$  $|S|\leq 2,$ such 
%that $V={\bf B}_{G}^{r}(S),$ then $\tw(G)\leq
%f_{2}(r,g).$
%\end{proposition}

The following proposition is a consequence of the result from \cite{Eppstein00} on the treewidth of graphs with bounded genus and diameter.
\begin{proposition}
\label{boundtwsurf2}
There exists   function $f_{1}:\Bbb{Z}^{+}\times \Bbb{Z}^{+}\rightarrow \Bbb{Z}^{+}$ such that if 
$G=(V,E)$ is a graph of Euler genus at most $g$ 
 such 
that $V={\bf B}_{G}^{r}(v) $ for some $v\in V$,  then $\tw(G)\leq
f_{1}(r,g).$
\end{proposition}

%An estimation of $f_{2}$ can be easily derived from~\cite[Lemma 6]{FominGT11}. In particular,
%it follows that $f_{2}(r,g)\leq 12\cdot (g+1)^{3/2}\cdot (8r+4)=O(rg^{3/2}).$
%\medskip

For the purposes of the proof of the next lemma, we permit the existence of multiple edges or loops  in the embedding.
Thus   contracting edges can create  multiple edges 
or loops which we do not delete.  
 We call a face {\em trivial} if it is incident with at most two edges.
We call a loop  {\em empty} if it is the boundary of some face of $G.$
 %\todo{define face somewhere here?done, face defs moved above}
%We also adapt the definition of the contraction operation so that multiple edges 
%or loops that are created after a contraction are not removed. 
%Moreover, 
%we permit deletion  of a loop {\em only} if it is empty.

A {\em  walk} 
of length $\lambda$   in a multigraph $G$  %embedded in a surface $\Sigma$ 
is a sequence ${\cal C}=v_{0}e_{1}v_{1} \cdots e_{\lambda} v_{\lambda}$ of alternating vertices 
and edges of $G$ such that for every $i\in\{1,\ldots,\lambda\},$   the 
vertices $v_{i-1}$ and $v_{i}$
are the endpoints of  edge $e_{i}$. Thus an edge or a vertex can appear many times in a walk.
If in the previous definition we additionally demand that $v_{0}=v_{\lambda}$, then the walk is a {\em closed walk}.
% A closed walk ${\cal C}$ of a graph $G$ embedded in a surface $\Sigma$ is  {\em contractible} if 
%every cycle of the graph $G_{\cal C}$ is contractible, {i.e. null-homotopic, in $\Sigma$}.
%In an embedded  graph, all closed walks of length smaller than its face-width 
%are contractible because  each cycle of $G$ corresponds to a noose of equal length. 

We are ready to proceed with the proof of the lemma.
%of the next  technical lemma.

%\begin{lemma}
%\label{highrep:region} 
%Let $r$ be a positive integer and let $G=(V,E)$ be a graph, embedded in a surface ${\Sigma}$  of Euler genus $g$ such that
%either $g=0,$ 
%or the face-width of $G$ is more than $6r+3.$ Let $S\subseteq V$, $|S|\leq k$, be 
%such that  ${{\bf B}}_{G}^{r}(S)=V.$ Then  $G$ admits an 
%$(\alpha k,\beta)$-protrusion decomposition,  
% where $\beta$ is a universal constant and $\alpha=c_{1} (r +g)$ 
% for some universal constant $c_{1}$.
% \end{lemma}

\begin{proof}[of Lemma~\ref{thm:decomgenus2}] 
Let us note that 
by adding edges we do not  
  increase  distances between  vertices. Thus by 
 Observation~\ref{obs:minprot}, we may assume that all the faces in the embedding of $G$ in $\Sigma$
are {\em triangular}, meaning that they are incident with at most 3 edges, and that $G$ is connected.

% Given an equivalence class 
%$\overline{e}$ of the $\sim$ relation, we define 
%$$E(\overline{e})=\{e\in \overline{e}\mid \mbox{$\forall e'\in\overline{e}$ there is at most one face incident to $e$ and $e'$}\}$$
%and notice that $E(\overline{e})$ contains at most two edges. 

For every $v\in S$, we  construct a breadth-first search tree  $T_v$   {of depth at most $r$}
rooted at $v$. Because { ${\bf B}_{G}^{r}(S)=V$,} we have  
that every vertex of $G$ is in some $T_v$ for some $v\in S.$ While some vertices can be  
  within distance $r$ from several vertices of $S$, by suitably modifying these trees, 
  we may assume that every vertex is assigned to exactly one tree. That way, the
vertex sets of the trees in ${\cal T}=\{T_v\mid v\in S\}$ form a partition
of $V.$

We denote by $H$ the graph obtained 
from $G$ after contracting all the edges of the trees in ${\cal T}.$
 Notice that $V(H)=S,$   
and  as $G$ is triangulated, every  face of $H$ is incident to at most $3$ edges.
We further simplify $H$ as follows. 
\begin{itemize}
\item As long as  there are two edges incident with a trivial 
face, we delete one of them;
\item As long as  there is an empty loop, we delete it.
\end{itemize}
We denote the resulting graph by $\tilde{H}$. Again, every face of $\tilde{H}$ is incident to at most $3$ edges. Also $V(\tilde{H})=S$.

Using Euler's formula for graphs embedded in surfaces, see e.g. \cite[(4.4)]{MoharT2001}, 
we derive that 
$\tilde{H}$ has at most $2k+2g-4$ faces and at most $3k+3g-6$ edges. 
The edges of $\tilde{H}$  can be seen as the  edges of $G$ which were not contracted or deleted during the construction of $\tilde{H}$. 
For  every  edge $\tilde{e}$ of $\tilde{H},$ we denote by  $e$ the corresponding edge of  $G.$

Let  $\tilde{e}$ be an edge of  $\tilde{H}$ with endpoints $u,v\in S.$   
Let $x_u$ and $x_v$ be the endpoints of the corresponding edge $e$ in $G.$ 
If $u=v,$ then $x_u$ and $x_v$ are   vertices of ${T}_v.$ If $u\neq v,$ then $x_u$  is a vertex of ${T}_u$ and $x_v$ is a vertex of $T_v.$ 
 In both cases, there are unique paths $P_{u,x_u}$ in $T_u$ and $P_{v,x_v}$ in $T_v$ from $u$ to $x_u,$ and from $v$ to $x_v$ correspondingly. Each of these paths is of length at most $r.$ 
 We set  $P_e = P_{u,x_u}\cup \{e\} \cup  P_{v,x_v}.$ Let us note that if $u=v,$ then $P_e$ is a closed  walk, and if $u\neq v,$ then it is a path.
The length of  $P_e$ is at most $2r+1.$ 
 
 Let $\tilde{G}$ be the graph obtained from $G$ by contracting for every edge $\tilde{e}$   of  $\tilde{H}$ all edges except $e$ in the corresponding walk $P_e$.  Thus besides $S$, the vertex set of $\tilde{G}$ contains all vertices of $G$ not covered by walks $P_e$. %Since we do not contract edges and loops from  $\tilde{H}$,  each such edge or  loop survives  in $\tilde{G}$. 
 By construction,     $\tilde{G}[S] \supseteq  \tilde{H}$. 
  We take the drawing of $\tilde{G}$ in $\Sigma$ and observe that 
   $\tilde{G}[S]$ contains the drawing of $\tilde{H}$ in $\Sigma$. In the drawings of  $\tilde{G}$ and  $\tilde{H}$, 
  every face $f$ of  $\tilde{H}$ covers a subset of vertices $X_f$ of  $\tilde{G}$. The set $X_f$ is  separated in $\tilde{G}$ by the vertices incident to $f$ from the remaining vertices of the graph $\tilde{G}$. 
% { \blue{ (FIGURE HERE?)}}
   
%   
%   We now construct a protrusion decomposition ${\cal P}$ of $G$ in a way that its protrusions 
%will roughly correspond to   faces    of $\tilde{H}.$ 
  In $\tilde{G}$, every vertex $v\not\in S$ belongs to some set $X_f$. Thus, in $G$, every vertex is either in some $X_f$ or belongs to some walk $P_e$.
  We define  vertex subset  $R_0$ of $G$, as the union of the vertices of all walks corresponding  to  edges of $\tilde{H}$, i.e. 
    $$R_0= \bigcup_{\tilde{e}\in E(\tilde{H})} V(P_e).$$
    
Sets $R_0$ and     $X_f$, $f\in \tilde{F}$, have the following properties. 
 \begin{claim}\label{claimR_0}$|R_0|\leq  k+2r(3k+3g-6)$.
 \end{claim}
 \begin{proof}[of Claim]
 There are at most $3k+3g-6$ edges in 
$\tilde{H}$ and each edge corresponds in $G$ to a walk of length at most 
$2r+1$  connecting vertices of $S$. There are at most $k$ vertices in $S$ and thus 
 $|R_0|\leq  k+2r(3k+3g-6)$.
\end{proof}

Let $C_1, C_2, \ldots, C_\ell$ be the connected components of $G \setminus R_0$.
We use the following properties of these connected components. 
 \begin{claim}\label{claimR_X}
 \begin{align}\label{eqn:gprimevertices}
\big{|}\{i~:~|N_G(C_i)| \geq 3\}\big{|} \leq 2|R_0| - 2g - 4,
\end{align}
\begin{align}\label{eqn:gprimeedges}
\sum_{\{i~:~|N_G(C_i)| \geq 3 \}} |N_G(C_i)| \leq 6|R_0| + 6g - 12.
\end{align}
\end{claim}
\begin{proof}[of Claim]
Make a new graph $G'$ from $G$ by deleting all components $C_i$ such that $|N(C_i)| < 3$, contracting each component $C_i$ with  $|N(C_i)| \geq 3$ to a single vertex, removing all edges between vertices in $R_0$, and removing double edges and self loops. Thus $G'$ is bipartite simple graph and therefore every face of $G'$ is incident to at least $4$ edges. This fact, together with Euler's formula  yields the claim.
Here  \eqref{eqn:gprimevertices} counts the number of vertices of $G'$ in the bipartition corresponding to components, while  ~\eqref{eqn:gprimeedges} counts the number of edges in $G'$.
\end{proof}

\begin{claim}\label{claim:compR0TwBound} For each connected component $C_i$ of $G \setminus R_0$, the treewidth of $G[N[C_i]]$ is at most
   $f_1(4r+2, g)$.
\end{claim}

\begin{proof}
By construction of $R_0$, the component $C_i$ is a subset of $X_f$ for some face $f$ of $\tilde{H}$. The face $f$ is incident to at most $3$ vertices, say $x$, $y$ and $z$. In the graph $\tilde{G}$,  the neighborhood of $X_f$ is a subset of $\{x,y,z\}$. Hence in the graph $G$, the set $N_G(X_f)$ is a subset of vertices which were contracted to  $x,y$, or $z$.  Thus, also for $C_i$ it holds that $N_G(C_i)$ is a subset of the vertices which were contracted to $x,y$, or $z$. 

For any vertex $u$ in $C_i$ there is a path on at most $r$ vertices starting in $u$ and ending in $S$. This path must contain a vertex $u'$ in $N_G(C_i)$, and from $u'$ we can reach $\{x,y,z\}$ in at most $r$ steps. It follows that from any vertex in $C_i$ we can reach  $\{x,y,z\}$ in at most $2r$ steps. Since $x$ can reach $y$ and $z$ in $2r+1$ steps it follows that $N[C_i]$ is covered by a ball of radius $4r+2$ centered at $x$. Then by Proposition~\ref{boundtwsurf2}, the treewidth of $G[N[C_i] ]$ is at most $f_1(4r+2,g)$.
\end{proof}

For each $i \leq \ell$ define $G_i = G[N(C_i)]$. By Claim~\ref{claim:compR0TwBound} we have that the treewidth of $G_i$ is at most $t =  f_1(4r+2, g)$. Next we    claim the following.

\begin{claim}\label{claimR_5} 
For every $i$, there exists a set $Y_i \subseteq V(G_i)$ such that
\begin{itemize}
     \item $N_G(C_i) \subseteq Y_i$,
     \item $|Y_i| \leq 2|N_G(C_i)|(t+1)$,
     \item Every connected component of $G_i \setminus Y_i$ has at most $2(t+1)$ neighbors in $Y_i$. 
\end{itemize}
\end{claim}
\begin{proof}[of Claim] The proof of this claim is almost identical to the proof of Lemma~\ref{lem:split}. Here the role of the set $Z$ is given to $N_G(C_i)$. We compute a nice tree decomposition of $G_i$ and mark all upper most forget nodes of the decomposition forgetting vertices of  $N(C_i)$. We keep marking each lowest common ancestor of marked nodes, as long as possible. The vertices contained in all marked bags form the set $Y_i$. 
\end{proof}

We use Claim~\ref{claimR_5} to find sets $Y_i$ for every $G_i$ and define the set  
\[
R = R_0 \cup \bigcup_{\{i~:~|N(C_i)| \geq 3\}} Y_i.
\]
We partition the remaining set of vertices $V(G)\setminus R$ into sets $Q_1, Q_2, \dots, Q_q$, where every $Q_i$ is the  union of connected components of $G\setminus R$ with the same neighborhood in $R$. We claim that ${\cal P}=(R, \{Q_i\}_{1\leq i \leq q})$  is the desired $(\alpha k,  \beta)$-protrusion decomposition of $G$. 

First, we have the following bound on $|R|$. 
\[
|R| \leq |R_0| + \sum_{\{i~:~|N(C_i)| \geq 3\}} |Y_i| \leq |R_0| + 2(t+1) \sum_{\{i~:~|N(C_i)| \geq 3\}} |N(C_i)| =O(k)
\]
Here the last bound follows from  \eqref{eqn:gprimeedges} together with the bound of Claim~\ref{claimR_0} that $|R_0| = O(k)$

There are at most $|R|$ sets $Q_i$ such that $|N(Q_i)| = 1$. By Euler's formula there are at most $3|R|+3g-6$ sets $Q_i$ with exactly two neighbors in $R$. Again, by Euler's formula, exactly as in ~\eqref{eqn:gprimevertices}, 
the number of sets $Q_i$ with at least three neighbors in $R$ is at most $2|R|+2g-4$. Hence $q \leq 6|R| +5g=O(k)$. 

By Claim~\ref{claimR_5}, we have that $|N(Q_i)| \leq 2(t+1)$ for every $i$. Furthermore, for every $i$ we have that each connected component of $G[Q_i]$ is in fact $C_j$ for some $j$, and hence by Claim~\ref{claim:compR0TwBound}, $G[Q_i]$ has treewidth at most $t$. Hence $G[N[Q_i]]$ is a protrusion with treewidth at most $3t+2$ and boundary size at most $2(t+1)$. This completes the proof of Lemma~\ref{thm:decomgenus2}.
\qed
\end{proof}

\subsection{Decomposition lemma for quasi-coverable problems}
%Proof of Lemma~\ref{lem:quasicompactcombcondition}}
\label{subsec:proofoftheoremquasi}
%\label{subsec:proofoftheoremquasi}
In this section we prove the following decomposition lemma. 
\begin{lemma}
\label{lem:quasicompactcombcondition}
Every $r$-quasi-coverable problem  has the weak protrusion  decomposition  property {\bf B}$^*\!.$
\end{lemma}

Given the definition of $r$-quasi-coverability,  Lemma~\ref{lem:quasicompactcombcondition} is a direct consequence of the following graph-theoretic result.

\begin{lemma}  
\label{lem:thmith2}
There exist functions $\zeta_{1}$ and $\zeta_{2}$ such that the following holds:
Let $r,g,p,$ and $k$ be non-negative integers and let 
$G=(V,E)$ be a graph embedded in a surface ${\Sigma}$   of Euler genus $g$ 
 such that 
\begin{itemize}
\item $G$ contains a   set $S$ of vertices,  where $|S|\leq k$ and $\tw(G\setminus {\bf R}_{G}^{r}(S))\leq r,$ and
\item for every $\lambda\leq \zeta_{1}(r,g),$ $G$ has no $\lambda$-protrusion of size at least $p.$
\end{itemize}
Then  $G$ has a   $(ck,c)$-protrusion decomposition, where $c=\zeta_{2}(g,r,p).$
\end{lemma}
Indeed, we set  $g=r$ in Lemma~\ref{lem:thmith2}. Then combinatorial propertry ${\bf B}^{*}$ holds for $c'= \zeta_{1}(r,g)$ and $g(x)=\zeta_{2}(r,r,x).$

\medskip

The rest of this section is devoted to the proof of Lemma~\ref{lem:thmith2}. 
Let us  outline first  the main ideas of the proof.
Let $S$ be a subset of $V$ of size $k$  such that removal of balls of  radius $r$ (in radial distance) around vertices of $S$ from $G,$ results in a graph  of treewidth at most $r.$ 
We enlarge the set $S$ by adding at most $k$ new vertices and we want the new set $S'$ to satisfy the following property:
\begin{itemize}
\item Balls of radius $\mu$ (in radial distance) around vertices of $S'$ cover all vertices of $G,$ where $\mu$ is a constant depending on $r,p$ and $g.$ 
%\item[$(ii)$] or for every vertex $v$ at distance $\geq r+p$ from $S',$ in the graph $G\setminus{\bf R}_{G}^{p}(v)$ there are at most two connected components containing vertices of $S'.$
\end{itemize}
If we succeed to find such a set $S',$ then we can use 
Lemma~\ref{thm:decomgenus2} to obtain a $(c k,c)$-protrusion decomposition of $G$ for some constant  $c.$
To find the required set $S',$ we  show how to construct a superset $S'$ of $S$ of size at most $2k,$ such that 
for every vertex $v$ at distance $\geq 2\mu$ from $S'$  in the graph $G\setminus{\bf B}_{G}^{\mu}(v)$ there are at most two connected components containing vertices of $S'.$ This construction is given in Lemma~\ref{lem:maketwo}. 
To prove that $S'$ is the required set, we have to prove that every vertex of $G$ is at radial distance $\mu$ from some vertex of $S'.$ The proof of this fact is based on the proof 
that in graphs embedded in a surface of bounded genus, two connected sets embedded at a large radial distance from each other and non-separable by ``small" separators, form an obstruction for having ``small" treewidth (Lemma~\ref{lema:sepzz}). Because the treewidth of the  graph  $G\setminus{\bf R}_{G}^{r}(S')$ is at most   $r,$ we obtain that if there is a  vertex $v$ at distance  $>\mu$ from $S',$ then
a ball of radius $p$ around this vertex should be separated from the remaining graph by a small separator. This yields that 
$G$ has a protrusion containing a ball of radius $p$ around $v,$ and thus of size at least $p.$ But by the assumption of the lemma, there is no such a protrusion. Thus every vertex 
$v$ is within distance $\leq \mu$ from $S'.$  

We proceed with the proof of Lemma~\ref{lem:thmith2}.

\medskip
\noindent\textbf{Constructing $S'$ from  $S.$}
Let $G$ be a graph, $H$ be a subgraph of $G$  and  $S\subseteq V(G).$ An {\em $S$-component}
of $H$ is a connected component of $H$  containing some of the vertices of $S.$
%We denote by $\delta_{G}(S)$ the set of all vertices of $G$ that are not 
%in $S$ and are adjacent with some vertex in $S.$

\begin{lemma}
\label{lem:maketwo}
Let $\mu$ be a positive integer, $G=(V,E)$ be a connected graph,  and  $S$ be a subset of $V.$ Then there is a set $S'\supseteq S$ such that 
\begin{itemize}
\item $|S'|\leq \max\{2|S|-2,1\},$ and 
\item
 for 
every $v\in V\setminus {\bf B}_{G}^{2\mu}(S'),$   graph $G\setminus {\bf B}_{G}^{\mu}(v)$
has at most two  $S'$-components.
\end{itemize}
\end{lemma}

\begin{proof}
We use induction on $|S|.$ As the lemma is obvious when $|S|\leq 2,$ we 
assume that  $|S|=k>2$ and that the lemma holds  for all sets $S$ of smaller sizes.
 Suppose that $G$ contains a vertex $u$ such that 
 $\dist_{G}(u,S)\geq 2\mu+1$ and  $G^-=G\setminus {\bf B}^{\mu}_{G}(u)$ has at least three $S$-components. (If there is no such a vertex $u$, we are done.) We denote these components 
by $C_{1},\ldots,C_{h},$ $h\geq 3$, and we denote by $C_{h+1},\ldots,C_{\ell},$  the  
connected components of $G^-$  not containing vertices from $S.$
For $\ i\in\{1,\ldots,\ell\},$ we define
$$S_{i}=(S\cap V(C_{i}))\cup \{u\},$$
{and}
$$G_{i}=G[{\bf B}_{G}^{\mu}(u)\cup V(C_{i})].$$ 
Notice that each $S_{i}$ is a vertex subset of  the connected graph $G_{i}$ and 
that $1\leq |S_{i}| \leq |S|-1=k-1$.
This means that  the induction hypothesis holds for $G_{i}$ and $S_{i}.$ Thus 
  for every $i\in\{1,\ldots,\ell\},$ there is   a set $S'_{i} \supseteq S_{i}$ such that  $|S_{i}'|\leq \max\{2|S_{i}|-2,1\}$,   and  
\begin{eqnarray}
\forall v\in V(G_{i})\setminus {\bf B}_{G_{i}}^{2\mu}(S_{i}'),  \mbox{ graph\ } 
 G_{i}\setminus {\bf B}_{G_{i}}^{\mu}(v) \mbox{\ has at most two  $S_{i}'$-components.}\label{eq:atmosttwo}
\end{eqnarray}
We now set $S'=\bigcup_{1\leq i \leq \ell}S_{i}'.$ Clearly, $S'\supseteq S.$ Notice also that $u$ appears in every $S_{i}',$ while each other vertex of $S'$
appears in exactly one of $S_{1}',\ldots,S_{h}'.$ Therefore,  
\begin{eqnarray*}
|S'| & = & (\sum_{i=1}^h|S_{i}'|)-(h-1)\\
  & \leq & 2\cdot (\sum_{i=1}^h |S_{i}|)-2h-h+1\\
%  & = & 2\cdot (\sum_{i\in\{1,\ldots,r\}}|S_{i}^-|+1)-3r+1\\
  & =& 2\cdot (\sum_{i=1}^h|S_{i}\setminus \{u\}|)+2h-3h+1\\
  & = & 2|S|-h+1 \leq  2k-2.
\end{eqnarray*}
(For the last inequality, we use  the assumption that $h\geq 3$.)

We claim that  
for every $v\in V \setminus {\bf B}_{G}^{2\mu}(S'),$   the graph 
 $G\setminus {\bf B}_{G}^{\mu}(v)$ has at most two $S'$-components.
 Without loss of generality, let us assume that $v$ belongs to the 
 connected component $C_1$ of $G^- =G\setminus {\bf B}^{\mu}_{G}(u)$. 
 By \eqref{eq:atmosttwo},  in the corresponding graph $G_1$, the subgraph
  $G_{1}\setminus {\bf B}_{G_{1}}^{\mu}(v)$  has at most two  $S_{1}'$-components, where $S'_1= V(G_1)\cap S',$ and  one of these components  contains $u.$ The distance from $u$ to $v$ is at least $2\mu +1$ and hence the  whole ball ${\bf B}_{G_{v}}^{\mu}(v)$ is contained in $C_1$. 
 Therefore every vertex $w\in S'\setminus S_1$ is connected with $u$ in $G$  by a path avoiding ${\bf B}_{G }^{\mu}(v).$
 %\todo{Why? Donot we need an explanation?} 
 Hence, $G \setminus {\bf B}_{G }^{\mu}(v)$  has at most two  $S'$-components.
\end{proof}

\medskip
\noindent\textbf{Treewidth obstructions.} 
The main result of this subsection is Lemma~\ref{lema:sepzz} which can be seen as an extension of the following result: if a graph of bounded genus has two vertices which are far apart (in the radial distance) and cannot be separated by a small separator, then the treewidth of  the graph is large.
However for the purposes of the proof,  we need an extension of this result for two ``radially" connected and non-separable vertex sets. 
%sets

%For the purposes of the proof we need the 

%
%Next we prove combinatorial results about the graph structures enforcing large
%treewidth. 
%
%

To prove  Lemma~\ref{lema:sepzz} we need several combinatorial results. 
We use the following proposition  from  %Malni{\v{c}} and Mohar ~
\cite{JuvanMohar} (see also  ~\cite[Proposition 4.2.7]{MoharT2001}).

\begin{proposition}
\label{prop:homotop}
Let $G$ be a graph embedded in a surface $\Sigma$ of Euler genus $g,$ $x,y\in V(G),$ 
and let  ${\cal P}$ be a collection of pairwise  
internally vertex disjoint paths from $x$ to $y$  such that no two of them are homotopic.
% \ff{If we need to define homotopic}
Then, $|{\cal P}|\leq h(g),$ where 
\[h(g)= \left\{\begin{array}{lll}
&g+1& \mbox{if $g\leq 1$}\\
&3g-2
& \mbox{if $g\geq 2.$}\\
\end{array}\right.\]
\end{proposition}

Let $G=(V,E)$ a graph and let $X,Y,$ and $Z$ be pairwise disjoint subsets of $V.$
We say that $Y$ {\em separates} $X$ and $Z$ if 
$X$ and $Z$ are in different connected components of 
$G\setminus Y.$ We say that $Y$ is a {\em minimal $(X,Z)$-separator}
if  no subset of $Y$ separates $X$ and $Z.$
For   $S\subseteq V,$ we say that   $S$ is {\em connected in $G$}
if $G[S]$ is a connected graph.

The following properties of minimal separators of connected vertex sets in triangulated graphs are important  for obtaining treewidth obstructions. 
 \begin{lemma}
\label{lema:triangsep}
Let  $G$ be a triangulated graph embedded in a surface $\Sigma$ with Euler genus $g$ and let $S$ be a minimal separator for connected vertex  subsets 
 $X_{1} $ and $X_{2}$ of $G.$
Then $S$ has at most $h(g)$ connected components.
\end{lemma}

\begin{proof}
Let $C_1, C_2, \dots, C_r$ be the connected components of $G\setminus S.$ Without loss of generality, we assume that $C_1$ contains $X_1$ and $C_2$ contains $X_2.$ For each component $C_i$ we select a vertex $x_i\in C_i,$ $i\in \{1, \dots, r\}.$
  We call the vertices in $S$ {\em separation vertices} and the vertices   
$ \{x_{1}, x_{2}, \ldots, x_r\}$ {\em satellite} vertices.
From $G$ we 
construct   graph $H$ by  exhaustively 
contracting or removing edges according to the following rules:
\begin{itemize}
\item We contract all edges except the edges with one endpoint being  a satellite vertex and the other endpoint a separation vertex. 
\item We  delete loops which are not surface separating, and as long as possible, we delete one of the multiple edges incident with a trivial  faces, 
i.e. face incident with two edges.
\end{itemize}
Notice that every connected component $C_i$  is contracted to a single vertex $x_i$  and every connected component of $G[S]$ is also contracted to a single vertex.  In addition,    each application of the above rules results  in a triangulated graph, thus 
$H$ is triangulated. Let $S'$ be the vertices of $H$ resulted in contracting of $G[S].$ The vertices of $S'$ form a minimal $(x_{1},x_{2})$-separator in $H,$ and thus each of $x_i,$ $i\in \{1,2\},$ is adjacent to all vertices of $S'.$ 
  Hence there exist 
$|S'|$ internally vertex disjoint paths of length two from 
$x_{1}$ to $x_{2}$ in $H.$ Because $H$ is triangulated,   
these $(x,y)$-paths are pairwise  non-homotopic, otherwise 
some edge in $H[S']$ could be further contracted or deleted. Combining this with 
Proposition~\ref{prop:homotop}, we deduce that $|S'|\leq h(g).$
The lemma now follows by observing that each connected component of 
$S$ shrinks to a single vertex of $S',$ therefore $S$ has $|S'|\leq h(g)$ connected components. 
\end{proof}

 We say that two vertex subsets $X,Y$ of graph $G$ {\em touch}
if either $X\cap Y\neq\emptyset$ or there exist an edge of $G$ with one endpoint in $X$ and the other in $Y.$
A {\em bramble} of $G$ is a collection ${\cal B}$
of mutually touching connected subsets of $V(G).$ 
The {\em order} of a bramble ${\cal B}$ is the minimum  size 
of a set $S$ that intersects all its elements.
The {\em bramble} number of $G$ is the maximum order a bramble 
of $G$ may have.

The following min-max characterization of treewidth 
was proved in~\cite{SeymourT93}.

\begin{proposition}
\label{lema:minmaxtreewidthbramble}
The treewidth of a graph is one less than its  bramble number.
\end{proposition}

We define   functions $f_{1},f_{2}$ such that
$f_{1}(x,y)=(x+1)y$ and $f_{2}(x,y)=x({(x+1)y\choose x+1})+1.$
The following lemma can be seen as a generalization of~\cite[(3.2)]{SeymourT93}.

\begin{lemma}
\label{lema:mutually}
Let $q,t$ be non-negative integers and let  $r_{1}=f_{1}(t,q)$ and 
$r_{2}=f_{2}(t,q).$
Let $G$ be a graph and let ${\cal X}=\{X_{1},\ldots,X_{r_{1}}\}$ be a collection of mutually  disjoint 
connected vertex sets of $G.$ Let also ${\cal Y}=\{Y_{1},\ldots,Y_{r_{2}}\}$ be a collection of 
mutually  disjoint vertex sets of $G,$ each with at most $q$ connected components
and such that for every $i \in\{1,\ldots,r_1\}$ and $j\in\{1,\ldots,r_2\},$ $X_{i}\cap Y_{j}\neq\emptyset.$
Then $\tw(G)\geq t.$
\end{lemma}

\begin{proof}
%Let $I=\{1,\ldots,r_{1}\}.$ 
For every set $Y_{j},$ 
 $j\in\{1,\ldots,r_{2}\},$ we select its connected component $Y_{j}'$  intersecting the largest number of  sets from  ${\cal X}.$ Because  every  $Y_{j}$ has at most $q$ connected components,  
set $Y_{j}'$ intersects at least $t+1=r_{1}/q$ sets from ${\cal X}.$ 

%it holds 
%that  $Y_{j}\cup\bigcup_{i\in\{1,\ldots,r_{1}\}} X_{i}$ has at most $q$ connected components.
%Among them, let $Y_{j}'$ be the one that intersects more sets in ${\cal X}.$ This means 
%that $Y_{j}'$ intersects at least $k+1=r_{1}/q$ sets in ${\cal X}.$ 

Let now $R$ be the intersection graph of  sets  ${\cal X}$
and ${\cal Y}'=\{Y_{1}',\ldots,Y_{r_{2}}'\}.$  Then $R$ is a bipartite graph 
with bipartition $({\cal X}, {\cal Y}')$ , and every vertex from ${\cal Y}'$ 
has degree $\geq t+1$ in $R.$ We remove edges from $R$ such that in the resulting 
graph all vertices of ${\cal Y}'$ have degree exactly $t+1.$ In the new graph 
the vertices from ${\cal Y}'$  have at most 
\[{|{\cal X}|\choose t+1}={(t+1)q \choose t+1}\] 
distinct neighbourhoods in ${\cal X}.$ Because  \[|{\cal Y}'|=|{\cal Y}|=t{(t+1)q \choose t+1}+1,\]
we deduce that there  should be at least $t+1$ vertices of  ${\cal Y}'$ with the same neighbourhood
in ${\cal X}.$ Let $I_{{\cal Y}}$ be the indices of these vertices in ${\cal Y}$
and let $I_{{\cal X}}$ be  the indices of their neighbours in ${\cal X}.$

It follows that for every $(i,j)\in I_{{\cal X}}\times I_{{\cal Y}},$ $X_{i}\cap Y_{j}'\neq\emptyset,$
and, as both $X_{i}$ and $Y_{j}'$ are connected, $X_{i}\cup Y_{j}'$ is also a connected set.
Moreover, because $|I_{\cal X}|=|I_{{\cal Y}}|=t+1,$ it follows that 
for every set $S$ of  $t$ vertices in $G,$ there are  $i\in I_{{\cal X}}$ and 
  $j\in I_{{\cal Y}}$ such that $S\cap (X_{i}\cup Y_{j}')=\emptyset.$
We can now conclude that 
the collection $\{X_{i}\cup Y_{j}'\mid (i,j)\in I_{\cal X}\times I_{{\cal Y}}\}\}$
is a bramble in $G$ of order $t+1.$ Therefore, the bramble number of $G$
is at least $t+1$ and the lemma follows from  Proposition~\ref{lema:minmaxtreewidthbramble}.
\end{proof}
%
%\begin{lemma}
%Let $G$ be a graph embedded in a surface $\Sigma$ of genus $g.$
%Let also $x,y$ be two vertices of $G$ connected by $k$ internally vertex disjoint 
%paths and such that their 
%radial distance in $G$ is at least $k.$ Then $\tw(G)\geq k.$
%\end{lemma}

%
%
%Let $G$ be a graph embedded in some surface $\Sigma$
%and let $S\subseteq V(G).$ We say that $S$ is {\em radially connected} of 

Let $G$ be a graph embedded in some surface $\Sigma.$
We define the {\em radial completion} of $G$ as the graph
obtained from drawing of $G$  in $\Sigma$  together with its radial graph $R_{G}.$ We denote the radial completion of $G$ by  $W_{G}.$ Let us remark that $W_G$ is triangulated and that   $R_{G}$ is a spanning subgraph of $W_{G}.$ %$\sed{This is not correct!}
Notice that every two adjacent vertices 
in $W_{G}$ have some common neighbour  in $R_{G}.$ This implies the following observation.

%STOPPED HERE

\begin{observation}
\label{lema:surfdist}
Let $G$ be a graph embedded in some surface $\Sigma.$ Then for every 
pair $x,y\in V(R_{G}),$ it holds that $\dist_{W_G}(x,y)\leq \dist_{R_{G}}(x,y)\leq 2\cdot \dist_{W_{G}}(x,y).$
\end{observation}

Loosely speaking, the following lemma says that 
in a graph of small treewidth which is embedded on a surface of fixed genus, every two connected sets 
will be either radially close or will be be separated by a small set.
Let $h$ be the function from  Lemma~\ref{lema:triangsep}, and  $f_1,$ $f_2$ be the functions defined before Lemma~\ref{lema:mutually}.

\begin{figure}[h]
\label{fig:lemt}
\begin{center}
 \scalebox{0.4}{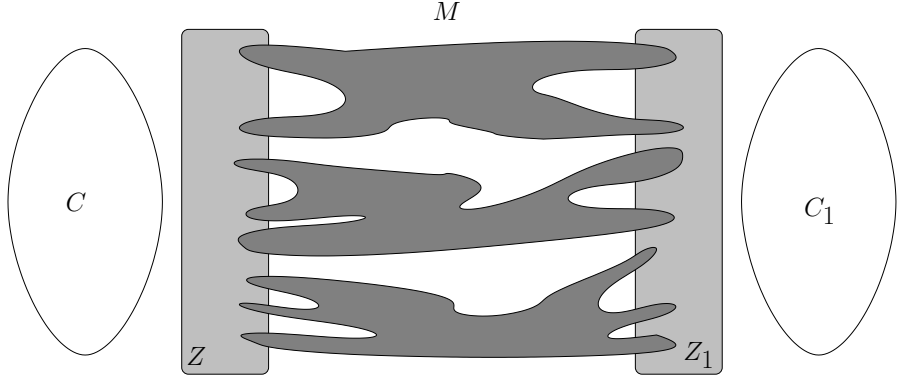}\end{center}
\caption{A visualization of the statement of Lemma~\ref{lema:sepzz}.}\end{figure}
%
% \begin{figure} 
%\begin{center}
%\includegraphics[scale=.2]{setM.pdf}
%\caption{Situation in Lemma~\ref{lema:sepzz}.}\label{fig:lem}
%\end{center}
%\end{figure}

\begin{lemma}
\label{lema:sepzz}
Let $G$ be a graph embedded in a surface $\Sigma$
of Euler genus $g,$  $t$ be a positive integer, 
and   $C,Z,Z_{1},C_{1}$ be disjoint subsets of $V(W_G)$
such that 
\begin{itemize}
\item $C$ and $C_{1}$ are connected  in $W_{G},$
\item $Z$ separates $C$ from $Z_{1}\cup C_{1}$
and $Z_{1}$ separates $C\cup Z$ from $C_{1}$ in $W_{G},$
\item $\dist_{W_{G}}(Z,Z_{1})\geq 3\cdot  f_{2}(t+1,h(g))+3,$ and
\item $G$ contains   $f_{1}(t+1,h(g))$ internally vertex-disjoint paths from $C\cap V(G)$ to $C_{1}\cap V(G).$
\end{itemize}
Then $\tw(G[V(M)\cap V(G)])> t,$ where 
 $M$ is the union of 
all  connected components of $W_G\setminus (Z\cup Z_{1})$
that have at least  one neighbor in  $Z$ and at least one neighbor in $Z_{1}.$ (See Fig. 1.)
\end{lemma}

\begin{proof}
We set $\mu=f_{1}(t+1,h(g))$ and $\lambda=f_{2}(t+1,h(g)).$
 Let  $P_{1},\ldots,P_{\mu}$ be   internally vertex-disjoint paths  in $G$ from $C\cap V(G)$ to $C_{1}\cap V(G).$  Each of these paths $P_i$ contains at least one  subpath with 
 one endpoint in $Z$ and the other in $Z_{1},$ and with all  internal vertices in $M.$
We denote   by $P_{1}',\ldots,P_{\mu'}'$ 
the set of such subpaths. Then $\mu'\geq \mu.$
 
 For $j\in\{1,\ldots,3\lambda+2\},$ let $A_j$ be the set of all vertices of $W_{G}$ that are within distance exactly $j$ 
from $Z$ and belonging to  $M.$ 
Notice that each $A_{j}$ is a $(Z,Z_{1})$-separator and thus also a 
$(C,C_{1})$-separator of $W_G.$  Clearly, each $A_{j}$ contains as a subset 
a minimal $(C,C_{1})$-separator $Y_{j}$ of $W_{G}.$ 
As each $Y_{j}$ is also a  $(Z,Z_{1})$-separator, 
it should contain at least one   internal vertex of every path in $P_{1}',\ldots,P_{\mu'}.$
Moreover, by its definition, $A_{j}$ should be a subset of $M.$

As $W_{G}$ is triangulated, by Lemma~\ref{lema:triangsep}, each $W_{G}[Y_{j}]$ contains at most 
$h(g)$ connected components. 
Recall that, by the definition of $W_{G},$ for each vertex $x\in V(W_{G})\setminus V(G),$ the graph induced by its neighborhood is a connected subgraph of $G.$
Using this fact, we obtain that $Y_{j}^+={\bf B}^{1}_{W_{G}}(Y_{j})\cap V(G)$ has also at most 
$h(g)$ connected components in $G$ for $j\in\{2,\ldots,3\lambda+1\}.$

Let $I=\{1,\ldots,\lambda\}$ and notice that, 
for any two distinct $h,l\in I,$ $Y_{3h}^+$ and $Y_{3l}^+$
are vertex-disjoint subgraphs of $G[V(M)\cap V(G)].$
For $j\in\{1,\ldots,\mu'\},$  we define $P_{j}''$ as the 
path obtained from $P'_{j}$ after removing its endpoints.
Observe  now  that $P_{1}'',\ldots,P_{\mu'}''$ are connected   vertex-disjoint 
subgraphs of $G[V(M)\cap V(G)],$ and each of these graphs intersect all graphs 
$Y_{3j}^+.$ 
 Applying Lemma~\ref{lema:mutually}
for   $\mu$  graphs from  $\{P_{1}'',\ldots,P_{\mu'}''\}$ 
and  $\lambda$ graphs from  $\{Y_{3j}^+\mid j\in I\},$  
we deduce  that $\tw(G[V(M)\cap V(G)])\geq t+1>t$ and the lemma 
follows. 
%\todo[inline]{some picture for this proof could be very helpful.}
\end{proof}

\medskip\noindent\textbf{Final step.} 
To conclude the proof of the main result of this section, we need the last lemma.
The following lemma essentially says that if 
$(G,k)$ is a YES-instance of a quasi-coverable problem $\Pi$
where $G$ has {no} big protrusions, then $G$ has an $r$-dominating set 
of size $O(k)$ for some $r$ that depends only on $\Pi$ and $g$
and therefore $(G,k)$ can   be treated as a 
YES-instance of a coverable problem.

\medskip

We define   function %$f_{3}$ and $f_{4}$ such  that 
$f_{3}(x,y)=2\cdot f_{1}(x+1,h(y+1))$
% and $f_{4}(x,y,w,z)=2(x+1+6f_{2}(y+1,c\cdot (w+1))+6+z),$ 
where $h$ is the function  
of Lemma~\ref{lema:triangsep}, and $f_1$ is the function defined before Lemma~\ref{lema:mutually}. 
\begin{lemma}
\label{lem:sizelessprotrusion}
Let $G=(V,E)$ be a graph embedded in a surface $\Sigma$ of Euler genus $g$
and let $p,t,$ and $r$ be non-negative  integers such that  
\begin{itemize}
\item there exists a set $S\subseteq V$ such that $\tw(G\setminus {\bf R}_{G}^{r}(S))\leq t$;
\item     for $\lambda \leq f_{3}(t,g)$, all $\lambda$-protrusions of $G$   are of size less than $p.$
\end{itemize}
Then there exist a set $S'\subseteq V$
and a constant $\mu$ (depending on $p,g,$ and $r$ only)
such that 
\begin{itemize}
\item $|S'|\leq 2|S| ,$ and 
\item ${\bf R}_{G}^{ \mu}(S')=V.$
\end{itemize}
\end{lemma}

\begin{proof} 

To prove the lemma, we prove a slightly different statement: Under the assumptions of the lemma, 
there is  a set $S'\subseteq V(W_G)$ such that $|S'|\leq 2|S|$ and   ${\bf B}_{W_G}^{ \mu}(S')=V(W_G).$ 
Then the statement of the lemma  can be 
  deduced from this alternative statement by   constructing set  $S_{\rm new}'$ as follows:
first set $S_{\rm new}'\leftarrow S'$ and then 
replace each vertex in $S'$
that does not belong to  $V(G)$ with one of its neighbors  from $V(G).$
It remains to observe that ${\bf R}_{G}^{ \mu +1}(S'_{\rm new})\supseteq {\bf B}_{W_G}^{\mu}(S').$ \medskip

We put  
$\mu =2p+2r + 2+2\mu',$ where $\mu'=3\cdot  f_{2}(t+1,h(g))+3,$
and  proceed with the proof of the above  alternative statement.
 We first apply Lemma~\ref{lem:maketwo} for $W_{G}$ and $S$    to obtain a set $S'\supseteq S$ of vertices,  where  $|S'|\leq 2|S| $ and such 
that for every $v\in W_{G}\setminus  {\bf B}_{W_{G}}^{2 \mu }(S'),$ graph 
$W_G\setminus {\bf B}_{W_{G}}^{ \mu}(v)$ has at most two $S'$-components. 
If  ${\bf B}_{W_{G}}^{2\mu}(S')=V(W_G),$ then we are done. Otherwise, let $v\in W_{G}\setminus  {\bf B}_{W_{G}}^{2 \mu }(S').$ Let $C_1, C_2$ be $S'$-components of 
$W_G\setminus {\bf B}_{W_{G}}^{ \mu  }(v)$ (one of these components can be an empty set), and let $S_i=C_i\cap S',$ $i\in\{1,2\}.$
We also define subgraphs of $W_G$ as follows,  
$W_1=W_G\setminus C_2$ and $W_2=W_G \setminus C_1$.

We claim that at least one of the sets $C_i$, $i\in\{1,2\},$ cannot be separated in $W_i$ from  
$C=B_{W_G}^{2p}(v)$  by a separator of size  at most $\lambda/2.$ Indeed, if it was the case, 
then  in $W_G$, $C$ is separable from  $C_1\cup C_2$, and thus from  
 ${\bf B}_{W_{G}}^{2r}(S')\subseteq C_1\cup C_2$ by a separator of size at most $\lambda.$ 
 By Observation~\ref{lema:surfdist}, this means that in   $G,$  
vertices $R_{G}^{p}(v)$ can be separated from ${\bf R}_{{G}}^{r}(S') $ by a separator of size at most $\lambda.$ Because   $\tw(G\setminus {\bf R}_{G}^{r}(S'))\leq t$ this yields that 
there is a  $\lambda$-protrusion   in $G$ containing  ${\bf R}_{G}^{p}(v).$ But 
$| {\bf R}_{G}^{p}(v)|\geq p,$  and thus the size of this protrusion is at least $p$ in $G,$ which  contradicts to the assumption of the lemma.

Without loss of generality, 
let  us assume that 
$C_1$ is  a $S'$-component of 
$W_G\setminus {\bf B}_{W_{G}}^{ \mu}(v)$ 
  that
  cannot be separated in $W_1$  from   $ C$ by a separator of size $\lambda/2$.  
  By    Menger's theorem, in graph $W_1$ there are $\lambda/2$ internally vertex-disjoint paths 
from $C $ to $C_1.$
We define $Z$ as the set of vertices at distance exactly $2p+1$ from $v$ in $W_1,$ and $Z_1$ as $N_{W_1}(C_1).$ Then $Z$ separates $C$ from $Z_1\cup C_1$ and $Z_1$ separates $C_1$ from $Z\cup C.$ 
The distance in $W_1$ between $Z$ and $Z_1$ is at least $\mu'.$
Let $M$ be the union of connected components of $W_1\setminus (Z_1\cup Z_2)$ having at least one neighbour in $Z$ and $Z_1$. 
By Lemma~\ref{lema:sepzz}, the treewidth of the subgraph $G_M$ of $G$ induced by $M\cap V(G)$ is more than $t$. On the other hand, every vertex of $M$ is at distance more than $r+1$ in $W_G$, and thus  at radial distance at least $r+1$ in $G$,  from each vertex of $S'$, and thus of $S$. Hence
$\tw(G_M)\leq \tw(G\setminus {\bf R}_{G}^{r}(S))$, which is at most $t$ by the assumption of the lemma. This contradiction concludes the proof of the lemma.
 \end{proof}
\begin{proof}[of Lemma~\ref{lem:thmith2}]
By applying Lemma~\ref{lem:sizelessprotrusion} for $r=t$ and $\zeta_{1}=f_{3},$ we have that 
$G$ contains a set of vertices $S'$ where $|S'|\leq 2k$ such that ${\bf R}_{G}^{\mu}(S')=V(G),$ where $\mu$ is the constant of Lemma~\ref{lem:sizelessprotrusion}. But then by Lemma~\ref{thm:decomgenus2},  $G$ has a
$(ck,c)$-protrusion decomposition for some $c$ depending on $g,r,$ and $p$ as required. \end{proof}

\section{Criteria for proving FII}
\label{sec:criterialforprovingfii}

To apply Theorem~\ref{thm:automata}, to prove that a specific 
parameterized problem on graphs admits a linear kernel
we have to show that it has FII. 
This property is not always easy to prove directly. In this section, we give some 
general criteria for establishing FII. 
These tools are used   in~Section~\ref{sec:implication}.
Early results that establish that problems have FII were obtained by
Bodlaender and de
Fluiter~\cite{BodlaenderF96a,BodlaendervA01a,Fluiter97}; another 
criterion for FII was given by van Rooij \cite[Section 11.2]{Roo11}.

%We denote by  ${\bf MinRep}(\psi,t)$   a set containing a representative (a boundaried structure of arity two)  for each equivalence class of $\equiv_{\sigma_\psi}$ with  the minimum number of  vertices in the graph of a structure.

\subsection{Strong monotonicity}
We first give a sufficient condition which implies that a large class of \pmm{} problems has FII. We prove it here for vertex versions of \pmm{} problems.
By ${\cal U}_{I}$  we denote the set of all boundaried structures of type $({\sf graph},{\sf vertex\ set})$, whose boundaried graph has label set $I$. 

\smallskip

 Let  $\Pi$ be a \pmin{} problem definable by some sentence $\psi.$
We say that a boundaried structure $(G',S')$ whose boundaried graph has label set $I$ is {\em $\psi$-feasible for some boundaried 
graph $G$ with label set $I$}  if there exist some $S\subseteq V(G)$ such that   $(G\oplus G',S\cup S')\models\psi.$
For a boundaried graph $G$ with label set $I$, 
we define the function $\zeta_G : {\cal U}_I \rightarrow \mathbb{Z}^+ \cup \{\infty\}$ as follows. For a structure $\alpha=(G',S')\in{\cal U}_{I}$ we set \smallskip
\begin{eqnarray}
\hspace{-3mm} \zeta_{G}(\alpha)\!=\!\left\{\begin{array}{lll}
&\hspace{-3mm} \min\{|S|\! \mid\! S\subseteq V(G) \wedge (G\oplus G',S\cup S')\models\psi\}& \mbox{\!\!if $\alpha$ is $\psi$-feasible for $G$}\\
& \infty& \!\!\mbox{otherwise}\label{eq:feasiblezeta}
%\\ & {\bf left}_{G}(y_{1}^{i,2l}) & \mbox{if $y_{1}^{i,2l}\in S$ and $i$ is odd}
\end{array}\right.
\end{eqnarray}

 Similarly, for $\Pi$  \pmax{}\ problems   we define\smallskip
 \begin{eqnarray*}
\hspace{-3mm} \zeta_{G}(\alpha)\!=\!\left\{\begin{array}{lll}
&\hspace{-3mm} \max\{|S|\!\mid\! S\subseteq V(G) \wedge (G\oplus G',S\cup S')\models\psi\}& \mbox{\!\!if $\alpha$ is $\psi$-feasible for $G$}\\
& -\infty& \!\!\mbox{otherwise}\label{eq:feasiblezetamax}
%\\ & {\bf left}_{G}(y_{1}^{i,2l}) & \mbox{if $y_{1}^{i,2l}\in S$ and $i$ is odd}
\end{array}\right.
\end{eqnarray*}
 
\begin{definition}
\label{def:minstrongmonmin}
A \pmin{} problem $\Pi$ is  \emph{strongly monotone} if there exists a function $f : \mathbb{Z}^+ \rightarrow \mathbb{Z}^+$ such that the following condition is  
satisfied. For every boundaried graph $G$ with label set $I$,  there exists a subset $W\subseteq V(G)$ such that for 
every $(G',S')\in {\cal U}_I$ such that 
%$S'$ does not intersect the boundary of $G$ and 
$\zeta_G(G',S')$ is finite, it holds that $(G\oplus G',W\cup S')\models \psi$  and $|W|\leq \zeta_G(G',S')+f(|I|).$ 
\end{definition}

For completeness we give below the maximization counterpart of Definition~\ref{def:minstrongmonmin}.

\begin{definition}
\label{def:minstrongmonmax}
A \pmax{} problem $\Pi$ is \emph{strongly monotone} if there exists a function $f : \mathbb{Z}^+ \rightarrow \mathbb{Z}^+$ such that the following condition is  
satisfied. For every boundaried graph $G$ with label set $I$ there exists a subset $W\subseteq V(G)$ such that for every $(G',S')\in {\cal U}_I$
such that %$S'$ does not intersect the boundary of $G$ and 
$\zeta_G(G',S')$ is finite,  it holds that  $(G\oplus G',W\cup S')\models \psi$  and $|W|\geq \zeta_G(G',S')-f(|I|).$ 
\end{definition}

\subsection{FII for \pmm\ problems}

\begin{lemma}
\label{lem:stronglymonotone}
Every strongly monotone \pmin{} and every strongly monotone \pmax{} problem has FII.\end{lemma}
\begin{proof}
We prove the lemma for a  \pmin{} problem; the proof for a \pmax{} problem is similar. Let $\Pi$ be a strongly monotone \pmin{} problem and 
%defined by some CMSO sentence $\psi,$ 
let $I\subseteq \Bbb{Z}^{+}.$  Let ${\bf MinRep}(\psi,I)$ be   a set containing a representative (a boundaried structure of arity two)  for each equivalence class of $\equiv_{\sigma_\psi}$ with  the minimum number of  vertices in the graph of a structure. For brevity we denote ${\bf MinRep}(\psi,I)$ by ${\cal S}$. 
%Let ${\cal S}_t$ be a set of representatives of each of the equivalent classes of $\equiv_{\psi,t}.$
%(given 
%in Definition~\ref{donfbf} above).
From Lemma~\ref{log_lem} we know that $|{\cal S}|$ 
is bounded by some function of $|\psi|$ and $|I|.$ \smallskip

Consider a boundaried graph $G$ with label set $I$ and define  $\zeta_G^{\cal S} : {\cal S} \rightarrow \mathbb{Z}^+ \cup \{\infty\}$ 
to be the function $\zeta_G$ with domain restricted to  ${\cal S}.$ 
Let $L_{G}^{\cal S}=\{\zeta_{G}^{\cal S}(\alpha)\mid \alpha\in{\cal S}\}\setminus \{\infty\}.$ 
%
%As $G\in{\cal F}_{t},$ \sed{Fix this!}
%%(see Definition~\ref{defhhds})
% the representative in ${\cal S}$ of the trivial 
% structure $(I_{t},\emptyset)$ is $\psi$-feasible for $G$ (here, \sed{The issue is here!}$I_{t}$ is the graph with $t$ isolated terminal vertices).
% Therefore, $L_G^{\cal S}\neq \emptyset.$
We first  argue
 that  if $f$ is the function in the definition of the strong monotonicity of $\Pi$ (i.e., Definition~\ref{def:minstrongmonmin}) and $L_{G}^{\cal S}\neq\emptyset,$
 then \begin{eqnarray}
 \max L_{G}^{\cal S}-\min L_{G}^{\cal S}\leq f(|I|)\label{eq:trivstronglmon}
 \end{eqnarray}
Since $\Pi$ is strongly monotone, there exists   $W\subseteq V(G)$ such that
 for every $(G',S')\in {\cal U}_I$ where $\zeta_G(G',S')\neq \infty,$ it holds that 
 \begin{eqnarray}
&& (G\oplus G',W\cup S')\models \psi \  \mbox{and } \label{eq:thiseqstrongmon}\\
&& |W|\leq \zeta_G(G',S')+f(|I|)\label{eq:zetastromon}
 \end{eqnarray}
 
 Let $\alpha=(G',S') \in {\cal S}$ such that $\zeta_G^{\cal S}(\alpha)\neq \infty.$ 
Then~\eqref{eq:thiseqstrongmon} implies that $\zeta_G^{\cal S}(\alpha) \leq |W|.$ 
This, together with~\eqref{eq:zetastromon},
yields that  $|W|-f(|I|) \leq \zeta_G^{\cal S}(\alpha) \leq |W|$ and~\eqref{eq:trivstronglmon} holds. 
Hence the 
minimum and the maximum finite values of  $\zeta_G^{\cal S}$ can 
differ by at most $f(|I|).$

We now assign for each boundaried graph $G$ with label set $I$ a {\em  signature} $\chi_{G}: {\cal S}\rightarrow \{0,\ldots,f(|I|),\infty\}$
in a way  that for each  $\alpha\in{\cal S},$ 
\begin{eqnarray}
\chi_{G}(\alpha) & = & \zeta_{G}^{\cal S}(\alpha)-\min L_{G}^{\cal S}\label{eq:boundazeta}
\end{eqnarray}
In~\eqref{eq:boundazeta}, we make the agreement
that infinite values remain infinite after subtracting an integer.
Notice that it is possible that in~\eqref{eq:boundazeta}
$\min L_{G}^{\cal S}$ may not exist and this happens 
in the extreme case where 
$L_{G}^{\cal S}=\emptyset.$  In such a case,  we set  $\chi_{G}(\alpha)=\infty$ for all $\alpha\in{\cal S}.$

We say that $G_{1}\sim G_{2}$ if and only if  $\chi_{G_{1}}=\chi_{G_{2}}$ and observe that $\sim$ is an equivalence
relation. Observe that the number of different  signatures of  boundaried graphs with label set $I$ is bounded 
by some function of $|\psi|$ and $|I|.$ Therefore, the same holds for the number of equivalent classes of $\sim.$
To prove that $\equiv_{\Pi}$ has FII, it is enough to prove that $\sim$ is a refinement of $\equiv_{\Pi},$
which means that 
%\sed{problem with $\equiv_{\Pi,t}$}
if $G_{1}\sim G_{2},$ then $G_1\equiv_{\Pi}G_{2}.$ For this, we claim that if  $G_{1}\sim G_{2},$ then 
there exists some constant $c\in\Bbb{Z}$ (depending on $G_1$ and $G_{2}$)
such that 
%\sed{define ${\cal F}_{t}$}
\begin{eqnarray}
\forall(F,k)\in {\cal F}\times \Bbb{Z} &&  (G_1 \oplus F, k) \in \Pi \Leftrightarrow (G_2 \oplus F, k+c) \in \Pi.\label{eq:transpoprodep}
\end{eqnarray}
To prove the above statement we first determine  the constant $c.$
As $G_{1}\sim G_{2},$ we have that
 $\chi_{G_{1}}=\chi_{G_{2}}.$
 In the extreme case where  $\chi_{G_1}(\alpha)=\chi_{G_2}(\alpha)=\infty$ for all $\alpha\in{\cal S},$~\eqref{eq:transpoprodep} holds trivially  for $c=0$ as  $\forall(F,k)\in {\cal F}\times \Bbb{Z}^+$ both sides of the equivalence are false
 (for completeness, recall that according to the way we defined parameterized problems,  $\forall(F,k)\in {\cal F}\times \Bbb{Z}^-$ both sides of the equivalence in~\eqref{eq:transpoprodep} have the same value). 
 From now onwards we assume that both $\min L_{G_{1}}^{\cal S}$ and $\min L_{G_{2}}^{\cal S}$ exist.
Therefore, from~\eqref{eq:boundazeta}, for each $\alpha\in {\cal S},$\
$\zeta_{G_{2}}^{\cal S}(\alpha)=\zeta_{G_{1}}^{\cal S}(\alpha)-\min L_{G_{1}}^{\cal S}+\min L_{G_{2}}^{\cal S}.$
We set $c=\min L_{G_{2}}^{\cal S}-\min L_{G_{1}}^{\cal S}.$ and we conclude that
\begin{eqnarray}
\forall\alpha\in{\cal S} & & \zeta_{G_{2}}^{\cal S}(\alpha)=\zeta_{G_{1}}^{\cal S}(\alpha)+c.\label{eq:againchoice}
\end{eqnarray}
Let $(F,k)\in {\cal F}\times \Bbb{Z}$ and assume that $(G_1 \oplus F, k) \in \Pi.$ This means 
that there exists a set $S\subseteq V(G_1\oplus F)$ such that  $|S|\leq k$ and 
\begin{eqnarray}
(G_{1}\oplus F,S) & \models  & \psi.\label{eq:subsetstransp}
\end{eqnarray}
Let  $S_{F}=S\cap V(F)$ and $S_{G_1}=S\setminus S_{F}$ and observe that 
\begin{eqnarray}
|S_{G_1}|+|S_{F}| & \leq  & k.\label{eq:plustwotransp}
\end{eqnarray}
We rewrite~\eqref{eq:subsetstransp} as follows:

\begin{eqnarray}
(G_{1},S_{G_{1}})\oplus (F,S_{F}) & \models & \psi. \label{eq:plustwotramore}
%(G_{1}\oplus F,S_{G_{1}}\cup S_{F}) & \models & \psi. \label{eq:plustwotramore}
 \end{eqnarray}

  Let $(F',S_F')\in{\cal S}$ be the representative 
of $(F,S_{F}).$ As $(F,S_{F})\equiv_{\sigma_\psi}(F',S_F'),$~\eqref{eq:plustwotramore} implies that  
\begin{eqnarray}
(G_{1},S_{G_{1}})\oplus   (F',S_F') & \models & \psi \nonumber \\ 
\iff (G_{1}\oplus F',S_{G_{1}} \cup S_F') & \models & \psi 
\label{eq:plusttwomorezeta}
\end{eqnarray}
From~\eqref{eq:feasiblezeta},~\eqref{eq:plusttwomorezeta}  implies that $\zeta_{G_{1}}(F',S_F')\leq |S_{G_{1}}|.$ From~\eqref{eq:againchoice}, we get $\zeta_{G_{2}}^{\cal S}(F',S_F')\leq |S_{G_{1}}|+c$ which, again from~\eqref{eq:feasiblezeta}, 
means that there exists  $S_{G_{2}},$ where 
\begin{eqnarray}
&& (G_{2}\oplus F',S_{G_{2}}\cup S_{F}')  \models  \psi \ \mbox{and}\label{eq:opluszetatransp}\\
& & |S_{G_{2}}|  \leq  |S_{G_{1}}|+c. \label{eq:opluszetatranspmore}
\end{eqnarray}

We rewrite~\eqref{eq:opluszetatransp} as follows:
\begin{eqnarray}
(G_{2},S_{G_{2}})\oplus (F',S_{F}' ) & \models & \psi. \label{eq:fortythree}
%(G_{1}\oplus F,S_{G_{1}}\cup S_{F}) & \models & \psi. \label{eq:plustwotramore}
 \end{eqnarray}

As $(F',S_{F}')\equiv_{\sigma_\psi} (F,S_{F}),$~\eqref{eq:fortythree} implies that 
\begin{eqnarray*} 
(G_{2},S_{G_{2}})\oplus (F,S_{F}) & \models & \psi \\
\iff (G_{2}\oplus F,S_{G_{2}}\cup S_{F}) & \models &  \psi.
\end{eqnarray*}
 Moreover, 
$|S_{G_2}\cup S_{F}|\leq |S_{G_{2}}|+|S_{F}|\leq^{\eqref{eq:opluszetatranspmore}}|S_{G_{1}}|+c+|S_{F}|\leq^{\eqref{eq:plustwotransp}} k+c.$
We conclude that $(G_{2}\oplus F,k+c)\in\Pi$ and we proved the one direction of~\eqref{eq:transpoprodep}. The other
direction is symmetric.
\end{proof}

\begin{remark}
In Definitions~\ref{def:minstrongmonmin} and~\ref{def:minstrongmonmax} 
we defined the notion of strong monotonicity for  \pmm\ problems
where $S$ is a subset of the vertices of the input graph. If instead 
we ask $S$ to be an edge subset then an analogue 
of Lemma~\ref{lem:stronglymonotone} can be proved  
%\smallskip \sef{What we do with the edge version?}
 in a similar manner.
\end{remark}

Let ${\cal G}$ be a graph class. We say that ${\cal G}$ is \term{CMSO{\em -definable}}
if there exist a sentence $\psi$ on graphs such that 
%\sed{Clarify how we nominate the scope of the formulas} 
${\cal G}=\{G\mid G\models \psi\}$ and, in such a  case,
we say that $\psi$ defines the class ${\cal G}.$
 Recall that,
given a  parameterized graph problem  $\Pi$ and a graph class ${\cal G},$ 
we denote by $\Pi\doublecap {\cal G}$ %\sed{Again this definition...} 
the problem obtained by 
removing from $\Pi$ all instances that 
encode graphs that do not belong to ${\cal G}.$

A necessary tool to adapt our results to problems on special graph
classes is the following. The proof follows  directly by the definitions.
% trivial application of Lemma~\ref{lem:stronglymonotone}.

%\sed{I removed the old ridiculously complicated proof. But still we might need to ad some argument here.}
\begin{lemma}
\label{lem:intersectclose}
Let $\Pi$ be a parameterized problem on graphs and let ${\cal G}$
be a CMSO-definable graph class.
Then if $\Pi$ has FII,  so does $\Pi \doublecap {\cal G}.$
\end{lemma}

%
%\begin{proof} Let $\psi$ be a sentence defining ${\cal G}$ and  \sfdd{Remove this joke from here! Also Lemma~\ref{lem:intersectionfii}}
%let $\psi'=\psi\wedge[S = S]$ (we just added a  trivial MSOL-definable condition about $S$ that is  true for every $S$).
%Clearly $\psi'$ defines a \pmin\ problem $\Pi_{\psi'}.$ We next prove that $\psi'$
%is strongly monotone for $f(t)=0.$ For this we consider 
%a $t$-boundaried graph  $G\in{\cal G}$ and we set $S=\emptyset.$
%It trivially follows that $\zeta_{G}(G',S')=0\Leftrightarrow G\oplus G'\in{\cal G}.$
%Therefore if $\zeta_{G}(G',S')\neq\infty$ then $(G\oplus G',S\cup S')\in\psi'$
%and $|S|=|\zeta_{G}(G',S')|+f(t).$ From Lemma~\ref{lem:stronglymonotone},
%$\Pi_{\psi'}$ has FII. The  lemma follows from 
%%\sed{$\Pi_{\psi}$ is a more general issue here on how we nominate problems of simple, annotated, and general type}
%Lemma~\ref{lem:intersectionfii} as $\Pi\doublecap {\cal G}=\Pi\cap \Pi_{\psi'}.$ 
%\end{proof}
%
%

 \section{Implications of our results}\label{sec:implication}
 %%%%%%%%%%%%%%%%%%%%
 %%FVF
 \newcommand{\fd}{{\sc $p$-$\mathcal{H}$-Deletion}}
 \newcommand{\pvc}{{\sc $p$-Vertex Cover}}
 \newcommand{\pfvs}{{\sc $p$-Feedback Vertex Set}}
\newcommand{\pdhs}{{\sc $p$-Diamond Hitting Set}} 
  \newcommand{\paou}{{\sc $p$-Almost Outerplanar}}
  \newcommand{\parbt}{{\sc $p$-Almost-$t$-bounded treewidth}} 
  \newcommand{\parbp}{{\sc $p$-Almost-$t$-bounded pathwidth}}
  \newcommand{\pfp}{{\sc $p$-${\cal H}$-Packing}}
  \newcommand{\pcyp}{{\sc $p$-Cycle Packing}}
  \newcommand{\psc}{{\sc $p$-${\cal S}$-Covering}}
  \newcommand{\psp}{{\sc $p$-${\cal S}$-Packing}}
%%%%%%%%%%%%%%%%%%%%%%%%%%%%%%input{meta_implications}

In this section we mention a few parameterized problems for which we can obtain either polynomial or linear 
kernel using Theorems~\ref{thm:cmsol},~\ref{thm:cmsolnotanotated}, and~\ref{thm:automata}. In Appendix we provide a full list of the problems amenable to  our approach.
%Problems for 
%which we can obtain either polynomial or linear kernels using our results 
%and the techniques of this section are mentioned in  Appendix.% \sef{Tables and stuff?}

\subsection{Preliminary tools}

All of our results concern problems defined 
on graphs of bounded genus.  Recall that we 
denote by ${\cal G}_{g}$ the class of all graphs of Euler genus at most $g.$ %, where $g\in\Bbb{Z}^{+}.$
In this way for every parameterized problem  $\Pi$ on graphs, we 
define the problem $\Pi_{g}=\Pi\doublecap {{\cal G}_{g}},$ that contains 
only YES-instances of $\Pi,$ encoding graphs of Euler genus at most $g.$
We need to distinguish the two variants $\Pi$ and $\Pi_{g}.$ The reason for this is that, in many cases, for some fixed value $g,$ $\Pi_{g}$ admits  a polynomial kernel
while the general version $\Pi$ is not even believed to be fixed parameter tractable.
A typical example is {\sc Planar Dominating Set} that admits a vertex kernel of size $67k$ while the general 
{\sc Dominating Set} problem is {\sc W[2]}-complete~\cite{DowneyF98}.

 The following lemma is a    direct consequence of the definition of coverability
 and quasi-coverability.

 \begin{lemma}
 \label{lem:subsetcomporquasicomp}
Let $\Pi_{1},\Pi_{2}$ be graph problems whose instances are of the form $(G,k).$
Then if $\Pi_{1}\subseteq \Pi_{2}$ and $\Pi_{2}$ is $r$-(quasi)-coverable, then 
so is $\Pi_{1}.$
\end{lemma} 

%
% \begin{lemma}
% \label{lem:compproduct}
%Let $\Pi$ be an $r$-(quasi)-compact problem  whose instances are of the form $(G,k).$
%\sef{check this!}
%Then for every $c\in\Bbb{Z}^{+},$
%the problem $c\cdot \Pi$ is a  $c\cdot r$-(quasi)-compact problem.
%\end{lemma} 

The next lemma is useful when we work on graphs of bounded genus. 
%with the bounded genus variant of a problem on graphs.  
 
\begin{lemma}
\label{lem:restrictiongenus}
Let $\Pi$ be a parameterized problem on graphs.
If $\Pi$ has FII, then for every $g\in\Bbb{Z}^{+},$ $\Pi_{g}$ has FII.
\end{lemma} 

%We define ${\cal G}_{g}$ as the set of all graphs with Euler genus at most $g.$

\begin{proof}
Let ${\cal O}_{g}$ be the set containing all minor-minimal 
elements of the class of graphs with Euler genus more than $g.$ 
According to the results of~\cite{Mohar99}, ${\cal O}_{g}$ is finite for each fixed  $g.$
Notice that ${\cal G}_{g}=\{G\mid \forall_{H\in{\cal O}_{g}} \ H\nleq_{\rm mn} G\}$ 
and as minor checking can be expressed in CMSO, %\sef{Reference for CMSO of minor checking?}
 the class 
${\cal G}_{g}$ is CMSO-definable. Therefore, the lemma follows from 
Lemma~\ref{lem:intersectclose}.
\end{proof}

\subsection{Covering minors}
\label{subsec:minorcovering}
%
%\ff{Why not Hitting Minors? {\bf sed$>$}: it can be {\em covering, hitting,} or {\em  touching}. Naomi Nishimura  told me once that {\em touching} is more politically correct than {\em hitting}. But I go bowl and prefer {\em covering}.}

A {\em minor-model} of a graph $H$ in a graph $G$
is a minimal subgraph $F$ of $G$ that contains $H$ as a minor. 
Notice that $H\leq_{\rm mn} G$ if and only if $G$ contains as a subgraph
some minor-model of $H.$

%
%\paragraph*{Minor Covering and Packing}
%

We give below a  generic problem that  subsumes many problems in itself. Let ${\cal H}$ be a finite set of connected  graphs
containing at least one planar graph.\medskip
%\sed{For FII it is important that all graphs are connected. For q-c we just want one of them to be planar.}

\begin{center}
\fbox{\begin{minipage}{10cm}
\noindent  \fd \\ %{\sc Vertex-${\cal H}$-Covering}\\ 
{\sl Input:} A graph $G$ and  $k\in\Bbb{Z}^{+}.$\\
{\sl Parameter:} $k.$\\ 
{\sl Question:} Is there  $S\subseteq V(G)$ such that $|S|\leq k$ and $G\setminus S$ does  \\ 
\phantom{Question:}\, not contain any of the graphs from  ${\cal H}$ as a minor?
\end{minipage}}
\end{center}
\noindent\smallskip
%We similarly define {\sc Edge-${\cal H}$-Covering} by demanding $S\subseteq E$ in the above definition.

\begin{lemma}
\label{fii:mincovpackddgssd}
If $\Pi=$\fd, then for every  $g\in\Bbb{Z}^{+},$\
$\Pi_{g}$ is quasi-coverable.
\end{lemma}

\begin{proof}
Let $(G,k)$ be a YES-instance for $\Pi_{g}.$
This means that there exists a set $S\subseteq V(G)$ of cardinality at most $k$  such that none of the 
graphs in  ${\cal H}$ is a minor of $G\setminus S.$ Let $H$ be a  planar 
graph in ${\cal H}.$ As $G\setminus S$ excludes $H$ as a minor and $H$ is planar, it follows from~\cite{RobertsonST94} that $\tw(G\setminus S)\leq c_{H}$ for some constant 
that depends only on $H.$ Set $r=\max\{g,c_{H}\}$ and take an embedding of $G$ in a surface of 
genus at most $g.$ Observe that $G\setminus {\bf R}_{G}^{r}(S)\subseteq G\setminus S,$ therefore, $\tw(G\setminus {\bf R}_{G}^{r}(S))\leq \tw(G\setminus S).$ Thus $\Pi_{g}$ has the $r$-quasi-coverability property
 for some 
$r$   depending on   $ H $ and $g.$\medskip
\end{proof}

%\sed{Put a comment on the planar graph : it is unnecessary}
\begin{lemma}
\label{fii:mincovpacsssskdd}
If $\Pi=$\fd, then for every  $g\in\Bbb{Z}^{+},$\
$\Pi_{g}$ has FII.
\end{lemma}

\begin{proof}
Let $\psi = [\forall {H\in{\cal H}}\  H\nleq_{\rm mn}(G\setminus S)].$
%and observe that $\Pi$ is a \pmin\ problem that is defined by sentence $\psi.$
As minor-checking is CMSO-definable, $\psi$ can be written as a CMSO sentence, hence $\Pi$ is a \pmin\ problem. 
We now prove that $\Pi$ has FII. By Lemma~\ref{lem:stronglymonotone} and~\ref{lem:restrictiongenus},
 it suffices to prove that  $\Pi$ is strongly monotone. 
 Let  %$f:\Bbb{Z}^{+}\rightarrow\Bbb{Z}^{+}$ where $f(t)=t$ and let
  $G$ be a boundaried graph with label set $I$ and the 
  boundary $\delta(G)=B.$ 
Let $S^-$ be a set of  minimum size such that  
  $(G\setminus B)\setminus S^-$ does not contain any of the graphs from  ${\cal H}$
as a minor and let $W=S^-\cup B.$ 

Let $(G',S')\in{\cal U}_{I}$ be  a $\psi$-feasible structure.
%For a feasible structure $(G',S')\in{\cal U}_{t},$ let   $S^{*}\subseteq V(G)$ be a set of minimum size such that 
%$(G\oplus G')\setminus (S^{*}\cup S')$
%contains no graph from  ${\cal H}$ as a minor. 
We first prove that $(G\oplus G',W\cup S')\models \psi.$
%We say that a graph $H$ is an {\em ${\cal H}$-minor} of $G$ if 
%$H\leq_{\rm mn} G$ an $H\in{\cal H}.$
For this, assume in contrary, that $R$ 
is %the vertex set of 
 a minor-model of some $H$ from ${\cal H}$ contained in 
  $(G\oplus G')\setminus (W\cup S').$
As $H$ is connected and $B$ is a separator of $G\oplus G',$
$R$ should be either a subgraph of $G\setminus W=(G\setminus B)\setminus S^{-},$ 
or a subgraph of $(G'\setminus B)\setminus S'.$  
The first case contradicts to the choice of $S^-.$
In the second case, $R$ would be a subgraph of 
$(G'\setminus B)\setminus S',$ which contradicts the feasibility of $(G',S').$
% \subseteq (G\oplus G')\setminus (S'\cup S'),$ again a contradiction.

We next prove that $|W|\leq \zeta_{G}(G',S')+f(|I|),$ where $f(|I|)=|I|.$
For $(G',S')\in{\cal U}_{I},$ let    $S^{*}\subseteq V(G)$ be a set of minimum size such that 
$(G\oplus G')\setminus (S^{*}\cup S')$
contains no graph from  ${\cal H}$ as a minor.  Thus $|S^*|=\zeta_{G}(G',S').$
Notice that  $G\setminus B$ does not contain vertices from $S'.$
Therefore for every $H\in {\cal H},$  every  minor-model $R$ of $H$  in $G\setminus B$
should be intersected by vertices from $S^{*}$---otherwise $R$ would also be 
a subgraph of $(G\oplus G')\setminus (S^{*}\cup S'),$ which is a contradiction.
By the choice of $S^{-},$ we have  $|S^{-}|\leq |S^{*}|.$  
We conclude that $|W|=|S^{-}\cup B|\leq |S^{-}|+|B|\leq |S^{*}|+|B|=\zeta_{G}(G',S')+f(|I|).$
\end{proof}

\fd{} contains various problems as a special case.
Some examples are presented below (all of them are parameterized by solution size $k$).
\begin{itemize}
\item  \pvc{} : In this problem given an input graph $G$ and a $k\in\Bbb{Z}^{+},$ the objective is to 
test whether it is possible to remove at most $k$ vertices from $G$ and 
obtain an edgeless graph. This problem is generated by taking ${\cal H}=\{K_{2}\}.$

\item \pfvs{} : In this problem given an input graph $G$ and a $k\in\Bbb{Z}^{+},$ the objective is to 
test whether it is possible to remove at most $k$ vertices from $G$ and 
obtain an acyclic graph. This problem is generated by taking ${\cal H}=\{K_{3}\}.$
\item 
\pdhs{} : In this problem given an input graph $G$ and a $k\in\Bbb{Z}^{+},$ the objective is to 
test whether it is possible to  remove at most $k$ vertices from $G$ and 
obtain a graph where no edge is contained in more than one cycle.  This problem is generated by taking ${\cal H}=\{K_{4}^{-}\}$ where $K_{4}^{-}$ is the graph obtained from a $K_{4}$ after removing an edge.

\item  \paou{} %\sed{is this name OK?},  
 : In this problem given an input graph $G$ and a $k\in\Bbb{Z}^{+},$ the objective is to 
test whether it is possible to remove at most $k$ vertices from $G$ and 
obtain an outerplanar  graph.   This problem is generated by taking ${\cal H}=\{K_{4},K_{2,3}\}.$
\item \parbt{}  %\textsc{Almost-$r$-bounded treewidth}  %\sed{is this name OK?}, 
: In this problem given an input graph $G$ and a $k\in\Bbb{Z}^{+},$ the objective is to 
test whether it is possible to remove at most $k$ vertices from $G$ and 
obtain a graph of treewidth bounded by some fixed constant $t.$ This problem is generated by taking ${\cal H}$ to be the set of minor minimal graphs with treewidth  $>t$ (from the  results in~\cite{RobertsonST94}, this set  always contains a connected planar graph).
\item \parbp{} %\textsc{Almost-$r$-bounded pathwidth} 
: In this problem given an input graph $G$ and a $k\in\Bbb{Z}^{+},$ the objective is to 
test whether it is possible to remove at most $k$ vertices from $G$ and 
obtain a graph of pathwidth bounded by some fixed constant $t.$
This problem is generated by taking ${\cal H}$ to be the set of minor minimal graphs with pathwidth  bigger than $t$.
% (from the  results in~\cite{BienstockRST91-Qu,Takahashi94-Mi}, this set contains always a tree).\smallskip
\end{itemize}
%
%A variant of the \sef{This is not correct for FII. Edge-strong monotonicity is not able to do anything here. We have to use special tricks to prove FII for edge variants.}
% {\sc Vertex-${\cal H}$-Covering} problem is the 
% {\sc Edge-${\cal H}$-Covering} where the set $S$ that is removed 
% is now an edge set. Counterparts of Lemmata~\ref{fii:mincovpackddgssd} and~\ref{fii:mincovpacsssskdd}
%  can be proved for these problems applying directly the definition of FII.\medskip

\subsection{Packing minors}
\label{packsingsminorfsd}

We consider the following problem that, in a sense, is dual to the one
%\sed{I guess we have nothing about packing of edge disjoint models...}
examined in Section~\ref{subsec:minorcovering}. Again, let ${\cal H}$ be a finite set of connected  graphs
containing at least one planar graph.\medskip

\begin{center}
\fbox{\begin{minipage}{12cm}
\noindent  \pfp \\%{\sc Vertex-${\cal H}$-Packing}\\
{\sl Input:} A graph $G$ and  $k\in\Bbb{Z}^{+}.$\\
{\sl Parameter:} $k.$\\ 
{\sl Question:} Does there exist $k$ vertex disjoint subgraphs $G_{1},\ldots,G_{k}$ of $G$ such\\
\phantom{Question:} \,\!\! that each of them  contains some graph from ${\cal H}$ as a minor.
\end{minipage}}
\end{center}\smallskip
\noindent
%We define {\sc Edge-${\cal H}$-Packing} by demanding that $G_{1},\ldots,G_{k}$ be {\em edge disjoint} 
%subgraphs of $G.$ 

For proving the quasi-coverability of  \pfp, we need to examine its relation to  \fd.

\begin{lemma}
\label{fii:mincovpackdd}
If $\Pi=$\pfp, then for every  $g\in\Bbb{Z}^{+},$\
$\Pi_{g}$ is quasi-coverable.\end{lemma}

\begin{proof}
 Given two  graphs $G$ and $H,$ we define  $\cov_H(G),$  as the minimum size of a set  $S\subseteq V(G)$ of vertices  such that $G \setminus S$ does not contain any minor model of $H.$ 

We also define 
 \begin{eqnarray}
\pack_H(G) & = & \max\{k\mid \mbox{$\exists$   partition $V_{1},\ldots,V_{k}$ of $V(G)$ such that} \nonumber \\
        	                      &  &\hspace{1.7cm} \forall_{i\in\{1,\ldots,k\}}\  G[V_{i}] \mbox{~is a minor-model of $H$}\}. \nonumber 
\end{eqnarray}

Let $H$ be a connected planar graph in ${\cal H}.$ 
To prove that  $\Pi_g$ is quasi-coverable,  we show that 
$\overline{\Pi}_g=((\Sigma^{*}\times \Bbb{Z}^{+})\setminus \Pi_{g})\doublecap {\cal G}_{g}$ 
has the quasy-coverability property. % for some constant $c$ that depends on $g$ 
%and the size of $H.$
In order to do so, we  prove that if $(G,k)\in\overline{\Pi}_g,$ i.e.,  $G\in{\cal G}_g$ and has no  ${\cal H}$-packing into $k$ sets, 
then $(G,ck)$ is a YES-instance for 
  $\Pi^{\rm hd}_g,$  where $\Pi^{\rm hd}=$\fd{},    for 
some constant $c$ that depends 
only on $g$ and $H.$  By Lemma~\ref{fii:mincovpackdd},\fd{}  is $r$-quasi-coverable, and thus $\overline{\Pi}_g$ would posses a quasi-coverability property.
  
%
%This will imply that $\Pi'_{g}\subseteq c\cdot \Pi^{\rm hd}_{g}$\sef{There is an issue here with the product definition}
%and the quasi-coverability of $\Pi_{g}$ will follow from Lemma~\ref{lem:subsetcomporquasicomp} and~\ref{lem:compproduct} and the quasi-coverability of $\Pi^{\rm hd}_{g}$ (Lemma~\ref{fii:mincovpackdd}).\smallskip
%

Suppose that $(G,k)\in\overline{\Pi}_g.$ % is a NO-instance of $\Pi$ where $G\in{\cal G}_g.$ 
 This implies that $\pack_H(G)< k.$
According to the Erd\H{o}s-P\'osa type of result of~\cite{FominST08imp}, for every two graphs 
$H$ and $W,$ where $H$ is planar and $W$ is any graph, there exists 
a constant $c_{H,W},$ depending only on $H$ and $W,$ such
that for every graph $G$ excluding $W$ as a minor,  
$\cov_H(G)\leq c_{H,W}\cdot \pack_{H}(G).$ Let $W$ be a graph of Euler
genus $g+1.$ As the class ${\cal G}_{g}$ is closed under taking of minors, we have that every graph in ${\cal G}_{g}$ excludes $W$ as a minor.
Applying the aforementioned result, we have that  ${\bf cov}_{H}\leq c_{H,W}\cdot k,$ therefore $(G,c\cdot k)$ is a YES-instance for $\Pi^{\rm hd}_{g}$ for 
some $c$ depending only on $H$ and $g,$
as required.  This implies that $\overline{\Pi}_g$  has a quasi-coverability property, hence  ${\Pi}_g$ is quasi-coverable.
\end{proof}

Notice that when ${\cal H}=\{{K_3}\},$  \pfp{}
 is the \pcyp{} problem. Here, given an input graph $G$ and a $k\in\Bbb{Z}^{+},$ the objective is to check 
whether $G$ contains  $k$ vertex-disjoint cycles. While the general problem has FII for 
 every choice of ${\cal H},$ we present the proof for this special case in order 
to clearly explain the machinery that we use for such type of problems.  
 After the end of the proof of Lemma~\ref{lem:cvpfiigenus}, we outline  how to extend the proof 
 for the general case.

\begin{lemma}
\label{lem:cvpfiigenus}
If $\Pi=$\pcyp, then for every  $g\in\Bbb{Z}^{+},$\
$\Pi_{g}$ has FII.\end{lemma}
 
 \begin{proof} By Lemma~\ref{lem:restrictiongenus}, it is sufficient to  prove  that $\Pi$ has FII.
Let $G$ be a boundaried graph with label set $I$ and with   boundary $\delta(G)=B^*$. The proof proceeds 
in three stages: the first stage defines some characteristic  of the problem 
that depends on the boundary of the input boundaried graph.  
%is particular to the boundary of every $t$-boundaried graph. 
The second  uses this characteristic  to define 
an equivalence relation on boundaried graphs that will have finite index, and the last one proves that this equivalence 
relation is a refinement of $\equiv_{\Pi}$ and  therefore has finitely many equivalence classes  
as well.

\medskip

\noindent{\it Characteristic.}
We define set  ${\cal R}$  as the set of all matchings $R$  (not necessarily maximal)   of a complete graph on the vertex set 
$B^*$.
% such that for each $R\in {\cal R},$ every connected component of $R$ is an edge. 
 Let us remark,  that matching  $R\in {\cal R}$ is not necessarily  a subgraph of $G$;  each graph in ${\cal R}$ corresponds 
to a  set of mutually disjoint pairs 
from $B^*.$ 
We define $\zeta_{G}: {\cal R}\rightarrow \Bbb{Z}^{+}$
so that, for every $R\in{\cal R},$ the value $\zeta_{G}(R)$ is the maximum number 
of cycles that can be contained in a subgraph $J$ of $G$ such that: 
\begin{itemize}
 \item $\Delta(J)\leq 2,$ and 
  \item  for every edge $\{x,y\} $ of $R,$  $J$ contains an 
$(x,y)$-path.
 \end{itemize}
 Let us remark that all $(x,y)$-paths of $J$ are internally vertex disjoint. 
%  
%  connected component $P,$ i.e. a path, of $R,$  $J$ contains internally vertex disjoint 
%$(x,y)$-paths for every $\{x,y\}\in E(G).$ 
In case such a graph 
$J$ does not exist, we set $\zeta_{G}(R)=-\infty.$
Function $\zeta_{G}$ can be seen as a way 
to encode the tables 
of a dynamic programming for \pcyp{} 
on graphs of treewidth at most $|I|.$ 
The proof that follows can be  seen as an
alternate way to prove that such a dynamic programming algorithm  uses  
tables whose sizes depend only on $|I|.$ 
\medskip

\noindent{\it Definition of equivalence.} %We set $L_G=\{\zeta_{G}(R)\mid R\in{\cal R}\}\setminus \{-\infty\}$
%and observe that  
Let $x$ be the maximum number of vertex disjoint cycles in $G.$ Thus 
%$x=\zeta_{G}(R_\emptyset)\geq 0,$ where $R_{\emptyset}$ is the empty graph, 
%and  
 for every $ {R\in{\cal R}},$ we have  $\zeta_{G}(R)\leq x.$ % ($x$ is the maximum number of disjoint cycles $G$ may have).
We define the {\em  signature} of $G$ as the function $\chi_{G}: {\cal R}\rightarrow \{-|I|,\ldots,0\}\cup\{-\infty\}$ such that 
\begin{eqnarray*}
\hspace{-0mm} \chi_{G}(R)\!=\!\left\{\begin{array}{lll}
&\hspace{-0mm} \zeta_{G}(R)-x & \mbox{if  $x-|I|\leq \zeta_{G}(R)\leq x$}\\
&- \infty& \mbox{otherwise}\label{eq:feasiblezetamaxcyclepack}
%\\ & {\bf left}_{G}(y_{1}^{i,2l}) & \mbox{if $y_{1}^{i,2l}\in S$ and $i$ is odd}
\end{array}\right.
\end{eqnarray*}
Notice that the number of different signatures is bounded by some function of $|I|.$
Given two boundaried graphs $G_{1}$ and $G_{2},$ 
we say that $G_{1}\sim G_{2}$ if and only if $\Lambda(G_1)=\Lambda(G_2)$  and $\chi_{G_{1}}=\chi_{G_{2}}.$
Clearly, for every $I\subseteq \mathbb{Z}^{+}$,  $\sim$ is an equivalence relation with finite number of equivalence classes.  \medskip

\noindent{\it Refinement proof.}
The result will follow if we prove that 
$\sim$ is a refinement of $\equiv_{\Pi}.$ For this we claim that if $G_{1}\sim G_{2}$
then $G_{1}\equiv_{\Pi} G_{2}$ or, equivalently, there is some constant $c$, depending on $G_1$ and $G_2,$ such that
\begin{eqnarray}
\forall (F,k)\in {\cal F} \times \Bbb{Z} & &  (G_1 \oplus F, k) \in \Pi\Leftrightarrow(G_2 \oplus F, k+c) \in \Pi.\label{eq:refinementcyclepack}
\end{eqnarray}

Suppose that $G_{1}\sim G_{2}.$ Let $(F,k)\in {\cal F}\times \Bbb{Z}$ such that $(G_1 \oplus F, k) \in \Pi.$ Our target is to prove that  $(G_2 \oplus F, k+c) \in \Pi.$ (The proof for other direction of~\eqref{eq:refinementcyclepack} is symmetric and thus omitted.) Let us also assume that 
$G_1$ and $G_2$ are boundaried graphs with label set $I$ and    $\delta(G_1)=B$. 

The fact that $(G_1 \oplus F, k) \in \Pi$ means that 
$G_1\oplus F$ contains a collection of $k$ disjoint cycles.
Let ${\cal C}$ be such a collection of maximum size in $G_1\oplus F.$ Clearly, $|{\cal C}|\geq k.$
We partition ${\cal C}$  into four sets ${\cal C}_{G_1},$ ${\cal C}_{B},$ ${\cal C}_{F}^{B},$ and ${\cal C}_{F},$ where
\begin{itemize} 

\item ${\cal C}_{G_1}$ are the cycles that are entirely 
inside $G_1,$ 
\item   ${\cal C}_{B}$ are the cycles of ${\cal C}$ that are not entirely in $G_{1}$ or $F,$
\item ${\cal C}_{F}^{B}$ are the cycles that are  entirely
inside $F$ and intersect the boundary $B,$ and 
\item ${\cal C}_{F}$ are the cycles that are  entirely
inside $F$ and do not  intersect  $B.$
\end{itemize}
Notice that $|{\cal C}_{B}|+ |{\cal C}_{F}^{B}|\leq |I|.$
%Clearly,  ${\cal C}^{*}\leq t.$
 Graph $G_1\cap (\bigcup_{C\in {\cal C}_{B}}C)$ is a collection of internally disjoint 
paths between pairs of terminals  in $B.$ By replacing each of these 
paths by edges, we create   graph $R\in {\cal R}.$ Graph $R$ represents  the possibility of linking the pairs corresponding to the edges in ${\cal R}$
by disjoint paths inside $G_{1}$ in a way that these paths are disjoint from the 
disjoint cycles in ${\cal C}_{G_{1}}.$

For $i\in\{1,2\},$  
let ${\cal C}^{*}_{i}$ be a maximum size collection of cycles in $G_{i},$ and let $x_{i}=|{\cal C}^{*}_{i}|.$
 Notice that $x_{1}$ and $x_{2}$
depend only on $G_{1}$ and $G_{2}.$ We claim that 
$x_1-|I|\leq |{\cal C}_{G_{1}}|.$ Indeed, 
${\cal C}^*={\cal C}^{*}_1\cup {\cal C}_{F}$ is also a cycle packing in $G_1\oplus F.$ If
 $|{\cal C}_{G_{1}}|< x_1-|I|=|{\cal C}^{*}_1|- |I|,$  then
$|{\cal C}^*|=|{\cal C}^{*}_1|+|{\cal C}_{F}|>|{\cal C}_{G_{1}}|+ |I| +|{\cal C}_{F}|\geq |{\cal C}_{G_{1}}|+ |{\cal C}_{B}|+ |{\cal C}_{F}^{B}|+|{\cal C}_{F}|=|C|,$  contradicting the maximality of ${\cal C}.$  

 We set $c=x_{2}-x_{1}.$
By the definition of $\zeta_{G},$ we have that $|{\cal C}_{G_1}|\leq \zeta_{G_1}(R)\leq x_{1}.$ We conclude that $x_1- |I| \leq \zeta_{G_1}(R)\leq x_1$ and thus $\chi_{G_{1}}(R)>-\infty.$
As $G_{1}\sim G_{2},$ we have that $\chi_{G_{1}}(R)=\chi_{G_{2}}(R),$
 and therefore $\zeta_{G_{2}}(R)=\zeta_{G_{1}}(R)-x_{1}+x_{2}=\zeta_{G_{1}}(R)+c\geq |{\cal C}_{G_{1}}|+c.$ This in turn, means that $G_{2}$ contains a collection of disjoint cycles ${\cal C}_{G_{2}}$ and $|{\cal C}_{G_{2}}|= \zeta_{G_{2}}(R)\geq |{\cal C}_{G_{1}}|+c$  and $|E(R)|$ internally vertex disjoint paths 
  that are also disjoint from the cycles in ${\cal C}_{G_{2}},$ 
 one for each pair of vertices represented by the edges of $R.$
 
 Notice now that if we take the union of these paths with the graph $F\cap (\bigcup_{C\in {\cal C}_{B}}C),$  we obtain  a collection ${\cal C}_{B}'$ of $|{\cal C}_{B}|$ vertex disjoint 
 cycles in $G_{2}\oplus F$ that are also disjoint with the cycles from  ${\cal C}_{G_{2}}.$ The cycles from  ${\cal C}_{G_{2}}\cup {\cal C}_{B}$
 are disjoint from cycles  ${\cal C}_{F}^{B}$ and ${\cal C}_{F}.$
 Therefore,  ${\cal C}_{G_{2}}\cup {\cal C}_{B}'\cup {\cal C}_{F}^{B}\cup {\cal C}_{F}$ is a collection of cycles in $G_{2}\oplus F$ that has size at least $|{\cal C}_{G_{1}}|+c+|{\cal C}_{B}|+|{\cal C}_{F}^{B}|+|{\cal C}_{F}|=k+c.$
 We conclude that $(G_{2}\oplus F,k+c)\in\Pi$ as required.\smallskip
 \end{proof}

The proof that, in general, \pfp{} has FII 
follows  the same line as the proof of Lemma~\ref{fii:mincovpackdd}.
Instead of cycles we have minor-models of graphs in ${\cal H}$ and instead of paths between terminals of the border, we 
have {\em partial models} 
that are parts of minor-models of graphs in ${\cal H}$ 
that are cropped by $G_{1}.$ The signature $\chi$ is now encodes 
all the ways such partial models might be ``rooted'' in the boundary.
This can be done by the ``folio'' structure introduced in~\cite{RobertsonS-GMXIII}
for doing dynamic programming for the minor checking problem 
and the disjoint paths problem on graphs of bounded treewidth.
Variants of folios have been used for similar purposes in~\cite{AdlerGK08comp,GroheKMW11find,KaminskiT12cont,FominLMS12}.
%\ff{\cite{GroheM12}???}

\subsection{Subgraph Covering and Packing} 
\label{lem:subcovpack}
Let ${\cal S}$ be a  finite set of connected graphs.   We define the following two general problems.

\begin{center}
\fbox{\begin{minipage}{12cm}
\noindent \psc \\ %{\sc Vertex-${\cal S}$-Covering}\\ 
{\sl Input:} A graph $G$ and   $k\in\Bbb{Z}^{+}.$\\
{\sl Parameter:} $k.$\\ 
{\sl Question:} Is there   a $S\subseteq V(G)$ such that $|S|\leq k$ and $G\setminus S$ contains\\ 
\phantom{Question:} no subgraph  isomorphic to a graph from  ${\cal S}$?
\end{minipage}}
\end{center}
\noindent
%We similarly define {\sc Edge-${\cal H}$-Covering} by demanding $S\subseteq E$ in the above definition.

\begin{center}
\fbox{\begin{minipage}{12cm}
\noindent \psp \\
{\sl Input:} A graph $G$ and   $k\in\Bbb{Z}^{+}.$\\
{\sl Parameter:} $k.$\\ 
{\sl Question:} Does there exist $k$ vertex disjoint subgraphs $G_{1},\ldots,G_{k}$ of $G$ such\\
\phantom{Question:} that each of them  contains a subgraph isomorphic to a graph in ${\cal S}$?
\end{minipage}}
\end{center}
\noindent
%We define {\sc Edge-${\cal H}$-Packing} by demanding that $G_{1},\ldots,G_{k}$ be {\em edge disjoint} 
%subgraphs of $G.$ 

%\noindent
%We define  {\sc Edge-${\cal S}$-Covering} by considering $S$ to be an edge set. %subgraphs of $G.$ 
%Also,  we define {\sc Edge-${\cal S}$-Packing} by demanding that $G_{1},\ldots,G_{k}$ are {\em edge disjoint} \sef{What and where we say about this?}
%subgraphs of $G.$ \smallskip\ff{come back to that and add that edge are some}

Let us remark that it is not true in general, that   if $\Pi=$\psc{} or $\Pi=$\psp, then $\Pi_{g}$ is coverable.
However, the problems become coverable  if we modify instances by applying  the following  simple preprocessing rule.
\medskip

\begin{center}
{\begin{minipage}{10.5cm}
{\sf Redundant Vertex  Rule}:   For a  graph  $G,$ while this is possible, delete a vertex   that does not belong to  any subgraph of $G$ isomorphic 
to any graph in ${\cal S}.$  
\end{minipage}
}
\end{center}
\medskip

A graph $G$ is  {\sf RV}-{\em ${\cal S}$-reduced} if each its vertex  belongs to a subgraph isomorphic to a graph in ${\cal S}$.
%the application of the above rule on $G$ does not change the graph.
 We denote by ${\cal R}({\cal S})$ the set
of all  {\sf RV}-{\em ${\cal S}$-reduced}  graphs.
%Given a problem $\Pi$ whose instances are of the form $(G,k),$
%we define   its restriction to reduced 
%graphs $\Pi^{\sf RV}=\Pi\doublecap {{\cal R}({\cal S})}.$
%

\begin{lemma}
\label{preprofss}
Let $\Pi$ be  either  \psc{}
  or \psp. There is a polynomial time algorithm transforming  $(G,k)\in \Pi_g$ into an equivalent instance  $(G',k)\in \Pi_g^{\sf RV}=\Pi_g\doublecap {{\cal R}({\cal S})}.$
%  
%  
%  
%There is a %n $O(n^{2})$ 
%polynomial-time reduction from $\Pi_{g}$ to $\Pi^{\sf RV}_{g}.$
\end{lemma}

\begin{proof}
%The lemma follows from the above discussion as
% the application of the  {\sf RV} rule always creates an equivalent instance and 
%no more than $O(n)$ applications for this rule are possible.
Let $s$ be the maximum diameter of a graph in ${\cal S}$ and let $G$ be a graph of genus $g.$
We can perform the {\sf Redundant Vertex  Rule} in $O(|V(G)|^2)$  time by checking  for every vertex $v\in V(G)$   if the subgraph $G^s(v)$ induced by ${\bf B}^{s}_G (v)$ has a subgraph isomorphic to a 
graph in $\cal S$ containing vertex $v.$ By Proposition~\ref{boundtwsurf2}, the treewidth of  $G^s(v)$  is bounded by some function of $s$ and $g$  only and thus  for every $v$ such a check can be performed in time  $O(|V(G)| ),$ see, e.g. \cite{Eppstein00}.
\end{proof}

We are now ready to prove the following lemma.

\begin{lemma}
\label{fii:subgraohcovpack}
Let $\Pi$ be   
\psc{}  or
\psp.    Then $\Pi^{\sf RV}_{g}$ is coverable.
\end{lemma}

\begin{proof}
Let $s$ be the maximum diameter of a graph in ${\cal S}$
and let $\Upsilon=$\psc.
Let $(G,k)$ be a YES-instance 
of $\Upsilon_{g}^{\sf RV}$ and let $S$ be a vertex  set of size  at most $k$
  such that each subgraph of $G$ that is isomorphic to some graph in ${\cal S}$  intersects $S.$ 
 Consider an embedding of $G$ in some surface of Euler genus at most $g.$
As $G\in {\cal R}({\cal S}),$  every vertex  in $G$ is within distance at most $s$ from $S.$ Therefore, ${\bf B}_G^{s}(S)=V(G).$ By Observation~\ref{obs:radialmetric},  ${\bf R}_G^{2s}(S)\supseteq {\bf B}_G^{s}(S)$ and  
thus $\Upsilon^{\sf RV}_{g}$ has the $r$-coverability property for
$r=2s.$  %$r=\max\{2s,g\}.$

Assume now that  $\Psi=$\psp. To prove the coverability of $\Psi^{\sf RV}_{g},$ we will prove that 
$\bar{\Psi}^{\sf RV}_{g}=((\Sigma^{*}\times \Bbb{Z}^{+})\setminus \Psi^{\sf RV}_{g})\doublecap {\cal G}_{g}$ 
has the $r$-coverability property. % for some constant $c$ that depends on $g$ 
%and ${\cal S}.$ 
 Let $c$   be  the maximum number of 
vertices in a graph of ${\cal S}.$
We claim that if $(G,k)$ is a NO-instance for $\Psi_g^{\sf RV},$ 
where $G\in{\cal G}_g,$ 
then $(G,ck)$ is a YES-instance of  $\Upsilon^{\sf RV}_{g}.$ 
Indeed,  as $(G,k)$ is a NO-instance,  
$G$  does not contain $k$ vertex  disjoint subgraphs from  ${\cal S}.$
A set $S$ of vertices of size $\leq k\cdot c$
 ``hitting"  all subgraphs of $G$  isomorphic to   graphs in ${\cal S}$ can be constructed by the following greedy procedure:
\begin{quote}
{\sl Initialize  $S=\emptyset$ and,   
as long as $G$ contains a subgraph that is isomorphic to some graph in ${\cal S},$ add all its vertices to $S$ and remove them from $G.$}
\end{quote}
Notice that 
the above procedure  cannot be applied more than $k-1$ times, otherwise
the removed graphs would constitute  
a vertex  packing of graphs of ${\cal S}$ in ${ G}.$ When the procedure cannot be applied anymore, the set $S$ intersects every 
subgraph of $G$ that is isomorphic to some graph from ${\cal S}$
and $|S|\leq c\cdot (k-1).$
Therefore $(G,ck)$ is a YES-instance of $\Upsilon^{\sf RV}_{g},$ which is already shown to be coverable. Now the coverability of 
 $\Psi_{g}^{\sf RV}$  follows from Lemma~\ref{lem:subsetcomporquasicomp}. 
%
%
%
%This would  imply 
%that $\Lambda_{g}\subseteq c\cdot \Upsilon^{\sf RV}_{g}$
%and then the  coverability of $\Psi_{g}^{\sf RV}$ will follow from Lemma~\ref{lem:subsetcomporquasicomp} and~\ref{lem:compproduct} \ff{$c\Pi$ issue} and the quasi-coverability of $\Upsilon^{\sf RV}_{g}.$\smallskip
%
%
%
%Let $(G,k)$ be a NO-instance of $\Psi^{\sf RV}$ where $G\in{\cal G}_g.$ 
\end{proof}

Using a  modification of the proof of Lemma~\ref{fii:mincovpacsssskdd},
%\sef{IS this some kind of absolute jargon? Are we sure? Yes it is a highly non trivial issue. Let's discuss it!}
it is possible to show  that\psc{} %and  {\sc Edge-${\cal S}$-Covering}
  has FII. The proof that  \psp{} %and  {\sc Edge-${\cal S}$-Packing} 
   has FII follows the same steps as in the proof of Lemma~\ref{lem:cvpfiigenus}. The only difference in all cases 
is that we work with subgraphs instead of minors.

\subsection{Domination and its variants} 
\label{domitsvarl}

Given two integers $r,q\in\Bbb{Z}^{+},$ a graph $G,$ and a set $S\subseteq V(G),$
we say that $S$ is a \emph{$(q,r)$-dominating set of $G$} if for every 
vertex $x$ in $V(G)\setminus S,$ there are at least $q$ vertices in $S$ within distance at most $r$ from $x.$ 
We define a series of problems related to domination. In all of them 
the input is a graph $G$ and a parameter $k\in\Bbb{Z}^{+}.$  We mention below the variants and the questions corresponding to each of them.

%
%\subsection{Domination and its Variants} 
%

%Given two integers $r,q\in\Bbb{Z}^{+},$ a graph $G,$ and a set $S\subseteq V(G),$
%we say that $S$ is a \emph{$(q,r)$-dominating set of $G$} if for every 
%vertex $x$ in $V(G)\setminus S,$ there are at least $q$ vertices in $S$ that are in distance at least $r$ from $x.$

\begin{itemize}
\item {\sc $p$-$r$-Dominating Set}: Is there  
a  $(1,r)$-dominating set $S$ of size at most $k$ in $G$? For $r=1$ the problem is known as  {\sc $p$-Dominating Set}.
 \item {\sc $p$-$q$-Threshold Dominating Set}:  Is there  
a  $(q,1)$-dominating set $S$ of size at most $k$  in $G$?
 \item   {\sc $p$-Efficient Dominating Set}:  Is there  
a  $(1,1)$-dominating set $S$ of size at most $k$ in $G$ such that $G[S]$ is edgeless (i.e. $S$ is an independent set) and each vertex from $V(G)\setminus S$ is adjacent to exactly one vertex in $S.$ This problem is also known as {\sc $p$-Perfect Code}.

\item  {\sc $p$-Connected Dominating Set}: 
 Is there   a $(1,1)$-dominating set $S$ of size at most $k$  in $G$  such that  $G[S]$ is connected?
 \end{itemize}

\begin{lemma}
\label{fii:domination}
If $\Pi$ is one of the following problems:
 {\sc $p$-$r$-Dominating Set},  {\sc $p$-$q$-Threshold Dominating Set}, {\sc $p$-Efficient Dominating Set}, then for every $g\in\Bbb{Z}^{+},$ $\Pi_{g}$ is coverable and has FII.
\end{lemma}
\begin{proof}
For all these problems, $\Pi_{g}$ is $2r$-coverable by definition because if $S$ % \ff{Is (2,2)-dominating set FII?}
 is a $(q,r)$-dominating set of $G$ and $G$ is embeddable in some 
surface of Euler genus at most $g$ then, by  Observation~\ref{obs:radialmetric}, ${\bf B}_{G}^{r}(S)\subseteq {\bf R}_{G}^{2r}(G).$

By Lemma~\ref{lem:restrictiongenus}, it is enough to prove that  each of the problems has FII.
We start from  {\sc $p$-$r$-Dominating Set}. Since {\sc $p$-$r$-Dominating Set}
 is a \pmin\ problem, by Lemma~\ref{lem:stronglymonotone}, it is enough to 
prove that it is strongly monotone.
For a boundaried graph $G$ with label set $I$ and boundary $\delta(G)=B,$  let $S''\subseteq V(G)$ be a minimum sized $r$-dominating set of $G.$ 
We put  $W=S''\cup B.$  
For a boundaried  structure
$(G',S')\in {\cal U}_I,$  let $S^{*}\subseteq V(G)$
be a set of minimum size such that $S^{*}\cup S'$ is an $r$-dominating set of $G\oplus G'.$
Thus $\zeta_{G}(G',S')=|S^{*}|.$
Observe  that   $S^{*}\cup B$ is an $r$-dominating set of $G,$  hence $|S''|\leq |S^{*}|+|B|.$  
Therefore, $|W|= |S''\cup B|\leq |S''|+|B|\leq |S^{*}|+2|I| = \zeta_{G}(G',S')+2|I|.$
Also observe that   $W\cup B$ is an $r$-dominating set of $G',$ and thus $W\cup S'$ is an $ r$-dominating set of $G\oplus G'.$
This implies that    
$(G\oplus G',S\cup S')\in \Pi$ and  the strong monotonicity of  {\sc $p$-$r$-Dominating Set} follows.

 The proof that {\sc $p$-$q$-Threshold Dominating Set} is strongly monotone is based on the same observations as the proof for  {\sc $p$-$r$-Dominating Set} and thus omitted. 
To prove that {\sc  $p$-Efficient Dominating Set} has FII, we use the fact that 
\begin{eqnarray*}
 \mbox{\sc  $p$-Efficient Dominating Set} & = &\mbox{\sc $p$-$1$-Dominating Set}\doublecap {\cal G}^{\rm eds}, 
\end{eqnarray*}
where  ${\cal G}^{\rm eds}$ is the class of all graphs that have an efficient dominating set. The equality follows from a theorem of~\cite{BangeBS88}, 
asserting that if a graph $G$ has an efficient dominating set, then the size of the minimum efficient dominating set is equal to the size of the minimum  dominating set of $G.$  As ${\cal G}^{\rm eds}$ is CMSO-definable, {\sc $p$-Efficient Dominating Set} has FII by  Lemma~\ref{lem:intersectclose}.
\end{proof}

In the remaining part of this subsection, we prove that
when $\Pi$ is {\sc $p$-Connected Dominating Set}, then  $\Pi_{g}$ is  coverable and has FII. For this we  first need some 
auxiliary definitions and results on connected domination.
Given a graph $G$ and a set $V(G)$ we say
that a dominating set $S$ is a {\em component-wise} connected dominating set 
of $G$ if  
for every connected component $C$ of $G,$ $C[S\cap V(C)]$ is connected. In particular, if $G$ is connected, then every component-wise dominating set of $G$ is also a connected dominating set of $G.$

We need the following proposition attributed to \cite{Duchet82}
\begin{proposition}
\label{lem:bb}
Let $G$ be a connected graph and let $Q$ be a dominating set of $G$ such that $G[Q]$ has at most $\rho$ connected components. Then 
there exists a set $Z\subseteq V(G)$ of size  at most $2\cdot (\rho-1)$  
 such that $Q\cup Z$ is a connected dominating set in $G.$
\end{proposition}
%
%\begin{proof}
%Let $T$ be a minimum size tree in $G$
%that meets all sets in the connected components of $G[Q],$ (this  a minimum size group Steiner tree).
%Notice that, by its minimality, such a tree cannot contain 
%a subpath consisting of more than $2$ internal vertices that do not belong in $Q.$
%Let $Z$ be the internal vertices of all these subpaths.
%By the definition of $Z,$ it easily follows that $|Z|\leq 2(\rho-1).$
%Clearly $V(T)$ is a connected dominating set of $G$
%and the lemma follows as $|V(T)|=Z\cup Q.$
%\end{proof}

\begin{lemma}
\label{lemma:dominmon2}
Let $G$ be a graph   and let $B$  be 
a subset of $G.$ Let also 
$R$ be  a component-wise connected dominating set of $G.$
Then there exists a set $S\supseteq R\cup B$ that is also a component-wise 
connected dominating set of $G$  and has at most $|R|+3|B|$ vertices.
\end{lemma}

\begin{proof}
Let ${\cal C}$ be the set of connected components of $G.$
For $C\in {\cal C},$  let $B_{C}=V(C)\cap B$ and $R_{C}=R\cap V(C).$
Observe that 
$C[B_{C}\cup R_{C}]$ cannot have more than $1+|B_{C}|$ connected 
components. By Proposition~\ref{lem:bb}, there exists 
a set $Z_{C}\subseteq V(C)$ such that $Z_{C}\cup R_{C}\cup B_{C}$ induces a connected subgraph of $C$  such that 
$|Z_{C}|\leq 2|B_{C}|.$ This means  that $|B_{C}\cup R_{C}\cup Z_{C}|\leq |R_{C}|+3|B_{C}|.$ Moreover, as $R_{C}$ is a dominating set of $C,$ the same holds for its superset $B_{C}\cup R_{C}\cup Z_{C}.$
Therefore, the set $S=\bigcup_{C\in{\cal C}}B_{C}\cup R_{C}\cup Z_{C}$ is a component-wise dominating set of $G$
that containing $B\cup R.$   
It is now easy to check that $|S|\leq |R|+3|B|.$
\end{proof}

%label set $I$ and boundary $\delta(G)=B,$
\begin{lemma}
\label{lem:dominmon1}
Let    $G$ and $G'$ be boundaried graphs with label set $I$ and boundary $\delta(G)=B$.  Let also $S^*\subseteq V(G)$ and $S'\subseteq V(G')$ such that $S^{*}\cup S'$ is a 
  component-wise connected dominating set of $G\oplus G'.$ Then $G$ contains a  component-wise connected dominating set $S^+$ of size at most $3|B|+|S^*|.$%\sed{This equally indexed is the correct! In other places the nomenclature may be wrong!}
\end{lemma}

\begin{proof}
We first prove  the lemma under the assumption  that $H=G\oplus G'$ is a connected graph. Let us remark that $G$ is not necessarily connected.
Notice that $Q=S^*\cup B$
is a dominating set of $G.$
Let $C_{1},\ldots,C_{\mu}$ be the connected components of $G$ and, for each $i\in\{1,\ldots,\mu\},$
let $Q_{i}^{1},\ldots, Q_{i}^{\delta_{i}}$ be the vertex sets 
of the connected components of $C_{i}[V(C_{i})\cap Q].$ 
We claim that  $\sum_{1\leq i\leq \mu}\delta_{i} \leq |B|+1.$
Indeed, if $S^*\cup S'$ does not intersect $B,$ then  since $H[S^*\cup S']$ is connected we have that $G[S^*\cup S']$ is connected  
and in this case $Q$ may have at most $|B|+1$ connected components, therefore 
$\sum_{1\leq i\leq \mu}\delta_{i} \leq |B|+1.$
In case  $S^*\cup S$ intersects $B,$ then each connected component of $Q$ should  contain at least one vertex of $B,$  and, again, we have
  $\sum_{1\leq i\leq \mu}\delta_{i} \leq |B|<|B|+1.$

We now apply Proposition~\ref{lem:bb} for the sets $Q_{i}^{1},\ldots, Q_{i}^{\delta_{i}}$ of the graph $C_{i},$  for each $i\in\{1,\ldots,\mu\}.$  That way we find, for every $i\in\{1,\ldots,\mu\},$ a collection of sets $Z_{1},\ldots,Z_{\mu},$ where $Z_{i}$ is a connected dominating set of $C_{i}.$
This means that $S^{+}=\bigcup_{1\leq i \leq \mu} Z_{i}$
is a component-wise connected dominating set of $G.$
By Proposition~\ref{lem:bb}, $|Z_{i}|  \leq  2(\delta_{i}-1)+|V(C_{i})\cap Q|.$
We now  have that: 
\begin{eqnarray*}
|S^+|  & = &  \sum_{i= 1}^{\mu} |Z_{i}|\\
& \leq &   \sum_{i= 1}^{\mu} 2(\delta_{i}-1)+\sum_{i= 1}^{\mu}|V(C_{i})\cap Q| \\
& \leq &  2|B|+|Q| 
= 3|B|+|S^*|
\end{eqnarray*}
as required. 

If $G\oplus G'$ is not a connected graph, then the required component-wise connected dominating set is the union of the component-wise connected dominating sets obtained if we apply the above proof for each of the connected components of $G\oplus G'.$
\end{proof}

We also need the following lemma. The proof is based on the definition of connected dominating set and  is omitted.
%\sed{I omit it because it is a mess to write it... But it is correct! I suggest we leave it like that except if a simple half page proof is found.}

\begin{lemma}
\label{lem:superbb}
 Let $G$ and $G'$ be boundaried graphs with label set $I$ and boundary $\delta(G)=B$ 
%with a common boundary $B$ 
such that  $C=G\oplus G'$ is connected.
Let also
$S^*\subseteq V(G)$ and $S'\subseteq V(G')$ be such that $S^*\cup S'$ is a connected dominating set of $C.$
Let $S\subseteq V(G)$ be a component-wise dominating set of $G$ such that  $B\subseteq S.$ Then $S\cup S'$ is a connected dominating set of $G\oplus G'.$
\end{lemma}

%
%\begin{proof}
%Notice that $Q=S_{1}\cup S_{2}$ is a dominating set of $G=G_{1}\oplus G_{2}.$ Let $C_{1},\ldots,C_{\gamma}$
%be the connected components of $G,$ let $Q_{i}=Q\cap V(C_{i})$ and let $B_{i}=Β\cap V(C_{i}),  i=1,\ldots,\gamma.$
%For each $i\in\{1,\ldots,\gamma\},$ the connectivity of $C_{i}$
%implies that
%the number of the connected components 
%of $G_{1}[C_{i}\cap V(G_{1})]$ plus the number of the connected components 
%of $G_{2}[C_{i}\cap V(G_{2})]$ cannot be more than $|B_{i}|+1.$
%As $Q_{i}$ indices a connected subgraph of $Q_{i},$ 
%it follows  that $C_{i}[Q_{i}]$ cannot have more than $|B_{i}|+1$
%connected components. Let $Q_{1}^i,\ldots,Q_{\rho}^{\delta_{i}}$
%be the vertex sets of the connected components of $C_{i}[Q\cap V(C_{i})].$
%From Lemma~\ref{lem:bb}, there exists 
%a set $Z_{i}\subseteq V(C_{i})$ such that $Q_i\cup Z_i$
%is a component-wise connected dominating set 
%of $C_{i}$ and where $|Z_i|\leq 2|\rho_i-1|.$
%Notice that $|Q_{i}\cup Z_{i}|\leq |Q_{i}|+2|B_{i}|$
%and that, if $Z=\bigcup_{i\in\{1,\ldots,\rho\}}Z_{i},$ then 
%$S=Q\cup Z$ is a component-wise dominating set of $G$ and that $|Q\cup Z|\leq |Q|+2\cdot \sum_{i\in\{1,\ldots,\rho\}}|B_{i}|=|S_{1}\cup S_{2}|+2|B|$ as required.
%\end{proof}

\begin{lemma}
\label{fii:dominationconnected}
If $\Pi=${\sc $p$-Connected Dominating Set}, then for every $g\in\Bbb{Z}^{+},$ $\Pi_{g}$ is coverable and has FII.
\end{lemma}

\begin{proof}
The coverability of  $\Pi_{g}$ is trivial.
%compact follows from Lemmata~\ref{fii:domination} and~\ref{lem:subsetcomporquasicomp} and the fact that $\Pi_{g}\subseteq \Pi^*_{g},$
%where $\Pi^*=${\sc 1-Dominating Set}.
  To show  that {\sc $p$-Connected Dominating Set} has FII, we define the following 
auxiliary problem:
\begin{eqnarray*}
\Pi'= && \{(G,k)\mid \mbox{$G$ has a component-wise connected dominating set $S$ }\}
\end{eqnarray*}
%In other words, for every connected component $C$ of $G,$ the part of  dominating set $S$ in $C$ is connected.
Notice that 
\mbox{\sc $p$-Connected Dominating Set} $=  \Pi'\doublecap {\cal G}_{\rm con},$
where ${\cal G}_{\rm con}$ is the class of all connected graphs.
Let us remark that ${\cal G}_{\rm con}$ is CMSO-definable
and  $\Pi'$ is a \pmin\  problem.

Let $G$ be a boundaried graph with label set $I$ and boundary $\delta(G)=B$. 
%\sed{Here we have to check globally the boundary issue}.
  Let $R$ be a minimum size component-wise dominating set of $G.$  By Lemma~\ref{lemma:dominmon2}, $G$ has  a component-wise connected dominating 
set $W$ that contains the boundary of $G$ ($B\subseteq W $ ) as a subset and  $|W|\leq |R|+3|I|.$

For a boundaried  structure
$(G',S')\in {\cal U}_I,$  let $S^{*}\subseteq V(G)$
be a set of minimum size subset of $G$ such that $S^{*}\cup S'$ is a component-wise connected dominating set of $G\oplus G'.$
Thus $\zeta_{G}(G',S')=|S^{*}|.$
From Lemma~\ref{lem:dominmon1}, $G$ 
contains a  component-wise connected dominating set 
$S^{+}$ of size at most  $|S^{*}|+3|I|.$ By the definition of $R,$
we have that $|R|\leq |S^{+}|\leq |S^{*}|+3|I|=\zeta_{G}(G',S')+3|I|,$ therefore $|S|\leq  |R|+3|I|\leq \zeta_{G}(G',S')+6|I|.$

In order to prove that $(G\oplus G',W\cup S')\in \Pi',$  we have to show that 
$W\cup S'$ is component-wise connected dominating set of $G\oplus G'.$  
%Let $B$ be the common boundary of $G$ and $G'$, that is,  $\delta(G)=\delta(G')=B$.  
Let ${\cal C}$ be the  set of the connected components of $G\oplus G',$ and for every $C\in{\cal C},$ we set 
$G_{C}=G[V(C)],$ $G_{C}'=G'[V(C)],$ $S_{C}^*=S^*\cap V(C),$ $W_{C}=W\cap V(C),$ $S_{C}'=S'\cap V(C),$ and $B_{C}=B\cap V(C).$ Notice that $C=G_{C}\oplus G'_{C}.$
As $S^{*}\cup S'$ is a component-wise dominating set of $G\oplus G',$ we  have that the set $S^{*}_{C}\cup S'_{C}$ is a connected dominating set of $C.$ Moreover, the fact that $S$ is a component-wise dominating set of $G,$ implies that $W_{C}$
is also a component-wise dominating set of $G_{C}.$ Recall that the boundary of $G$ is contained in $W,$ therefore 
$B\subseteq W$ and this  implies that $B_{C}\subseteq W_{C}.$
From Lemma~\ref{lem:superbb}, $W_{C}\cup S_{C}'$ is a connected dominating set of $C.$ Therefore, $W\cup S'=\bigcup_{C\in{\cal C}}W_{C}\cup S'_{C}$ is a component-wise connected dominating set of $G\oplus G'$ as required.
\end{proof}

Using ideas similar to those in the proof of Lemma~\ref{fii:domination}, it is possible to prove 
that other problems such as {\sc $p$-Connected Vertex Cover}, {\sc $p$-Edge Dominating Set}, or {\sc$p$- Cycle Domination} have FII.
%\sef{This need a decision on where the handwaving goes... Edge variants are dangerous!}

\subsection{Scattered sets}
\label{lemgppodk}
Given an $r\in\Bbb{Z}^{+},$ a  graph $G,$ and a set $S\subseteq V(G),$ we say that $S$ 
is an {\em $r$-independent set} if every two vertices in $S$ have distance greater than
$r.$

We consider the following problem:

\begin{center}
\fbox{\begin{minipage}{10cm}
\noindent {\sc $p$-$r$-Scattered Set}\\
{\sl Input:} A graph $G$  and a $k\in\Bbb{Z}^{+}.$\\ 
{\sl Parameter:} $k$.\\ 
{\sl Question:} Is there an $r$-independent set in $G$ of size at least $k$?
\end{minipage}}
\end{center}

\begin{lemma}
\label{plllr8ujd}
For every positive integer $r,$ and every $g\in \Bbb{Z}^+,$  if $\Pi^{r}=\mbox{\sc $r$-Scatte}\-
\mbox{\sc red Set},$ then $\Pi_{g}^{r}$ is coverable.
\end{lemma}

\begin{proof}
To prove the coverability of $\Pi^{r}_{g},$ we will prove that 
$\Psi_{g}=((\Sigma^{*}\times \Bbb{Z}^{+})\setminus \Pi^r_{g})\doublecap {\cal G}_{g}$ 
has the $r$-coverability property for some constant $c$ that depends on $g$ 
and $r.$  Let $(G,k)$ be a NO-instance of $\Pi_{g}^{r}.$
This means that $G$ does not contain any $r$-independent set of size $k.$
According to the result in~\cite{Dvorak2011}, $G$ has an $r$-dominating 
set of size $c\cdot k$ where $c$ is a constant depending on the Euler genus of $G$ (actually, the result of~\cite{Dvorak2011} holds for much more general classes of sparse
graphs that include graphs of bounded Euler genus). Recall that, from Observation~\ref{obs:radialmetric}, given an embedding of $G$ in a surface of Euler genus $\leq g,$ we have that ${\bf R}_{G}^{2r}\subseteq {\bf B}_{G}^{r}(S),$ therefore $\Psi_{g}$ has the $c$-coverability 
property for $c=\max\{r,g\}.$
\end{proof}

We present in details the proof of the following lemma as it is based
on slightly different ideas than the one used in Lemma~\ref{lem:cvpfiigenus}.

\begin{lemma}
\label{lema:scatfii}
For every positive integer $r,$ if $\Pi^{r}=\mbox{\sc $p$-$r$-Scattered Set},$ then $\Pi_{g}$ has FII.
\end{lemma}
\begin{proof} Using Lemma~\ref{lem:restrictiongenus}, we prove instead  that $\Pi^r$ has FII.  
%Let  $$\Upsilon^{r}=\{(G,k)\mid \mbox{$G$ has an $r$-independent set}\}.$$
%As $\Pi^{r}=\Upsilon^{2r},$  the result follows from Lemma~\ref{lem:intersectclose} if we prove  that $\Upsilon^{r}$ has FII, for every positive integer $r.$ 
Below we prove this fact by adapting the three-stage machinery of the proof of Lemma~\ref{lem:cvpfiigenus}.\smallskip

\noindent{\it Characteristic.}
Let $G$ be a boundaried graph with label set $I$ and the boundary $\delta(G)=B$.  Furthermore, let 
$\ell_{G}~:~I\times I \rightarrow \{0,\ldots,r\}$ be a function that for $i,j\in I$ defines 
\[\ell_G(i,j)=  \min \Big\{\dist_{G}\Big(\lambda^{-1}(i),\lambda^{-1}(j)\Big),r \Big\}. \]
That is, the shortest distance in $G$ between  $\lambda^{-1}(i)$ and $\lambda^{-1}(j)$ if it is at most $r$ and if it is more than $r$ then 
$\ell_G(i,j)$ is  $r$ itself. 
%  for every $\{i,j\}\in I,$ the distance in $G'=G\setminus S$ between $\lambda^{-1}(i)$ and $\lambda^{-1}(j)$ is at least   $f(i,j)+1.$ That is, 
%$\dist_{G'}(\lambda^{-1}(i),\lambda^{-1}(j))\geq f(i,j)+1$. 
% as boundary and  with a vertex labelling $\lambda:B\rightarrow I,$ where $I=\{1,\ldots,t\}.$
Let also ${\cal S}$ be the set containing all functions 
mapping the integers of $I$ to integers in $\{0,\ldots,r\}\cup\{\infty\}.$
Given a $\sigma\in{\cal S},$ we define 
$\zeta_{G}(\sigma)$ as the maximum size of an $r$-independent set $S$
in $G$ with the property  that for every $i\in I,$ 
the distance in $G$ between $\lambda^{-1}(i)$ and every vertex in $S$ is at least 
$\sigma(i).$ As the empty set is always such a set, it holds that $\forall_{\sigma\in{\cal S}}\  \zeta_{G}(\sigma)\geq 0.$ 
%Also, by the definition of $\zeta_{G},$ 
%it holds that  $\sigma_1\geq \sigma_2\Rightarrow \zeta_{G}(\sigma_{1})\leq \zeta_{G}(\sigma_{2})$ 
%(where $\sigma_1\geq \sigma_2$ \sed{How much we use these observations?}
%means that $\forall_{b\in I}\ \sigma_{G}(b)\geq \sigma_{G}(b)).$
\medskip

\noindent{\it Definition of equivalence.}
%Let $G$ be a $t$-boundaried graph with $B$ as a boundary and  with a vertex labelling $\lambda:B\rightarrow I,$ where $I=\{1,\ldots,t\}.$
Let  $\sigma^{(0)}\in{\cal S}$  such that $\forall_{i\in \lambda(B)}\ \sigma^{(0)}(i)=0.$ We also set $x_{G}=\zeta_{G}(\sigma^{(0)}).$
We have that $\forall_{\sigma\in {\cal S}}\ 
%\zeta_{G}(\sigma^{(r)})\leq 
\zeta_{G}(\sigma)\leq x_{G}.$  
%Let now $S^{(0)}$ be an $r$-independent set of $G$  where $\zeta_{G}(\sigma^{(0)})=|S^{(0)}|.$  
We define a function $\chi_{G}: {\cal S}\rightarrow\{-\infty\}\cup\{-2t,\ldots,0\}$ as follows:
\begin{eqnarray*}
\hspace{-0mm} \chi_{G}(\sigma)\!=\!\left\{\begin{array}{lll}
&\hspace{-0mm} \zeta_{G}(\sigma)-x_{G} & \mbox{if  $x_{G}-2t\leq \zeta_{G}(\sigma)\leq x_{G}$}\\
&- \infty& \mbox{otherwise}\label{eq:feasiblezetamaxrset}
%\\ & {\bf left}_{G}(y_{1}^{i,2l}) & \mbox{if $y_{1}^{i,2l}\in S$ and $i$ is odd}
\end{array}\right.
\end{eqnarray*}
Given two boundaried graphs $G_{1}$ and $G_{2},$ we say that $G_{1}\sim G_{2}$ if $\Lambda(G_1)=\Lambda(G_2)$, $\ell_{G_1}=\ell_{G_2}$ 
and $\chi_{G_{1}}=\chi_{G_{2}}.$  Notice that for every finite $I\subseteq \mathbb{Z}^{+}$,  $\sim$ is an equivalence relation with 
finitely many equivalence classes.  
\medskip

\noindent{\it Refinement proof.}
The result will follow if we prove 
 that $\equiv_{\Pi^{r}}$ is a refinement of $\sim.$  
 For this we claim that if $G_{1}\sim G_{2}$
then $G_{1}\equiv_{\Pi^{r}} G_{2}$ or,  equivalently, that there is some constant $c$, depending on $G_1$ and $G_2,$ such that
\begin{eqnarray}
\forall (F,k)\in \cal{F} \times \Bbb{Z} & &  (G_1 \oplus F, k) \in \Pi^{r}\Leftrightarrow(G_2 \oplus F, k+c) \in \Pi^{r}.\label{eq:scatrefinement}
\end{eqnarray}
\noindent 
Suppose that $G_{1}\sim G_{2}.$  This implies that $\Lambda(G_1)=\Lambda(G_2)$. Let  $\Lambda(G_1)=\Lambda(G_2)=I$ and $|I|=t$.  
Let $(F,k)\in  \cal{F} \times \Bbb{Z}$ such that $(G_1 \oplus F, k) \in \Pi^{r}.$ Our target is to prove that  $(G_2 \oplus F, k+c) \in \Pi^{r}$ (the other direction of~\eqref{eq:scatrefinement} is symmetric). 

The fact that $(G_1 \oplus F, k) \in \Pi^{r}$ means that  $(G_1\oplus F)$ contains  an $r$-independent set $S$ where $|S|\geq k.$ 
%Let $B$ be the common boundary of $\deltaG_{1}$ and $F$ and  
Let $B$ be the  boundary of  $G_1$, that is, $\delta(G_{1})=B$  and  let $S_{1}=S\cap V(G_{1})$  and $S_{F}=S\setminus S_1.$ 
Let also $\lambda_{1}$ and $\lambda_{2}$ be the labelings of boundaries of $G_{1}$ and $G_{2}$, respectively. 
We define $\sigma$ as follows: for $i\in I$  
%; if $i\in \lambda_{1}(B)$ we 
set $\sigma(i)$ to be the minimum distance 
of a vertex of $S_{1}$ from $\lambda^{-1}_{1}(i)$
in $G_{1}$.  
%and if $i\not\in \lambda_{1}(B),$ then we set $\sigma(i)=\infty.$
By the definition of $\zeta_{G_{1}},$ we have 
that $\zeta_{G_{1}}(\sigma)\geq |S_1|.$
Before we proceed, we need to prove the following 
claim:\smallskip

\noindent{\em Claim:} $|S_{1}|\geq x_{G_{1}}-2t.$
Let  $S'_{1}$
be an $r$-independent set of $G_{1}$ such that $|S'_{1}|=x_{G_{1}}.$
Mark in $S'_{1}$ all vertices that are within distance at most $\lfloor\frac{r}{2}\rfloor$ from $B$ and denote by $S_{1}^{*}$
the set of the non-marked vertices of $S_{1}'.$ Notice that $S_{1}^{*}$ is an $r$-independent set of $G_1.$
The proof of the 
claim is a consequence of the following two subclaims:\smallskip

\noindent{\em Subclaim 1:}  $|S_{1}^{*}|\geq x_{G_1}-t.$
For this it is enough to prove that no more than $|B|$
vertices can  be marked from $S'_{1}.$
Indeed if this is not the case, then there should exist
two vertices $x$ and $y$ in $S'_{1}$ that are within
distance at most $\lfloor\frac{r}{2}\rfloor$ from some vertex $z$
of $B.$ Then the distance between $x$ and $y$ should be less 
than $2\cdot \lfloor\frac{r}{2}\rfloor\leq r,$ a contradiction to the fact that $S'_{1}$ is an $r$-independent set of $G_{1}.$ \smallskip

\noindent{\em Subclaim 2:} $|S_{1}|\geq |S_{1}^{*}|-t.$
For this, we mark in $S$ the vertices of $G_{1}\oplus F$ that are within distance at most 
$\lfloor\frac{r}{2}\rfloor$ from some vertex of $B.$
As above, the marked vertices cannot be more than $|B|.$
Let $S^{-}$ be the set obtained from $S$ after removing the marked vertices. Notice that $|S^{-}|\geq |S|-t,$
therefore $|S^{-}\cap V(G_{1})|+|S^{-}\setminus V(G_{1})|\geq |S|-t.$ Notice that $S^{-}\cap V(G_{1})$ is an $r$-independent 
set of $G_{1},$ therefore $|S^{-}\cap V(G_{1})|\leq x_{G}.$
Notice that  $S^{*}_{1}\cup (S^{-}\setminus V(G_{1}))$
is an $r$-independent set of $G_{1}\oplus F.$ Indeed if there 
are two vertices $x\in S_{1}^{*}$ and $y\in S^{-}\setminus V(G_{1})$ within distance $r,$ then either 
$x$ or $y$ would be within distance $\lfloor\frac{r}{2}\rfloor$
from some vertex in $B,$ a contradiction.
We obtain that $|S^{*}_1|+|S^{-}\setminus V(G_{1})|=|S^{*}
_{1}\cup (S^{-}\setminus V(G_{1}))|\leq |S|\leq |S^{-}|+t=|S^{-}\cap V(G_{1})|+|S^{-}\setminus V(G_{1})|+t$
and therefore, $|S^{*}_1|\leq |S^{-}\cap V(G_{1})|+t\leq |S_{1}|+t.$\medskip

\medskip 

We just proved that $\zeta_{G_{1}}(\sigma)\geq |S_{1}|\geq x_{G_{1}}-2t.$
This means that $\chi_{G}(\sigma)>-\infty.$ As $G_{1}\sim G_{2},$ we have that $\ell_{G_1}=\ell_{G_2}$ and  $\chi_{G_{1}}(\sigma)=\chi_{G_{2}}(\sigma).$  
By the definition of $\chi_{G},$ we obtain that 
$\zeta_{G_{2}}(\sigma)=\zeta_{G_{1}}(\sigma)-\zeta_{G_1}(\sigma^{(0)})+
\zeta_{G_2}(\sigma^{(0)})=\zeta_{G_{1}}(\sigma)+c\geq 
|S_{G_{1}}|+c$ where $c$ is a constant depending 
only on $G_{1}$ and $G_{2}.$ This implies that, 
there exists an $r$-independent set $S_{G_{2}}$ in $G_{2}$
with  least  $|S_{G_{1}}|+c$ vertices and 
for every $i\in \lambda_{2}(B),$  the distance in $G_{2}$ between $\lambda^{-1}_{2}(i)$ and 
the vertices  in $S_{2}$ is at least 
$\sigma(i).$
%As $B$ is a separator of $G_{2}\oplus F,$ all the 
%vertices of $S_{G_{2}}$ should be  within distance more than $r$ from the vertices of $S_{F}.$
The facts that $\ell_{G_1}=\ell_{G_2}$ and  $\chi_{G_{1}}(\sigma)=\chi_{G_{2}}(\sigma)$ together imply that $S_{G_{2}}\cup S_{F}$ is an $r$-independent set of $G_{2}\oplus F$ of size $|S_{G_{2}}\cup S_{F}|=|S_{G_{2}}|+|S_{F}|\geq |S_{G_{1}}|+|S_{F}|+c\geq   |S_1|+|S_{F}|+c \geq k+c.$ We conclude that $(G_{2},k+c)\in \Pi^{r},$ as required.
\end{proof}

\subsection{Problems on Directed Graphs} 
\label{directedproblems}
 Our results also apply to problems on directed graphs whose underlying undirected graph is of bounded 
genus. In this direction we mention three problems considered in the literature. 
In all cases the input is a directed graph $D=(V,A)$ where $V$ is the set of its vertices and 
$A$ is the set of its directed edges (i.e., $A\subseteq V\times V$).

\begin{itemize}
\item {\sc $p$-Directed Domination} \cite{AlberDN06}: Is there a  subset $S\subseteq V$  of size at most $k$ such that for very vertex 
$u\in V \setminus S$ there is a vertex $v\in S$ such that $(u,v)\in A$?  Such a set $S$ is called a {\em directed dominating set} of $D.$  
\item {\sc $p$-Independent Directed 
Domination}%
\footnote{In literature it is known as  ``{\sc $p$-Kernels}''.  We call it differently here 
to avoid confusion with problem kernels.}~\cite{GutinKLY05}: Is there a subset $S\subseteq V$ of 
size at most $k$ such that $S$ is an 
independent set 
and for every vertex $u\in V \setminus S$ 
there is a vertex $v\in S$ such that $(u,v)\in A$? 
\item  {\sc $p$-Maximum Internal Out-branching}~\cite{abs-0801-1979}:
Does $D$ contain  a directed  rooted  spanning tree, an out-branching, 
%(with all arcs directed outwards from the vertices) 
 with at least $k$ internal vertices?
\end{itemize}

In order to formally state our results, we extend the notion of coverability   to directed graphs by applying the definitions to their underlying undirected graphs.

\begin{lemma}
\label{fii:dirgraph}%\sef{why not the undirected independent dominating set? somehow we have to put comment here!}
The following statements hold:
\begin{itemize}
\item Let $\Pi$ be either {\sc $p$-Independent Directed Domination}, or   {\sc   $p$-Maximum Internal Out-branching}. Then $\Pi_{g}$ is a coverable \pmin{}\ problem.
\item Let $\Pi$ be {\sc $p$-Directed Domination}. Then $\Pi_{g}$ is a coverable problem and has FII.
%\item If $\Pi$={\sc   Minimum Leaf Out-branching}, then $\Pi_{g}$ is a  compact \pmin{}\ problem.
\end{itemize}
%
%,  {\sc Directed Domination} is compact and has FII and 
%$\Pi$={\sc Minimum Leaf Out-branching} is a \pmax{} problem and $\overline{\Pi}$ is compact. 
\end{lemma}
\begin{proof} Problems
{\sc $p$-Independent Directed Domination} and  {\sc $p$-Directed 
Domination} can easily be seen to be \pmin{} problems while 
{\sc   $p$-Maximum Internal Out-branching}  can be proved to be 
a \pmax{} problem. The strong monotonicity of  {\sc $p$-Directed 
Domination}  %\sed{Are we sure about s-m of directed dom set?} 
can be proved 
%\sef{We claim that FII definition and  proofs on this extend also for directed case. Shall we accept  that we do not have a proof?}
%by
 using the same arguments as in the proof of Lemma~\ref{fii:domination}.
%and taking in mind that Observations~\ref{obs:dominmon1} and~\ref{obs:dominmon2}
%hold also for directed domination. 
 This, together with Lemmata~\ref{lem:stronglymonotone} and~
\ref{lem:restrictiongenus},
implies that   for $\Pi$={\sc $p$-Directed Domination},  $\Pi_{g}$ has FII.

{\sc $p$-Independent Directed Domination} and {\sc $p$-Directed Domination} are coverable by definition.
Let $\Pi$={\sc   $p$-Maximum Internal Out-branching}. 
We claim that if $(D,k)\not\in \Pi,$ then the underlying undirected graph of  $D$
has a dominating set of size at most $k-1.$  
For this let $k_{0}=\max \{k'\mid (D,k')\in\Pi\}$ and observe that 
$k_0<k.$ Moreover, it also holds 
that $(D,k_{0})\in\Pi$ while $(D,k_{0}+1)\not\in \Pi.$
These two facts together imply that $D$ has 
a rooted directed spanning tree 
with {\em exactly} $k_0$ internal vertices and   all other 
vertices of $D$ being its leaves. These internal vertices form a  dominating 
set for the underlying undirected graph of  $D$. 
As $k_0<k,$  the underlying undirected graph of  $D$ has a dominating set of size at most $k-1.$ 
%We conclude that $(\Sigma^{*}\times \Bbb{Z}^{+}\setminus \Pi_{g})\doublecap {\cal G}_{g}\subseteq
%\Pi^{*}$ where  $\Pi^{*}=\mbox{\sc Directed Sominating Set}.$
Then  the coverability of $\Pi_{g}$ follows from the coverability of {\sc $p$-Dominating Set} and Lemma~\ref{lem:subsetcomporquasicomp}.  
\end{proof}

\subsection{A direct proof of FII for a minimization problem}
\label{thisssspecs}
Although Lemma~\ref{lem:stronglymonotone} is very useful for showing that a concrete problem has FII, sometimes a minimization problem may have FII even though it may not be strongly monotone. For an example, consider the following problem.
Let $s\geq 3$ be an integer. 
%\sef{Make more general comment of edge-variants!}
\medskip

\begin{center}
\fbox{\begin{minipage}{12cm}
\noindent{\sc $s$-Cycle Transversal} \\
{\sl Input:} A graph $G$ and a $k\in\Bbb{Z}^{+}.$\\
{\sl Parameter:} $k$\\ 
{\sl Question:} Is there an  edge subset $S \subseteq E(G)$ such that 
 $G'=G\setminus S$  does not contain\\
\phantom{Question:}   any cycle of length at most $s$ (i.e. $G'$ has girth more than $s$)?
\end{minipage}}
\end{center}
\medskip
%
%. For a fixed constant $s \geq 3,$ the problem is defined as follows. Input is a graph $G=(V,E)$ together with an integer $k,$ the parameter. The question is whether there exist an edge subset $S \subseteq E$ such that $G' = (V, E \setminus S)$ does not contain any cycle of length at most $s.$ In this section, let $\Pi$ be the {\sc $s$-Cycle Transversal} problem, and  $G \setminus S$ for the edge set $S$ means deleting the edges in $S$ from $G.$

Notice that for each integer $s\geq 3,$ the above problem is the edge deletion counterpart
of  {\sc Edge-${\cal S}$-Covering} when ${\cal S}$ contains the cycles of size 
 at least $3$ and at most $s.$ 
% However, in Section~\ref{lem:subcovpack}\sef{I guess this is not correct as we have nothing on the FII of  {\sc Edge-${\cal S}$-Covering}}
% we mentioned that this problem has FII and the proof follows by adapting the strong
% monotonicity proof of~\ref{fii:mincovpacsssskdd}. 

\begin{lemma}
\label{lem:scycletrans} 
If $\Pi^s=${\sc $s$-Cycle Transversal}, then $\Pi_{g}^s$  has FII.
\end{lemma}

\begin{proof}Using Lemma~\ref{lem:restrictiongenus}, we prove instead  that $\Pi^s$ has FII.
We present the proof in three stages, as we did in the cases of Lemmata~\ref{lem:cvpfiigenus} and~\ref{lema:scatfii}.\medskip

\noindent{\em Characteristic.} Let $G$ be a boundaried graph with label set $I$ and the boundary $\delta(G)=B$.  Let $|I|=t$. 
We use the term $s$-cycle for a cycle of length at most $s$. Let
$X$ be the set of unordered pairs of distinct indices in $I$ and  
%We also set $I_{B}=\lambda(B)$ and define $X_{B}$ as 
%the set of unordered pairs of distinct indices in $I_{B}.$
%We define  ${\cal X}$ as the set of all different subsets of $X.$
 ${\cal H}$ be the set containing all functions 
from $X$ to $\{0,\ldots,s\}.$ 
We define the function $\zeta_{G}:{\cal H} \rightarrow \Bbb{Z}^{+}$
such that, given a function $f\in{\cal H}$,   
$\zeta_{G}(f)$ is the size of a minimum 
set of edges $S$ in $G$ such that the following hold:
\begin{itemize}
%\item for every pair $x,y\in B,$ if  $\{\lambda(x),\lambda(y)\}\in X_{B}\cap Z,$ then $\{x,y\}\in S\cap X,$
\item  the graph $G\setminus S$ has girth $>s,$ and
\item  for every $\{i,j\}\in I,$ the distance in $G'=G\setminus S$ between $\lambda^{-1}(i)$ and $\lambda^{-1}(j)$ is at least   $f(i,j)+1.$ That is, 
$\dist_{G'}(\lambda^{-1}(i),\lambda^{-1}(j))\geq f(i,j)+1$. 
\end{itemize}
In case  a set satisfying the above conditions does not exist, we set $\zeta_{G}(f)=\infty.$\medskip

\noindent{\em Definition of equivalence.} We denote by  $f^{\rm min}$ the function in ${\cal H}$
where, for all $\{i,j\}\in X,$ $f^{\rm min}(\{i,j\})=0.$ 
% and given $f_{1},f_{2}\in{\cal F}$ we say that $f_{1}\leq f_{2}$ if $\forall_{p\in X} \ f_{1}(p)\leq f_{2}(p).$%
Notice that $ \zeta_{G}(f^{\min})<\infty$ (just take $S=E(G)$).
We set $x_{G}=\zeta_{G}(f^{\min}).$ The definition of $\zeta_{G}$ implies that 
\begin{eqnarray}
\forall f\in {\cal H}\  && 
 x_{G}\leq \zeta_{G}(f)\label{eq:defiscycle}
%\forall Z\in {\cal X},f_{1},f_{2}\in{\cal F} && f_{1}\leq f_{2}\Rightarrow \zeta{G}(f_{1},Z)\geq \zeta_{G}(f_{2},Z)\label{eq:defiscyclemore}
\end{eqnarray}
%We now define $\chi_{G}:{\cal F} \rightarrow \Bbb{Z}^{+}$ such that for all $f\in {\cal H},$\  
%$\chi_{G}(f)=\zeta_{G}(f)-x_{G}.$
We 
now define the {\em  signature} of $G$ as the function $\chi_{G}: {\cal H}\rightarrow \{0,\ldots,3{t \choose 2}\}\cup\{\infty\}$,  where
\begin{eqnarray}
\hspace{-0mm} \chi_{G}(f)\!=\!\left\{\begin{array}{lll}
&\hspace{-0mm} \zeta_{G}(f)-x_{G} & \mbox{if  $x_{G}\leq \zeta_{G}(f)\leq x_{G}+3{t \choose 2}$}\\
& \infty& \mbox{otherwise}\label{eq:feasiblezetamaxscycletra}
%\\ & {\bf left}_{G}(y_{1}^{i,2l}) & \mbox{if $y_{1}^{i,2l}\in S$ and $i$ is odd}
\end{array}\right.
\end{eqnarray}
We say that $G_{1}\sim G_{2}$ if $\Lambda(G_1)=\Lambda(G_2)$ and $\chi_{G_{1}}=\chi_{G_{2}}.$
Notice that the number of different signatures is bounded by some function of $t$ and $s.$
%Given two $t$-boundary graphs $G_{1}$ and $G_{2}$
%we say that $G_{1}\sim G_{2}$ if and only if  $\chi_{G_{1}}=\chi_{G_{2}}.$
Clearly, for every $I\subseteq \mathbb{Z}^{+}$,   $\sim$ is an equivalent relation with finitely many equivalence classes.  \medskip

\noindent{\it Refinement proof.}
The result will follow if we prove that 
$\sim$ is a refinement of $\equiv_{\Pi}.$ For this we claim that if $G_{1}\sim G_{2}$
then $G_{1}\equiv_{\Pi} G_{2}$ or, equivalently, 
that there is some constant $c$, depending on $G_1$ and $G_2,$ such that
\begin{eqnarray}
\forall (F,k)\in {\cal F} \times \Bbb{Z} & &  (G_1 \oplus F, k) \in \Pi\Leftrightarrow(G_2 \oplus F, k+c) \in \Pi.\label{eq:refequivdepscycle}
\end{eqnarray}

Suppose that $G_{1}\sim G_{2}.$ Let  $(F,k)\in {\cal F}\times \Bbb{Z}$ such that $(G_1 \oplus F, k) \in \Pi.$ Our target is to prove that  $(G_2 \oplus F, k+c) \in \Pi$ (the other direction of~\eqref{eq:refequivdepscycle} is symmetric and is omitted).
\smallskip

The fact that $(G_1 \oplus F, k) \in \Pi,$ means that there is a set $S\subseteq E(G_1 \oplus F)$ of edges 
such that all cycles in $(G_{1}\oplus F)\setminus S$ have length $>s.$  
Recall that $\lambda_G$ is an injective labelling from the boundary of the graph to $I$. We denote by $\lambda_{1},\lambda_{2}$ and 
$\lambda_{F}$ the labelings of the boundaried graphs $G_{1},$ $G_{2},$ and $F$  respectively. Let $B=\lambda_{1}^{-1}(\Lambda(G_1)\cap \Lambda(F))$ and $B'=\lambda_{2}^{-1}(\Lambda(G_2)\cap \Lambda(F))$. 
Since $G_1$, $G_2$ and $F$ are boundaried graphs with label set $I$ we have that   $|B|,|B'| = |I|= t.$
%
%the common boundary $B=\lambda_{1}(\Lambda(G_1)\cap \Lambda(F))=\lambda_{F}(\Lambda(G_1)\cap \Lambda(F))$. Similarly, let $B'$ denote the common boundary between  $G_{2}$) and $F$ and observe that $|B|,|B'| = |I|= t.$
%
%Let $B$ (resp. $B'$) be common boundary of $G_{1}$ (resp. $G_{2}$) and $F$
%and observe that $|B|,|B'|\leq t.$
Let also $S_{G_1}=E(G_{1})\cap S$ and  $S_{F}=E(F)\cap S.$
The set ${\cal C}$ of $s$-cycles in $G_{1}\cup F$ 
is partitioned into three sets: 
\begin{itemize}
\item ${\cal C}_{1}$ are the cycles in ${\cal C}$ that are entirely inside $G_{1},$
\item ${\cal C}_{F}$ are the cycles in ${\cal C}$ that are entirely inside $F,$ and
\item ${\cal C}_{B}$ are the cycles in ${\cal C}$ that contain both edges that are not in $G_{1}$
and edges that are not in $F,$ i.e., ${\cal C}_{B}={\cal C}\setminus ({\cal C}_{G_{1}}\cup {\cal C}_{F}).$
\end{itemize}
Observe that $S_{F}$
intersects all $s$-cycles in  ${\cal C}_{F}$ and the set
 $S_{G_1}$ intersects 
all $s$-cycles in ${\cal C}_{1}.$ 
Observe that $S_{G_{1}}\cap S_{F}$ contains only edges with both endpoints in  $B,$ therefore $|S_{G_{1}}\cap S_{F}|\leq {t \choose 2}.$ This implies that 
\begin{eqnarray}
|S_{G_{1}}| + |S_{F}| - {t \choose 2} & \leq & |S|.\label{eq:morescycles}
\end{eqnarray}
Recall that $x_{G_{1}}=\zeta_{G_{1}}(f^{\min}).$  We prove the following claim. Let $x_{G_1}$ denote the cardinality 
of a minimum sized subset of $E(G_{1})$ intersecting all $s$-cycles in $G_{1}.$

\medskip

\noindent{\em Claim:}  $|S_{G_1}|\leq x_{G_1}+3{t\choose 2}.$\smallskip

\noindent{\em Proof of Claim:}  
Let $S^{*}_{G_1}$ be  a minimum size subset of $E(G_{1})$ intersecting all $s$-cycles in $G_{1}.$ By definition, $|S_{G_1}^{*}|=x_{G_1}.$
Notice that the set $S_{G_1}^*\cup S_{F}$ meets all cycles in ${\cal C}_{1}\cup {\cal C}_{F}.$
Let ${\cal C}_{B}^{\bullet}$ be the cycles of ${\cal C}_{B}$ that are not met by $S_{G_1}^*\cup S_{F}.$

Our  first aim is to find a set $S_{B}$ of at most $2{t \choose 2}$
edges that interest all cycles of ${\cal C}_{B}^{\bullet}.$ 
Observe that each cycle in ${\cal C}_{B}^{\bullet}$
meets at least two vertices in $B.$ 
Let $W$ be the set of pairs in $X$
that are met by the cycles in ${\cal C}_{B}^{\bullet}.$
For each pair $p=\{x,y\},$ we denote by ${\cal Q}_{p}^{\rm left}$  (resp, ${\cal Q}_{p}^{\rm right}$) the 
set of all $(x,y)$-paths in $G_{1}$ that belong to cycles in ${\cal C}_{B}^{\bullet}.$
We claim that for each $p=\{x,y\}$ where $x,y\in B,$ 
at most one of the $(x,y)$-paths in ${\cal Q}_{p}^{\rm left}$
can have length at most $s/2.$ Suppose in contrary 
that $P_{1},P_{2}$ are two $(x,y)$-paths 
of $G_{1}$ of length $\leq s/2.$ The union of $P_{1}$ and $P_{2}$  
contains a cycle $C_{x,y}$ that is entirely in $G_{1}$. By the definition of ${\cal C}_{B}^{\bullet},$ we have that 
$C_{x,y}$  does not contain any edge $e$ from $S^*_{G_1}.$  This contradicts the fact that $S^{*}_{G_1}$ 
intersects all $s$-cycles in $G_{1}.$  
 %Therefore $C_{x,y}\in {\cal C}_{G_{1}}$ and this 
%cycle should contain at least one edge $e$ from $S^*_{G_1}.$
%
%W.l.o.g. let $e$ be an edge of $P_{1}$ (otherwise apply the same argument for $P_{2}$) and let 
%$C_{e}$ be the cycle of ${\cal C}_{B}^{\bullet}$
%that contains $P_{1}.$ 
%By the definition of ${\cal C}_{B}^{\bullet},$  none of its cycles of  contain
%edges from $S_{G_1}^*,$ a contradiction. 
Therefore,  for each $p=\{x,y\}$ where $x,y\in B,$
at most one, say $Q_{p}^{\rm right},$ of the $(x,y)$-paths in ${\cal Q}_{p}^{\rm right}$
can have length at most $s/2.$ 
Using the same arguments on $F,$ instead of $G_{1},$
it follows that  for each $p=\{x,y\}$ where $x,y\in B,$
at most one, say ${Q}_{p}^{\rm left},$ of the $(x,y)$-paths in ${\cal Q}_{p}^{\rm left}$
can have length at most $s/2.$ \smallskip

We now 
construct the set $S_{B}$ by adding to 
it, for each pair $p\in X,$ one edge from the $Q_{p}^{\rm right}$
and one edge from $Q_{p}^{\rm left}.$ As there are at most ${t\choose 2}$ pairs in $X,$ we 
obtain that $|S_{B}|\leq 2{t \choose 2}.$  We next prove 
that $S_{B}$ meets all cycles in ${\cal C}_{B}^{\bullet}.$
For this, let $C$ be a cycle in ${\cal C}_{B}^{\bullet}.$
Clearly, there are at least two internally vertex-disjoint paths 
contained in $C$ (these two paths may not contain all the vertices on $C$) that are entirely inside $G_{1}$ or $F$
and have their endpoints in $B.$ Since $C$ is an $s$-cycle, we have that 
at least one, say $Q,$ of these paths should have length $\leq s/2.$
Let $x$ and $y$ be the endpoints of $Q$ and $p=\{x,y\}.$ Clearly, 
$Q$ belongs in one of  ${\cal Q}_p^{\rm left}$ or 
${\cal Q}_{p}^{\rm right}.$ W.l.o.g., suppose that 
$Q$ belongs  in  ${\cal Q}_{p}^{\rm left}.$
As $Q$ has length at most $s/2,$ then $Q$ is the unique 
path in ${\cal Q}_{p}^{\rm left}$ that has such a length. 
By its construction,
$S_{B}$ intersects $Q$
and, as $Q$ is a path of $C,$ $S_{B}$ 
intersects $C$ as well.

We just proved that $S_{B}$ intersects all $s$-cycles in ${\cal C}'_{B}$ and contains  at most $2{t \choose 2}$ edges.
This implies that $S_{G_1}^*\cup  S_{B}\cup S_{F}$ is intersecting
all $s$-cycles in ${\cal C}.$ By 
the definition of $S,$ 
%and the fact that $S_{B}\cap (S^{*}_{G_{1}}\cup S_{F})=\emptyset,$ 
we have that 
 $|S|\leq |S_{G_1}^*\cup S_{B}\cup S_{F}|\leq |S^{*}_{G_{1}}|+|S_{B}|+|S_{F}|.$
 Therefore, $|S_{G_{1}}| + |S_{F}| - {t \choose 2} \leq^{\eqref{eq:morescycles}} |S|\leq |S_{G_1}^{*}|+|S_{B}|+|S_{F}|\leq x_{G_{1}}+2{t \choose 2}+|S_{F}|.$
We conclude that $|S_{G_1}|\leq x_{G_1}+2{t \choose 2}+{t \choose 2}$ and the claim follows.  $\Box$
%\end{proof}
%\qed
\medskip

%We set up the set $Z$ containing all pairs $\{i,j\}\in X$
%where $\{\lambda_{1}^{-1}(i),\lambda_{1}^{-1}(j)\}$ is an edge of $S_{G_{1}}\cap S_{F}.$ 
%, therefore $|Z|=|S_{G_{1}}\cap S_{F}|.$

 For every pair $\{i,j\}\in X,$\  let $s(i,j)$ be 
equal to $s$ minus the distance between $\lambda_{F}^{-1}(i)$ and $\lambda_{F}^{-1}(j)$ in $F.$   We define  the function $f\in {\cal F}$ 
as follows. For every pair $\{i,j\}\in X,$ if $\{\lambda_{1}^{-1}(i),\lambda_{1}^{-1}(j)\}$ is an edge of $S_{G_{1}}\cap S_{F}$ then define 
\[f(i,j)=\max\{1,s(i,j),\}\]
else define $f(i,j)=s(i,j)$. The choice of $f$ and the definition of $\zeta_{G_1},$ imply that 
\begin{eqnarray}
\zeta_{G_1}(f)\leq |S_{G_{1}}|.\label{eq:zetalesszeta}
\end{eqnarray} 

From~\eqref{eq:defiscycle} we have that $x_{G_{1}}\leq \zeta_{G_{1}}(f).$
Moreover, from~\eqref{eq:zetalesszeta} and  the above claim, we obtain $\zeta_{G_{1}}(f)\leq x_{G_{1}}+3{t \choose 2}.$
By~\eqref{eq:feasiblezetamaxscycletra}, $\chi_{G_{1}}(f)=\zeta_{G_{1}}(f)-x_{G_{1}}.$
%eq:feasiblezetamaxscycletra
Recall now that $G_{1}\sim G_{2},$ hence $\chi_{G_{2}}(f)=\chi_{G_{2}}(f).$  
This means that $\zeta_{G_{2}}(f)=\zeta_{G_{1}}(f)+c,$ where 
$c=x_{G_{2}}-x_{G_{1}},$ and clearly $c$ depends only on $G_{1}$ and $G_{2}.$

Let $S_{G_{2}}$ be a subset of $E(G_{2})$ such that 
$\zeta_{G_{2}}(f)=|S_{G_{2}}|.$ By the definition of $\zeta_{G_2},$
$S_{G_{2}}$ has the following properties:
\begin{enumerate}
%\item\label{enum:first} for every pair $x,y\in B',$ if  $\{\lambda_2(x),\lambda_2(y)\}\in X_{B'}\cap Z,$ then $\{x,y\}\in S_{G_{2}}\cap X,$
\item[(A)] %\label{enum:second}  
the graph $G_2\setminus S_{G_2}$ has girth $>s,$ and
\item[(B)]%\label{enum:third} 
  for every $\{i,j\}\in X,$ the distance in $G_{2}\setminus S_{G_{2}}$ between $\lambda^{-1}_{2}(i)$ and $\lambda^{-1}_{2}(j)$ is at least   $f(i,j)+1.$
\end{enumerate}
%
%
%property 
%that $Z$ is the set of edges of $S_{G_{2}}$ that have endpoints in $B$
%and that for every pair $\{x,y\}\in X,$
%the distance in $G_{2}\setminus S_{G_{2}}$ between $x$ and $y$ is at least $f(\{x,y\}).$
By the definition of $f,$ and Properties~(A) and (B), %\eqref{enum:second} and~\eqref{enum:third}, 
 all $s$-cycles in $G_{2}\oplus F$ that 
are not entirely in $F$ are intersected by $S_{G_{2}}.$ Hence,
$S'=S_{G_{2}}\cup S_{F}$ intersects all cycles in $G_{2}\oplus F.$ Moreover,  by the definition of $f$ we obtain that $S_{G_{1}}\cap S_{F}\subseteq S_{G_{2}}.$ This implies that 
$S'=S_{G_{2}}\cup S_{F}=S_{G_{2}}\cup  (S_{G_{1}}\cap S_{F})\cup (S_{F}\setminus (S_{G_{1}}\cap S_{F}))=S_{G_{2}}\cup  (S_{F}\setminus (S_{G_{1}}\cap S_{F})).$

We now have that
$|S'|\leq |S_{G_{2}}|+|S_{F}\setminus (S_{G_{1}}\cap S_{F})|=\zeta_{G_{2}}(f)+|S_{F}\setminus (S_{G_{1}}\cap S_{F})|=\zeta_{G_{1}}(f)+c+|S_{F}\setminus (S_{G_{1}}\cap S_{F})|\leq^{\eqref{eq:zetalesszeta}} |S_{G_{1}}|+|S_{F}\setminus (S_{G_{1}}\cap S_{F})|+c=|S_{G_{1}}\cup S_{F}|+c=|S|+c\leq k+c.$
Therefore $(G_{2}\oplus F,k+c)\in \Pi$ and  the  lemma follows.  
\end{proof}

\subsection{Summary of consequences of our results}

In this section, we discuss some of the consequences of our main meta-algorithmic results, namely 
Theorem~\ref{thm:automata} and Theorem~\ref{thm:cmsol}.

We start with the consequences of Theorem~\ref{thm:automata} to minimization problems that have FII.
%{\sc $p$-min-CMSO} problems.

\begin{corollary}
\label{cor00}
If $g\in\Bbb{Z}^{+}$ and if $\Pi$ is one of the following problems:
\pvc, %(FII(SM)+Co=linear)
\pfvs, % (FII(SM)+QC=linear)%
{\sc Almost Outperplanar},
\pdhs,  \parbt, \parbp,
 %\textsc{Almost-$r$-bounded treewidth},
 % \textsc{Almost-$r$-bounded pathwidth},
\fd,
{\sc $p$-Edge Dominating Set},   % (FII(SM)+Co=linear)
{\sc $p$-Minimum-Vertex Feedback Edge Set}, %(FII(SM)+QC=linear) 
{\sc $p$-Dominating Set}, %(FII(SM)+Co=linear) 
{\sc $p$-$r$-Dominating Set}, %(FII(SM)+Co=linear)
{\sc $p$-$q$-Threshold Dominating Set}, %(FII(SM)+Co=linear)  
{\sc $p$-Efficient Dominating Set},  %(FII(but not SM)+Co=linear)
{\sc $p$- Connected Dominating Set}, %(FII(SM)+Co=linear)
{\sc $p$-Connected Vertex Cover}, %(FII(SM)+Co=linear) 
{\sc $p$-Cycle Domination}, %(FII(SM)+Qc) 
{\sc $p$-Directed Domination},  %(FII(SM)+Co=linear) 
{\sc $p$-${\cal S}$-Co\-vering}, %(FII(SM)+Co preprocessing) 
{\sc $p$-Minimum Partition Into Cli\-ques}, %(FII(SM)+Co).
{\sc $p$-Edge Clique Cover},
and {\sc $p$-$s$-Cycle Trans\-versal}, %(FII(but not SM) +  Compact after preprocessing  remove all vertices not appearing in some cycle of length s = linear)
then $\Pi_{g}$ admits  a linear kernel.
\end{corollary}

\begin{proof}
The definitions of \pvc, %(FII(SM)+Co=linear)
\pfvs, % (FII(SM)+QC=linear)%
\paou,
\pdhs,  \parbt, \parbp{}
 %\textsc{Almost-$r$-bounded treewidth},
%  \textsc{Al\-most-$r$-bounded pathwidth},
have been given in Subsection~\ref{subsec:minorcovering}
and all of them are special cases of the \fd{} problem.
They all have  FII  because of Lemma~\ref{fii:mincovpacsssskdd} 
and the quasi-coverability of $\Pi_{g}$
follows from Lemma~\ref{fii:mincovpackddgssd}. 
We remark  that  not all of these problems are coverable.
% except from 
%{\sc Vertex Cover} are compact.
%then the resulting problem is compact as a vertex cover of such a graph is also a dominating set.
 %However, coverability can follow directly  in the case of  vertex cover, restricted to connected graphs,  as the vertex cover   of a connected graph is also a dominating set.

{\sc $p$-Edge Dominating Set} asks whether a graph $G$ contains a set $F$ of at most $k$ edges such that every other edge 
shares a common endpoint with some edge in $F$. The coverability of $\Pi_{g}$ follows by the fact that the endpoints of the 
edges in $F$ form a dominating set of $G$. Moreover, the {\sc $p$-Edge Dominating Set} 
problem can be easily expressed as a \pmin{}  
problem (with edge quantification)
and the proof of its strong monotonicity is similar to the one of Lemma~\ref{fii:domination}. Therefore it has FII as well.
Using similar arguments one can prove that if $\Pi$={\sc Minimum-Vertex Feedback Edge Set} -- 
given an undirected graph $G$ and a positive integer $k$ the task is to find a spanning tree $T$ of $G$ in which at most 
$k$ vertices have a degree smaller than in $G$, then $\Pi_{g}$ is quasi-coverable (however, it is not coverable). Moreover, {\sc Minimum-Vertex Feedback Edge Set}  has 
FII because it can be expressed as a \pmin{}   
problem and can be proved to be strongly monotone with a proof that uses the ideas of Lemma~\ref{fii:domination}.

{\sc $p$-Dominating Set}, %(FII(SM)+Co=linear) 
{\sc $p$-$r$-Dominating Set}, %(FII(SM)+Co=linear)
{\sc $p$-$q$-Threshold Dominating Set}, %(FII(SM)+Co=linear)  
{\sc $p$-Efficient Dominating Set},  are defined in Subsection~\ref{domitsvarl}. All these problems are coverable 
and have FII because of Lemma~\ref{fii:domination}.
Notice that  for the  first three  problems the FII property follows by  expressing them as \pmin{}  
problems and proving that 
are are strongly monotone. However, {\sc $p$-Efficient Dominating Set} is {\sl not} strongly monotone and the proof 
that it has FII uses a different idea.

{\sc $p$-Connected Dominating Set} is also defined  in Subsection~\ref{domitsvarl}. The coverability of $\Pi_{g}$ 
and the FII property is proved in
Lemma~\ref{fii:dominationconnected}. Using similar ideas, the same results can be proved also for {\sc Connected Vertex Cover}.
%(FII(SM)+Co=linear) 

The {\sc Cycle Domination} problem asks whether a graph $G$ contains a set $S$ of at most $k$ vertices 
such that the removal  of $S$ together with its neighbours from $G$  results in an acyclic graph. This problem can be 
seen as a common extension of \pfvs{} and {\sc $p$-Dominating Set}. $\Pi_{g}$ can be proven to be quasi-coverable
with arguments similar to those in the case of \pfvs{} ({\sc $p$-Cycle Domination} is not a coverable problem). 
The problem is easily expressible as a \pmin{} 
problem and the proof that it is strongly monotone is a blend of the ideas  of the proofs 
of Lemmata~\ref{fii:mincovpacsssskdd} and~\ref{fii:domination}.

{\sc $p$-Directed Domination} is defined in Subsection~\ref{directedproblems}. The coverability and the FII property of $\Pi_{g}$
are proved in Lemma~\ref{fii:dirgraph}.
%
%{\sc Vertex-${\cal H}$-Covering} has been defined in Subsection~\ref{subsec:minorcovering}. The  quasi-coverability and the 
%FII property of $\Pi_{g}$ 
%follow from Lemmata~\ref{fii:mincovpackddgssd} and \ref{fii:mincovpacsssskdd} respectively. Notice that {\sc Vertex-${\cal H}$-Covering} is not a compact problem.

\psc{} has been defined in Subsection~\ref{lem:subcovpack}.
The existence of a linear kernel for this problem makes use of the {\sf Redundant Vertex  Rule} (Lemma~\ref{preprofss}),
Lemma~\ref{fii:subgraohcovpack} (for coverability) and the ideas in the proof of Lemma~\ref{fii:mincovpacsssskdd} (for the FII property).

The {\sc $p$-Minimum Partition Into Cli\-ques} problem asks whether the vertex set of a graph $G$ 
scan be partitioned into at most $k$ sets each inducing a clique in $G$ (in other words, we are asking for a $k$-coloring 
of the complement of $G$).  Let $S$ be a set containing a vertex from each clique. 
Notice that $S$ is  a dominating set of $G$. Therefore, $\Pi_{g}$ is a coverable problem. 
To prove that it also has FII, one needs to express it as a \pmin{}  
problem and then to use arguments {similar} to those of Lemma~\ref{fii:domination}
in order to prove that it is strongly monotone.

The {\sc $p$-Edge Clique Cover} asks whether a graph $G$ contains 
a collection of at most $k$ cliques such that for every edge of $G$, both its endpoints
belongs to some of those cliques. We observe first that $\Pi_{g}$ is quasi-coverable. To see this, just notice that if we consider a set with one 
vertex from each such clique, then the removal of the closed neighbourhood of this set from $G$ results to an edgeless graph.
The proof that the problem has FII is omitted in this paper.

Finally, {\sc $p$-$s$-Cycle Transversal} has been defined in Section~\ref{thisssspecs}. While this problem is not strongly monotone, it has 
FII because of Lemma~\ref{lem:scycletrans}. To prove that it has a linear kernel, one needs first to apply to its instances 
the following preprocessing 
routine:  {\sl remove each vertex that does not appear in some cycle of $G$ of length $\leq s$}.  This routine can be seen
as  a special case of the {\sf Redundant Vertex  Rule} presented in Subsection~\ref{lem:subcovpack} and, with a proof similar to the 
one of Lemma~\ref{preprofss}, one can show that it produces equivalent instances. Under these circumstances, 
the coverability of $\Pi_{g}$ can be proved following the arguments of Lemma~\ref{fii:subgraohcovpack}.
\end{proof}

We continue with the consequences of Theorem~\ref{thm:automata} to maximization 
%{\sc $p$-max-CMSO} 
problems that have FII.

\begin{corollary}
\label{cor11}
If $g\in\Bbb{Z}^{+}$ and if $\Pi$ is one of the following problems:
{\sc $p$-$r$-Scattered Set}, %(FII(but not SM)+Co=linear)
{\sc $p$-Independent Set}, %(FII(SM)+Co=linear)  (scattered set)
{\sc $p$-Induced Matching}, %(FII(but not SM)+Co=linear) (scattered set)
{\sc $p$-Triangle Edge Packing},   %(FII(SM)+Co after preprocessing = linear)
{\sc $p$-Maximum Internal Spanning Tree}, %(FII(but not SM)+Co=linear)   (dominating set)
{\sc $p$-Maximum Full-Degree Spanning Tree}, %(FII(not SM)+Co)
\pcyp, %(FII(but not SM)+Qc = linear)
\pfp, %(FII(but not SM)+Qc)
{\sc $p$-Triangle Vertex Packing}, %(FII(SM)+Co after preprocessing = linear)
\psp, %(FII(but not SM)+Co-preproccesing)
and {\sc $p$-Edge Cycle Packing},
then $\Pi_{g}$ admits  a linear kernel.
\end{corollary}

\begin{proof}
The {\sc $p$-$r$-Scattered Set} problem has been defined in Subsection~\ref{lemgppodk}.
The coverability of $\Pi_{g}^{r}$ is proved in Lemma~\ref{plllr8ujd}, while the problem has FII because 
of Lemma~\ref{lema:scatfii}. We stress that the {\sc $p$-$r$-Scattered Set} problem is, in general, not a strongly monotone
problem. The {\sc $p$-Independent Set} problem asks whether a graph $G$ contains a set of at least $k$ mutually non-adjacent vertices.
If $\Pi$={\sc $p$-Independent Set}, then $\Pi_{g}$ is coverable using an argument that is very similar to the one of Lemma~\ref{plllr8ujd}.
Similarly, one may use the arguments of Lemma~\ref{lema:scatfii} to prove that the problem has FII. Alternatively, one may
express {\sc $p$-Independent Set} as a \pmax{} 
problem and then prove that it is strongly monotone. 

The {\sc $p$-Induced Matching} problem asks whether a graph $G$ contains a set of at least $k$ edges 
such that no vertex in $G$ has as neighbours endpoints of more than one edges in this set.
The problem is quasi-coverable because every NO-instance without isolated vertices 
has a  $(1,3)$-dominating of size at most $k$.
Moreover, the FII property uses ideas of the proof of~\ref{lema:scatfii}. 
We stress that  {\sc $p$-Induced Matching} {is not} a strongly monotone problem.

The  {\sc $p$-Triangle Edge Packing} problem asks whether a graph $G$ contains at least $k$ triangles
such that no two of them have any edge in common. The existence of a linear kernel for this problem makes use of the 
{\sf Redundant Vertex  Rule} and is based in suitable adaptations of the proofs of  
Lemma~\ref{fii:subgraohcovpack} (for coverability) and Lemma~\ref{fii:mincovpacsssskdd} (for the FII property).

%extends to cliques

The {\sc $p$-Maximum Internal Spanning Tree} problem asks whether a graph $G$ has a spanning tree with at least $k$ internal vertices.
The coverability of $\Pi_{g}$ follows by observing that a NO-instance  has a connected dominating set of less than $k$ vertices. The problem is {not strongly monotone} and proving that it has FII requires a {direct proof} that we omit in this paper.

The {\sc $p$-Maximum Full-Degree Spanning Tree} problem asks whether a graph $G$ has a spanning tree $T$ containing  at least $k$ vertices of full degree (a vertex $v$ of $T$ has {\em full degree}  if $N_{T}(v)=N_{G}(v)$). 
Clearly, a NO-instance of $\Pi_{g}$ cannot have a 2-independent set of size at least $k$, otherwise
we grow can a spanning tree with $\geq k$ full-degree vertices by 
starting from the neighbourhoods of the vertices in such a set. 
But then, using the arguments of the proof of Lemma~\ref{plllr8ujd}, $G$ has a dominating set of size $c\cdot k$ where $c$ is a constant 
that depends on the Euler genus $g$ of $G$. This implies the coverability  of $\Pi_{g}$.
For the FII property we only mention that the problem 
is {not strongly monotone} and a specialized proof is required that is omitted in this paper.

The \pcyp, %(FII(but not SM)+Qc = linear)
asks whether a graph contains at least $k$ mutually vertex disjoint cycles. 
This is a special case of the \pfp{} problem
where ${\cal H}=\{K_{3}\}$. For both problems, the quasi-coverability of $\Pi_{g}$ follows 
from Lemma~\ref{fii:mincovpackdd}. The FII property of  \pcyp{} follows from Lemma~\ref{lem:cvpfiigenus} and this proof can be extended for the general case of the  \pfp{} problem, as mentioned in the end of Subsection~\ref{packsingsminorfsd}.
Notice that both problems are neither strongly monotone nor coverable.

The {\sc $p$-Triangle Vertex Packing} problem asks whether a graph $G$ contains a set of at least $k$ triangles 
where no two such triangles share some common vertex. 
 {\sc $p$-Triangle Vertex Packing} is a special case of the 
\psp{} problem where ${\cal S}=\{K_{3}\}$.
The existence of a linear kernel for these problem makes use of the {\sf Redundant Vertex  Rule} (Lemma~\ref{preprofss}),
Lemma~\ref{fii:subgraohcovpack} (for coverability) and the ideas in the proof of Lemma~\ref{lem:cvpfiigenus} (for the FII property).

{\sc $p$-Edge Cycle Packing} asks whether a graph $G$ contains 
a collection of  at least $k$ mutually edge-disjoint cycles. 
To prove the quasi-coverability of $\Pi_{g}$ observe that a NO-instance,
cannot contain a collection of $k$ vertex disjoint cycles. But then, 
by the application of Erd\H{o}s-P\'osa property on bounded genus 
graphs (see, e.g.~\cite{FominST08imp,KloksLL02newa})
$G$ contains a set of at most $c\cdot k$ vertices meeting all the cycles of $G$, where $c$ is a constant 
depending on the Euler genus $g$ of $G$.
The proof that the problem has FII is omitted.
\end{proof}

Corollaries~\ref{cor00} and~\ref{cor11} unify and generalize results presented in~\cite{AlberDN06,AlberFN04,BodlaenderP08,BodlaenderPT08,ChenFKX07,FT04ICALP,GN07ICALP,GuoNW06,KanjPXS08,LokshtanovMS09,MoserS07,Xia:2010pi}. \medskip

We conclude this subsection with some consequences of Theorem~\ref{thm:cmsol}
for problems that do not have  FII.
%
%By applying Theorem~~\ref{thm:cmsol}, Corollary~\ref{thm:cmsolnotanotated}, and Lemmata~\ref{fii:mincovpack}, \ref{fii:subgraohcovpack}, \ref{fii:domination}, and \ref{fii:dirgraph}, we get the 
%following corollary for problems which do not have FII.
 
\begin{corollary}     
\label{metacor2}
If $g\in\Bbb{Z}^{+}$ and if $\Pi$ is one of the following problems:
%{\tt  {\sc Minimum Clique Cover},} % (not-FII, CMSO, compact)
{\sc $p$-Independent Dominating Set}, % (not-FII, CMSO, compact)
{\sc $p$-Acyclic Dominating Set}, 
{\sc $p$-Independent Directed Domination}, %(not-FII, CMSO, compact)
{\sc $p$-Maximum Internal Out-branching},  %(not-FII, CMSO, compact)
{\sc $p$-Odd Set}, %(needs counting!, not-FII, CMSO, compact)
%{\sc Edge-${\cal H}$-Packing}, {\sc Edge-${\cal H}$-Covering}, 
and {\sc $p$-Edge-${\cal S}$-Cove\-ring}, % (not-FII, CMSO, compact after preprocessing) 
 %
%{\sc Edge-${\cal S}$-Packing}, %(not-FII (in general), NOT CMSO, compact after preprocessing) 
%
%\sef{What about this?} 
then $\Pi_{g}$ admits  a polynomial kernel.
\end{corollary}
%%%%%%%%%%%%%%%%%%%%%%%%%%%%%%input{meta_implications}

\begin{proof}
The {\sc $p$-Independent Dominating Set} problem asks whether a graph $G$ contains 
a dominating set of at most $k$ mutually non-adjacent vertices. 
The  {\sc $p$-Acyclic Dominating Set} problem asks whether a graph $G$ contains 
a dominating set $S$ of at most $k$ vertices  such that $G[S]$ is acyclic.  
While these problems {do not have  FII}, they can be both expressed as \pmin{}  
problems and are obviously coverable.

Problems {\sc $p$-Independent Directed Domination} and {\sc $p$-Maximum Internal Out-branch\-ing} have been defined in Subsection~\ref{directedproblems} and they { do not have FII}. 
According to Lemma~\ref{fii:dirgraph}, in both cases, $\Pi_{g}$ is a coverable  \pmin{}\ problem.

The {\sc $p$-Odd Set} problem asks whether a graph $G$ contains a  set $S$ of at most $k$ vertices 
such that for every vertex of $G$, the number of its neighbors in $S$ is odd.  
Clearly, such  a set is a dominating set, therefore $\Pi_{g}$ is coverable. 
{\sc $p$-Odd Set}  {does not have  FII}. However, it can be expressed as a \pmin{} 
problem (notice that here we {have} to use the ``counting'' expressive power of CMSO).

Given  some fixed finite collection  of graphs  ${\cal S}$,
the {\sc $p$-Edge-${\cal S}$-Covering} problem asks whether a graph $G$ contains a set of at most $k$ 
edges meeting every subgraph of $G$ that is isomorphic to a graph in ${\cal S}$.
%The {\sc Edge-${\cal S}$-Packing} problem asks whether a graph $G$ contains a collection of at lest $k$ 
%subgraphs each isomorphic
%to a graph of some fixed collection  ${\cal S}$ such that no two such subgraphs share a common edge.
For this problem, a linear kernel requires the application of the 
{\sf Redundant Vertex  Rule}. The coverability of $\Pi_{g}$ follows {similarly} 
to the proof of  Lemma~\ref{fii:subgraohcovpack}.  {\sc Edge-${\cal S}$-Covering} does not 
have, in general, FII  (while it    has FII when  if ${\cal S}$ contains only cliques). However, it is possible 
to formulate it as a \pmin{}  problem.% or a $p$-max-CMSO problem respectively.}
\end{proof}

Concluding this section, we mention that there are several problems that do not satisfy  
 the conditions of Theorems~\ref{thm:automata} and~\ref{thm:cmsol}.

Apart from the problems mentioned in Corollary~\ref{cor11}, other 
examples of  $p$-max-CMSO problems that do not have  FII  are 
{\sc $p$-Maximum Cut}, {\sc $p$-Longest Path}, and {\sc $p$-Longest Cycle}, see~\cite{Fluiter97}.
Notice that
{\sc $p$-Maximum Cut} is (trivially) quasi-coverable, 
while {\sc $p$-Longest Path} and {\sc $p$-Longest Cycle} are not.
In fact, {\sc $p$-Maximum Cut} admits a trivial $2k$ kernel on 
general graphs while  {\sc $p$-Longest Path}, and {\sc $p$-Longest Cycle} do not admit polynomial 
kernels unless ${\sf coNP} \subseteq  {\sf NP}/{\sf poly}$  \cite{BDFH08}.

As an example of a problem that has FII but it is neither coverable or quasi-coverable, 
we mention  {\sc $p$-Hamiltonian Path Completion} (asking whether 
the addition of at most $k$ edges in a graph can make it Hamiltonian).
This problem can be expressed as a \pmin{}  and it is {possible} to 
prove that it is strongly monotone. Therefore, it has FII. However, none of 
our results apply on this problem as it is not quasi-coverable. In fact, 
{\sc $p$-Hamiltonian Path Completion} cannot have a 
kernel, unless {\sf P}={\sf NP}, as such a kernelization algorithm, 
for $k=1$, would be a polynomial algorithm for the {\sc Hamiltotonian Path} Problem.

%\subsection{Relaxing Definitions and Extending the Applicability}
\section{Open Problems and Further Directions}\label{sec:conclusion}
This paper gives  the first   meta-theorems on kernelization, %as called by Grohe and Kreutzer~\cite{Grohe07logic,Kreutzer2011}, 
 where logical and combinatorial properties of problems lead to kernels of polynomial or linear sizes. Our results are quite general in the sense that they can be applied to a large number of combinatorial problems on graphs on fixed surfaces and generalize a large collection of known results. Still, there are several directions in which our results could possibly be extended. We conclude  with some new problems and further research directions opened by our results.

%\paragraph*{Further extensions} 
\myparagraph{Further extensions.} 
The first natural question for further research is if our logical and combinatorial properties can be extended to larger classes of problems. The property that problems should satisfy some kind of coverability or quasi-coverability cannot be omitted. For instance, even though the problem of finding a path of length $k$ is expressible in first order logic, it does not admit a polynomial kernel on planar graphs, unless 
${\sf coNP} \subseteq  {\sf NP}/{\sf poly}$  \cite{BDFH08}.  
%the polynomial hierarchy
%collapses to the third level, a collapse which is deemed unlikely~\cite{BDFH08}. 
An interesting question for further research is  
 \begin{itemize}
 \setlength{\itemsep}{-1pt}
% \item Can we replace the coverability condition with the weaker notion of {\em quasi-coverability} in Theorem~\ref{thm:cmsol}?
 \item 
 Do all quasi-coverable CMSO problems admit a linear kernel on graphs of bounded genus? 
 \end{itemize}
 This question is interesting even restricting ourselves to planar graphs. 
%
%Another interesting question concerns constructive meta-kernelization theorems. Our meta-kernelization theorems are pure existence results, we prove that there are kernelization algorithms but we do not know how to construct them.  Obtaining of constructive meta-theorems is an important challenge. 

\medskip 

It is very natural to ask whether our results can be extended to more general classes of graphs. The most natural candidates for such extensions are graphs of bounded local-treewidth~\cite{FrickG01-dec} and graphs of bounded expansion~\cite{NesetrilM08}. 
The first step in this direction is done in \cite{F.V.Fomin:2010oq}. 

\myparagraph{Practical considerations.} 
Our meta-theorems provide simple criteria to decide whether a problem admits a polynomial or linear kernel on graphs of bounded genus.
 %Of sec:avariodcourctheoe, 
 It is expected that for concrete problems, tailor-made kernels will 
have much smaller constant factors, than what would follow from a direct application of our results.
However, our approach might be useful for computer aided design of kernelization algorithms:  
%with a Myhill-Nerode approach, 
a computer program can in some cases output a set of rules that transform each 
protrusion to a minimum size representative and estimate the obtained kernel size.
This seems an interesting and far from trivial algorithm-engineering problem. 
In general, finding linear kernels {\em with reasonably small constant factors}
for concrete problems on planar graphs or graphs with small genus remains
a worthy topic of further research.

%\paragraph*{Some concrete open problems}
\myparagraph{Some concrete open problems.}
We conclude with some concrete problems that cannot be resolved by our approach. These include \textsc{$p$-Directed Feedback Vertex Set}~\cite{ChenLLSR08}  and \textsc{$p$-Odd Cycle Transversal}~\cite{RSV04} to name a few. All these problems are expressible in CMSO but none of them are known to be quasi-coverable. 
For  \textsc{$p$-Directed Feedback Vertex Set} no polynomial kernel is known even  on planar graphs. For 
\textsc{$p$-Odd Cycle Transversal} a randomized kernel for general graphs was obtained recently in 
 \cite{KratschW12} but existence of a deterministic kernel even on planar graphs is open. 

%\paragraph*{Impact}
\myparagraph{Impact.}
The protrusion replacement technique for kernelization was 
 introduced in the preliminary conference version of this paper \cite{H.Bodlaender:2009ng} 
appears to be useful in different algorithmic approaches. They were used to obtain kernels for a wide set of  bidimensional problems on 
$H$-minor-free graphs \cite{F.V.Fomin:2010oq,FominLST12}, vertex removal problems on general  and unit disc graphs
\cite{FominLMPS11}, and problems on graphs excluding a fixed graph as a topological minor \cite{FominLST12a,KimLPR12a}. It was also used in the design of fast parameterized algorithms 
and approximation algorithms \cite{FominLRS11,FominLS12,FominLMS12,JoretPSST11,KimPG12,KimLPR12a} 
 \medskip \medskip   \medskip \medskip
 
%\cite{FominLRS11,FominLS12}
%\subsection*{Acknowledgments.}
\myparagraph{Acknowledgements.}~We thank Jiong Guo, Ge Xia, and Yong Zhang for sending us the full versions of~\cite{GN07ICALP} and \cite{Xia:2010pi}. We also thank the anonymous reviewers of FOCS'09 and J. ACM  for their valuable comments on previous versions of this paper. 
% on the conference proceedings version of this article.

%
%%%%%%%%%%%%%%%%%%%%%%%%%%%%%%input{meta_open}

%\newpage

%%%%%%%%%%%%%%%%%%%%%%%%%%%%%%%%
%\bibliographystyle{acmtrans}

%\todo[inline]{Clean up the bibliography.}
%
%\bibliographystyle{plain}
%%\bibliographystyle{plain}
%\bibliography{metakernels_extended}
%

\newpage
\begin{appendix}
%%%%%%%%%%%%%%%%%%%%%%%%%%%%%%%%

%%%%%%%%%%%%%%%input{meta_appendix}
\section{Problem Compendium}\vspace{-1mm}\vspace{-1mm}
\label{apx:probcomp}

In this compendium we present the kernelization status of all problems that have been mentioned in this paper.
%We refer to problem compendium given in~\cite{DowneyF98} or the compendium of parameterized problems 
%provided at 
%\\
%\texttt{http://bravo.ce.uniroma2.it/home/cesati/research/compendium/} for the definitions 
%of problems given below. 

\vspace{-1mm}\vspace{-1mm}\vspace{-1mm}\subsection{Minimization problems that have FII and are quasi-coverable -- {\em linear kernels for graphs of bounded genus}.}
\label{subsec:probthatavefii}
\pvc, %(FII(SM)+Co=linear)
\pfvs, % (FII(SM)+QC=linear)%
%{\sc $p$Almost Outperplanar},
\paou, 
\pdhs,  \parbt,  \parbp,
% \textsc{Almost-$r$-bounded treewidth},
 % \textsc{Almost-$r$-bounded path\-wi\-dth},
\fd,
{\sc $p$-Edge Dominating Set},   % (FII(SM)+Co=linear)
{\sc $p$-Minimum-Vertex Feedback Edge Set}, %(FII(SM)+QC=linear) 
{\sc $p$-Dominating Set}, %(FII(SM)+Co=linear) 
{\sc $p$-$r$-Dominating Set}, %(FII(SM)+Co=linear)
{\sc $p$-$q$-Threshold Dominating Set}, %(FII(SM)+Co=linear)  
{\sc $p$-Efficient Dominating Set}$^{*}$,  %(FII(but not SM)+Co=linear)
{\sc $p$-Connected Dominating Set}, %(FII(SM)+Co=linear)
{\sc $p$-Connected Vertex Cover}, %(FII(SM)+Co=linear) 
{\sc $p$-Cycle Domination}, %(FII(SM)+Qc) 
{\sc $p$-Directed Domination},  %(FII(SM)+Co=linear) 
\psc, 
%{\sc $p$-${\cal S}$-Co\-vering}$^+$, %(FII(SM)+Co preprocessing) 
%
{\sc $p$-Minimum Partition Into Cli\-ques}, %(FII(SM)+Co).
{\sc $p$-Edge Clique Co\-ver}$^*$,
and {\sc $p$-$s$-Cycle Trans\-versal}$^*$.

\vspace{-1mm}\vspace{-1mm}\vspace{-1mm}\subsection{Maximization problems that have FII and are quasi-coverable -- {\em linear kernels for graphs of bounded genus}.}
 \label{subsec:probthatavefiibutnotquasi}
{\sc $p$-$r$-Scattered Set}$^*$, %(FII(but not SM)+Co=linear)
{\sc $p$-Independent Set}, %(FII(SM)+Co=linear)  (scattered set)
{\sc $p$-Induced Matching}$^*$, %(FII(but not SM)+Co=linear) (scattered set)
{\sc $p$-Triangle Edge Packing}$^+$,   %(FII(SM)+Co after preprocessing = linear)
{\sc $p$-Maximum Internal Spanning Tree}$^{*}$, %(FII(but not SM)+Co=linear)   (dominating set)
{\sc $p$-Maximum Full-Degree Spanning Tree}$^{*}$, %(FII(not SM)+Co)
\pcyp$^{*}$, %(FII(but not SM)+Qc = linear)
\pfp$^{*}$, %(FII(but not SM)+Qc)
{\sc $p$-Triangle Vertex Packing}$^+$, %(FII(SM)+Co after preprocessing = linear)
\psp$^+$,
and {\sc $p$-Edge Cycle Packing}$^{*}$,\medskip

For all problems with an asterisk ``$^*$'', a direct proof that they have FII is required. For the rest, FII property follow 
by expressing them as a {\sc $p$-min/max-CMSO} problem and proving strong monotonicity.
For the problems with a cross ``$^+$'', the linear kernel assumes the application of some preprocessing routine.

\vspace{-1mm}\vspace{-1mm}\vspace{-1mm}\subsection{Problems that do not have FII and are coverable {\sc $p$-min/max-CMSO} -- {\em polynomial kernels for graphs of bounded genus}.}
{\sc $p$-Independent Dominating Set}, % (not-FII, CMSO, coverable)
{\sc $p$-Acyclic Dominating Set},
{\sc $p$-Independent Directed Domination}, %(not-FII, CMSO, coverable)
{\sc $p$-Maximum Internal Out-branching},  %(not-FII, CMSO, coverable)
{\sc $p$-Odd Set}, %(needs counting!, not-FII, CMSO, coverable)
%{\sc Edge-${\cal H}$-Packing}, {\sc Edge-${\cal H}$-Covering}, 
and {\sc $p$-Edge-${\cal S}$-Cove\-ring}.

\vspace{-1mm}\vspace{-1mm}\vspace{-1mm}\subsection{A problem that has FII but  is not  
quasi-coverable.}
 \label{subsec:problfiinoquasi}
{\sc $p$-Hamiltonian Path Completion}.  
%{\sc Maximum Leaf Spanning Tree},    {\sc Maximum Leaf Out-Branching},  
 
\vspace{-1mm}\vspace{-1mm}\vspace{-1mm}\subsection{A quasi-coverable problem that   has no FII.}
\label{pnfii}
{\sc $p$-Maximum Cut}.

\vspace{-1mm}\vspace{-1mm}\vspace{-1mm}\subsection{ Problems that do not have FII and they are not quasi-coverable.}
\label{pnfiinotqcov}
{\sc $p$-Longest Path} and  {\sc $p$-Longest Cycle}. 
%
%\noindent
%The following lemma is shown in~\cite{Fluiter97}.
%\begin{lemma}{\rm \cite{Fluiter97}}
% {\sc Longest Path}, {\sc Longest Cycle}, {\sc Maximum Cut} and {\sc Minimum Covering by Cliques} do not 
%have  FII.  
%\end{lemma}

\end{appendix}
%%%%%%%%%%%%%%%input{meta_appendix}
\end{document}